\documentclass[12pt]{article}
\usepackage{jheppub}

\pdfoutput=1

\usepackage{amsmath,bbm,array,amsfonts,graphicx,wrapfig,lscape,float,mathtools,multirow,longtable}
\usepackage[dvipsnames]{xcolor}
\usepackage{array}
\usepackage{adjustbox}

\newcommand{\be}{\begin{equation}}
\newcommand{\ee}{\end{equation}}
\newcommand{\beq}{\begin{equation}}
\newcommand{\beql}[1]{\begin{equation}\label{#1}}
\newcommand{\eeq}{\end{equation}}
\newcommand{\ba}{\begin{array}}
\newcommand{\ea}{\end{array}}
\newcommand{\bea}{\begin{eqnarray}}
\newcommand{\beal}[1]{\begin{eqnarray}\label{#1}}
\newcommand{\eea}{\end{eqnarray}}
\newcommand{\ben}{\begin{enumerate}}
\newcommand{\een}{\end{enumerate}}
\newcommand{\bean}{\begin{eqnarray*}}
\newcommand{\eean}{\end{eqnarray*}}
\newcommand{\eref}[1]{(\ref{#1})}
\newcommand{\sref}[1]{\S\ref{#1}}
\newcommand{\tref}[1]{Table~\ref{#1}}
\newcommand{\nn}{\nonumber}

\newcommand{\fref}[1]{Figure \ref{#1}}
\newcommand{\btab}[1]{\begin{tabular}{#1}}
\newcommand{\etab}{\end{tabular}}

\newcommand{\comment}[1]{}

\newcommand{\qed}{\nobreak \ifvmode \relax \else
      \ifdim\lastskip<1.5em \hskip-\lastskip
      \hskip1.5em plus0em minus0.5em \fi \nobreak
      \vrule height0.75em width0.5em depth0.25em\fi}

\definecolor{darkspringgreen}{rgb}{0.09, 0.45, 0.27}
\definecolor{forestgreen}{rgb}{0.13, 0.55, 0.13}

\usepackage{array}
\usepackage{physics}
\usepackage{subcaption}
\usepackage{tikz,tikz-3dplot}
\usepackage[colorlinks=true]{hyperref}
\newcolumntype{C}[1]{>{\centering\let\newline\\\arraybackslash\hspace{0pt}}m{#1}}
\definecolor{yellow2}{rgb}{0.98, 0.80, 0.20}

\title{Classification and Birational Equivalence \\ of Dimer Integrable Systems \\ for Reflexive Polygons} 
\author[a]{Minsung Kho,}
\author[b]{Norton Lee,}
\author[a,c]{Rak-Kyeong Seong}

\affiliation[a]{
Department of Mathematical Sciences, and 
${}^{c}$Department of Physics,\\ 
Ulsan National Institute of Science and Technology,\\
50 UNIST-gil, Ulsan 44919, South Korea
}

\affiliation[b]{
Center for Geometry and Physics, Institute for Basic Science (IBS),\\
Pohang 37673, South Korea
}

\emailAdd{minsung@unist.ac.kr}
\emailAdd{norton.lee@ibs.re.kr}
\emailAdd{seong@unist.ac.kr}

\preprint{
\begin{flushright}
UNIST-MTH-25-RS-05 \\
CGP25011
\end{flushright}
}

\abstract{
Brane tilings are bipartite periodic graphs on the 2-torus and realize a
large family of $4d$ $\mathcal{N}=1$ supersymmetric gauge theories corresponding to toric Calabi-Yau 3-folds.
We present a complete classification 
of dimer integrable systems corresponding to the 30 brane tilings whose toric Calabi-Yau 3-folds are given by the 16 reflexive polygons in 2 dimensions.
For each dimer integrable system associated to a reflexive polygon, 
we present the Casimirs, the single Hamiltonian built from 1-loops, 
the spectral curve, and the Poisson commutation relations.
We also identify all birational equivalences between 
dimer integrable systems in this classification by presenting 
the birational transformations that match the Casimirs and the Hamiltonians as well as the spectral curves and Poisson structures between equivalent dimer integrable systems.
In total, we identify 16 pairs of birationally equivalent dimer integrable systems
which combined with Seiberg duality between the corresponding brane tilings form 5 distinct equivalence classes.
Echoing phenomena observed for brane brick models 
realizing a family of $2d$ $(0,2)$ supersymmetric gauge theories corresponding to toric Calabi-Yau 4-folds, 
we illustrate that deformations of brane tilings, including mass deformations, correspond to the birational transformations we discover in this work, and leave invariant
the number of generators of the mesonic moduli space as well as the corresponding $U(1)_R$-refined Hilbert series. 
}

\begin{document}

\maketitle

%%%%%%%%%%%%%%%%%%%%%%%%%%%%%%%%%%%%%%%%%%%%%%%%%%%%%%%%%%%%%
%%%%%%%%%%%%%%%%%%%%%%%%%%%%%%%%%%%%%%%%%%%%%%%%%%%%%%%%%%%%%

%=================================================================
\section{Introduction}
%=================================================================

\textbf{Brane tilings} \cite{Hanany:2005ve, Franco:2005j,Franco:2005sm} as bipartite periodic graphs on a 2-torus form 
a large family of $4d$ $\mathcal{N}=1$ supersymmetric gauge theories
corresponding to toric Calabi-Yau 3-folds.
The $4d$ $\mathcal{N}=1$ supersymmetric gauge theories are worldvolume theories 
of a stack of D3-branes probing the associated toric Calabi-Yau 3-folds. 
By Goncharov and Kenyon, brane tilings have also been shown to correspond to an equally large family of integrable systems, 
now known as \textbf{dimer integrable systems} \cite{ goncharov2012dimersclusterintegrablesystems, Eager:2011dp}.
The bipartite graph on the 2-torus, 
also known as a dimer in the literature \cite{KENYON1997591, kenyon2003introductiondimermodel},
not only encodes
the $4d$ $\mathcal{N}=1$ theory
as well as the Type IIB brane configuration in string theory that realizes it, 
but also 
the Casimirs, Hamiltonians, the spectral curve and the Poisson commutation relations of the underlying dimer integrable system. 

\begin{figure}[http!!]
\begin{center}
\resizebox{0.8\hsize}{!}{
\includegraphics{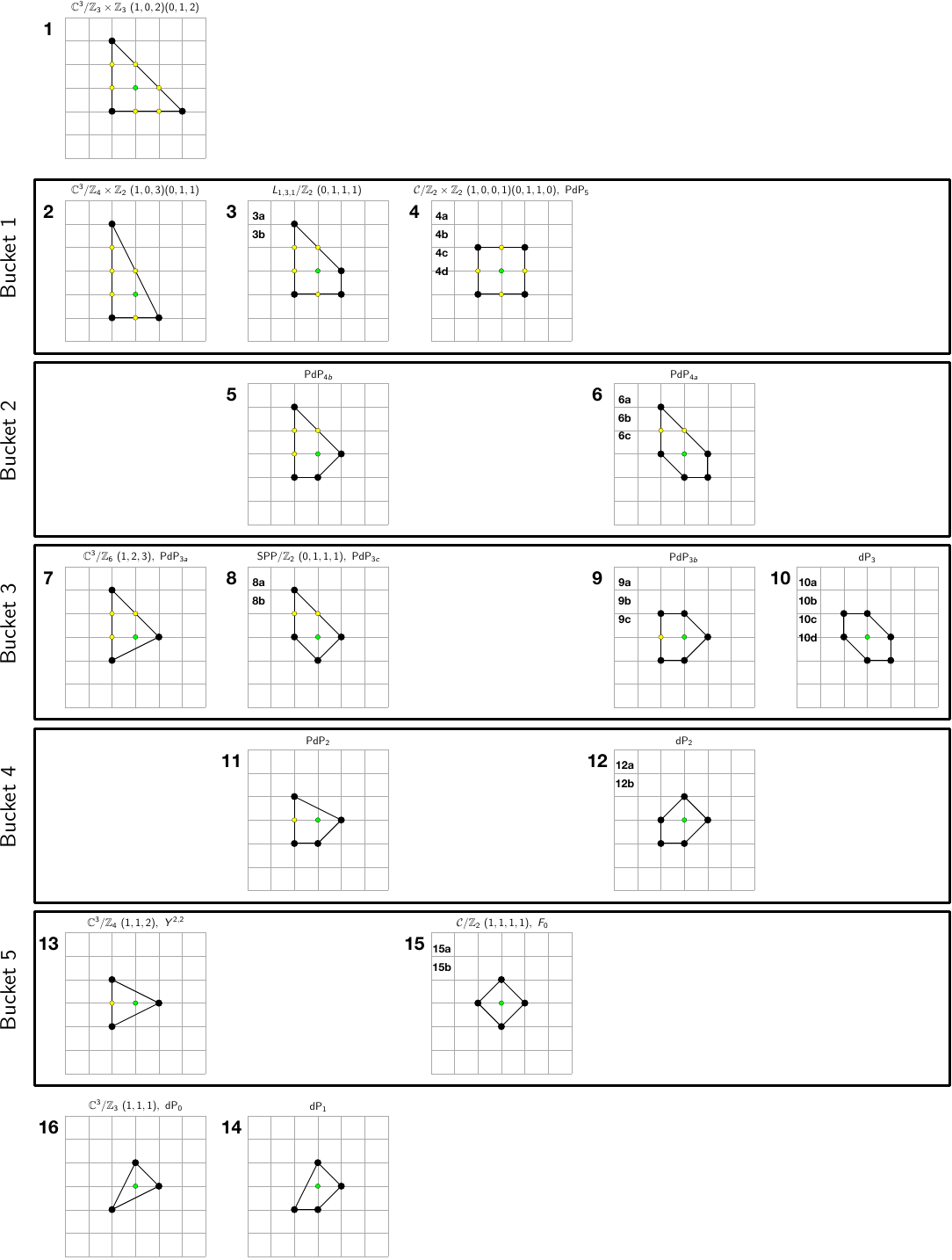}
}
\caption{
The 16 reflexive polygons in 2 dimensions with labels corresponding to the associated 30 brane tilings classified in \cite{Hanany:2012hi}.
Birational transformations between toric Calabi-Yau 3-folds correspond to birational equivalence between the associated dimer integrable systems.
Combined with Seiberg duality, we identify 5 equivalence classes called buckets amongst the 30 brane tilings and dimer integrable systems classified in this work.
}
\label{fig_bucketsummary}
 \end{center}
 \end{figure}

Various aspects of dimer integrable systems have been studied \cite{Bershtein:2017swf, Marshakov:2019vnz, Lee:2023wbf,  Lee:2024bqg, Bershtein:2024lvd, Kho:2025fmp, Lee:2025pee} since the initial work by Goncharov and Kenyon.
However, there has not been a systematic attempt in classifying dimer integrable systems 
as it is the case for brane tilings and corresponding $4d$ $\mathcal{N}=1$ supersymmetric gauge theories \cite{Hanany:2012hi}.
For brane tilings, 
one of the largest collections
has been obtained through the classification based on toric Calabi-Yau 3-folds
whose toric diagrams \cite{fulton1993introduction,Leung:1997tw, He:2017gam, Krefl:2017yox, Choi:2023rqg, Bao:2024nyu} are one of the \textbf{16 reflexive polygons} in $\mathbb{Z}^2$.
Reflexive polytopes \cite{1993alg.geom.10003B, 1994alg.geom..2002B} are convex lattice polytopes with a single interior point as the origin
and have been classified up to dimension 4 by Kreuzer and Skarke in \cite{Kreuzer:1998vb, Kreuzer:2000qv,Kreuzer:2000xy}.
The classification in \cite{Hanany:2012hi}
is based on the 16 reflexive polygons in dimension 2 shown in \fref{fig_bucketsummary}
that correspond to 16 toric Calabi-Yau 3-folds,
including the zeroth Hirzebruch surface \cite{Feng:2001xr, Hanany:2005ve, Franco:2005j}, the del Pezzo surfaces \cite{Feng:2000mi, Feng:2001xr, Feng:2002fv,Feng:2002zw}
and certain abelian orbifolds of $\mathbb{C}^3$ \cite{Davey:2010px,Hanany:2010ne}.
The classification in \cite{Hanany:2012hi}
resulted in 30 brane tilings associated to these 16 toric Calabi-Yau 3-folds.
There are more brane tilings and associated $4d$ $\mathcal{N}=1$ supersymmetric gauge theories 
than toric Calabi-Yau 3-folds
due to the fact that some brane tilings and $4d$ $\mathcal{N}=1$ theories
correspond to the same toric Calabi-Yau 3-fold
due to Seiberg duality \cite{Seiberg:1994pq}, 
which is also known as toric duality in this context \cite{Feng:2000mi,Feng:2001xr,Feng:2002zw, Feng:2001bn,Beasley:2001zp}.

Based on this classification of brane tilings for
toric Calabi-Yau 3-folds with reflexive polygons as toric diagrams, 
the following work has the aim to identify the corresponding dimer integrable systems.
For the 30 brane tilings, 
we identify 30 dimer integrable systems
with their corresponding 
Casimirs and Hamiltonians,
the spectral curve and the Poisson
commutation relations.\footnote{A subset of these dimer integrable systems are studied in \cite{Bershtein:2024lvd}.}
Here we note that because
these dimer integrable systems correspond to toric Calabi-Yau 3-folds with reflexive toric diagrams, 
the dimer integrable systems possess only one Hamiltonian associated to the single interior point of the toric diagrams.

%--------------------------------------------
%--------------------------------------------

As part of this classification, 
we also identify the complete collection of birational equivalences
between dimer integrable systems corresponding to the reflexive polygons in 2 dimensions.
As observed in \cite{Kho:2025fmp},
when two toric Calabi-Yau 3-folds
with their corresponding toric varieties 
are related by a birational transformation \cite{galkin2010mutations, 2012SIGMA...8..047I, akhtar2012minkowski, 2018arXiv180100013B, 2021RSPSA.47710584C, Ghim:2024asj,Ghim:2025zhs}, 
then the associated brane tilings define dimer integrable systems, which are birationally equivalent to each other.
Under what is now known as \textbf{birational equivalence} between dimer integrable systems,
the birational transformation 
identifies the Casimirs and Hamiltonians
as well as the spectral curve and the Poisson commutation relations between the two birationally equivalent dimer integrable systems.
In the following work, we identify out of the 30 dimer integrable systems in our classification
in total 16 pairs of birationally equivalent dimer integrable systems. 

We note here that besides birational equivalence, 
dimer integrable systems
can also be equivalent
when the corresponding brane tilings and the $4d$ $\mathcal{N}=1$ supersymmetric gauge theories
are related by
Seiberg duality \cite{Seiberg:1994pq,Feng:2000mi,Feng:2001xr,Feng:2002zw, Feng:2001bn,Beasley:2001zp}.
Under Seiberg duality, 
the bipartite period graph of the brane tiling undergoes a local deformation
also referred to as a spider move or urban renewal \cite{goncharov2012dimersclusterintegrablesystems, CIUCU199834, 1999math......3025K}.
Under this local mutation of the periodic bipartite graph, 
the associated dimer integrable system undergoes a canonical transformation
that leaves the integrable system and the corresponding Poisson moduli space invariant \cite{goncharov2012dimersclusterintegrablesystems}.
When equivalence due to Seiberg duality is combined with birational equivalence between dimer integrable systems, 
we are able to identify 
5 distinct equivalence classes amongst the 30 dimer integrable systems classified in this work.
We refer to these equivalence classes as \textbf{buckets} \cite{akhtar2012minkowski}.
These are illustrated with the corresponding reflexive toric diagrams in \fref{fig_bucketsummary}.

Birational transformations
between toric Calabi-Yau 4-folds
and associated $2d$ $(0,2)$ supersymmetric gauge theories given by brane brick models \cite{Franco:2015tna, Franco:2015tya, Franco:2016nwv, Franco:2016qxh, Franco:2022gvl, Kho:2023dcm}
have been studied extensively in \cite{Ghim:2024asj, Ghim:2025zhs}.
In particular, it has been shown that 
mass deformation between brane brick models \cite{Franco:2023tyf} realizing $2d$ $(0,2)$ supersymmetric gauge theories
can be identified with a birational transformation between the corresponding toric Calabi-Yau 4-folds.
This result has been recently extended to relevant deformations of brane brick models in \cite{Carcamo:2025shw}.
In this work, we see that 
brane tilings related by deformations \cite{2022SIGMA..18..030H, Franco:2023flw, Cremonesi:2023psg}, 
including mass deformations \cite{Klebanov:1998hh,Gubser:1998ia, Bianchi:2014qma},
correspond to
birationally equivalent dimer integrable systems.

As observed for brane brick models corresponding to toric Calabi-Yau 4-folds \cite{Ghim:2024asj, Ghim:2025zhs}
as well as in the context of generalized toric polygons (GTPs) \cite{Arias-Tamargo:2024fjt}, 
we observe as part of our classification that
brane tilings and dimer integrable systems that are related by a birational transformation
have the same Hilbert series \cite{Benvenuti:2006qr, Hanany:2006uc, Butti:2007jv, Feng:2007ur, Hanany:2007zz} of the mesonic moduli space of the associated abelian $4d$ $\mathcal{N}=1$ supersymmetric gauge theory \cite{Feng:2000mi},
when the Hilbert series is refined under a $U(1)_R$ symmetry
that gives generators of the mesonic moduli space the same $U(1)_R$ charge
and gives the superpotentials of the brane tilings $U(1)_R$ charge $2$. 
Moreover, as observed for brane brick models in \cite{Ghim:2024asj, Ghim:2025zhs},
we also confirm in this work that 
the mesonic moduli spaces have the same number of generators 
for brane tilings and dimer integrable systems that are related by a birational transformation.
\\

Our work is organized as follows. 
Section \sref{sec:02}
gives a brief overview about brane tilings and the corresponding family of $4d$ $\mathcal{N}=1$ supersymmetric gauge theories.
The section also reviews the moduli spaces of the abelian $4d$ $\mathcal{N}=1$ theories, including the mesonic moduli space \cite{Feng:2000mi} and
the master space \cite{Forcella:2008bb, Forcella:2008eh, Hanany:2010z},
and then summarizes the family of toric Calabi-Yau 3-folds whose toric diagrams are reflexive polygons.
While discussing the moduli spaces, the section also gives an overview of Seiberg duality that preserves the mesonic moduli space, 
also referred to as toric duality \cite{Seiberg:1994pq,Feng:2000mi,Feng:2001xr,Feng:2002zw, Feng:2001bn,Beasley:2001zp}, 
as well as specular duality \cite{Hanany:2012vc} which preserves the master space.
The section then gives a detailed review on how brane tilings define dimer integrable systems
and the observation in \cite{Kho:2025fmp}
on how dimer integrable systems can be equivalent under birational transformations between the corresponding toric Calabi-Yau 3-folds.
Sections \sref{sec:03} to \sref{sec:18}
summarize the 30 brane tilings and the corresponding dimer integrable systems corresponding to the 16 reflexive polygons, giving explicit 
expressions for the Casimirs, Hamiltonians, the spectral curve and the Poisson commutation relations for each of the dimer integrable systems.
Sections \sref{sec:19} to \sref{sec:23}
summarize the birational transformations 
that map between equivalent dimer integrable systems in the classification.
The sections explicitly show the mapping between 
the Casimirs, Hamiltonians, the spectral curve and the Poisson commutation relations of the birationally equivalent dimer integrable systems.
Moreover, the sections are organized in terms of buckets 
containing dimer integrable systems that are birationally equivalent and
also dimer integrable systems that are equivalent under Seiberg duality of the corresponding brane tilings.
For completeness, 
these sections also summarize how within the buckets
brane tilings that are related by birational transformations all share the same
number of generators of the mesonic moduli space of the associated $4d$ $\mathcal{N}=1$ theories.
Moreover, the sections illustrate how the Hilbert series of the mesonic moduli space refined only under the $U(1)_R$ symmetry
is invariant within each of the buckets. 
We conclude our work in section \sref{sec:24} with an overview of our results as well as an overview 
on the correspondence 
between birational transformations of toric Calabi-Yau 3-folds, 
deformations of brane tilings, birational equivalence between dimer integrable systems, 
and Hanany-Witten moves for $(p,q)$ webs and corresponding $5d$ $\mathcal{N}=1$ theories.
\\

%=================================================================

\section{Background \label{sec:02}}

%=================================================================
%=================================================================
\subsection{Brane tilings and $4d$ $\mathcal{N}=1$ Quiver Gauge Theories}
\label{sec_branetilings}
%=================================================================

A \textbf{brane tiling} \cite{Hanany:2005ve, Franco:2005j,Franco:2005sm}, 
also known as a \textbf{dimer model} \cite{KENYON1997591, kenyon2003introductiondimermodel},
is a periodic bipartite graph on a 2-torus $T^2$. 
The bipartite graph consists of black and white nodes where edges connect nodes of opposite color. 
Brane tilings realize a family of $4d$ $\mathcal{N}=1$ gauge theories, which are worldvolume theories of 
D3-branes probing a toric Calabi-Yau 3-fold \cite{Douglas:1997de, Douglas:1996sw, Feng:2000mi, Feng:2001xr}.

%------------
\begin{table}[ht!]
\begin{center}
\begin{tabular}{|c|cccc|cccccc|}
\hline
\; & 0 & 1 & 2 & 3 & 4 & 5 & 6 & 7 & 8 & 9 \\
\hline \hline
D3 & $\times$ & $\times$ & $\times$ & $\times$ & $\cdot$ & $\cdot$ & $\cdot$ & $\cdot$ & $\cdot$ & $\cdot$\\
CY3 & $\cdot$ & $\cdot$ & $\cdot$ & $\cdot$ &$\times$&$\times$&$\times$& $\times$& $\times$ & $\times$ \\
\hline
\end{tabular}
\caption{
The D3-branes probing a toric Calabi-Yau 3-fold.
The worldvolume theory on the probe D3-branes is a $4d$ $\mathcal{N}=1$ supersymmetric gauge theory
given by a brane tiling.
\label{tab_01}
}
\end{center}
\end{table}
%------------

The probe D3-branes at the Calabi-Yau singularity, as summarized in \tref{tab_01}, 
become under T-duality
D5-branes suspended between a NS5-brane wrapping a 2-torus $T^2$.
In this Type IIB brane configuration,
the D5-branes extend along the $(012345)$ directions, and the NS5-brane extends along the $(0123)$ directions and wraps a holomorphic curve $\Sigma$ defined in terms of directions $(4567)$ as summarized in \tref{tab_02}. 
The holomorphic curve $\Sigma$ is given by,
\beal{es01a01}
\Sigma : P(x,y)=0~\text{for}~x,y \in \mathbb{C}^*
~,~
\eea
where the complex coordinates $x$ and $y$ come from the directions $(45)$ and $(67)$, respectively.
$P(x,y)$ in \eref{es01a01} is known as a \textbf{Newton polynomial} given by the toric diagram $\Delta$ of a toric Calabi-Yau 3-fold.
The Newton polynomial for a toric diagram $\Delta$ is defined as follows,
\beal{es01a01b}
P(x,y) = \sum_{(n_x,n_y) \in \Delta} c_{(n_x,n_y)} x^{n_x} y^{n_y} ~,~
\eea
where the sum is over vertices in $\Delta$ with coordinates $(n_x,n_y) \in \mathbb{Z}^2$. 
The coefficients are associated to complex structure moduli 
in the corresponding mirror Calabi-Yau and are chosen to be in $c_{(n_x,n_y)} \in \mathbb{C}^*$ \cite{Hori:2000ck, Hori:2000kt, Hori:2003ic}.

%------------
\begin{table}[htb!]
\begin{center}
\begin{tabular}{|c|cccc|cccc|cc|}
\hline
\; & 0 & 1 & 2 & 3 & 4 & 5 & 6 & 7 & 8 & 9 \\
\hline \hline
D5 & $\times$ & $\times$ & $\times$ & $\times$ & $\times$ & $\cdot$ & $\times$ & $\cdot$ & $\cdot$ & $\cdot$\\
NS5 & $\times$ & $\times$ & $\times$ & $\times$ &\multicolumn{4}{c|}{------$\Sigma$------} & $\cdot$ & $\cdot$ \\
\hline
\end{tabular}
\caption{
The Type IIB brane configuration given by a brane tiling, 
consisting of D5-branes suspended between a NS5-brane wrapping a holomorphic curve $\Sigma$.
\label{tab_02}
}
\end{center}
\end{table} 
%------------

The brane tiling as a bipartite graph on a 2-torus $T^2$ represents the Type IIB brane configuration in \tref{tab_02}. 
In the following paragraph, we summarize the dictionary that translates between the bipartite graph on $T^2$ and the corresponding $4d$ $\mathcal{N}=1$ quiver gauge theory:

%---------------------------------
\begin{itemize}

\item \textbf{Faces} correspond to $U(N)_i$ gauge groups of the $4d$ $\mathcal{N}=1$ gauge theory. 
The faces are all even-sided because of the bipartite nature of the brane tiling on $T^2$.
This also implies that the number of fundamental and anti-fundamental chiral multiples $X_{ij}$ associated to a gauge group $U(N)_i$ is always the same.

\item \textbf{Edges} correspond to bifundamental chiral multiplets $X_{ij}$ of the $4d$ $\mathcal{N}=1$ supersymmetric gauge theory. 
Every chiral field $X_{ij}$ in the brane tiling transforms under the bifundamental representation of associated gauge groups $U(N)_i$ and $U(N)_j$, which correspond to the adjacent faces of the edge associated to $X_{ij}$ in the bipartite graph on $T^2$.

\item \textbf{White (Black) nodes} correspond to positive (negative) monomial terms in the superpotential of the associated $4d$ $\mathcal{N}=1$ gauge theory. 
The monomial terms corresponding to white (black) nodes are given by products of chiral fields, which are associated with the edges that connect to the given white (black) nodes in a clockwise (anti-clockwise) orientation. 
This orientation along white (black) nodes determines the bifundamental representation of chiral fields associated to the connected edges.
This ensures that the monomial product of chiral fields corresponding to the white (black) node is gauge-invariant. 

\end{itemize}
%---------------------------------

The Newton polynomial of $\Delta$ is also given by the permanent of the
\textbf{Kasteleyn matrix} \cite{kasteleyn1967graph}
of the brane tiling.
The Kasteleyn matrix $K$ for a brane tiling is a $n \times n$ square matrix, where $n$ is the number of white nodes which is the same as the number of black nodes in a brane tiling. 
Here, white nodes $w_j$ and black nodes $b_k$ are labelled by $j,k=1, \dots, n$.
The colouring of nodes in the brane tiling allows us to assign also an orientation \textit{along} the edges from a white node to a black node.
Under this orientation, an edge $e_{jk} = (w_j, b_k)$ 
can be assigned a winding number $h(e_{jk}) = (h_1,h_2) \in \mathbb{Z}^2$ along the two independent $S^1$ cycles on $T^2$.
Based on the winding number assignment on edges of the brane tiling, 
the elements of the Kasteleyn matrix are given by, 
\beal{es01a02}
K_{w_j,b_k} (x, y) = \sum_{e_{jk}=(w_j,b_k)}e_{jk} ~x^{h_1(e_{jk})} y^{h_2(e_{jk})}  ~,~ 
\eea
where $x$ and $y$ are the fugacities for the winding numbers.
The permanent of the Kasteleyn matrix, 
\beal{es01a03}
\text{perm}~ K(x,y)
=
P(x,y)
~,~
\eea
gives the Newton polynomial defined in \eref{es01a01b}.
We note here that the particular form of the Newton polynomial depends on the $GL(2,\mathbb{Z})$ frame chosen for the toric diagram $\Delta$, or equivalently the choice of the fundamental domain in the brane tiling that determines the winding number of edges. 
We also note that the coefficients $c_{(n_x, n_y)}$ in the Newton polynomial in \eref{es01a03} correspond to products of 
edge variables $e_{jk}$, which themselves are associated to chiral fields in the $4d$ $\mathcal{N}=1$ supersymmetric gauge theory. 
These products of edge variables correspond to a particular subset of edges in the brane tiling associated to each vertex in the toric diagram $\Delta$, which are known as perfect matchings. 
\\

A \textbf{perfect matching} $p_{a}$ \cite{Hanany:2005ve, Franco:2005j,Franco:2005sm, KENYON1997591, kenyon2003introductiondimermodel} is a collection of edges in a brane tiling that covers all white and black nodes in the bipartite graph precisely once. 
All perfect matchings for a brane tiling are summarized in a $\abs{E}$ $\times$ $c$ matrix, which we call the perfect matching matrix $P$. 
Here, $\abs{E}$ and $c$ indicate the number of edges and perfect matchings, respectively.
For simplicity, we label here the edges $e_k$ and the corresponding chiral fields $X_k$ in the brane tiling with a single index $k = 1, \dots, \abs{E}$.
Then, the entries in a perfect matching matrix $P$ are given by,
\beal{es01a04}
P_{k a} = \begin{cases}
1 & ~~\text{if}~ e_{k} \in p_{a}\nn\\
0 & ~~\text{if}~ e_{k} \notin p_{a}
\end{cases}
~.~
\eea
The perfect matchings correspond to gauged linear sigma model (GLSM) fields \cite{Witten:1993yc},
and can be used to express 
each bifundamental chiral field $X_k$ as a product of perfect matchings as follows,
\beal{es01a05}
X_k=\prod_{a}(p_{a})^{P_{k a}}~,~
\eea
We note here that the F-term constraints from the superpotential $W$ of the $4d$ $\mathcal{N}=1$ theory automatically satisfy the relations in \eref{es01a05}. 
\\

The space of gauge invariant operators satisfying the F- and D--terms of the $4d$ $\mathcal{N}=1$ supersymmetric quiver gauge theory 
is known as the \textbf{mesonic moduli space} \cite{Feng:2000mi, Feng:2001xr}.
For an abelian $4d$ $\mathcal{N}=1$ theory with $U(1)$ gauge groups,
the mesonic moduli space is precisely the probed toric Calabi-Yau 3-fold.
It is defined as follows, 
\beal{es01a06a}
\mathcal{M}^{mes}
= 
\text{Spec} 
\left(
\mathbb{C}[X_{ij}]
/ \mathcal{I}_{\partial_{W}}
\right) 
// U(1)^{G-1}
~,~
\eea
where $\mathbb{C}[X_{ij}]$ is the coordinate ring formed by the chiral fields $X_{ij}$ of the $4d$ $\mathcal{N}=1$ theory 
and 
$\mathcal{I}_{\partial_{W}}$ is the irreducible component of the ideal formed by the F-terms of the form
$\partial_{X_{ij}} W = 0$. 
The F-terms are binomial due to the bipartite nature of the brane tiling
and $\mathcal{I}_{\partial_{W}}$ forms a binomial ideal giving a \textbf{toric variety} \cite{fulton1993introduction,Leung:1997tw}. 
We also note that $G$ is the total number of $U(1)$ gauge groups in the abelian $4d$ $\mathcal{N}=1$ theory, where an overall $U(1)$ decouples, and $i,j=1,\dots, G$ are the gauge group labels. 
When we remove the quotient by the $U(1)$ gauge groups in \eref{es01a06a}, 
we remain with the space of chiral fields $X_{ij}$ subject to the F-terms of the $4d$ $\mathcal{N}=1$ theory, 
\beal{es01a06b}
\mathcal{F}^\flat_\text{Irr}
= 
\text{Spec} ~
\mathbb{C}[X_{ij}]
/ 
\mathcal{I}_{\partial_{W}}
~,~
\eea
which is known as the \textbf{master space} \cite{Forcella:2008bb, Forcella:2008eh, Hanany:2010z} of the brane tiling.

In terms of the GLSM fields corresponding to perfect matchings of the brane tiling, we
can express the master space and the mesonic moduli space as the following symplectic quotients, 
\beal{es01a06c}
\mathcal{F}^\flat_\text{Irr}
& = &
\text{Spec} ~
\mathbb{C}[p_1, \dots, p_c]
// Q_F
~,~
\nn\\
\mathcal{M}^{mes}
& = &
\text{Spec} 
\left(
\mathbb{C}[p_1, \dots, p_c]
// Q_F
\right) // Q_D
~,~
\eea
where the 
F-term and D-term constraints 
are given as $U(1)$ charges on the GLSM fields $p_a$,
which are summarized in the $Q_F$ and $Q_D$ charge matrices, respectively.
The computation of the $Q_F$ and $Q_D$ charge matrices 
using the perfect matching matrix $P$ 
follows what is known as the forward algorithm for brane tilings \cite{Feng:2000mi, Feng:2001xr}.
\\

%=================================================================
\subsection{Reflexive Polygons, Toric Duality and Specular Duality \label{sec_reflexive}}
%=================================================================

%------------------------------------------------------------------------------------------------------------------
\begin{figure}[http!!]
\begin{center}
\resizebox{0.88\hsize}{!}{
\includegraphics{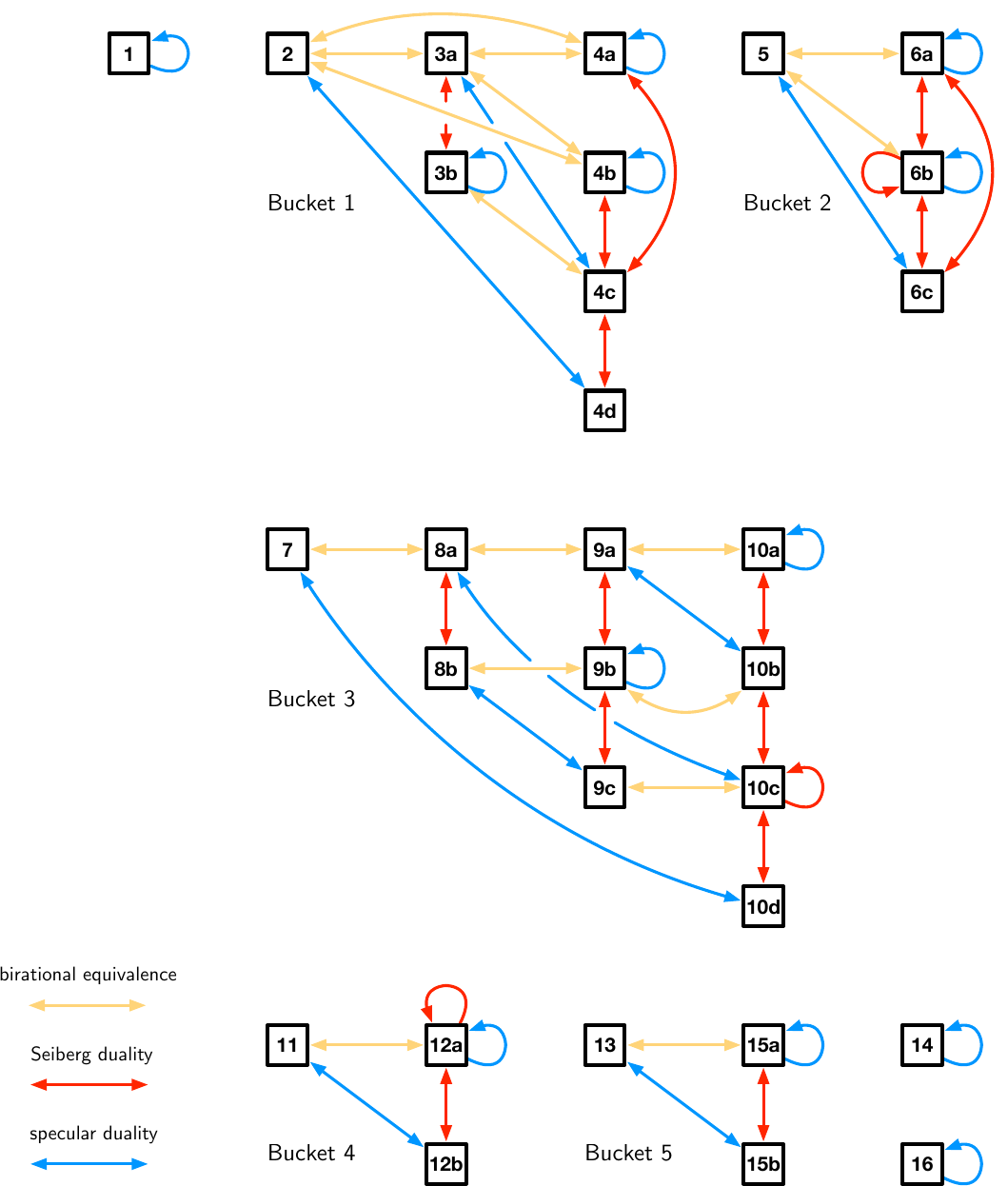}
}
\vspace{0.3cm}
\caption{
The 30 brane tilings corresponding to the 16 reflexive polygons in dimension 2 are related by Seiberg duality (red), specular duality (blue), and birational transformations (yellow).
Under Seiberg duality and under birational transformations, the associated dimer integrable systems are equivalent and form equivalence classes, which we call buckets. 
The labels correspond to the 30 brane tilings classified in \cite{Hanany:2012hi}
with the corresponding 16 reflexive toric diagrams given in \fref{fig_bucketsummary}.
}
\label{fig_02}
 \end{center}
 \end{figure}
%------------------------------------------------------------------------------------------------------------------

In this paper, we mainly focus on a special family of brane tilings associated to \textbf{reflexive polygons} in $\mathbb{Z}^2$.
It is known based on the classification in \cite{Hanany:2012hi} that there are 30 distinct brane tilings corresponding to the toric Calabi-Yau 3-folds whose toric diagrams \cite{fulton1993introduction,Leung:1997tw, He:2017gam, Krefl:2017yox, Choi:2023rqg, Bao:2024nyu}
are one of the 16 reflexive polygons in $\mathbb{Z}^2$.
There are more brane tilings because some of them correspond to the same toric Calabi-Yau 3-fold due to \textbf{Seiberg duality} between the corresponding $4d$ $\mathcal{N}=1$ theories \cite{Seiberg:1994pq}.
This correspondence in the context of toric Calabi-Yau 3-folds associated to $4d$ $\mathcal{N}=1$ theories
is also known as \textbf{toric duality} \cite{Feng:2000mi,Feng:2001xr,Feng:2002zw, Feng:2001bn,Beasley:2001zp}. 

%------------------------------------------------------------------------------------------------------------------
\begin{table}[http!!]
\begin{center}
\begin{tabular}{|c|c|}
\hline
$d$ & number of reflexive polytopes
\\
\hline
1 & 1
\\
2 & 16
\\
3 & 4319
\\
4 & 473800776
\\
\hline
\end{tabular}
\caption{The number of reflexive polytopes in dimension $d \leq 4$ \cite{Kreuzer:1998vb, Kreuzer:2000qv,Kreuzer:2000xy}.}
\label{tab_reflexive}
 \end{center}
 \end{table}
%------------------------------------------------------------------------------------------------------------------

A convex $d$-dimensional lattice polytope $\Delta$ is \textbf{reflexive} if its dual polytope $\Delta^\circ$ defined as 
\beal{es01a07}
\Delta^\circ = 
\{
\mathbf{u} \in \mathbb{Z}^d 
~|~ 
\mathbf{u} \cdot \mathbf{v} \geq -1,~ \forall\mathbf{v} \in \Delta
\}~,~
\eea
is also a convex lattice polytope in $\mathbb{Z}^d$ \cite{1993alg.geom.10003B, 1994alg.geom..2002B}. 
Due to a classification by Kreuzer and Skarke \cite{Kreuzer:1998vb, Kreuzer:2000qv,Kreuzer:2000xy},
it is known up to lattice dimension 4 that there are finitely many reflexive polytopes up to $GL(d, \mathbb{Z})$ equivalence.
\tref{tab_reflexive} summarizes the number of reflexive polytopes up to dimension $4$.
\\

\begin{figure}[http!!]
\begin{center}
\resizebox{0.75\hsize}{!}{
\includegraphics{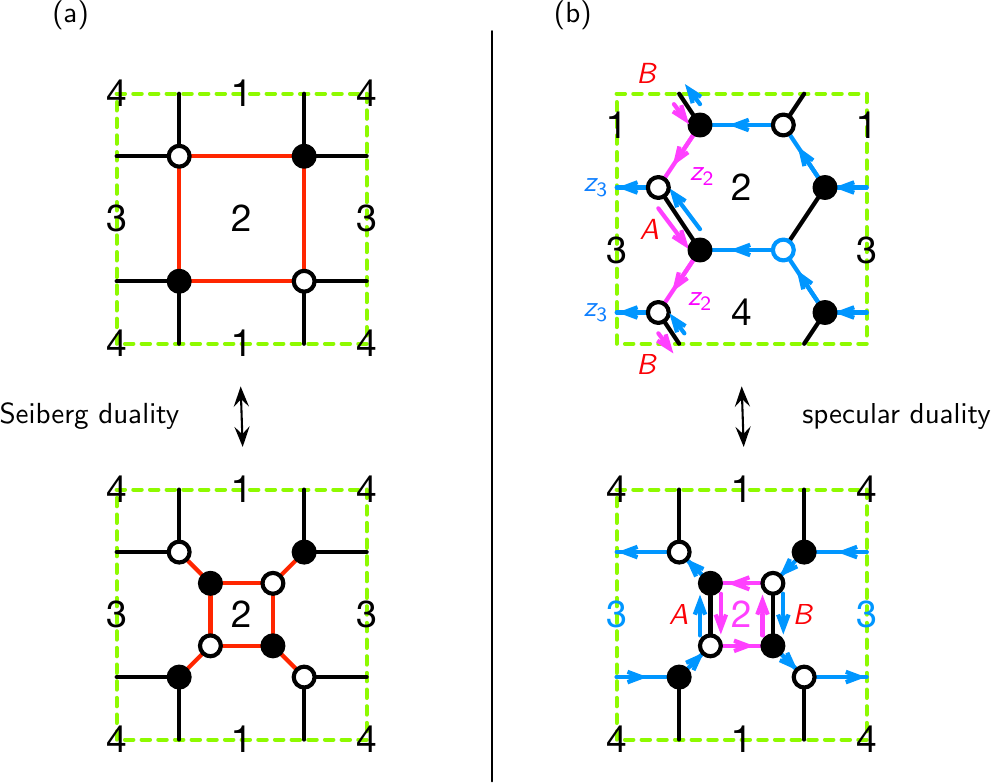}
}
\vspace{0.3cm}
\caption{
(a) Seiberg duality on brane tilings \cite{Feng:2000mi,Feng:2001xr,Feng:2002zw, Feng:2001bn,Beasley:2001zp} is a local deformation of the bipartite graph on the 2-torus acting on square faces, which is also known as urban renewal or spider moves \cite{goncharov2012dimersclusterintegrablesystems, CIUCU199834, 1999math......3025K}. 
(b) Specular duality on brane tilings \cite{Hanany:2012vc} swaps directed paths along edges corresponding to zig-zag paths with 
directed paths around faces and vice versa while preserving intersections between these paths.
}
\label{fig_duality}
 \end{center}
 \end{figure}

As illustrated in \fref{fig_bucketsummary}, there are 16 reflexive polygons up to $GL(2,\mathbb{Z})$ equivalence in $\mathbb{Z}^2$. 
The 30 brane tilings corresponding to these 16 reflexive polytopes have been fully classified in \cite{Hanany:2012hi}.
Under Seiberg duality, 
multiple brane tilings and the corresponding $4d$ $\mathcal{N}=1$ theories can correspond to the same toric Calabi-Yau 3-fold as summarized in \fref{fig_02}.
Seiberg duality can be interpreted as a local mutation of the bipartite graph on $T^2$,
which is also referred to as \textbf{urban renewal} or \textbf{spider move} \cite{goncharov2012dimersclusterintegrablesystems, CIUCU199834, 1999math......3025K}.
The brane tilings and $4d$ $\mathcal{N}=1$ theories
corresponding to the same toric Calabi-Yau 3-fold
are referred to as \textbf{toric phases} \cite{Feng:2000mi,Feng:2001xr,Feng:2002zw, Feng:2001bn,Beasley:2001zp}.
\fref{fig_duality}(a) illustrates the local mutation on the brane tiling that identifies the two toric phases corresponding to the cone over the zeroth Hirzebruch surface $F_0$, whose toric diagram is one of the 16 reflexive polygons in $\mathbb{Z}^2$.
The two toric phases are referred to as Model 15a and 15b in \fref{fig_bucketsummary} and \fref{fig_02}. 

%------------------------------------------------------------------------------------------------------------------
%------------------------------------------------------------------------------------------------------------------

The rich combinatorial structure of brane tilings led to the discovery of a new correspondence in \cite{Hanany:2012vc}
now known as \textbf{specular duality}. 
This new correspondence 
identifies brane tilings and the associated abelian $4d$ $\mathcal{N}=1$ theories
that have the same master space as defined in \eref{es01a06b}.
Like Seiberg duality, 
specular duality can be interpreted as a deformation of the bipartite graph on a 2-torus $T^2$, 
where for reflexive toric diagrams the resulting bipartite graph is again on a 2-torus $T^2$.
Specular duality swaps the roles played by zig-zag paths and faces in a brane tiling as illustrated in \fref{fig_duality}(b).
\\

%=================================================================
\subsection{Dimer Integrable Systems}
%=================================================================

In the following section, we review various aspects of integrable systems corresponding to brane tilings
and bipartite graphs on $T^2$ that were introduced by Goncharov and Kenyon in \cite{goncharov2012dimersclusterintegrablesystems, Eager:2011dp}.
Every consistent brane tiling on a 2-torus defines such a \textbf{dimer integrable system}
whose Casimirs and Hamiltonians as well as the spectral curve and the Poisson commutation relations 
are encoded in the bipartite graph on $T^2$.

\paragraph{Edge Variables and Perfect Matching Weights.}
In order to review dimer integrable systems
and how they are encoded in a brane tiling, 
we first recall that every edge in the bipartite graph on $T^2$
is associated to a bifundamental chiral field $X_{ij}$ of the $4d$ $\mathcal{N}=1$ theory, where the indices $i,j$ label 
the gauge groups of the $4d$ $\mathcal{N}=1$ theory associated to the faces of the brane tiling.

Equivalently, we can label each edge by an \textbf{edge variable}
$e_{jk} = (w_j, b_k)$, 
where now $j$ labels white nodes
$w_j$ and $k$ labels black nodes $b_k$ of the brane tiling.
As in \cite{Kho:2025fmp}, 
we also introduce \textbf{directed edge variables} $e_{jk}^+$ and $e_{jk}^-$,
which indicate respectively whether one moves along an edge from a white node to a black node, 
or from a black node to a white node, 
\beal{es02a00}
e_{jk}^+ ~:~ w_j \rightarrow b_k
~,~
e_{jk}^- ~:~ b_k \rightarrow w_j 
~.~
\eea
Here, we set the convention $e_{jk}^+ \equiv e_{jk}$.
As illustrated in \cite{Kho:2025fmp}, 
directed edge variables $e_{jk}^\pm$ allow us to express connected paths along edges in the brane tiling as a sequence of directed edge variables that alternate between white and black nodes.
When these connected paths along edges are closed, it is argued in \cite{Jejjala:2010vb, Hanany:2015tgh, Kho:2025fmp} that they form permutations of directed edge variables $e_{jk}^\pm$ in the permutation group $S_{2|E|}$, where $|E|$ is the number edges in the brane tiling.

%------------------------------------------------------------------------------------------------------------------
\begin{figure}[http!!]
\begin{center}
\resizebox{0.88\hsize}{!}{
\includegraphics{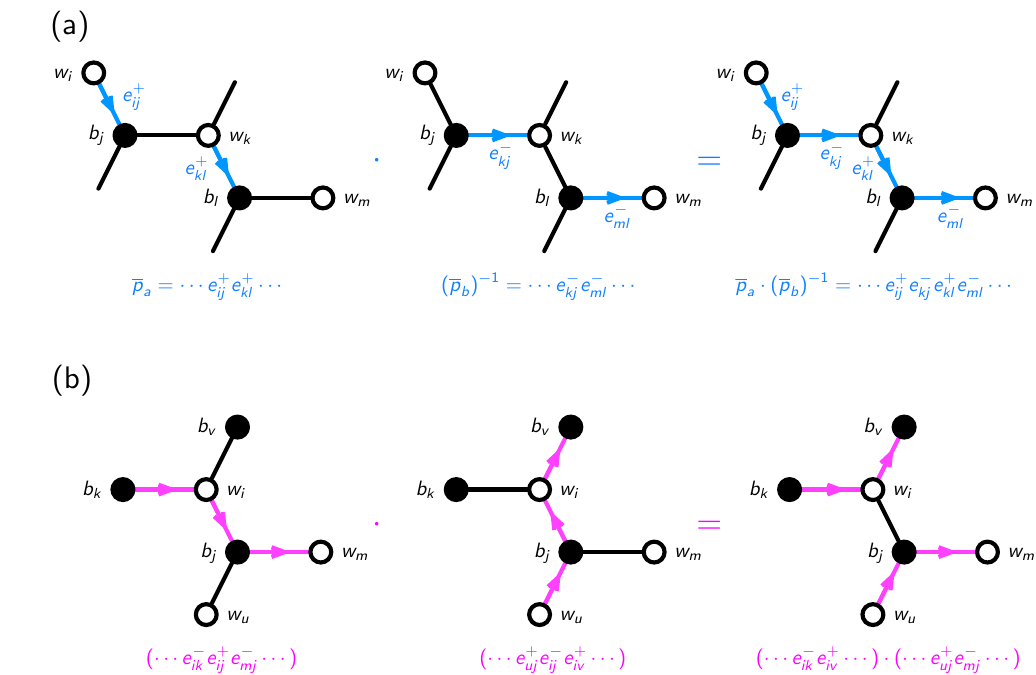}
}
\vspace{0.3cm}
\caption{
(a) A product of perfect matchings weights $\overline{p}_a$ and $(\overline{p}_b)^{-1}$
in terms of directed edge variables, 
and (b) a product of closed directed paths given by permutations in $S_{2|E|}$ with a cancellation between a pair of directed edges.
}
\label{fig_directedpaths}
 \end{center}
 \end{figure}
%------------------------------------------------------------------------------------------------------------------

We can also make use of directed edge variables $e_{jk}^\pm$
in order to introduce \textbf{perfect matching weights}
$\overline{p}_a$ associated to a perfect matching $p_a$
of a brane tiling,
\beal{es02a01}
\overline{p}_a
= \prod_{e_{jk} \in p_a} e_{jk}^+
~,~
(\overline{p}_a)^{-1}
= \prod_{e_{jk} \in p_a} e_{jk}^-
~.~
\eea
Defining perfect matching weights $\overline{p}_a$ in terms of directed edge variables allows us 
to introduce a product of perfect matching weights that can be associated to a directed path along edges of the brane tiling.
Taking
\beal{es02a02}
\overline{p}_a
=
\cdots 
e_{ij}^+
e_{kl}^+
\cdots
~,~
(\overline{p}_b)^{-1}
=
\cdots
e_{kj}^-
e_{ml}^-
\cdots
~,~
\eea
we define the following product of perfect matching weights, 
\beal{es02a03}
\overline{p}_a \cdot (\overline{p}_b)^{-1}
\equiv 
\cdots 
e_{ij}^+
e_{kj}^-
e_{kl}^+
e_{ml}^-
\cdots ~,~
\eea
where we see that under the product
we obtain a directed connected path along the edges of a brane tiling
alternating between white and black nodes as illustrated in \fref{fig_directedpaths}.

The convention used in \cite{Kho:2025fmp} is that all directed paths along edges in a brane tiling 
alternate between white and black nodes and can be therefore expressed as an alternating sequence of directed edge variables $e_{jk}^+$ and $e_{lk}^-$.
Moreover, when the connected paths are closed, then the directed edge variables $e_{jk}^\pm$
form permutation tuples of the permutation group $S_{2|E|}$.
Taking two permutation tuples in $S_{2|E|}$ in terms of $e_{jk}^\pm$, 
we identify the product between the permutation tuples to be as follows, 
\beal{es02a05}
(\cdots e_{ik}^{-}~ e_{ij}^+ ~ e_{mj}^- ~\cdots)
\cdot
(\cdots e_{uj}^{+}~ e_{ij}^- ~ e_{iv}^+ ~\cdots)
=
(\cdots e_{ik}^{-} ~ e_{iv}^+ ~\cdots)
\cdot
(\cdots e_{uj}^{+} ~ e_{mj}^- ~\cdots)
~,~
\nn\\
\eea
giving a new pair of closed paths
with certain edge variables cancelling each other under the following identities, 
\beal{es02a06}
(e_{jk}^\pm)^{-1}
= e_{jk}^\mp
~,~
e_{jk}^+ \cdot e_{jk}^- = 1
~.~
\eea

%------------------------------------------------------------------------------------------------------------------
\begin{figure}[http!!]
\begin{center}
\resizebox{0.6\hsize}{!}{
\includegraphics{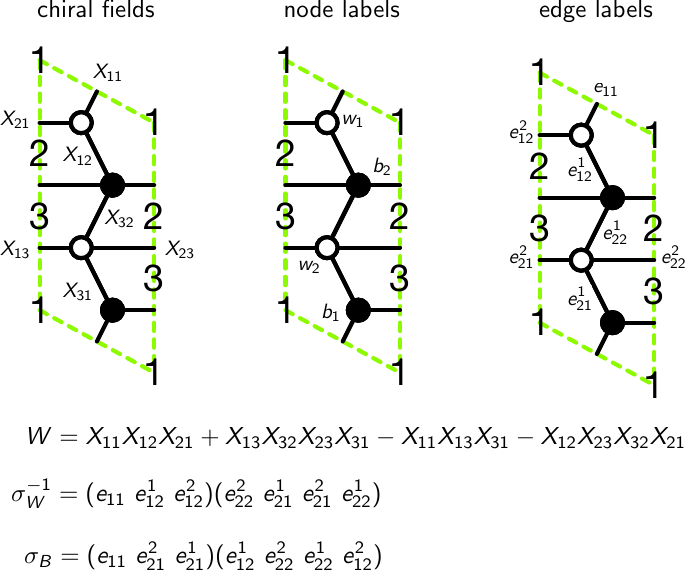}
}
\vspace{0.3cm}
\caption{The brane tiling for the suspended pinch point (SPP) with chiral fields $X_{ij}$, node labels $w_j$ and $b_k$, and edge labels $e_{jk}$.
The superpotential $W$ and the corresponding permutation tuples $\sigma_W^{-1}$ and $\sigma_B$ in terms of edge labels are also shown.
}
\label{fig_spp}
 \end{center}
 \end{figure}
%------------------------------------------------------------------------------------------------------------------

Examples of closed directed paths in a brane tiling are \textbf{zig-zag paths} and \textbf{face paths} that go around the boundary edges of a face in the brane tiling. 
By first using edge variables $e_{jk}$ instead of chiral fields $X_{ij}$, 
we are able to rewrite the superpotential $W$ of the brane tiling as a pair of permutation tuples $\sigma_B, \sigma_W \in S_{|E|}$ \cite{Jejjala:2010vb, Hanany:2015tgh}, 
where 
$\sigma_B$ contains a cycle for every black node in the brane tiling associated to a negative term in $W$,
while $\sigma_W^{-1}$ has a cycle for every white node in the brane tiling associated to a positive term in $W$ as reviewed in section \sref{sec_branetilings}.
These cycles in $\sigma_B$ and $\sigma_W^{-1}$ follow the clockwise and anti-clockwise orientation around white and black nodes in the brane tilings, respectively. 
Let us illustrate this for the brane tiling for the suspended pinch point (SPP) shown in \fref{fig_spp}, 
whose superpotential $W$
is as follows, 
\beal{es02a07}
W = X_{11} X_{12} X_{21} + X_{13} X_{32} X_{23} X_{31}
- X_{11} X_{13} X_{31} - X_{12} X_{23} X_{32} X_{21}
~.~
\eea
The corresponding permutation tuples in terms of edge variables $e_{jk}$ are given by,
\beal{es02a08}
\sigma_W^{-1} =
(e_{11}~e_{12}^1~e_{12}^2) 
(e_{22}^2~e_{21}^1~e_{21}^2~e_{22}^1)
~,~
\sigma_B =
(e_{11}~e_{21}^2~e_{21}^1) 
(e_{12}^1~e_{22}^2~e_{22}^1~e_{12}^2)
~.~
\eea

%------------------------------------------------------------------------------------------------------------------
\begin{figure}[http!!]
\begin{center}
\resizebox{0.85\hsize}{!}{
\includegraphics{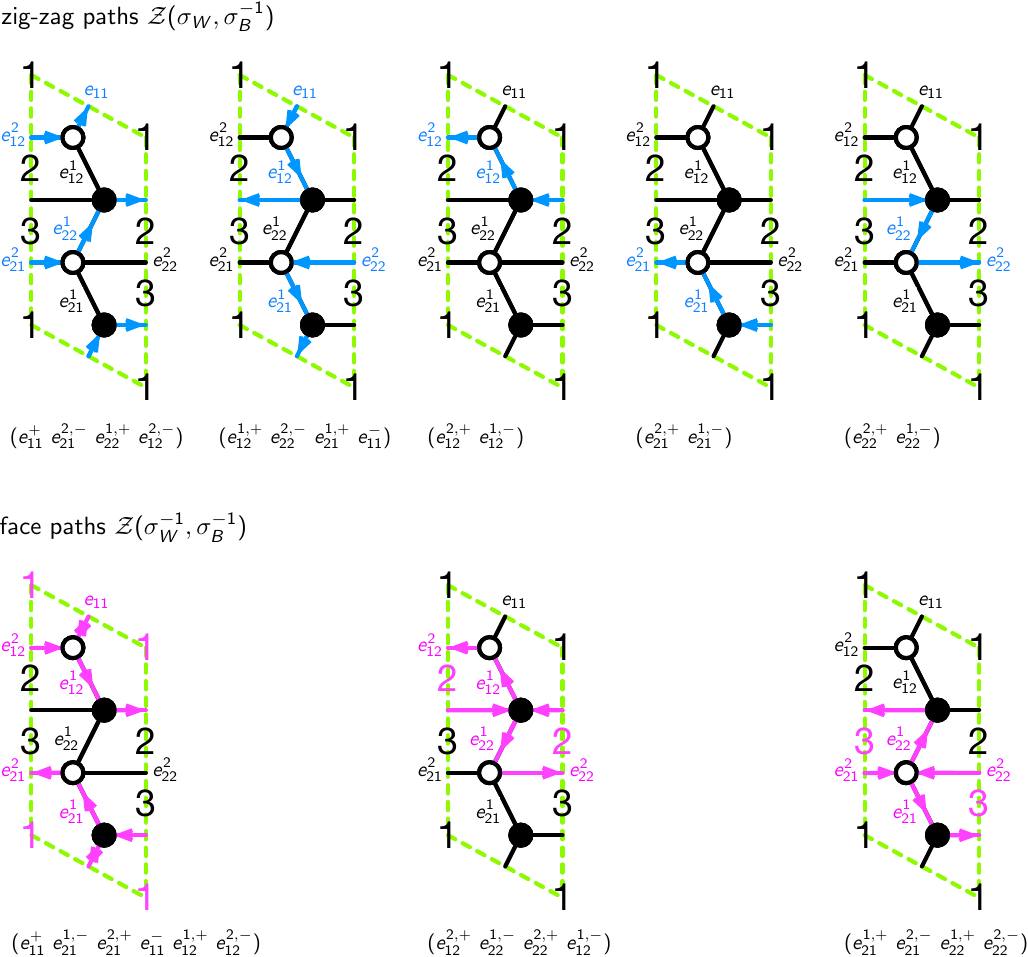}
}
\vspace{0.3cm}
\caption{The brane tiling for the suspended pinch point (SPP) with zig-zag paths given by $\Sigma_e \cdot (\sigma_W^{-1})^+ \cdot (\sigma_B)^-$ and face paths given by $\Sigma_e \cdot (\sigma_W^{-1})^+ \cdot (\sigma_B^{-1})^-$.
}
\label{fig_paths}
 \end{center}
 \end{figure}
%------------------------------------------------------------------------------------------------------------------

In terms of the permutation tuples $\sigma_W, \sigma_B \in S_{|E|}$, we can define the following permutations in $S_{2|E|}$ in terms of directed edge variables $e_{jk}^\pm$, 
\beal{es02a09}
\Sigma_z = 
(\sigma_W^{-1})^+ \cdot
(\sigma_B)^-
~,~
\Sigma_f = 
(\sigma_W^{-1})^+ \cdot
(\sigma_B^{-1})^-
~,~
\Sigma_e = \prod_{e_{jk}} (e_{jk}^+ ~ e_{jk}^-)
~,~
\eea
where here in $(\sigma)^+$
all edge variables $e_{jk}$ become directed edge variables $e_{jk}^+$, 
while in $(\sigma)^-$ all edge variables $e_{jk}$ become directed edge variables $e_{jk}^-$.
Using these permutations in $S_{2|E|}$, we can write permutations that encode the zig-zag paths in the brane tiling \cite{Jejjala:2010vb, Hanany:2015tgh} as follows, 
\beal{es02a10}
\Sigma_e
\cdot
\Sigma_z
=
\prod_{e_{jk}} (e_{jk}^+ ~ e_{jk}^-)
\cdot
(\sigma_W^{-1})^+ \cdot
(\sigma_B)^-
~,~
\eea
while the permutations that encode the directed paths
around boundary edges of faces in the brane tiling are given by,
\beal{es02a11}
\Sigma_e
\cdot
\Sigma_f
=
 \prod_{e_{jk}} (e_{jk}^+ ~ e_{jk}^-)
\cdot
(\sigma_W^{-1})^+ \cdot
(\sigma_B^{-1})^-
~.~
\eea
For the SPP example with $\sigma_W^{-1}$ and $\sigma_B$ given in \eref{es02a08}, 
all distinct zig-zag paths are given by,
\beal{es02a12}
\Sigma_e
\cdot
\Sigma_z
=
(e_{11}^{+}~e_{21}^{2,-}~e_{22}^{1,+}~e_{12}^{2,-})
(e_{12}^{1,+}~e_{22}^{2,-}~e_{21}^{1,+}~e_{11}^{-})
(e_{12}^{2,+}~e_{12}^{1,-})
(e_{21}^{2,+}~e_{21}^{1,-})
(e_{22}^{2,+}~e_{22}^{1,-})
~,~
\nn\\
\eea
where every cycle corresponds to a closed zig-zag path in the brane tiling.
Similarly, the directed paths around the 3 faces of the SPP brane tiling are given by,
\beal{es02a13}
\Sigma_e
\cdot
\Sigma_f
=
(e_{11}^{+}~e_{21}^{1,-}~e_{21}^{2,+}~e_{11}^{-}~e_{12}^{1,+}~e_{12}^{2,-})
(e_{12}^{2,+}~e_{22}^{1,-}~e_{22}^{2,+}~e_{12}^{1,-})
(e_{21}^{1,+}~e_{21}^{2,-}~e_{22}^{1,+}~e_{22}^{2,-})
~,~
\nn\\
\eea
where we can see that directed paths go anti-clockwise around each of the faces as illustrated in \fref{fig_paths}.
\\

\paragraph{Casimirs, $1$-loops and the Spectral Curve.}
The Kasteleyn matrix defined in \eref{es01a02}
is written in terms of edge variables $e_{jk}$.
By taking all the edge variables to be positively oriented such that, 
\beal{es02a15}
K_{w_j, b_k}^{+}(x,y) = \sum_{e_{jk}=(w_j, b_k)} e_{jk}^+ ~x^{h_1(e_{jk})} y^{h_2(e_{jk})}
~,~
\eea
the characteristic polynomial from the permanent then takes the form, 
\beal{es02a16}
P^{+}(x,y) = \text{perm} ~K_{w_j, b_k}^{+}(x,y) = \mathop{\sum_{p_{(n_x, n_y)}}}_{(n_x,n_y)\in \Delta}~ \overline{p}_{(n_x, n_y)}~ x^{n_x} y^{n_y}
~,~
\eea
where $\overline{p}_{(n_x,n_y)}$ is the perfect matching weight defined in \eref{es02a01}
corresponding to perfect matching $p_{(n_x,n_y)}$ in the brane tiling associated to vertex $(n_x,n_y)$ in the toric diagram $\Delta$.
Here, we note that multiple perfect matchings can correspond to the same vertex in the toric diagram
and the sum in \eref{es02a16} goes over all of them. 

By choosing a \textbf{reference perfect matching} $p_0$, 
we can factor out its corresponding weight from the characteristic polynomial in \eref{es02a16} to obtain, 
\beal{es02a20}
P^{+}(x,y) 
= 
\overline{p}_0
\cdot 
\mathop{\sum_{p_{(n_x, n_y)}}}_{(n_x,n_y)\in \Delta}~
\delta_{(n_x,n_y)}~
x^{n_x} y^{n_y}
~,~
\eea
where
\beal{es02a21}
\delta_{(n_x,n_y)}
=
(\overline{p}_0)^{-1} \cdot
\overline{p}_{(n_x, n_y)}
~,~
\eea
are identified 
as \textbf{Casimirs}
of the associated dimer integrable system
given by the brane tiling. 
For the origin $(0,0) \in \Delta$, 
which can be chosen by an overall shift of the toric diagram
such that it becomes the unique internal vertex for reflexive polygons, 
we have multiple associated perfect matchings.
The corresponding perfect matching weights give the \textbf{Hamiltonian} of the dimer integrable system.
The expression for the Hamiltonian is as follows, 
\beal{es02a22}
\delta_{(0,0)}
\equiv H = 
\sum_{p_{(0,0)}^{u}}
\gamma_u
~,~
\eea
where
\beal{es02a23}
\gamma_u
=
(\overline{p}_0)^{-1}
\cdot
\overline{p}_{(0,0)}^{u}
~,~
\eea
are the \textbf{1-loops} of the dimer integrable system.
Here, 
the sum in \eref{es02a22} is over all perfect matchings associated to the interior point $(0,0)$
of the reflexive toric diagram, 
where $u=1, \dots, m$
labels the perfect matchings with $m$ being the multiplicity of the interior vertex. 
$\overline{p}_{(0,0)}^{u}$ is the weight of the $u$-th perfect matching associated to the interior vertex $(0,0)$. 
By factorizing out the reference perfect matching weight $\overline{p}_0$, 
these perfect matching weights are then identified with the
1-loops $\gamma_u$ in \eref{es02a23}, 
with the sum over all 1-loops associated to the unique interior point of the toric diagram corresponding to the Hamiltonian $H$ of the dimer integrable system.
In the case of brane tilings corresponding to non-reflexive toric diagrams with $N_i$ internal vertices, 
there would be $N_i$ Hamiltonians of the form $H_1, \dots, H_{N_i}$
where the $n$-th Hamiltonian is given by the sum over $n$-loops.

Accordingly, for brane tilings associated with toric Calabi-Yau 3-folds with reflexive toric diagrams,
we can 
write down the general form of the \textbf{spectral curve} $\Sigma$
of the dimer integrable system following the factorized form of the characteristic polynomial in \eref{es02a20},
\beal{es02a24}
\Sigma ~:~
H + 
\mathop{
\sum_{
(n_x,n_y)\in \Delta
}
}_{
(n_x,n_y) \neq (0,0)
}~
\delta_{(n_x,n_y)}~
x^{n_x} y^{n_y}
= 0
~,~
\eea
where the single Hamiltonian $H$ takes the form given in \eref{es02a22}.
In terms of the Newton polynomial $P^{+}(x,y)$ defined in \eref{es02a20}, 
we can express the spectral curve of the dimer integrable system as follows, 
\beal{es02a24b}
\Sigma ~:~
(\overline{p}_0)^{-1} \cdot P^{+}(x,y) 
= 0
~.~
\eea
\\

\paragraph{Poisson Commutation Relations.}
We can define Poisson commutation relations between directed closed paths, 
also referred to as loops, in the dimer integrable systems.
For oriented face paths $f_i$, the Poisson commutation relations are given by,
\beal{es02a25}
\{f_i , f_{j}\}
= I_{i,  j} ~f_i f_{j} 
~,~
\eea
where $I_{i, j}$
is the number of arrows from node $i$ to $j$
minus the number of arrows from node $j$ to $i$ in the quiver diagram of the corresponding brane tiling.

We note here that since brane tilings are embedded on a 2-torus $T^2$ and the fact that brane tilings are bipartite making every face even-sided, 
the face paths satisfy the following overall constraint relation,
\beal{es02a26}
\prod_{i=1}^G
f_i
= \text{constant}
~,~
\eea
where $G$ is the number of faces corresponding to the number of gauge groups in the brane tiling.
Moreover,  
the face paths $f_i$ form with the zig-zag paths $z_{r}$ non-trivial relations.
Since in this work, we only consider reflexive polygons as toric diagrams, 
the dimer integrable systems only have a single Hamiltonian with \textbf{canonical variables} $e^P$ and $e^Q$.
The face paths $f_i$ of the corresponding dimer integrable system can be expressed in terms of these canonical variables.
We also note that the face paths $f_i$ correspond to cluster variables of the quiver in the corresponding brane tiling \cite{goncharov2012dimersclusterintegrablesystems, Marshakov:2012kv, Fock:2014ifa}. 

%------------------------------------------------------------------------------------------------------------------
\begin{figure}[http!!]
\begin{center}
\resizebox{0.9\hsize}{!}{
\includegraphics{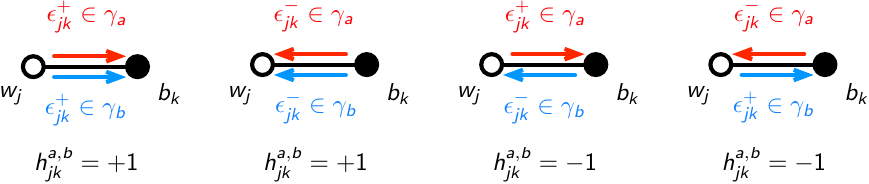}
}
\caption{
The possible directed intersections between directed paths $\gamma_a$ and $\gamma_b$
at edge $e_{jk}$
with the corresponding directed intersection number $h_{jk}^{a,b}$. 
}
\label{fig_poissonint}
 \end{center}
 \end{figure}
%------------------------------------------------------------------------------------------------------------------

The associated oriented 1-loops $\gamma_u$ of the single Hamiltonian $H$ 
can be written in terms of face paths and zig-zag paths, 
as well as the canonical variables $e^P$ and $e^Q$.
In general, they satisfy 
the following Poisson commutation relations of the form,
\beal{es02a34}
\{
\gamma_u , \gamma_{u^\prime}
\}
=
\frac{\partial \gamma_u}{\partial P} \cdot
\frac{\partial \gamma_{u^\prime}}{\partial Q}
-
\frac{\partial \gamma_u}{\partial Q} \cdot
\frac{\partial \gamma_{u^\prime}}{\partial P}
~.~
\eea
These Poisson commutation relations can be rewritten in terms of the original 1-loops 
$\gamma_u$ and $\gamma_{u^\prime}$ 
as
follows,
\beal{es02a35}
\{
\gamma_u , \gamma_{u^\prime}
\}
=
\epsilon_{u, u^\prime} ~\gamma_u  \gamma_{u^\prime}
~,~
\eea
where 
\beal{es02a36}
\epsilon_{u, u^\prime}  = 
\sum_{e_{jk} \in \gamma_u, \gamma_{u}^\prime}
h_{jk}^{u, u^\prime} 
\eea
is the ordered intersection number between 1-loops $\gamma_u$ and $\gamma_{u^\prime}$.
The sum in \eref{es02a36} is over all common edges $e_{jk}$ between $\gamma_u$ and $\gamma_{u^\prime}$,
and $h_{jk}^{u,u^\prime}$ is the directed intersection number
at edge $e_{jk}$ where $\gamma_u$ and $\gamma_{u^\prime}$ intersect.
The different values that $h_{jk}^{u,u^\prime}$ can have at a particular intersection at edge $e_{jk}$ are given in \fref{fig_poissonint} with the corresponding illustrations of the directed intersections.

The Poisson commutation relations between 1-loops can also be written in terms of a \textbf{commutation matrix} $C$ 
as follows, 
\beal{es02a37}
\{
\gamma_u , \gamma_{u^\prime}
\}
= C_{u, u^\prime}~ \gamma_u \gamma_{u^\prime}
~,~
\eea
where $C_{u, u^\prime} \in \mathbb{Z}$ are elements of the commutation matrix.
The commutation matrices
are presented in the following classification for all 30 dimer integrable systems corresponding to reflexive polygons in $\mathbb{Z}^2$.
\\

%=================================================================
\subsection{Birational Transformations on the Dimer Integrable Systems}
%=================================================================

Birational transformations 
have been studied extensively in \cite{akhtar2012minkowski, 2018arXiv180100013B, 2021RSPSA.47710584C}
in order to identify birationally equivalent toric Fano 3-folds.
In \cite{Ghim:2024asj, Ghim:2025zhs}, this equivalence has 
been interpreted as a correspondence between $2d$ $(0,2)$ supersymmetric gauge theories
associated with toric Fano 3-folds and more generally toric Calabi-Yau 4-folds 
realized by brane brick models \cite{Franco:2015tna, Franco:2015tya, Franco:2016nwv, Franco:2016qxh, Franco:2022gvl, Kho:2023dcm}.

In this work, we focus on birational transformations that relate 
toric Calabi-Yau 3-folds whose toric diagrams are given by the 16 reflexive polygons as summarized in section \sref{sec_reflexive}.
As observed in \cite{Kho:2025fmp},
when two of these
toric Calabi-Yau 3-folds are related
by a birational transformation,
they are associated to a pair of brane tilings on the 2-torus 
that define dimer integrable systems which are birationally equivalent.
Under this equivalence, 
it is shown in \cite{Kho:2025fmp} 
that the Casimirs and Hamiltonians
as well as the Poisson commutation relations of the 
integrable systems are identified to each other by the birational transformation.
The spectral curves
as defined in \eref{es02a24}
are also mapped to each other by the birational transformation, 
making the transformation a true equivalence between two distinct dimer integrable systems. 

Given the Newton polynomial $P(x,y)$ as defined in \eref{es01a01b}
for the toric diagram $\Delta$ of a toric Calabi-Yau 3-fold, 
we can expand it in the following form, 
\beal{es02a40}
P(x,y) = \sum_{m=a}^{b}
C_m(y) x^m ~,~
\eea
where $a<0$ and $b>0$
and $C_m(y)$ are sub-polynomials of $P(x,y)$ for $a \leq m \leq b$.
Using this expanded form of the Newton polynomial, 
we can define a \textbf{birational transformation} $\varphi_A$ \cite{galkin2010mutations, 2012SIGMA...8..047I, akhtar2012minkowski, 2018arXiv180100013B, 2021RSPSA.47710584C} 
that acts on the coordinates $x,y \in \mathbb{C}^*$ of $P(x,y)$ as follows, 
\beal{es02a50}
\varphi_A ~:~ (x,y) \mapsto \left(A(y) x, y \right)
~,~
\eea
where $A(y)$ is a polynomial chosen such that 
$A(y)^{-m}$ is a polynomial divisor of $C_m(y)$
in the expansion in \eref{es02a40}
for $a \leq m \leq -1$. 
By calling the new Newton polynomial $P^\vee(x,y)$
with toric diagram $\Delta^\vee$, 
the toric varieties associated to the original toric diagram $\Delta$ and 
the new toric diagram $\Delta^\vee$
are known to be birationally equivalent to each other \cite{galkin2010mutations, 2012SIGMA...8..047I, akhtar2012minkowski, 2018arXiv180100013B, 2021RSPSA.47710584C}.
This birational equivalence exists if the birational map in \eref{es02a50}
applies to at least one chosen $GL(2,\mathbb{Z})$ frame or choice of origin in the $\mathbb{Z}^2$
lattice for the toric diagrams $\Delta$ and $\Delta^\vee$.
Given that the birational transformation $\varphi_A$
only exists for certain $GL(2,\mathbb{Z})$ frames of a given toric diagram, 
we can generalize the expression of the birational transformation in \eref{es02a50} to,
\beal{es02a51}
\varphi_{A;M;N}
= M \circ \varphi_A \circ N ~,~
\eea
in order to include the $GL(2,\mathbb{Z})$ transformations $M$ and $N$ 
on the coordinates $x,y$ in $P(x,y)$. 

Under such birational transformations, 
dimer integrable systems associated to
brane tilings corresponding to $\Delta$ and $\Delta^\vee$
are \textbf{birationally equivalent} to each other, as observed in \cite{Kho:2025fmp}.
This means that the birational map $\varphi_{A;M;N}$
acts on the spectral curve of the dimer integrable system as follows, 
\beal{es02a55}
\varphi_{A;M;N} 
\Sigma = \Sigma^\vee
~.~
\eea
In other words, the spectral curves
are mapped to each other by the birational transformation from $\Delta$ to $\Delta^\vee$. 
As a result of this, 
the Hamiltonian $H$
and the 1-loops $\gamma_u$ 
are identified to each other
between the two birationally equivalent dimer integrable systems.
This in turn leads to identifications between the
Poisson commutation relations 
as well as
relations 
between zig-zag paths and face paths 
of the birationally equivalent dimer integrable systems.
These relations form a canonical transformation between the birationally equivalent dimer integrable systems.
\\

In the following work, we classify all birational equivalences 
between dimer integrable systems
that correspond to the 30 brane tilings associated to toric Calabi-Yau 3-folds whose toric diagrams are one of the 16 reflexive polygons.
\fref{fig_02} summarizes the classification of all birational equivalences between 
the 30 dimer integrable systems corresponding to the 16 reflexive polygons.
While presenting the explicit birational maps that define these equivalences, 
we also present the relations between zig-zag paths and face paths
as well as the associated canonical variables that lead to 
the identifications of the
Hamiltonians, spectral curves and 1-loops between the equivalent dimer integrable systems.
\\

%=================================================================

%=================================================================

%=================================================================
\section{Model 1: $\mathbb{C}^3/\mathbb{Z}_3 \times \mathbb{Z}_3$ $(1,0,2)(0,1,2)$ \label{sec:03}}
%=================================================================

%------------------------------------------------------------------------------------------------------------------
\begin{figure}[H]
\begin{center}
\resizebox{0.9\hsize}{!}{
\includegraphics{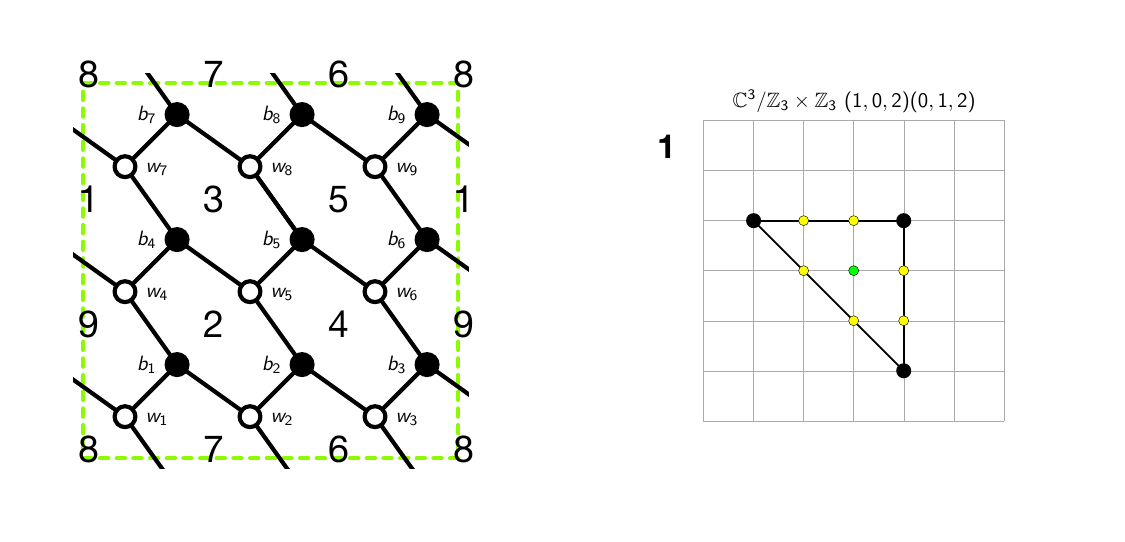}
}
\vspace{-0.5cm}
\caption{The brane tiling and toric diagram of Model 1.}
\label{mf_01}
 \end{center}
 \end{figure}
%------------------------------------------------------------------------------------------------------------------

The brane tiling for Model 1 can be expressed in terms of the following pair of permutation tuples
\beal{es10a01}
\sigma_B &=& (e_{11}\ e_{21}\ e_{41}) \ (e_{22}\ e_{32}\ e_{52}) \ (e_{13}\ e_{63}\ e_{33}) \ (e_{44}\ e_{54}\ e_{74}) \ (e_{55}\ e_{65}\ e_{85})\nn
\\&& (e_{46}\ e_{96}\ e_{66}) \ (e_{17}\ e_{77}\ e_{87}) \ (e_{28}\ e_{88}\ e_{98}) \ (e_{39}\ e_{99}\ e_{79})~, \nn
\\
\sigma_W^{-1} &=& (e_{11}\ e_{17}\ e_{13}) \ (e_{21}\ e_{22}\ e_{28}) \ (e_{32}\ e_{33}\ e_{39}) \ (e_{41}\ e_{46}\ e_{44}) \ (e_{52}\ e_{54}\ e_{55}) \nn
\\&& (e_{63}\ e_{65}\ e_{66})\ (e_{74}\ e_{79}\ e_{77}) \ (e_{85}\ e_{87}\ e_{88}) \ (e_{96}\ e_{98}\ e_{99})~,
\eea
which correspond to black and white nodes of the brane tiling, respectively.\\
 
The brane tiling for Model 1 has 9 zig-zag paths given by, 
\beal{es10a03}
&
z_1 = (e_ {74}^{+}~ e_ {44}^{-}~ e_ {41}^{+}~ e_ {11}^{-}~ e_ {17}^{+}~ e_{77}^{-})~,~\nn
z_2 = (e_ {46}^{+}~ e_ {96}^{-}~ e_ {98}^{+}~ e_ {28}^{-}~ e_ {21}^{+}~ e_{41}^{-})~,~\nn
&
\\
&
z_3 = (e_ {99}^{+}~ e_ {79}^{-}~ e_ {77}^{+}~ e_ {87}^{-}~ e_ {88}^{+}~ e_{98}^{-})~,~\nn
z_4 = (e_ {44}^{+}~ e_ {54}^{-}~ e_ {55}^{+}~ e_ {65}^{-}~ e_ {66}^{+}~ e_{46}^{-})~,~\nn
&
\\
&
z_5 = (e_ {52}^{+}~ e_ {22}^{-}~ e_ {28}^{+}~ e_ {88}^{-}~ e_ {85}^{+}~ e_{55}^{-})~,~\nn
z_6 = (e_ {79}^{+}~ e_ {39}^{-}~ e_ {32}^{+}~ e_ {52}^{-}~ e_ {54}^{+}~ e_{74}^{-})~,~\nn
&
\\
&
z_7 = (e_ {96}^{+}~ e_ {66}^{-}~ e_ {63}^{+}~ e_ {33}^{-}~ e_ {39}^{+}~ e_{99}^{-})~,~\nn
z_8 = (e_ {87}^{+}~ e_ {17}^{-}~ e_ {13}^{+}~ e_ {63}^{-}~ e_ {65}^{+}~ e_{85}^{-})~,~\nn
&
\\
&
z_9=(e_{22}^{+}~ e_{32}^{-}~ e_{33}^{+}~ e_{13}^{-}~ e_{11}^{+}~e_{21}^{-})~,~
\eea
and 9 face paths given by, 
\beal{es10a04}
&
f_1 = (e_ {44}^{+}~ e_ {74}^{-}~ e_ {79}^{+}~ e_ {99}^{-}~ e_ {96}^{+}~ e_{46}^{-})~,~\nn
f_2 = (e_ {54}^{+}~ e_ {44}^{-}~ e_ {41}^{+}~ e_ {21}^{-}~ e_ {22}^{+}~ e_{52}^{-})~,~\nn
&
\\
&
f_3 = (e_ {74}^{+}~ e_ {54}^{-}~ e_ {55}^{+}~ e_ {85}^{-}~ e_ {87}^{+}~ e_{77}^{-})~,~\nn
f_4 = (e_ {52}^{+}~ e_ {32}^{-}~ e_ {33}^{+}~ e_ {63}^{-}~ e_ {65}^{+}~ e_{55}^{-})~,~\nn
&
\\
&
f_5 = (e_ {85}^{+}~ e_ {65}^{-}~ e_ {66}^{+}~ e_ {96}^{-}~ e_ {98}^{+}~ e_{88}^{-})~,~\nn
f_6 = (e_ {99}^{+}~ e_ {39}^{-}~ e_ {32}^{+}~ e_ {22}^{-}~ e_ {28}^{+}~ e_{98}^{-})~,~\nn
&
\\
&
f_7 = (e_ {21}^{+}~ e_ {11}^{-}~ e_ {17}^{+}~ e_ {87}^{-}~ e_ {88}^{+}~ e_{28}^{-})~,~\nn
f_8 = (e_ {77}^{+}~ e_ {17}^{-}~ e_ {13}^{+}~ e_ {33}^{-}~ e_ {39}^{+}~ e_{79}^{-})~,~\nn
&
\\
&
f_9 = (e_ {46}^{+}~ e_ {66}^{-}~ e_ {63}^{+}~ e_ {13}^{-}~ e_ {11}^{+}~ e_{41}^{-})~,~
\eea
which satisfy the following constraints,
\beal{es10a05}
&
f_1^{-1} f_7 =  z_1 z_2 z_3~,~
f_2 f_8^{-1} = z_1 z_6 z_9~,~
f_3 f_9^{-1} = z_1 z_4 z_8~,~ 
&
\nn\\
&
f_4 f_7^{-1} = z_5 z_8 z_9~,~ 
f_5^{-1} f_8 = z_3 z_7 z_8~,~
f_6^{-1} f_9 = z_2 z_7 z_9~,~ 
&
\nn\\
&
f_7 f_8 f_9 = z_2 z_3 z_7~,~ 
f_1 f_2 f_3 f_4 f_5 f_6 f_7 f_8 f_9=1~.~ 
&
\eea
The face paths can be written in terms of the canonical variables as follows, 
\beal{es10a05_1}
&
f_1= e^{-Q-P} z_1^{-1} z_7~,~ 
f_2= e^{Q} z_1 z_6 z_9 ~,~ 
f_3= e^{P} z_1 z_4 z_8 ~,~ 
\nn
&
\\
&
f_4= e^{-Q-P} z_1^{-1} z_4^{-1} z_6^{-1} ~,~ 
f_5= e^Q z_3^{-1} z_7^{-1} z_8^{-1} ~,~ 
f_6= e^P z_2^{-1} z_7^{-1} z_9^{-1} ~,~ 
\nn
&
\\
&
f_7= e^{-Q-P} z_2 z_3 z_7 ~,~ 
f_8=e^Q~,~ 
f_9=e^P~.~
&
\eea

The Kasteleyn matrix of the brane tiling for Model 1 in \fref{mf_01} is given by,
\beal{es10a06}
K = 
\left(
\ba{c|ccccccccc}
\; & b_1 & b_2 & b_3 & b_4 & b_5 & b_6 & b_7 & b_8 & b_9 \\
\hline
w_1 &e_{11} & 0 & e_{13} x^{-1} & 0 & 0 & 0 & e_{17} y^{-1} & 0 & 0 \\
w_2 &e_{21} & e_{22} & 0 & 0 & 0 & 0 & 0 & e_{28} y^{-1} & 0 \\
w_3 &0 & e_{32} & e_{33} & 0 & 0 & 0 & 0 & 0 & e_{39} y^{-1} \\
w_4 &e_{41} & 0 & 0 & e_{44} & 0 & e_{46} x^{-1} & 0 & 0 & 0 \\
w_5 &0 & e_{52} & 0 & e_{54} & e_{55} & 0 & 0 & 0 & 0 \\
w_6 &0 & 0 & e_{63} & 0 & e_{65} & e_{66} & 0 & 0 & 0 \\
w_7 &0 & 0 & 0 & e_{74} & 0 & 0 & e_{77} & 0 & e_{79}x^{-1} \\
w_8 &0 & 0 & 0 & 0 & e_{85} & 0 & e_{87} & e_{88} & 0 \\
w_9 &0 & 0 & 0 & 0 & 0 & e_{96} & 0 & e_{98} & e_{99} \\
\ea
\right)~.~
\eea
By taking the permanent of the Kasteleyn matrix,
we obtain the spectral curve of the dimer integrable system for Model 1 as follows,
\beal{es10a07}
&&
0 = \text{perm}~K=\overline{p}_0  \cdot x^{-1} y^{-1} \cdot  \big[\delta_{(-2,1)}x^{-2} y+\delta_{(-1,0)} x^{-1}+\delta_{(-1,1)}x^{-1}y
\nn\\&&
\hspace{1cm}
+\delta_{(0,-1)}y^{-1}+\delta_{(0,1)} y+\delta_{(1,-2)}x y^{-2}+\delta_{(1,-1)}x y^{-1}+\delta_{(1,0)} x+\delta_{(1,1)}x y +H\big]~,~
\nn\\
\eea
where $\overline{p}_0= e_{13}^{+} e_{21}^{+} e_{32}^{+} e_{46}^{+} e_{54}^{+} e_{65}^{+} e_{79}^{+} e_{87}^{+} e_{98}^{+}$.
The Casimirs $\delta_{(m,n)}$ in \eref{es10a07} can be expressed in terms of the
zig-zag paths in \eref{es10a03} as follows, 
\beal{es10a08}
&
\delta_{(-2,1)}=1~,~
\delta_{(-1,0)}=z_2^{-1}+z_6^{-1}+z_8^{-1}~,~
\delta_{(-1,1)}=z_3+z_4+z_9~,~
&
\nn
\\
&
\delta_{(0,-1)}=z_2^{-1} z_6^{-1}+z_2^{-1} z_8^{-1}+z_6^{-1} z_8^{-1}~,~
\delta_{(0,1)}=z_3 z_4+z_3 z_9+z_4 z_9~,~
&
\nn
\\
&
\delta_{(1,-2)}= z_1 z_3 z_4 z_5 z_7 z_9~,~
\delta_{(1,-1)}=z_3 z_4 z_9 (z_1 z_5+z_1 z_7+z_5 z_7)~,~
&
\nn
\\
&
\delta_{(1,0)}=z_3 z_4 z_9 (z_1+z_5+z_7)~,~
\delta_{(1,1)}=z_3 z_4 z_9~.~
&
\eea
This leads to the following form of the spectral curve for Model 1, 
\beal{es10a09}
&&
\Sigma~:~
\frac{y}{x^2}+\Big[\frac{1}{z_2}+\frac{1}{z_6}+\frac{1}{z_8}+(z_3+z_4+z_9)y \Big]\frac{1}{x}+\Big(\frac{1}{z_2 z_6}+\frac{1}{z_2 z_8}+\frac{1}{z_6 z_8} \Big)\frac{1}{y}
\nn\\&&\hspace{1cm}
+(z_3 z_4 + z_3 z_9 + z_4 z_9)y+z_3 z_4 z_9 \Big[z_1 z_5 z_7 \frac{1}{y^2} + (z_1 z_5+z_1 z_7+z_5 z_7) \frac{1}{y}
\nn\\&&\hspace{1cm}
+(z_1+z_5+z_7)+y\Big]x+H 
= 0 
~.~
\eea

The Hamiltonian is a sum over all 21 1-loops $\gamma_i$,
\beal{es10a10}
H=\sum_{i=1}^{21} \gamma_i~,~
\eea
where the 1-loops $\gamma_i$ can be expressed in terms of zig-zag paths and face paths as follows, 
\beal{es10a11}
&
\gamma_1 = z_1 z_4 z_9 f_8 f_9~,~
\gamma_2 = z_1 z_4 z_9 f_8 ~,~ 
\gamma_3 = z_8^{-1} z_9 f_9^{-1} ~,~
\gamma_4 =  z_1^{-1} z_6^{-1} z_8^{-1} f_8^{-1} f_9^{-1} ~,~
&
\nn
\\
&
\gamma_5 =z_8^{-1} z_9 f_8 ~,~ 
\gamma_6 =z_8^{-1} z_9 f_8 f_9^{-1}  ~,~
\gamma_7 = z_1^{-1} z_6^{-1} z_8^{-1} f_9^{-1}  ~,~
\gamma_8 = z_1^{-1} z_2^{-1} z_6^{-1} f_8^{-1}~,~ 
&
\nn
\\
&
\gamma_9 =  z_2^{-1} z_4 f_9~,~ 
\gamma_{10} =  z_3 z_7 z_9 f_8^{-1} f_9^{-1} ~,~
\gamma_{11} = z_1 z_3 z_4 f_1 f_4 f_9 ~,~
\gamma_{12} = z_1 z_3 z_4 f_1 f_9~,~ 
&
\nn
\\
&
\gamma_{13} =  z_1 z_3 z_9 f_1 f_8~,~ 
\gamma_{14} =  z_3 z_6^{-1} f_1 ~,~
\gamma_{15} = z_2^{-1} z_6^{-1} z_7^{-1} f_9 ~,~
\gamma_{16} = z_4 z_6^{-1} f_2 f_6 f_9~,~ 
&
\nn
\\
&
\gamma_{17} = z_4 z_6^{-1} f_2 f_6~,~ 
\gamma_{18} = z_1^{-1} z_6^{-1} z_8^{-1} f_2 f_6 ~,~
\gamma_{19} = z_1^{-1} z_6^{-1} z_8^{-1} f_2 f_6 f_9^{-1} ~,~
&
\nn
\\
&
\gamma_{20} = z_2^{-1} z_6^{-1} z_7^{-1} f_7 f_9~,~ 
\gamma_{21} = z_1 z_3 z_4 f_9~.~ 
\eea

The commutation matrix $C$ for Model 1 takes the following form, 
\beal{es10a12}
&&
C =
\resizebox{0.75\textwidth}{!}{$
\left(
\begin{array}{c|ccccccccccccccccccccc}
& \gamma_1&\gamma_2&\gamma_3&\gamma_4&\gamma_5&\gamma_6&\gamma_7&\gamma_8&\gamma_9&\gamma_{10}&\gamma_{11}&\gamma_{12}&\gamma_{13}&\gamma_{14}&\gamma_{15}&\gamma_{16}&\gamma_{17}&\gamma_{18}&\gamma_{19}&\gamma_{20}&\gamma_{21} \\ \hline
\gamma_1&0 & -1 & -1 & 0 & -1 & -2 & -1 & 1 & 1 & 0 & 1 & 1 & -1 &0 & 1 & 1 & 0 & 0 & -1 & 1 & 1 \\
\gamma_2&1 & 0 & -1 & -1 & 0 & -1 & -1 & 0 & 1 & -1 & -1 & 0 & -1& -1 & 1 & 2 & 1 & 1 & 0 & 0 & 1 \\
\gamma_3& 1 & 1 & 0 & -1 & 1 & 1 & 0 & -1 & 0 & -1 & -2 & -1 & 0 &-1 & 0 & 1 & 1 & 1 & 1 & -1 & 0 \\
\gamma_4& 0 & 1 & 1 & 0 & 1 & 2 & 1 & -1 & -1 & 0 & -1 & -1 & 1 & 0& -1 & -1 & 0 & 0 & 1 & -1 & -1 \\
\gamma_5& 1 & 0 & -1 & -1 & 0 & -1 & -1 & 0 & 1 & -1 & -1 & 0 & -1& -1 & 1 & 2 & 1 & 1 & 0 & 0 & 1 \\
\gamma_6& 2 & 1 & -1 & -2 & 1 & 0 & -1 & -1 & 1 & -2 & -3 & -1 & -1& -2 & 1 & 3 & 2 & 2 & 1 & -1 & 1 \\
\gamma_7& 1 & 1 & 0 & -1 & 1 & 1 & 0 & -1 & 0 & -1 & -2 & -1 & 0 &-1 & 0 & 1 & 1 & 1 & 1 & -1 & 0 \\
\gamma_8& -1 & 0 & 1 & 1 & 0 & 1 & 1 & 0 & -1 & 1 & 1 & 0 & 1 & 1 &-1 & -2 & -1 & -1 & 0 & 0 & -1 \\
\gamma_9& -1 & -1 & 0 & 1 & -1 & -1 & 0 & 1 & 0 & 1 & 2 & 1 & 0 & 1& 0 & -1 & -1 & -1 & -1 & 1 & 0 \\
\gamma_{10}& 0 & 1 & 1 & 0 & 1 & 2 & 1 & -1 & -1 & 0 & -1 & -1 & 1 & 0& -1 & -1 & 0 & 0 & 1 & -1 & -1 \\
\gamma_{11}& -1 & 1 & 2 & 1 & 1 & 3 & 2 & -1 & -2 & 1 & 0 & -1 & 2 & 1& -2 & -3 & -1 & -1 & 1 & -1 & -2 \\
\gamma_{12}& -1 & 0 & 1 & 1 & 0 & 1 & 1 & 0 & -1 & 1 & 1 & 0 & 1 & 1 &-1 & -2 & -1 & -1 & 0 & 0 & -1 \\
\gamma_{13}& 1 & 1 & 0 & -1 & 1 & 1 & 0 & -1 & 0 & -1 & -2 & -1 & 0 &-1 & 0 & 1 & 1 & 1 & 1 & -1 & 0 \\
\gamma_{14}& 0 & 1 & 1 & 0 & 1 & 2 & 1 & -1 & -1 & 0 & -1 & -1 & 1 & 0& -1 & -1 & 0 & 0 & 1 & -1 & -1 \\
\gamma_{15}& -1 & -1 & 0 & 1 & -1 & -1 & 0 & 1 & 0 & 1 & 2 & 1 & 0 & 1& 0 & -1 & -1 & -1 & -1 & 1 & 0 \\
\gamma_{16}& -1 & -2 & -1 & 1 & -2 & -3 & -1 & 2 & 1 & 1 & 3 & 2 & -1 & 1 & 1 & 0 & -1 & -1 & -2 & 2 & 1 \\
\gamma_{17}& 0 & -1 & -1 & 0 & -1 & -2 & -1 & 1 & 1 & 0 & 1 & 1 & -1 &0 & 1 & 1 & 0 & 0 & -1 & 1 & 1 \\
\gamma_{18}& 0 & -1 & -1 & 0 & -1 & -2 & -1 & 1 & 1 & 0 & 1 & 1 & -1 & 0 & 1 & 1 & 0 & 0 & -1 & 1 & 1 \\
\gamma_{19}& 1 & 0 & -1 & -1 & 0 & -1 & -1 & 0 & 1 & -1 & -1 & 0 & -1& -1 & 1 & 2 & 1 & 1 & 0 & 0 & 1 \\
\gamma_{20}& -1 & 0 & 1 & 1 & 0 & 1 & 1 & 0 & -1 & 1 & 1 & 0 & 1 & 1 &-1 & -2 & -1 & -1 & 0 & 0 & -1 \\
\gamma_{21}& -1 & -1 & 0 & 1 & -1 & -1 & 0 & 1 & 0 & 1 & 2 & 1 & 0 & 1& 0 & -1 & -1 & -1 & -1 & 1 & 0 \\
\end{array}
\right)
$}
.~
\eea
Satisfying the commutation relations given by the commutation matrix above, 
the 1-loops can be written in terms of the canonical variables as follows, 
\beal{es10a13}
&
\gamma_1 = e^{Q+P} z_1 z_4 z_9~,~
\gamma_2 = e^Q z_1 z_4 z_9  ~,~ 
\gamma_3 = e^{-P} z_8^{-1} z_9  ~,~
&
\nn\\
&
\gamma_4 =  e^{-Q-P} z_1^{-1} z_6^{-1} z_8^{-1}  ~,~
\gamma_5 =e^Q z_8^{-1} z_9  ~,~ 
\gamma_6 =e^{Q-P} z_8^{-1} z_9  ~,~
&
\nn\\
&
\gamma_7 = e^{-P} z_1^{-1} z_6^{-1} z_8^{-1}   ~,~
\gamma_8 = e^{-Q} z_1^{-1} z_2^{-1} z_6^{-1} ~,~ 
\gamma_9 =  e^P z_2^{-1} z_4~,~ 
&
\nn\\
&
\gamma_{10} = e^{-Q-P} z_3 z_7 z_9  ~,~
\gamma_{11} = e^{-2Q-P} z_1^{-1} z_3 z_6^{-1} z_7  ~,~
\gamma_{12} = e^{-Q} z_3 z_4 z_7~,~ 
&
\nn\\
&
\gamma_{13} = e^{-P} z_3 z_7 z_9 ~,~ 
\gamma_{14} =  e^{-Q-P} z_1^{-1} z_3 z_6^{-1} z_7  ~,~
\gamma_{15} = e^P z_2^{-1} z_6^{-1} z_7^{-1}  ~,~
&
\nn\\
&
\gamma_{16} = e^{Q+2P} z_1 z_2^{-1} z_4 z_7^{-1} ~,~ 
\gamma_{17} =e^{Q+P} z_1 z_2^{-1} z_4 z_7^{-1} ~,~ 
\gamma_{18} = e^{Q+P} z_2^{-1} z_7^{-1} z_8^{-1}~,~
&
\nn\\
&
\gamma_{19} =e^{Q} z_2^{-1} z_7^{-1} z_8^{-1} ~,~
\gamma_{20} = e^{-Q} z_3 z_6^{-1} ~,~ 
\gamma_{21} = e^{P} z_1 z_3 z_4 ~.~ 
&
\eea
\\

%=================================================================
\section{Model 2: $\mathbb{C}^3/\mathbb{Z}_4 \times \mathbb{Z}_2$ $(1,0,3)(0,1,1)$}
%=================================================================

%------------------------------------------------------------------------------------------------------------------
\begin{figure}[H]
\begin{center}
\resizebox{0.9\hsize}{!}{
\includegraphics{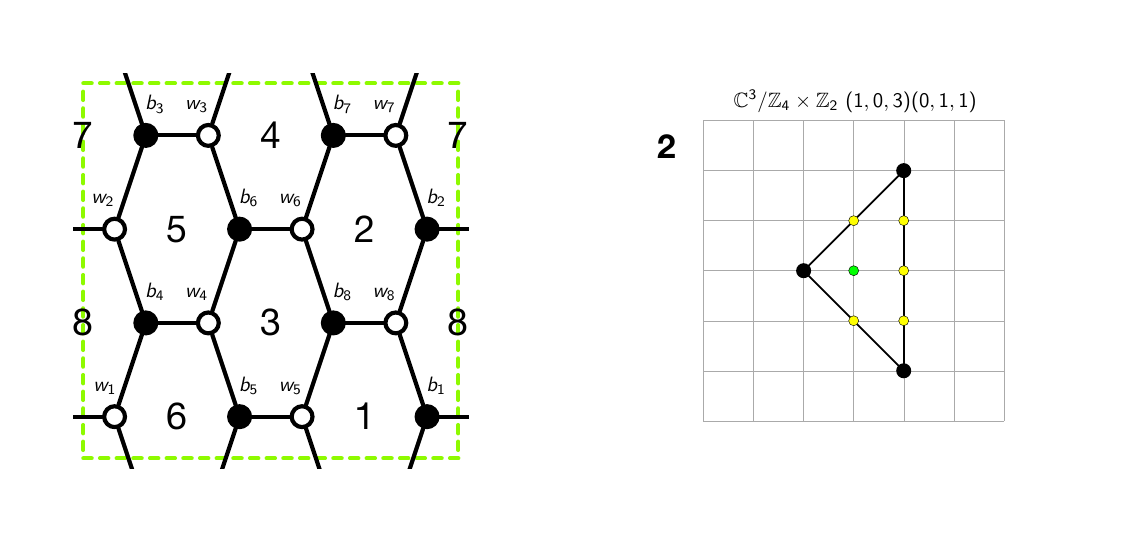}
}
\vspace{-0.5cm}
\caption{The brane tiling and toric diagram of Model 2.}
\label{mf_02}
 \end{center}
 \end{figure}
%------------------------------------------------------------------------------------------------------------------

The brane tiling for Model 2 can be expressed in terms of the following pair of permutation tuples
\beal{es11a01}
\sigma_B &=& (e_{11}\ e_{81}\ e_{71})\ (e_{22}\ e_{72}\ e_{82})\ (e_{13}\ e_{23}\ e_{33})\ (e_{14}\ e_{44}\ e_{24})\  (e_{35}\ e_{55}\ e_{45})
\nn\\&&
(e_{36}\ e_{46}\ e_{66})\ (e_{57}\ e_{67}\ e_{77})\ (e_{58}\ e_{88}\ e_{68})\ ,
\nn\\
\sigma_W^{-1} &=& (e_{11} \ e_{14} \ e_{13}) \ (e_{22} \ e_{23} \ e_{24}) \ (e_{33} \ e_{35} \ e_{36}) \ (e_{44} \ e_{46} \ e_{45}) \ (e_{55} \ e_{58} \ e_{57}) 
\nn\\&&
(e_{66} \ e_{67} \ e_{68}) \ (e_{71} \ e_{72} \ e_{77}) \ (e_{81} \ e_{88}\ e_{82} )\ , 
\eea
which correspond to 
black and white nodes of the brane tiling, respectively.\\

The brane tiling for Model 2 has  8 zig-zag paths given by,
\beal{es11a03}
&
z_1 = (e_{14}^{+}~e_{44}^{-}~e_{46}^{+}~e_{66}^{-}~e_{67}^{+}~e_{77}^{-}~e_{71}^{+}~e_{11}^{-})~,~
&
\nn\\
&
z_2 = (e_{23}^{+}~e_{33}^{-}~e_{35}^{+}~e_{55}^{-}~e_{58}^{+}~e_{88}^{-}~e_{82}^{+}~e_{22}^{-})~,~
&
\nn\\
&
z_3 = (e_{11}^{+}~e_{81}^{-}~e_{88}^{+}~e_{68}^{-}~e_{66}^{+}~e_{36}^{-}~e_{33}^{+}~e_{13}^{-} )~,~
&
\nn\\
&
z_4 = (e_{22}^{+}~e_{72}^{-}~e_{77}^{+}~e_{57}^{-}~e_{55}^{+}~e_{45}^{-}~e_{44}^{+}~e_{24}^{-} )~,~
&
\nn\\
&
z_5 = (e_{13}^{+}~e_{23}^{-}~e_{24}^{+}~e_{14}^{-})~,~
z_6 = (e_{36}^{+}~e_{46}^{-}~e_{45}^{+}~e_{35}^{-})~,~
&
\nn\\
&
z_7 = (e_{57}^{+}~e_{67}^{-}~e_{68}^{+}~e_{58}^{-})~,~
z_8 = (e_{72}^{+}~e_{82}^{-}~e_{81}^{+}~e_{71}^{-})~,~
&
\eea
and face paths given by, 
\beal{es11a04}
&
f_1 = (e_{57}^{+}~e_{77}^{-}~e_{71}^{+}~e_{81}^{-}~e_{88}^{+}~e_{58}^{-})~,~
f_2 = (e_{68}^{+}~e_{88}^{-}~e_{82}^{+}~e_{72}^{-}~e_{77}^{+}~e_{67}^{-})~,~
&
\nn\\
&
f_3 = (e_{45}^{+}~e_{55}^{-}~e_{58}^{+}~e_{68}^{-}~e_{66}^{+}~e_{46}^{-})~,~
f_4 = (e_{36}^{+}~e_{66}^{-}~e_{67}^{+}~e_{57}^{-}~e_{55}^{+}~e_{35}^{-})~,~
&
\nn\\
&
f_5 = (e_{24}^{+}~e_{44}^{-}~e_{46}^{+}~e_{36}^{-}~e_{33}^{+}~e_{23}^{-})~,~
f_6 = (e_{13}^{+}~e_{33}^{-}~e_{35}^{+}~e_{45}^{-}~e_{44}^{+}~e_{14}^{-})~,~
&
\nn\\
&
f_7 = (e_{11}^{+}~e_{71}^{-}~e_{72}^{+}~e_{22}^{-}~e_{23}^{+}~e_{13}^{-})~,~
f_8 = (e_{22}^{+}~e_{82}^{-}~e_{81}^{+}~e_{11}^{-}~e_{14}^{+}~e_{24}^{-})~,~
&
\eea
which satisfy the following constraints, 
\beal{es11a05}
&
f_1 f_6^{-1} = z_1 z_3 z_6 z_7,~
f_2 f_5^{-1} = z_2 z_4 z_6 z_7 ~,~
f_3 f_8^{-1} = z_2 z_3 z_5 z_6~,~ 
&
\nn\\
&
f_4 f_7^{-1} =z_1 z_4 z_5 z_6~,~ 
f_5 f_6 = z_5 z_6^{-1}~,~
f_7 f_8 = z_5^{-1} z_8~,~ 
f_1 f_2 f_3 f_4 f_5 f_6 f_7 f_8 =1~.~ 
&
\eea
The face paths can be written in terms of the canonical variables as follows, 
\beal{es11a05_1}
&
f_1= e^{-P} z_1 z_3 z_5 z_7~,~ 
f_2= e^{P}  z_2 z_4 z_6 z_7~,~ 
f_3= e^{-Q} z_2 z_3 z_6 z_8~,~ 
&
\nn\\
&
f_4= e^{Q} z_1 z_4 z_5 z_6~,~
f_5= e^{P} ~,~ 
f_6= e^{-P} z_5 z_6^{-1}~,~ 
f_7= e^{Q} ~,~ 
f_8=e^{-Q} z_5^{-1} z_8~.~  
&
\eea

The Kasteleyn matrix of the brane tiling for Model 2 in \fref{mf_02} is 
given by, 
\beal{es11a06}
K = 
\left(
\ba{c|cccccccc}
\; & b_1 & b_2 & b_3 & b_4 & b_5 & b_6 & b_7 & b_8
\\
\hline
w_1    &   e_{11} x^{-1} & 0 & e_{13} y^{-1} & e_{14} & 0 & 0 & 0 & 0 
\\
w_2.   &  0 & e_{22} x^{-1} & e_{23} & e_{24} & 0 & 0 & 0 & 0 
\\
w_3    &  0 & 0 & e_{33} & 0 & e_{35} y & e_{36} & 0 & 0 
\\
w_4    &  0 & 0 & 0 & e_{44} & e_{45} & e_{46} & 0 & 0 
\\
w_5    &  0 & 0 & 0 & 0 & e_{55} & 0 & e_{57} y^{-1} & e_{58}
\\
w_6    &  0 & 0 & 0 & 0 & 0 & e_{66} & e_{67} & e_{68} 
\\
w_7    &  e_{71} y & e_{72} & 0 & 0 & 0 & 0 & e_{77} & 0 
\\
w_{8} &  e_{81} & e_{82} & 0 & 0 & 0 & 0 & 0 & e_{88} 
\\
\ea
\right)
~.~
\eea
The permanent of the Kasteleyn matrix gives the spectral curve of 
the dimer integrable system for Model 2, given by
\beal{es11a07}
&&
0 = \text{perm}~K=\overline{p}_0
\cdot x^{-1} \cdot \Big[
\delta_{(-1,0)} \frac{1}{x} + \delta_{(0,-1)} \frac{1}{y}+ \delta_{(0,1)} y+ \delta_{(1,-2)} \frac{x}{y^2}
\nn\\&& \hspace{1cm}
+ \delta_{(1,-1)} \frac{x}{y}+ \delta_{(1,0)} x+ \delta_{(1,1)} x y+ \delta_{(1,2)} x y^2+ H\Big]
\eea
where $\overline{p}_0= e_{11}^{+} e_{22}^{+} e_{33}^{+} e_{44}^{+} e_{55}^{+} e_{66}^{+} e_{77}^{+} e_{88}^{+}$.
The Casimirs $\delta_{(m,n)}$ in \eref{es11a07} can be expressed in terms of
the zig-zag paths in \eref{es11a03} as follows, 
\beal{es11a08}
&
\delta_{(-1,0)} = 1 ~,~
\delta_{(0,-1)} = z_3^{-1} + z_4^{-1} ~,~ 
\delta_{(0,1)}= z_1 + z_2~.~
&
\nn\\
&
\delta_{(1,-2)} = z_3^{-1} z_4^{-1} ~,~
\delta_{(1,-1)} =z_3^{-1} z_4^{-1} (z_5^{-1} + z_6^{-1} + z_7^{-1} + z_8^{-1})~,~
&
\nn\\
&
\delta_{(1,0)}= z_1 z_2 (z_5 z_6 + z_5 z_7 + z_5 z_8 + z_6 z_7 + z_6 z_8 + z_7 z_8)~,~
&
\nn\\
&
\delta_{(1,1)}=z_1 z_2 (z_5 + z_6 + z_7 + z_8)~,~
\delta_{(1,2)} = z_1 z_2 ~.~
&
\eea
This allows us to express the spectral curve for Model 2 as follows, 
\beal{es11a09}
&&
\Sigma~:~ \Big(\frac{z_5}{y}  + 1\Big)\Big(\frac{z_6}{y}  + 1\Big) \Big(\frac{z_7}{y}  + 1\Big) \Big(\frac{z_8}{y}  + 1\Big) z_1 z_2 x y^2
+ \Big(\frac{1}{z_3} + \frac{1}{z_4}\Big) \frac{1}{y} 
\nn\\
&&
\hspace{1cm} + (z_1 + z_2) y+ \frac{1}{x} + H  =0 
~.~
\eea

The Hamiltonian is 
given by the sum over all 12 1-loops  $\gamma_i$,
\beal{es11a10}
H=\sum_{i=1}^{12} \gamma_i~,~
\eea
where the 1-loops $\gamma_i$
can be expressed in terms a combination of zig-zag paths and face paths as shown below, 
\beal{es11a11}
&
\gamma_1 = z_1 z_8 f_2 ~,~
\gamma_2 = z_1 z_8 f_2 f_3~,~
\gamma_3 = z_1 z_8  f_1 f_2 f_3~,~
\gamma_4 = z_1 z_8 f_2 f_3 f_6~,~
&
\nn\\
&
\gamma_5 = z_1 z_8 f_1 f_2 f_3 f_6~,~
\gamma_6 = z_1 z_8 f_5^{-1} f_7^{-1} f_8^{-1}~,~
\gamma_7 = z_1 z_8 f_4^{-1} f_5^{-1} f_8^{-1}~,~
&
\nn\\
&
\gamma_8 = z_1 z_8 f_5^{-1} f_8^{-1}~,~
\gamma_9 = z_1 z_8 f_2 f_5^{-1} f_8^{-1}~,~
\gamma_{10} = z_1 z_8 f_8^{-1}~,~
&
\nn\\
&
\gamma_{11} = z_1 z_8 f_2 f_8^{-1}~,~
\gamma_{12} = z_1 z_8 f_2 f_3 f_8^{-1}~.~
\eea

The commutation matrix $C$ for 
Model 2 is given by, 
\beal{es11a12}
&&
C=
\left(
\ba{c|cccccccccccc}
\; & \gamma_1
& \gamma_2
& \gamma_3
& \gamma_4 
& \gamma_5  
& \gamma_6 
& \gamma_7 
& \gamma_8 
& \gamma_9 
& \gamma_{10} 
& \gamma_{11} 
& \gamma_{12}
\\
\hline
\gamma_1 &  0 & 1 & 1 & 1 & 1 & 0 & 0 & -1 & -1 & -1 & -1 & 0 \\
\gamma_2 & -1 & 0 & 1 & 1 & 2 & 1 & 1 & 0 & -1 & -1 & -2 & -1 \\  
\gamma_3 & -1 & -1 & 0 & 0 & 1 & 1 & 1 & 1 & 0 & 0 & -1 & -1 \\
\gamma_4 & -1 & -1 & 0 & 0 & 1 & 1 & 1 & 1 & 0 & 0 & -1 & -1 \\
\gamma_5 & -1 & -2 & -1 & -1 & 0 & 1 & 1 & 2 & 1 & 1 & 0 & -1 \\
\gamma_6 & 0 & -1 & -1 & -1 & -1 & 0 & 0 & 1 & 1 & 1 & 1 & 0 \\
\gamma_7 & 0 & -1 & -1 & -1 & -1 & 0 & 0 & 1 & 1 & 1 & 1 & 0 \\
\gamma_8 & 1 & 0 & -1 & -1 & -2 & -1 & -1 & 0 & 1 & 1 & 2 & 1 \\
\gamma_9 & 1 & 1 & 0 & 0 & -1 & -1 & -1 & -1 & 0 & 0 & 1 & 1 \\
\gamma_{10} & 1 & 1 & 0 & 0 & -1 & -1 & -1 & -1 & 0 & 0 & 1 & 1\\
\gamma_{11} & 1 & 2 & 1 & 1 & 0 & -1 & -1 & -2 & -1 & -1 & 0 & 1 \\
\gamma_{12} & 0 & 1 & 1 & 1 & 1 & 0 & 0 & -1 & -1 & -1 & -1 & 0
\ea
\right)
~.~
\eea
The commutation relations are in terms of the 1 -loops, which
can be written in terms of the canonical variables as follows,
\beal{es11a13}
&
\gamma_1 = e^{P} z_3^{-1} z_5^{-1}~,~
\gamma_2 = e^{-Q+P} z_2 z_5^{-1} z_6 z_8~,~
\gamma_3 = e^{-Q} z_4^{-1} z_5^{-1}~,~
\gamma_4 =  e^{-Q} z_2 z_8 ~,~
&
\nn\\
&
\gamma_5 =e^{-Q-P} z_4^{-1} z_6^{-1}~,~
\gamma_6 =e^{-P} z_1 z_5~,~
\gamma_7 = e^{-P} z_4^{-1} z_6^{-1}~,~
\gamma_8 = e^{Q-P} z_1 z_5~,~ 
&
\nn\\
&
\gamma_9 =  e^{Q} z_3^{-1} z_8^{-1}~,~ 
\gamma_{10} = e^{Q}  z_1 z_5 ~,~
\gamma_{11} = e^{Q+P} z_3^{-1} z_8^{-1}~,~ 
\gamma_{12} = e^{P} z_2 z_6~.~
&
\eea
\\

%=================================================================
\section{Model 3: $L_{1,3,1}/\mathbb{Z}_2$ $(0,1,1,1)$}
%=================================================================

%=================================================================
\subsection{Model 3a}
%=================================================================

%------------------------------------------------------------------------------------------------------------------
\begin{figure}[H]
\begin{center}
\resizebox{0.9\hsize}{!}{
\includegraphics{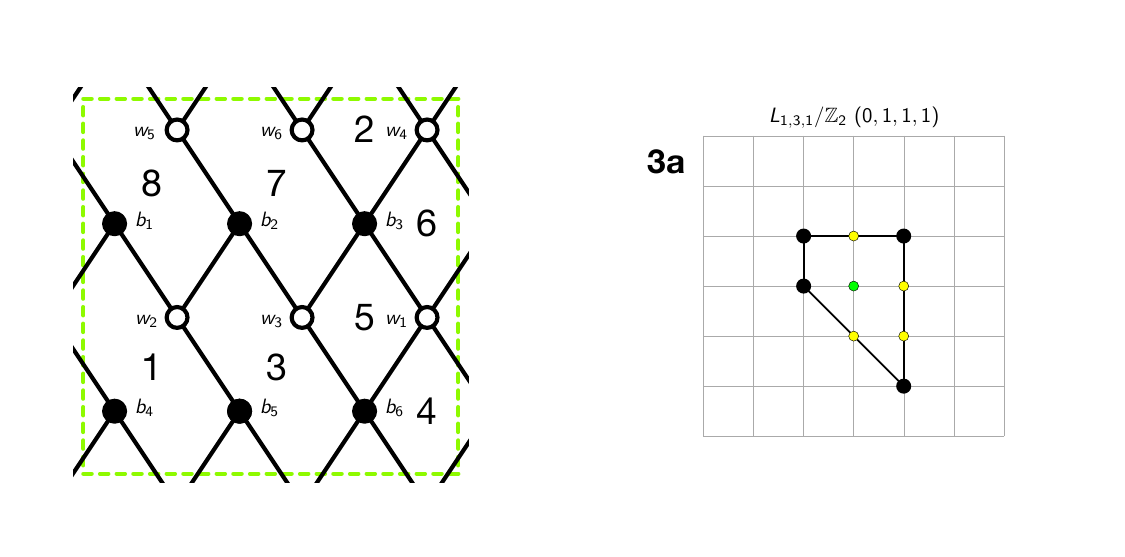}
}
\vspace{-0.5cm}
\caption{The brane tiling and toric diagram of Model 3a.}
\label{mf_03A}
 \end{center}
 \end{figure}
%------------------------------------------------------------------------------------------------------------------

The brane tiling for Model 3a can be expressed in terms of the following pair of permutation tuples
\beal{es12a01}
\sigma_B &=& (e_{11} \ e_{21} \ e_{41}) \ (e_{22} \ e_{32} \ e_{52}) \ (e_{13} \ e_{43} \ e_{63} \ e_{33} ) \ (e_{14} \ e_{44} \ e_{54}) 
\nn\\
&&
(e_{25} \ e_{55} \ e_{65}) \ (e_{16} \ e_{36} \ e_{66} \ e_{46} )  ~,~ 
\nn\\
\sigma_W^{-1} &=& (e_{11} \ e_{14} \ e_{16} \ e_{13})\ (e_{21} \ e_{22} \ e_{25}) \ (e_{32} \ e_{33} \ e_{36}) \ (e_{41} \ e_{43} \ e_{46} \ e_{44})
\nn\\
&&
(e_{52} \ e_{54} \ e_{55}) \ ( e_{63}\ e_{65}\ e_{66} )~,~
\eea
which are associated with black and white nodes in the brane tiling, respectively.\\

The brane tiling for Model 3a has  8 zig-zag paths given by,
\beal{es12a03}
&
z_1 = (e_ {54}^{+}~ e_ {14}^{-}~ e_ {16}^{+}~ e_ {36}^{-}~ e_ {32}^{+}~ e_{52}^{-})~,~
z_2 = (e_ {25}^{+}~ e_ {55}^{-}~ e_ {52}^{+}~ e_ {22}^{-})~,~
&
\nn\\
&
z_3 = (e_ {11}^{+}~ e_ {21}^{-}~ e_ {22}^{+}~ e_ {32}^{-}~ e_ {33}^{+}~ e_{13}^{-})~,~
z_4 = (e_ {55}^{+}~ e_ {65}^{-}~ e_ {66}^{+}~ e_ {46}^{-}~ e_ {44}^{+}~ e_{54}^{-})~,~
&
\nn\\
&
z_5 = (e_ {21}^{+}~ e_ {41}^{-}~ e_ {43}^{+}~ e_ {63}^{-}~ e_ {65}^{+}~ e_{25}^{-})~,~
z_6 = (e_ {63}^{+}~ e_ {33}^{-}~ e_ {36}^{+}~ e_ {66}^{-})~,~
&
\nn\\
&
z_7 = (e_ {46}^{+}~ e_ {16}^{-}~ e_ {13}^{+}~ e_ {43}^{-})~,~
z_8 = (e_ {14}^{+}~ e_ {44}^{-}~ e_ {41}^{+}~ e_ {11}^{-})~,~
&
\eea
and 8 face paths given by, 
\beal{es12a04}
&
f_1 = (e_ {14}^{+}~ e_ {54}^{-}~ e_ {55}^{+}~ e_ {25}^{-}~ e_ {21}^{+}~ e_{11}^{-})~,~
f_2 = (e_ {46}^{+}~ e_ {66}^{-}~ e_ {63}^{+}~ e_ {43}^{-})~,~
&
\nn\\
&
f_3 = (e_ {25}^{+}~ e_ {65}^{-}~ e_ {66}^{+}~ e_ {36}^{-}~ e_ {32}^{+}~ e_{22}^{-})~,~
f_4 = (e_ {16}^{+}~ e_ {46}^{-}~ e_ {44}^{+}~ e_ {14}^{-})~,~
&
\nn\\
&
f_5 = (e_ {36}^{+}~ e_ {16}^{-}~ e_ {13}^{+}~ e_ {33}^{-})~,~
f_6 = (e_ {11}^{+}~ e_ {41}^{-}~ e_ {43}^{+}~ e_ {13}^{-})~,~
&
\nn\\
&
f_7 = (e_ {65}^{+}~ e_ {55}^{-}~ e_ {52}^{+}~ e_ {32}^{-}~ e_ {33}^{+}~ e_{63}^{-})~,~
f_8 = (e_ {54}^{+}~ e_ {44}^{-}~ e_ {41}^{+}~ e_ {21}^{-}~ e_ {22}^{+}~ e_{52}^{-})~,~
&
\eea
which satisfy the following constraints, 
\beal{es12a05}
&
f_1 f_2 ^{-1} = z_4 z_5 z_8~,~
f_2 f_8 = z_2^{-1} z_4^{-1} z_5^{-1}~,~
f_3 f_6^{-1}=z_3^{-1} z_5^{-1} z_6^{-1}~,~
&
\nn\\
& 
f_4  f_7^{-1} = z_1 z_4 z_6~,
f_5 f_8^{-1} = z_1^{-1} z_3^{-1} z_8^{-1}~,~
f_6 f_7 = z_2 z_3 z_5~,~
f_1 f_2 f_3 f_4 f_5 f_6 f_7 f_8 =1~.~ 
&
\eea
The face paths can be written in terms of the canonical variables as shown below, 
\beal{es12a05_1}
&
f_1=e^{Q} z_4 z_5 z_8~,~ 
f_2= e^{Q}~,~ 
f_3= e^{-P} z_2 z_6^{-1}~,~ 
f_4= e^{P} z_1 z_4 z_6 ~,~ 
&
\nn\\
&
f_5= e^{-Q} z_6 z_7 ~,~ 
f_6= e^{-P} z_2 z_3 z_5~,~ 
f_7= e^{P} ~,~ 
f_8=e^{-Q} z_2^{-1} z_4^{-1} z_5^{-1}~.~ 
&
\eea

The Kasteleyn matrix of the brane tiling for Model 3a in \fref{mf_03A} is 
given by, 
\beal{es12a06}
K = 
\left(
\ba{c|cccccc}
\; & b_1 & b_2 & b_3 & b_4 & b_5 & b_6 
\\
\hline
w_1   &   e_{11} x & 0 & e_{13} & e_{14} x & 0 & e_{16}
\\
w_2   &  e_{21} & e_{22} & 0 & 0 & e_{25} & 0
\\
w_3   &  0 & e_{32} & e_{33} & 0 & 0 & e_{36}
\\
w_4   &  e_{41} x & 0 & e_{43} & e_{44} xy & 0 & e_{46} y
\\
w_5   &  0 & e_{52} & 0 & e_{54} y & e_{55} y & 0
\\
w_6   &  0 & 0 & e_{63} & 0 & e_{65}y & e_{66}y
\\
\ea
\right)
~.~
\eea
The permanent of the Kasteleyn matrix 
gives the spectral curve of the 
dimer integrable system for Model 3a as shown below, 
\beal{es12a07}
&&
0 = \text{perm}~K=\overline{p}_0 \cdot xy^2
\cdot \Big[
\delta_{(-1,0)}\frac{1}{x} + \delta_{(-1,1)}\frac{y}{x} + \delta_{(0,-1)}\frac{1}{y} + \delta_{(0,1)}y
\nn\\
&& \hspace{1cm}
+\delta_{(1,-2)}\frac{x}{y^2}+\delta_{(1,-1)}\frac{x}{y}+\delta_{(1,0)} x+\delta_{(1,1)} xy+
 H\Big]
 ~,~
\eea
where $\overline{p}_0= e_{11}^{+} e_{22}^{+} e_{33}^{+} e_{44}^{+} e_{55}^{+} e_{66}^{+}$.
The Casimirs $\delta_{(m,n)}$ in \eref{es12a07} can be expressed in terms of
zig-zag paths in \eref{es12a03} as follows, 
\beal{es12a08}
&
\delta_{(-1,0)} = z_3^{-1} z_4^{-1} z_7^{-1}~,~ 
\delta_{(-1,1)} = z_3^{-1} z_4^{-1} ~,~
\delta_{(0,-1)}= z_2 z_6 z_8 (z_1+z_5)~.~
&
\nn\\
&
\delta_{(0,1)} = z_3^{-1} + z_4^{-1} ~,~
\delta_{(1,-2)} =z_2 z_6 z_8~,~
\delta_{(1,-1)}= z_2 z_6+ z_2 z_8+z_6 z_8~,~
&
\nn\\
&
\delta_{(1,0)}=z_2+z_6+z_8~,~
\delta_{(1,1)} = 1 ~.~
&
\eea
Accordingly, we can express the spectral curve for Model 3a
as follows, 
\beal{es12a09}
&&
\Sigma~:~
\Big(\frac{1}{z_3}+\frac{1}{z_4} \Big) y+\frac{(1+z_7 y)}{z_3 z_4 z_7} \frac{1}{x}+(z_1+z_5) z_2 z_6 z_8 \frac{1}{y}
\nn\\
&&
\hspace{1cm}
+\Big(1+\frac{z_2}{y}\Big) \Big(1+\frac{z_6}{y}\Big) \Big(1+\frac{z_8}{y}\Big)xy+H
= 0 
~.~
\eea

The Hamiltonian is a sum over all 12 1-loops $\gamma_i$
given by, 
\beal{es12a10}
H=\sum_{i=1}^{12} \gamma_i~,~
\eea
where the 1-loops $\gamma_i$
can be expressed in terms  of zig-zag paths and face paths as follows,
\beal{es12a11}
&
\gamma_1 = z_3^{-1} z_8 f_6 ~,~
\gamma_2 = z_2 z_4^{-1} f_1 f_7^{-1}~,~
\gamma_3 = z_1 z_2 z_6 f_1~,~
&
\nn\\
&
\gamma_4 = z_2 z_4^{-1} f_1~,~
\gamma_5 = z_2 z_4^{-1} f_1 f_4~,~
\gamma_6 = z_1^{-1} z_2 z_3^{-1} z_4^{-1} z_8^{-1} f_1 f_4 f_8~,~
&
\nn\\
&
\gamma_7 = z_1 z_2 z_6 f_1 f_7 f_8~,~
\gamma_8 = z_4^{-1} z_8 f_4 f_5~,~
\gamma_9 = z_2 z_3 z_4^{-1} z_5 z_8 f_4 f_5 f_7^{-1}~,~
&
\nn\\
&
\gamma_{10} = z_1 z_2 z_8 f_5 ~,~
\gamma_{11} = z_2 z_4^{-1} z_6^{-1} z_7^{-1} f_5 f_7^{-1}~,~
\gamma_{12} = z_4^{-1} z_6 f_3~.~
\eea

The commutation matrix $C$ for Model 3a takes the following form,
\beal{es12a12}
&&
C=
\left(
\ba{c|cccccccccccc}
\; & \gamma_1
& \gamma_2
& \gamma_3
& \gamma_4 
& \gamma_5  
& \gamma_6 
& \gamma_7 
& \gamma_8 
& \gamma_9 
& \gamma_{10} 
& \gamma_{11} 
& \gamma_{12}
\\
\hline
\gamma_1      & 0 & 1 & 1 & 1 & 1 & 0 & 0 & -1 & -1 & -1 & -1 & 0 \\
\gamma_2      & -1 & 0 & 1 & 1 & 2 & 1 & 1 & 0 & -1 & -1 & -2 & -1 \\
\gamma_3      & -1 & -1 & 0 & 0 & 1 & 1 & 1 & 1 & 0 & 0 & -1 & -1 \\
\gamma_4      & -1 & -1 & 0 & 0 & 1 & 1 & 1 & 1 & 0 & 0 & -1 & -1 \\
\gamma_5      & -1 & -2 & -1 & -1 & 0 & 1 & 1 & 2 & 1 & 1 & 0 & -1 \\
\gamma_6      & 0 & -1 & -1 & -1 & -1 & 0 & 0 & 1 & 1 & 1 & 1 & 0 \\
\gamma_7      & 0 & -1 & -1 & -1 & -1 & 0 & 0 & 1 & 1 & 1 & 1 & 0 \\
\gamma_8      & 1 & 0 & -1 & -1 & -2 & -1 & -1 & 0 & 1 & 1 & 2 & 1 \\
\gamma_9      & 1 & 1 & 0 & 0 & -1 & -1 & -1 & -1 & 0 & 0 & 1 & 1 \\
\gamma_{10} & 1 & 1 & 0 & 0 & -1 & -1 & -1 & -1 & 0 & 0 & 1 & 1 \\
\gamma_{11} &  1 & 2 & 1 & 1 & 0 & -1 & -1 & -2 & -1 & -1 & 0 & 1 \\
\gamma_{12} & 0 & 1 & 1 & 1 & 1 & 0 & 0 & -1 & -1 & -1 & -1 & 0 \\
\ea
\right)
~.~
\eea
The 1-loops, 
which satisfy the commutation relations given by the above commutation matrix, 
can be written in terms of the canonical variables as follows, 
\beal{es12a13}
&
\gamma_1 = e^{-P} z_2 z_5 z_8~,~
\gamma_2 = e^{Q-P}  z_2 z_5 z_8~,~
\gamma_3 =  e^{Q} z_3^{-1} z_7^{-1}~,~ 
&
\nn\\
&
\gamma_4 = e^{Q} z_2 z_5 z_8~,~
\gamma_5 = e^{Q+P} z_3^{-1} z_7^{-1}~,~
\gamma_6 =e^{P} z_3^{-1} z_6~,~
&
\nn\\
&
\gamma_7 = e^{P} z_1 z_6 z_8~,~ 
\gamma_8 =e^{-Q+P} z_1 z_6^2 z_7 z_8~,~ 
\gamma_9 =  e^{-Q} z_4^{-1} z_6~,~
&
\nn\\
&
\gamma_{10} = e^{-Q} z_3^{-1} z_4^{-1} z_5^{-1}~,~
\gamma_{11} = e^{-Q-P} z_2 z_4^{-1}~,~ 
\gamma_{12} = e^{-P} z_2 z_4^{-1}~.~ 
&
\eea
\\

%=================================================================
\subsection{Model 3b}
%=================================================================
%------------------------------------------------------------------------------------------------------------------
\begin{figure}[H]
\begin{center}
\resizebox{0.9\hsize}{!}{
\includegraphics{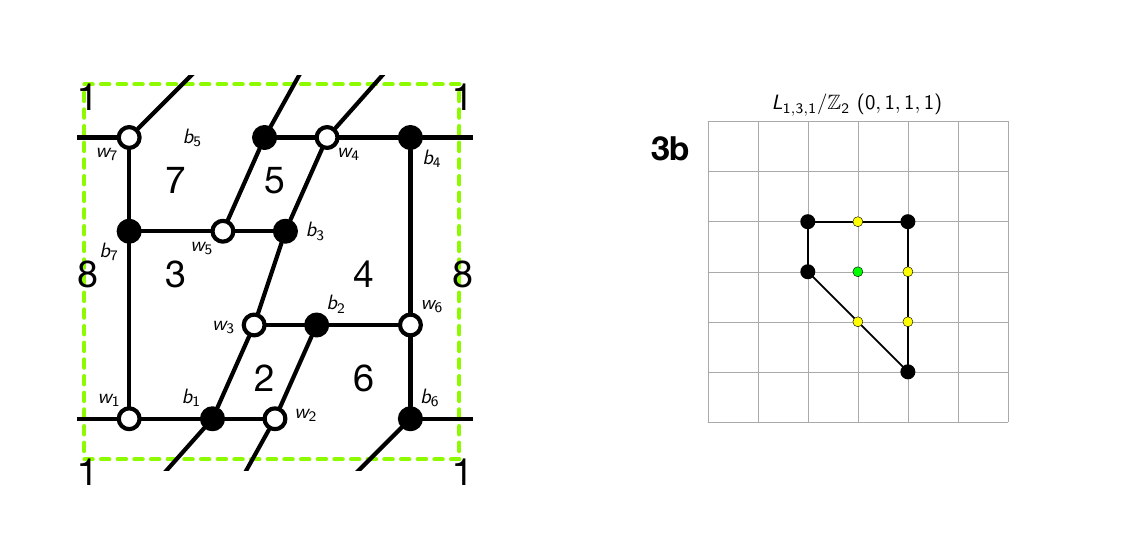}
}
\vspace{-0.5cm}
\caption{The brane tiling and toric diagram of Model 3b.}
\label{mf_03B}
 \end{center}
 \end{figure}
%------------------------------------------------------------------------------------------------------------------

The brane tiling for Model 3b can be expressed in terms of the following pair of permutation tuples
\beal{es12b01}
\sigma_B &=& (e_{11}\ e_{71}\ e_{21}\ e_{31})\ (e_{22}\ e_{62}\ e_{32}) \ (e_{33}\ e_{43}\ e_{53})\ (e_{44}\ e_{64}\ e_{74})
\nn\\
&&
(e_{25}\ e_{55}\ e_{45})\ (e_{16}\ e_{66}\ e_{46})\ (e_{17}\ e_{57}\ e_{77})~,~
\nn\\
\sigma_W^{-1} &=& (e_{11}\ e_{16}\ e_{17})\ (e_{21}\ e_{22}\ e_{25})\ (e_{31}\ e_{33}\ e_{32})\ (e_{43}\ e_{45}\ e_{46}\ e_{44})
\nn\\
&&
(e_{53}\ e_{57}\ e_{55})\ (e_{62}\ e_{64}\ e_{66})\ (e_{71}\ e_{77}\ e_{74})~,~
\eea
which
are associated with black and white nodes in the brane tiling, respectively.\\

The brane tiling for Model 3b has 8 zig-zag paths given by, 
\beal{es12b03}
&
z_1 = (e_ {74}^{+}~ e_ {44}^{-}~ e_ {43}^{+}~ e_ {53}^{-}~ e_ {57}^{+}~ e_{77}^{-})~,~
z_2 = (e_ {11}^{+}~ e_ {71}^{-}~ e_ {77}^{+}~ e_ {17}^{-})~,~
&
\nn\\
&
z_3 = (e_ {46}^{+}~ e_ {16}^{-}~ e_ {17}^{+}~ e_ {57}^{-}~ e_ {55}^{+}~ e_{45}^{-})~,~
z_4 = (e_ {21}^{+}~ e_ {31}^{-}~ e_ {33}^{+}~ e_ {43}^{-}~ e_ {45}^{+}~ e_{25}^{-})~,~
&
\nn\\\
&
z_5 = (e_ {32}^{+}~ e_ {22}^{-}~ e_ {25}^{+}~ e_ {55}^{-}~ e_ {53}^{+}~ e_{33}^{-})~,~
z_6 = (e_ {16}^{+}~ e_ {66}^{-}~ e_ {62}^{+}~ e_ {32}^{-}~ e_ {31}^{+}~ e_{11}^{-})~,~
&
\nn\\
&
z_7 = (e_ {44}^{+}~ e_ {64}^{-}~ e_ {66}^{+}~ e_ {46}^{-})~,~
z_8 = (e_ {71}^{+}~ e_ {21}^{-}~ e_ {22}^{+}~ e_ {62}^{-}~ e_ {64}^{+}~ e_{74}^{-})~,~
&
\eea
and 9 face paths given by, 
\beal{es12b04}
&
f_1 = (e_ {44}^{+}~ e_ {74}^{-}~ e_ {71}^{+}~ e_ {11}^{-}~ e_ {16}^{+}~ e_{46}^{-})~,~
f_2 = (e_ {31}^{+}~ e_ {21}^{-}~ e_ {22}^{+}~ e_ {32}^{-})~,~
&
\nn\\
&
f_3 = (e_ {11}^{+}~ e_ {31}^{-}~ e_ {33}^{+}~ e_ {53}^{-}~ e_ {57}^{+}~ e_{17}^{-})~,~
f_4 = (e_ {32}^{+}~ e_ {62}^{-}~ e_ {64}^{+}~ e_ {44}^{-}~ e_ {43}^{+}~ e_{33}^{-})~,~
&
\nn\\\
&
f_5 = (e_ {53}^{+}~ e_ {43}^{-}~ e_ {45}^{+}~ e_ {55}^{-})~,~
f_6 = (e_ {46}^{+}~ e_ {66}^{-}~ e_ {62}^{+}~ e_ {22}^{-}~ e_ {25}^{+}~ e_{45}^{-})~,~
&
\nn\\\
&
f_7 = (e_ {21}^{+}~ e_ {71}^{-}~ e_ {77}^{+}~ e_ {57}^{-}~ e_ {55}^{+}~ e_{25}^{-})~,~
f_8 = (e_ {74}^{+}~ e_ {64}^{-}~ e_ {66}^{+}~ e_ {16}^{-}~ e_ {17}^{+}~ e_{77}^{-})~,~
&
\eea
which are under the following constraints, 
\beal{es12b05}
&
f_2 f_8^{-1} = z_2 z_6 z_8~,~ 
f_3^{-1} f_6 =  z_3 z_5 z_6~,~ 
f_4 f_7^{-1}= z_1 z_5 z_8~,~
f_5^{-1} f_8 = z_1 z_3 z_7~,~ 
&
\nn\\
&
f_6 f_7 f_8 = z_3 z_8^{-1}~,~
f_1^{-1} f_6 f_7 =  z_2 z_3 z_7^{-1} z_8^{-1}~,~
f_1 f_2 f_3 f_4 f_5 f_6 f_7 f_8 =1~.~ 
&
\eea
The face paths 
can be written in terms of the canonical variables as follows, 
\beal{es12b05_1}
&
f_1=e^{-P} z_2^{-1} z_7~,~ 
f_2= e^{P} z_2 z_6 z_8~,~ 
f_3= e^{-Q-P} z_5^{-1} z_6^{-1} z_8^{-1}~,~ 
f_4= e^{Q} z_1 z_5 z_8 ~,~ 
&
\nn\\
&
f_5= e^{P} z_1^{-1} z_3^{-1} z_7^{-1}~,~ 
f_6= e^{-Q-P} z_3 z_8^{-1}~,~ 
f_7= e^{Q} ~,~ 
f_8=e^{P} ~.~ 
&
\eea

The Kasteleyn matrix of the brane tiling for Model 3b in \fref{mf_03B}
takes the following form, 
\beal{es12b06}
K = 
\left(
\ba{c|ccccccc}
\; & b_1 & b_2 & b_3 & b_4 & b_5 & b_6 & b_7
\\
\hline
w_1   &   e_{11}  & 0 & 0 &0& 0 & e_{16} x^{-1} & e_{17}
\\
w_2   &  e_{21} & e_{22} & 0 & 0 & e_{25} y^{-1} & 0 & 0
\\
w_3   &  e_{31} & e_{32} & e_{33} & 0 & 0 & 0 & 0
\\
w_4   &  0 & 0 & e_{43} & e_{44} & e_{45} & e_{46} y & 0
\\
w_5   &  0 & 0 & e_{53} & 0 & e_{55} & 0 & e_{57}
\\
w_6   &  0 & e_{62} & 0 & e_{64} & 0 & e_{66} & 0
\\
w_7 & e_{71} y & 0 & 0 & e_{74} x^{-1} & 0 & 0 & e_{77} 
\ea
\right)
~.~
\eea
By taking the permanent of the Kasteleyn matrix in 
\eref{es12b06} with a $GL(2,\mathbb{Z})$ transformation $M : (x,y) \mapsto (x,\frac{1}{y})$, 
we obtain the following spectral curve of the dimer integrable system for Model 3b, 
\beal{es12b07}
&&
0= \overline{p}_0 \cdot x^{-1}
\cdot \Big[
\delta_{(-1,1)}\frac{y}{x} + \delta_{(-1,0)}\frac{1}{x} + \delta_{(0,-1)}\frac{1}{y} + \delta_{(0,1)}y
\nn\\
&& 
\hspace{2cm}
+\delta_{(1,-2)}\frac{x}{y^2}+\delta_{(1,-1)}\frac{x}{y}+\delta_{(1,0)} x+\delta_{(1,1)} xy+
 H\Big]
 ~,~
\eea
where $\overline{p}_0= e_{17}^{+} e_{22}^{+} e_{33}^{+} e_{46}^{+} e_{55}^{+} e_{64}^{+} e_{71}^{+}$.
The Casimirs $\delta_{(m,n)}$ in \eref{es12b07} can be expressed in terms of
the zig-zag paths in \eref{es12b03} as follows, 
\beal{es12b08}
&
\delta_{(-1,1)} = z_3^{-1} z_4^{-1} z_8^{-1} ~,~
\delta_{(-1,0)} = z_3^{-1} z_8^{-1}~,~ 
\delta_{(0,-1)}= z_3^{-1}+z_8^{-1}~.~
&
\nn\\
&
\delta_{(0,1)} = z_2 z_5 z_7 (z_1+z_6)~,~
\delta_{(1,-2)} =1~,~
\delta_{(1,-1)}= z_2+z_5+z_7~,~
&
\nn\\
&
\delta_{(1,0)}=z_2 z_5+z_2 z_7+z_5 z_7~,~
\delta_{(1,1)} = z_2 z_5 z_7 ~.~
&
\eea
This leads to the following form of the spectral curve for Model 3b,   
\beal{es12b09}
&&
\Sigma~:~
\Big(\frac{1}{y}+z_2\Big)\Big(\frac{1}{y}+z_5\Big)\Big(\frac{1}{y}+z_7\Big) x y+\Big(\frac{1}{z_1 z_4}+\frac{1}{z_4 z_6} +\frac{1}{z_4 x}\Big)\frac{y}{z_3 z_8}
\nn\\
&& 
\hspace{1cm}
+\frac{1}{z_3 y} + \Big(\frac{1}{z_3}+ \frac{x}{y} \Big) \frac{1}{z_8 x} +H
= 0 
~.~
\eea

The Hamiltonian is a sum over all 14 1-loops $\gamma_i$
given by, 
\beal{es12b10}
H=\sum_{i=1}^{14} \gamma_i~,~
\eea
where the 1-loops $\gamma_i$
can be expressed in terms of zig-zag paths and face paths as follows, 
\beal{es12b11}
&
\gamma_1 = z_5 z_6 z_7 f_3 ~,~
\gamma_2 = z_2 z_8^{-1} f_1 ~,~
\gamma_3 = z_7 z_8^{-1} f_2 f_4 f_8^{-1}~,~
\gamma_4 = z_2 z_8^{-1} f_1 f_4~,~
&
\nn\\
&
\gamma_5 = z_3^{-1} z_7 f_4 f_6~,~
\gamma_6 = z_1^{-1} z_3^{-1} z_8^{-1} f_4~,~
\gamma_7 = z_1^{-1} z_2 z_3^{-1} z_6 f_4 f_8~,~
\gamma_8 = z_7 z_8^{-1} f_4~,~
&
\nn\\
&
\gamma_9 = z_4 z_5 z_7 z_8^{-1} f_2 f_4~,~
\gamma_{10} =z_7 z_8^{-1} f_2 f_4~,~
\gamma_{11} = z_2 z_4 z_7 z_8^{-1} f_2 f_4 f_7^{-1}~,~
\gamma_{12} = z_4^{-1} z_8^{-1} f_3 f_5~,~
&
\nn\\
&
\gamma_{13} = z_1^{-1} z_3^{-1} z_8^{-1} f_2 f_4 f_8~,~
\gamma_{14} = z_5 z_8^{-1} f_2~.~
\eea

The commutation matrix $C$ for Model 3b
has the following form, 
\beal{es12b12}
&&
C=
\left(
\ba{c|cccccccccccccc}
\; & \gamma_1
& \gamma_2
& \gamma_3
& \gamma_4 
& \gamma_5  
& \gamma_6 
& \gamma_7 
& \gamma_8 
& \gamma_9 
& \gamma_{10} 
& \gamma_{11} 
& \gamma_{12}
& \gamma_{13}
& \gamma_{14}
\\
\hline
\gamma_1      & 0 & 1 & 1 & 2 & 1 & 1 & 0 & 1 & 0 & 0 & -1 & -1 & -1 & -1 \\
\gamma_2      &  -1 & 0 & 1 & 1 & 0 & 1 & 1 & 1 & 1 & 1 & 0 & -1 & 1 & 0 \\
\gamma_3      &  -1 & -1 & 0 & -1 & -1 & 0 & 1 & 0 & 1 & 1 & 1 & 0 & 2 & 1 \\
\gamma_4      &  -2 & -1 & 1 & 0 & -1 & 1 & 2 & 1 & 2 & 2 & 1 & -1 & 3 & 1 \\
\gamma_5      &  -1 & 0 & 1 & 1 & 0 & 1 & 1 & 1 & 1 & 1 & 0 & -1 & 1 & 0 \\
\gamma_6      &  -1 & -1 & 0 & -1 & -1 & 0 & 1 & 0 & 1 & 1 & 1 & 0 & 2 & 1 \\
\gamma_7      &  0 & -1 & -1 & -2 & -1 & -1 & 0 & -1 & 0 & 0 & 1 & 1 & 1 & 1 \\
\gamma_8      &  -1 & -1 & 0 & -1 & -1 & 0 & 1 & 0 & 1 & 1 & 1 & 0 & 2 & 1 \\
\gamma_9      &  0 & -1 & -1 & -2 & -1 & -1 & 0 & -1 & 0 & 0 & 1 & 1 & 1 & 1 \\
\gamma_{10} &  0 & -1 & -1 & -2 & -1 & -1 & 0 & -1 & 0 & 0 & 1 & 1 & 1 & 1 \\
\gamma_{11} &   1 & 0 & -1 & -1 & 0 & -1 & -1 & -1 & -1 & -1 & 0 & 1 & -1 & 0 \\
\gamma_{12} &  1 & 1 & 0 & 1 & 1 & 0 & -1 & 0 & -1 & -1 & -1 & 0 & -2 & -1 \\
\gamma_{13} &  1 & -1 & -2 & -3 & -1 & -2 & -1 & -2 & -1 & -1 & 1 & 2 & 0 & 1 \\
\gamma_{14} &  1 & 0 & -1 & -1 & 0 & -1 & -1 & -1 & -1 & -1 & 0 & 1 & -1 & 0 \\
\ea
\right)
~.~
\eea
The 1-loops
forming the commutation relations
can be written in terms of the canonical variables as follows, 
\beal{es12b13}
&
\gamma_1 = e^{-Q-P} z_7 z_8^{-1}~,~
\gamma_2 = e^{-P} z_7 z_8^{-1}~,~ 
\gamma_3 = e^{Q} z_3^{-1} z_4^{-1}~,~
\gamma_4 =  e^{Q-P} z_1 z_5 z_7~,~
&
\nn\\
&
\gamma_5 =e^{-P} z_1 z_5 z_7~,~ 
\gamma_6 =e^{Q} z_3^{-1} z_5~,~
\gamma_7 = e^{Q+P} z_2 z_3^{-1} z_5 z_6 z_8~,~
\gamma_8 = e^{Q}  z_1 z_5 z_7 ~,~ 
&
\nn\\
&
\gamma_9 =  e^{Q+P} z_3^{-1} z_5~,~ 
\gamma_{10} = e^{Q+P} z_3^{-1} z_4^{-1}~,~
\gamma_{11} = e^{P}  z_2 z_3^{-1}~,~
\gamma_{12} = e^{-Q}  z_2 z_8^{-1}~,~ 
&
\nn\\
&
\gamma_{13}=  e^{Q+2P} z_2 z_3^{-1} z_5 z_6 z_8~,~ 
\gamma_{14} = e^{P} z_2 z_5 z_6~.~
&
\eea
\\

%=================================================================
\section{Model 4: $\mathcal{C}/\mathbb{Z}_2 \times \mathbb{Z}_2$ $(1,0,0,1)(0,1,1,0)$, $\text{PdP}_5$}
%=================================================================

%=================================================================
\subsection{Model 4a}
%=================================================================
%------------------------------------------------------------------------------------------------------------------
\begin{figure}[H]
\begin{center}
\resizebox{0.9\hsize}{!}{
\includegraphics{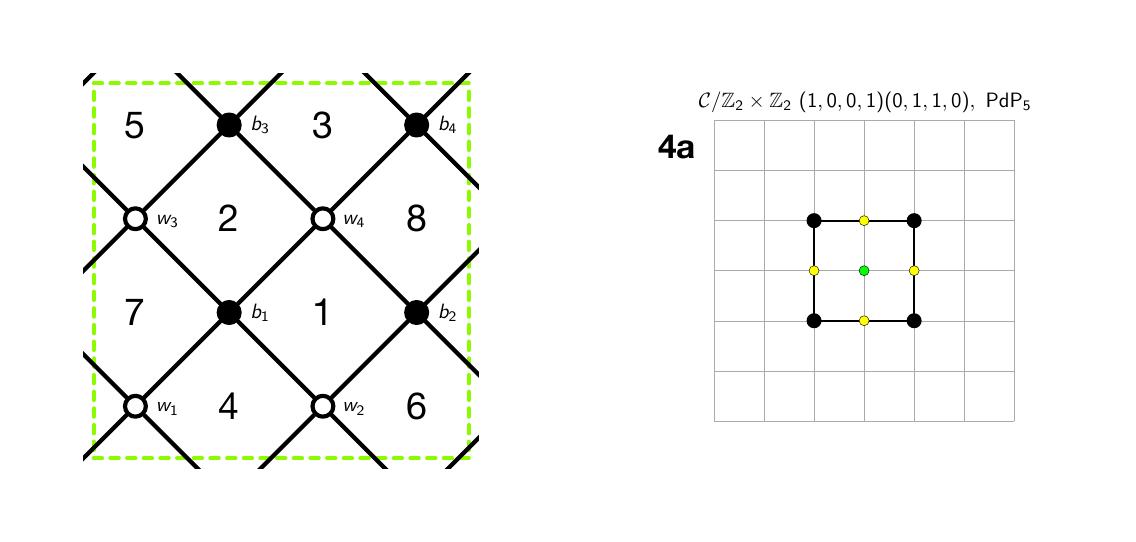}
}
\vspace{-0.5cm}
\caption{The brane tiling and toric diagram of Model 4a.}
\label{mf_04A}
 \end{center}
 \end{figure}
%------------------------------------------------------------------------------------------------------------------

The brane tiling for Model 4a can be expressed in terms of the following pair of permutation tuples
\beal{es13a01}
\sigma_B &=& (e_{11}\ e_{21}\ e_{41}\ e_{31})\ (e_{12}\ e_{32}\ e_{42}\ e_{22})\ (e_{13}\ e_{33}\ e_{43}\ e_{23})\ (e_{14}\ e_{24}\ e_{44}\ e_{34})~,~
\nn\\
\sigma_W^{-1} &=& (e_{11}\ e_{13}\ e_{14}\ e_{12})\ (e_{21}\ e_{22}\ e_{24}\ e_{23})\ (e_{31}\ e_{32}\ e_{34}\ e_{33})\ (e_{41}\ e_{43}\ e_{44}\ e_{42})~,~
\eea
which correspond to the black and white nodes in the brane tiling, respectively.\\

The brane tiling for Model 4a has 8 zig-zag paths
given by, 
\beal{es13a03}
&
z_1 = (e_{11}^{+}~e_{21}^{-}~e_{22}^{+}~e_{12}^{-})~,~
z_2 = (e_{32}^{+}~e_{42}^{-} ~e_{41}^{+}~e_{31}^{-})~,~
&
\nn\\
&
z_3 = (e_{33}^{+}~e_{43}^{-}~e_{44}^{+}~e_{34}^{-})~,~
z_4 = (e_{14}^{+}~e_{24}^{-} ~e_{23}^{+}~e_{13}^{-})~,~
&
\nn\\
&
z_5 = (e_{13}^{+}~e_{33}^{-}~e_{31}^{+}~e_{11}^{-})~,~
z_6 = (e_{21}^{+}~e_{41}^{-}~e_{43}^{+}~e_{23}^{-})~,~
&
\nn\\
&
z_7 = (e_{24}^{+}~e_{44}^{-}~e_{42}^{+}~e_{22}^{-})~,~
z_8 = (e_{12}^{+}~e_{32}^{-}~e_{34}^{+}~e_{14}^{-})~,~
&
\eea
and 8 face paths given by, 
\beal{es13a04}
&
f_1 = (e_{22}^{+}~e_{42}^{-}~e_{41}^{+}~e_{21}^{-})~,~
f_2 = (e_{43}^{+}~e_{33}^{-}~e_{31}^{+}~e_{41}^{-})~,~
&
\nn\\
&
f_3 = (e_{44}^{+}~e_{24}^{-}~e_{23}^{+}~e_{43}^{-})~,~
f_4 = (e_{21}^{+}~e_{11}^{-}~e_{13}^{+}~e_{23}^{-})~,~
&
\nn\\
&
f_5 = (e_{33}^{+}~e_{13}^{-}~e_{14}^{+}~e_{34}^{-})~,~
f_6 = (e_{12}^{+}~e_{22}^{-}~e_{24}^{+}~e_{14}^{-})~,~
&
\nn\\
&
f_7 = (e_{11}^{+}~e_{31}^{-}~e_{32}^{+}~e_{12}^{-})~,~
f_8 = (e_{34}^{+}~e_{44}^{-}~e_{42}^{+}~e_{32}^{-})~,~&
\eea
which satisfy the following constraints, 
\beal{es13a05}
&
f_6 f_8=z_7 z_8~,~
f_5 f_7=z_5^{-1} z_8^{-1} ~,~
f_4 f_8^{-1} =z_2 z_3 z_5 z_6~,~
f_3 f_7^{-1} =z_3 z_4 z_5 z_8~,~
&
\nn\\
&
f_2 f_6^{-1} =z_1 z_4 z_5 z_6~,~
f_1 f_5^{-1} =z_1 z_2 z_5 z_8~,~
f_1 f_2 f_3 f_4 f_5 f_6 f_7 f_8=1 ~.~
&
\eea
The face paths can be written in terms of the canonical variables as follows, 
\beal{es13a05_1}
&
f_1=e^{Q}~,~ 
f_2= e^{P} ~,~ 
f_3= e^{-Q} z_6^{-1} z_7^{-1} ~,~ 
f_4= e^{-P} z_5 z_6~,~ 
&
\nn\\
&
f_5= e^{Q} z_3 z_4 z_6 z_7 ~,~ 
f_6= e^{P} z_2 z_3 z_7 z_8~,~ 
f_7= e^{-Q} z_1 z_2 ~,~ 
f_8=e^{-P} z_2^{-1} z_3^{-1}~.~ 
&
\eea

The Kasteleyn matrix of the brane tiling for Model 4a in \fref{mf_04A} is 
given by, 
\beal{es13a06}
K = 
\left(
\ba{c|cccc}
\; & b_1 & b_2 & b_3 & b_4 
\\
\hline
w_1 & e_{11} & e_{12} x^{-1} & e_{13} y^{-1} & e_{14} x^{-1} y^{-1}
\\
w_2 & e_{21}  & e_{22} & e_{23} y^{-1} & e_{24} y^{-1}
\\
w_3 & e_{31}  & e_{32} x^{-1} & e_{33} & e_{34} x^{-1}
\\
w_4 & e_{41}  & e_{42} & e_{43} & e_{44}
\ea
\right)
~.~
\eea
The permanent of the Kasteleyn matrix in \eref{es13a06}
gives us the spectral curve of the dimer integrable system for Model 4a as follows, 
\beal{es13a07}
0 = \text{perm}~K&=&\overline{p}_0
\cdot  x^{-1} y^{-1} \cdot
\Big[
\delta_{(-1,-1)} \frac{1}{x y} 
+ \delta_{(-1,1)} \frac{y}{x} 
+ \delta_{(1,-1)} \frac{x}{y} 
+ \delta_{(1,1)} xy
\nn\\
&& 
+ \delta_{(-1,0)} \frac{1}{x} 
+ \delta_{(0,-1)} \frac{1}{y} 
+ \delta_{(1,0)} x 
+ \delta_{(0,1)} y
+ H
\Big]
~,~
\eea
where $\overline{p}_0= e_{14}^{+} e_{21}^{+} e_{32}^{+} e_{43}^{+}$.
The Casimirs $\delta_{(m,n)}$ in \eref{es13a07} 
can be expressed in terms of the zig-zag paths in \eref{es13a03} as follows, 
\beal{es13a08}
&
\delta_{(-1,-1)} = z_6^{-1} ~,~
\delta_{(-1,1)} = z_8 ~,~
\delta_{(1,-1)} = z_2^{-1} z_4^{-1} z_6^{-1} ~,~
\delta_{(1,1)} = z_1 z_3 z_8 ~,~
&
\nn\\
&
\delta_{(-1,0)} = 1+z_6^{-1} z_8  ~,~
\delta_{(0,-1)} = z_2^{-1} z_6^{-1}+ z_4^{-1} z_6^{-1}~,~
&
\nn\\
&
\delta_{(1,0)} = z_1 z_3 z_7 z_8+z_1 z_3 z_5 z_8 ~,~
\delta_{(0,1)} = z_1 z_8+z_3 z_8 ~.~
&
\eea
This leads to the following form for the spectral curve of Model 4a, 
\beal{es13a09}
&&
\Sigma~:~ \Big(\frac{1}{y} + z_6\Big) \Big(\frac{1}{y}+z_8\Big)  \frac{y}{z_6  x}
+\Big(\frac{1}{z_2} + \frac{1}{z_4}\Big) \frac{1}{z_6 y}
\nn\\
&&
\hspace{1cm}
+ \Big(\frac{1}{y}+ \frac{1}{z_5}\Big) \Big(\frac{1}{y}+\frac{1}{z_7}\Big) \frac{ x y }{z_2 z_4 z_6 } 
+(z_1 + z_3) z_8 y + H 
= 0 
~.~
\eea

The Hamiltonian is a sum over all 12 1-loops $\gamma_i$,
\beal{es13a10}
H=\sum_{i=1}^{12} \gamma_i~,~
\eea
where the 1-loops $\gamma_i$ can be expressed in terms of zig-zag paths and face paths as follows,
\beal{es13a11}
&
\gamma_1 = z_1 f_6 ~,~
\gamma_2 = z_1 f_3 f_6 ~,~
\gamma_3 = z_1 f_3 f_6 f_8 ~,~
\gamma_4 = z_1 f_3 f_4 f_6 ~,~
&
\nn\\
&
\gamma_5 = z_1 f_3 f_4 f_6 f_8 ~,~
\gamma_6 = z_1 f_1 f_3 f_4 f_6 f_8 ~,~
\gamma_7 = z_1 f_3 f_4 f_5 f_6 f_8 ~,~
&
\nn\\
&
\gamma_8 = z_1 f_1 f_3 f_4 f_5 f_6 f_8 ~,~
\gamma_9 = z_1 f_1 f_3 f_4 f_5 f_6^2 f_8 ~,~
\gamma_{10} = z_1 f_1 f_2 f_3 f_4 f_5 f_6 f_8 ~,~
&
\nn\\
&
\gamma_{11} = z_1 f_1 f_2 f_3 f_4 f_5 f_6^2 f_8 ~,~
\gamma_{12} = z_1 f_1 f_2 f_3^2 f_4 f_5 f_6^2 f_8 ~.~
\eea

The commutation matrix $C$ for Model 4a 
takes the following form,
\beal{es13a12}
&&
C=
\left(
\ba{c|cccccccccccc}
\; & \gamma_1
& \gamma_2
& \gamma_3
& \gamma_4 
& \gamma_5  
& \gamma_6 
& \gamma_7 
& \gamma_8 
& \gamma_9 
& \gamma_{10} 
& \gamma_{11} 
& \gamma_{12}
\\
\hline
\gamma_1 &  0 & 1 & 1 & 1 & 1 & 0 & 0 & -1 & -1 & -1 & -1 & 0 \\
\gamma_2 & -1 & 0 & 1 & 1 & 2 & 1 & 1 & 0 & -1 & -1 & -2 & -1 \\  
\gamma_3 & -1 & -1 & 0 & 0 & 1 & 1 & 1 & 1 & 0 & 0 & -1 & -1 \\
\gamma_4 & -1 & -1 & 0 & 0 & 1 & 1 & 1 & 1 & 0 & 0 & -1 & -1 \\
\gamma_5 & -1 & -2 & -1 & -1 & 0 & 1 & 1 & 2 & 1 & 1 & 0 & -1 \\
\gamma_6 & 0 & -1 & -1 & -1 & -1 & 0 & 0 & 1 & 1 & 1 & 1 & 0 \\
\gamma_7 & 0 & -1 & -1 & -1 & -1 & 0 & 0 & 1 & 1 & 1 & 1 & 0 \\
\gamma_8 & 1 & 0 & -1 & -1 & -2 & -1 & -1 & 0 & 1 & 1 & 2 & 1 \\
\gamma_9 & 1 & 1 & 0 & 0 & -1 & -1 & -1 & -1 & 0 & 0 & 1 & 1 \\
\gamma_{10} & 1 & 1 & 0 & 0 & -1 & -1 & -1 & -1 & 0 & 0 & 1 & 1\\
\gamma_{11} & 1 & 2 & 1 & 1 & 0 & -1 & -1 & -2 & -1 & -1 & 0 & 1 \\
\gamma_{12} & 0 & 1 & 1 & 1 & 1 & 0 & 0 & -1 & -1 & -1 & -1 & 0
\ea
\right)
~,~
\eea
where the 1-loops satisfying the commutation relations
can be written in terms of the canonical variables as follows, 
\beal{es13a13}
&
\gamma_1 = e^{P} z_4^{-1} z_5^{-1} z_6^{-1}~,~
\gamma_2 = e^{-Q+P} z_1 z_2 z_3 z_6^{-1} z_8~,~ 
\gamma_3 = e^{-Q} z_1 z_6^{-1} z_8~,~
&
\nn\\
&
\gamma_4 =  e^{-Q} z_4^{-1} z_6^{-1} z_7^{-1}~,~
\gamma_5 =e^{-Q-P} z_1 z_5 z_8~,~ 
\gamma_6 =e^{-P} z_1 z_5 z_8~,~
&
\nn\\
&
\gamma_7 = e^{-P} z_2^{-1}~,~
\gamma_8 = e^{Q-P}  z_2^{-1} ~,~ 
\gamma_9 =  e^{Q} z_3 z_7 z_8~,~ 
&
\nn\\
&
\gamma_{10} = e^{Q} z_2^{-1}~,~
\gamma_{11} = e^{Q+P}  z_3 z_7 z_8~,~
\gamma_{12} = e^{P}  z_3 z_6^{-1} z_8~.~ 
&
\eea
\\

%=================================================================
\subsection{Model 4b}
%=================================================================
%------------------------------------------------------------------------------------------------------------------
\begin{figure}[H]
\begin{center}
\resizebox{0.9\hsize}{!}{
\includegraphics{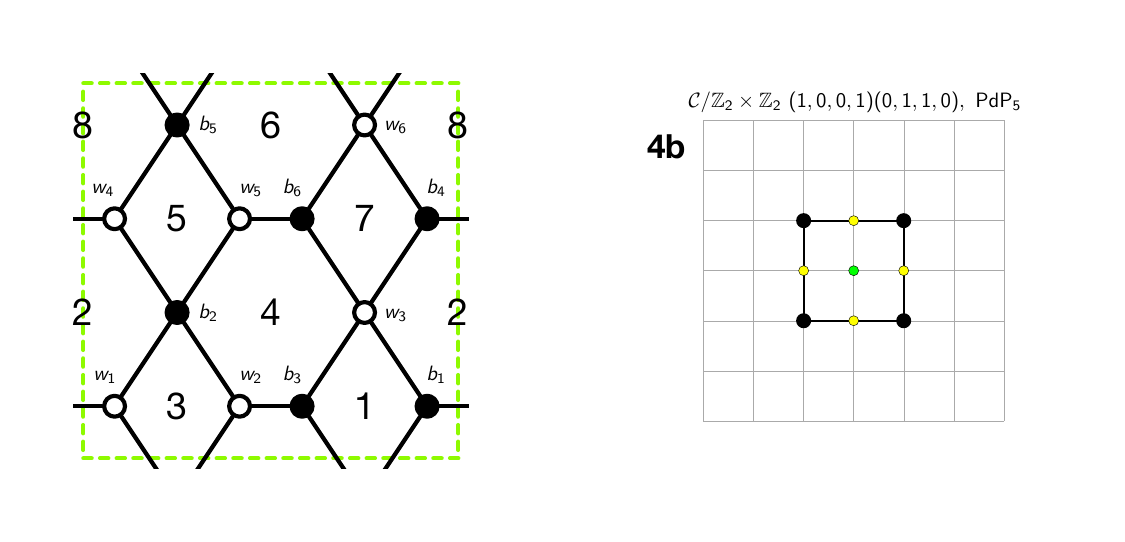}
}
\vspace{-0.5cm}
\caption{The brane tiling and toric diagram of Model 4b.}
\label{mf_04B}
 \end{center}
 \end{figure}
%------------------------------------------------------------------------------------------------------------------

The brane tiling for Model 4b can be expressed in terms of the following pair of permutation tuples
\beal{es13b01}
\sigma_B &=& (e_{11}\ e_{31}\ e_{61})\ (e_{12}\ e_{22}\ e_{52}\ e_{42})\ (e_{23}\ e_{63}\ e_{33})\ (e_{34}\ e_{44}\ e_{64})
\nn\\
&& (e_{15}\ e_{45}\ e_{55}\ e_{25})\ (e_{36}\ e_{66}\ e_{56})~,~
\nn\\
\sigma_W^{-1} &=& (e_{12}\ e_{15}\ e_{11})\ (e_{22}\ e_{23}\ e_{25})\ (e_{31}\ e_{33}\ e_{36}\ e_{34})\ (e_{42}\ e_{44}\ e_{45})
\nn\\
&& (e_{52}\ e_{55}\ e_{56})\ (e_{61}\ e_{64}\ e_{66}\ e_{63})~.~
\eea
The above permutation tuples correspond to black and white nodes in the brane tiling, respectively.\\
 
The brane tiling for Model 4b has 8 zig-zag paths given by, 
\beal{es13b03}
&
z_1 = (e_ {63}^{+}~ e_ {33}^{-}~ e_ {36}^{+}~ e_ {66}^{-})~,~
z_2 = (e_ {31}^{+}~ e_ {61}^{-}~ e_ {64}^{+}~ e_ {34}^{-})~,~
&
\nn\\
&
z_3 = (e_ {33}^{+}~ e_ {23}^{-}~ e_ {25}^{+}~ e_ {15}^{-}~ e_ {11}^{+}~ e_{31}^{-})~,~
z_4 = (e_ {42}^{+}~ e_ {12}^{-}~ e_ {15}^{+}~ e_ {45}^{-})~,~
&
\nn\\
&
z_5 = (e_ {34}^{+}~ e_ {44}^{-}~ e_ {45}^{+}~ e_ {55}^{-}~ e_ {56}^{+}~ e_{36}^{-})~,~
z_6 = (e_ {61}^{+}~ e_ {11}^{-}~ e_ {12}^{+}~ e_ {22}^{-}~ e_ {23}^{+}~ e_{63}^{-})~,~
&
\nn\\
&
z_7 = (e_ {22}^{+}~ e_ {52}^{-}~ e_ {55}^{+}~ e_ {25}^{-})~,~
z_8 = (e_ {44}^{+}~ e_ {64}^{-}~ e_ {66}^{+}~ e_ {56}^{-}~ e_ {52}^{+}~ e_{42}^{-})~,~
&
\eea
and 8 face paths given by, 
\beal{es13b04}
&
f_1 = (e_{33}^{+}~ e_{63}^{-}~ e_{61}^{+}~ e_{31}^{-})~,~
f_2 = (e_{31}^{+}~ e_{11}^{-}~ e_{12}^{+}~ e_{42}^{-}~ e_{44}^{+}~ e_{34}^{-})~,~
&
\nn\\
&
f_3 = (e_{22}^{+}~ e_{12}^{-}~ e_{15}^{+}~ e_{25}^{-})~,~
f_4 = (e_{52}^{+}~ e_{22}^{-}~ e_{23}^{+}~ e_{33}^{-}~ e_{36}^{+}~ e_{56}^{-})~,~
&
\nn\\
&
f_5 = (e_{42}^{+}~ e_{52}^{-}~ e_{55}^{+}~ e_{45}^{-})~,~
f_6 = (e_{63}^{+}~ e_{23}^{-}~ e_{25}^{+}~ e_{55}^{-}~ e_{56}^{+}~ e_{66}^{-})~,~
&
\nn\\
&
f_7 = (e_{34}^{+}~ e_{64}^{-}~ e_{66}^{+}~ e_{36}^{-})~,~
f_8 = (e_{11}^{+}~ e_{61}^{-}~ e_{64}^{+}~ e_{44}^{-}~ e_{45}^{+}~ e_{15}^{-})~,~
&
\eea
which satisfy the following constraints, 
\beal{es13b05}
&
f_1 f_3^{-1}= z_3 z_6~,~
f_2 f_6^{-1}= z_2 z_6 z_7 z_8~,~
f_3 f_7 = z_4 z_5 z_7 z_8~,~ 
f_4 f_8^{-1}= z_1 z_4 z_6 z_8~,~
&
\nn\\
& 
f_5^{-1} f_7=  z_5 z_8~,~ 
f_6 f_8 = z_1 z_2 z_3 z_5~,~ 
f_1 f_2 f_3 f_4 f_5 f_6 f_7 f_8=1~.~
&
\eea
The face paths can be written in terms of the canonical variables as follows, 
\beal{es13b05_1}
&
f_1=e^{P} z_3 z_6~,~ 
f_2= e^{Q} z_2 z_6 z_7 z_8~,~ 
f_3= e^{P} ~,~ 
f_4= e^{-Q} z_1 z_7^{-1}~,~ 
&
\nn\\
&
f_5= e^{-P} z_4 z_7 ~,~ 
f_6= e^{Q} ~,~ 
f_7= e^{-P} z_4 z_5 z_7 z_8~,~ 
f_8=e^{-Q} z_1 z_2 z_3 z_5~.~ 
\eea

The Kasteleyn matrix of the brane tiling for Model 4b in \fref{mf_04B} takes the following form, 
\beal{es13b06}
K = 
\left(
\ba{c|cccccc}
\; & b_1 & b_2 & b_3 & b_4 & b_5 & b_6 
\\
\hline
w_1 & e_{11} x^{-1} & e_{12} & 0 & 0 & e_{15} y^{-1} & 0
\\
w_2 & 0 & e_{22} & e_{23} & 0 & e_{25} y^{-1} & 0
\\
w_3 & e_{31} & 0 & e_{33} & e_{34} & 0 & e_{36}
\\
w_4 & 0 & e_{42} & 0 & e_{44} x^{-1} & e_{45} & 0
\\
w_5 & 0 & e_{52} & 0 & 0 & e_{55} & e_{56}
\\
w_6 &e_{61} y & 0 & e_{63} y & e_{64} & 0 & e_{66}
\ea
\right)
~.~
\eea
By taking the permanent of the Kasteleyn matrix,
we obtain the spectral curve of the dimer integrable system for Model 4b as follows,
\beal{es13b07}
&&
0 = \text{perm}~K=\overline{p}_0 
\cdot x^{-1} \cdot
\Big[
\delta_{(-1,-1)} \frac{1}{x y} 
+ \delta_{(-1,1)} \frac{y}{x} 
+ \delta_{(1,-1)} \frac{x}{y} 
\nn\\
&&  \hspace{1cm}
+ \delta_{(1,1)} xy
+ \delta_{(-1,0)} \frac{1}{x} 
+ \delta_{(0,-1)} \frac{1}{y} 
+ \delta_{(1,0)} x 
+ \delta_{(0,1)} y
+ H
\Big]
~,~
\eea
where $\overline{p}_0= e_{12}^{+} e_{23}^{+} e_{34}^{+} e_{45}^{+} e_{56}^{+} e_{61}^{+}$.
The Casimirs $\delta_{(m,n)}$ in \eref{es13b07} can be expressed in terms of the zig-zag paths in \eref{es13b03} as follows, 
\beal{es13b08}
&
\delta_{(-1,-1)} = z_2 z_3 z_4 z_8 ~,~
\delta_{(-1,0)} = z_5^{-1} z_6^{-1} z_7^{-1}+z_1^{-1} z_5^{-1} z_6^{-1} ~,~
\delta_{(-1,1)} = z_5^{-1} z_6^{-1}~,~
&
\nn\\
&
\delta_{(0,-1)} =z_2 z_3 z_4 + z_2 z_4 z_8~,~
\delta_{(0,1)} = z_5^{-1} + z_6^{-1} ~,~
\delta_{(1,-1)} = z_2 z_4~,~
&
\nn\\
&
\delta_{(1,0)} = z_2 + z_4 ~,~
\delta_{(1,1)} = 1~.~
&
\eea
This leads to the following form of the spectral curve for Model 4b,   
\beal{es13b09}
&&
\Sigma~:~
(z_2 + z_4) x + 
 z_2 z_4 \frac{x}{y} + \Big(\frac{1}{z_5}+\frac{1}{z_6}\Big) y + (z_2 z_3 z_4 + z_2 z_4 z_8)\frac{1}{y} 
\nn\\ 
&& 
\hspace{1cm}
 + \frac{1}{z_5 z_6} \frac{y}{x} + \Big(\frac{1}{z_5 z_6 z_7}+\frac{1}{z_1 z_5 z_6}\Big)\frac{1}{x} +z_2 z_3 z_4 z_8 \frac{1}{xy}+xy+ H 
 = 0 
~.~
\eea

The Hamiltonian is a sum over all 12 1-loops $\gamma_i$,
\beal{es13b10}
H=\sum_{i=1}^{12} \gamma_i~,~
\eea
where the 1-loops $\gamma_i$'s can be expressed in terms of zig-zag paths and face paths as follows,
\beal{es13b11}
&
\gamma_1 = z_1 z_2 z_4 z_8 f_1~,~
\gamma_2 = z_3^{-1} z_5^{-1} z_6^{-1} z_7^{-1} f_1 f_8~,~
\gamma_3 = z_1 z_2 z_4 z_8 f_1 f_7 f_8~,~
&
\nn\\
&
\gamma_4 = z_1 z_2 z_4 z_5^{-1} f_1 f_7 f_8~,~
\gamma_5 = z_4 z_8 f_5 f_8~,~
\gamma_6 = z_2 z_8 f_5 ~,~
&
\nn\\
&
\gamma_7 = z_6^{-1} z_7^{-1} f_5~,~
\gamma_8 = z_2 z_4^{-1} z_6^{-1} z_7^{-1} f_5 f_8^{-1}~,~
\gamma_9 = z_3 z_4 f_2~,~
&
\nn\\
&
\gamma_{10} = z_2 z_5 z_6^{-1} z_8 f_5 f_7^{-1} f_8^{-1}~,~
\gamma_{11} = z_2 z_6^{-1} f_1 f_8^{-1}~,~
\gamma_{12} = z_2 z_6^{-1} f_1~,~
\eea

The commutation matrix $C$ for Model 4b
takes the following form, 
\beal{es13b12}
&&
C=
\left(
\ba{c|cccccccccccc}
\; & \gamma_1
& \gamma_2
& \gamma_3
& \gamma_4 
& \gamma_5  
& \gamma_6 
& \gamma_7 
& \gamma_8 
& \gamma_9 
& \gamma_{10} 
& \gamma_{11} 
& \gamma_{12}
\\
\hline
\gamma_1 &     0 & 1 & 1 & 1 & 1 & 0 & 0 & -1 & -1 & -1 & -1 & 0 \\
\gamma_2 &     -1 & 0 & 1 & 1 & 2 & 1 & 1 & 0 & -1 & -1 & -2 & -1 \\
\gamma_3 &     -1 & -1 & 0 & 0 & 1 & 1 & 1 & 1 & 0 & 0 & -1 & -1 \\
\gamma_4 &     -1 & -1 & 0 & 0 & 1 & 1 & 1 & 1 & 0 & 0 & -1 & -1 \\
\gamma_5 &     -1 & -2 & -1 & -1 & 0 & 1 & 1 & 2 & 1 & 1 & 0 & -1 \\
\gamma_6 &     0 & -1 & -1 & -1 & -1 & 0 & 0 & 1 & 1 & 1 & 1 & 0 \\
\gamma_7 &     0 & -1 & -1 & -1 & -1 & 0 & 0 & 1 & 1 & 1 & 1 & 0 \\
\gamma_8 &     1 & 0 & -1 & -1 & -2 & -1 & -1 & 0 & 1 & 1 & 2 & 1 \\
\gamma_9 &     1 & 1 & 0 & 0 & -1 & -1 & -1 & -1 & 0 & 0 & 1 & 1 \\
\gamma_{10} & 1 & 1 & 0 & 0 & -1 & -1 & -1 & -1 & 0 & 0 & 1 & 1 \\
\gamma_{11} & 1 & 2 & 1 & 1 & 0 & -1 & -1 & -2 & -1 & -1 & 0 & 1 \\
\gamma_{12} & 0 & 1 & 1 & 1 & 1 & 0 & 0 & -1 & -1 & -1 & -1 & 0 \\
\ea
\right)
~.~
\eea
The 1-loops satisfying the commutation relations
can be written in terms of the canonical variables as follows, 
\beal{es13b13}
&
\gamma_1 =  e^{P} z_5^{-1} z_7^{-1}~,~ 
\gamma_2 = e^{-Q+P}  z_1 z_2 z_3 z_7^{-1} ~,~ 
\gamma_3 = e^{-Q} z_6^{-1} z_7^{-1}~,~
&
\nn\\
&
\gamma_4 =e^{-Q} z_1 z_2 z_3 z_4~,~
\gamma_5 =e^{-Q-P} z_4 z_6^{-1}~,~ 
\gamma_6 =  e^{-P}z_2 z_4 z_7 z_8~,~
&
\nn\\
&
\gamma_7 = e^{-P} z_4 z_6^{-1}~,~
\gamma_8 = e^{Q-P} z_2 z_4 z_7 z_8~,~ 
\gamma_9 = e^{Q} z_1^{-1} z_5^{-1}~,~
&
\nn\\
&
\gamma_{10} = e^{Q}  z_2 z_4 z_7 z_8~,~ 
\gamma_{11} = e^{Q+P}  z_1^{-1} z_5^{-1}~,~
\gamma_{12} = e^{P} z_2 z_3~.~
\eea
\\

%=================================================================
\subsection{Model 4c}
%=================================================================
%------------------------------------------------------------------------------------------------------------------
\begin{figure}[H]
\begin{center}
\resizebox{0.9\hsize}{!}{
\includegraphics{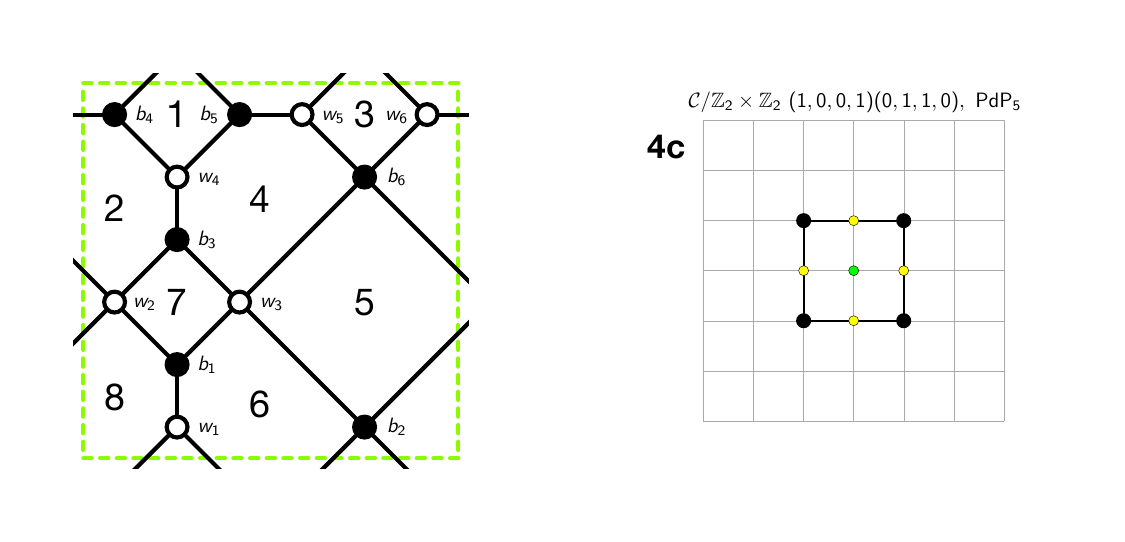}
}
\vspace{-0.5cm}
\caption{The brane tiling and toric diagram of Model 4c.}
\label{mf_04C}
 \end{center}
 \end{figure}
%------------------------------------------------------------------------------------------------------------------

The brane tiling for Model 4c can be expressed in terms of the following pair of permutation tuples
\beal{es13c01}
\sigma_B &=& (e_{11}\ e_{31}\ e_{21})\ (e_{22}\ e_{32}\ e_{52}\ e_{62})\ (e_{23}\ e_{33}\ e_{43})\ (e_{14}\ e_{64}\ e_{44})
\nn\\
&&
 (e_{15}\ e_{45}\ e_{55})\ (e_{26}\ e_{66}\ e_{56}\ e_{36})~,~
\nn\\
\sigma_W^{-1} &=& (e_{11}\ e_{15}\ e_{14})\ (e_{21}\ e_{22}\ e_{26}\ e_{23})\ (e_{31}\ e_{33}\ e_{36}\ e_{32})\ (e_{43}\ e_{44}\ e_{45})
\nn\\
&&
(e_{52}\ e_{56}\ e_{55})\ (e_{62}\ e_{64}\ e_{66})~,~
\eea
which correspond to the black and white nodes in the brane tiling, respectively.
 
The brane tiling for Model 4c has  8 zig-zag paths given by, 
\beal{es13c03}
&
z_1 = (e_ {15}^{+}~ e_ {45}^{-}~ e_ {43}^{+}~ e_ {23}^{-}~ e_ {21}^{+}~ e_{11}^{-})~,~
z_2 = (e_ {44}^{+}~ e_ {14}^{-}~ e_ {11}^{+}~ e_ {31}^{-}~ e_ {33}^{+}~ e_{43}^{-})~,~
&
\nn\\
&
z_3 = (e_ {45}^{+}~ e_ {55}^{-}~ e_ {52}^{+}~ e_ {62}^{-}~ e_ {64}^{+}~ e_{44}^{-})~,~
z_4 = (e_ {26}^{+}~ e_ {66}^{-}~ e_ {62}^{+}~ e_ {22}^{-})~,~
&
\nn\\
&
z_5 = (e_ {14}^{+}~ e_ {64}^{-}~ e_ {66}^{+}~ e_ {56}^{-}~ e_ {55}^{+}~ e_{15}^{-})~,~
z_6 = (e_ {56}^{+}~ e_ {36}^{-}~ e_ {32}^{+}~ e_ {52}^{-})~,~
&
\nn\\
&
z_7 = (e_ {23}^{+}~ e_ {33}^{-}~ e_ {36}^{+}~ e_ {26}^{-})~,~
z_8 = (e_ {22}^{+}~ e_ {32}^{-}~ e_ {31}^{+}~ e_ {21}^{-})~,~
&
\eea
and 8 face paths given by, 
\beal{es13c04}
&
f_1 = (e_{45}^{+}~ e_{15}^{-}~ e_{14}^{+}~ e_{44}^{-})~,~
f_2 = (e_{44}^{+}~ e_{64}^{-}~ e_{66}^{+}~ e_{26}^{-}~ e_{23}^{+}~ e_{43}^{-})~,~
&
\nn\\
&
f_3 = (e_{56}^{+}~ e_{66}^{-}~ e_{62}^{+}~ e_{52}^{-})~,~
f_4 = (e_{43}^{+}~ e_{33}^{-}~ e_{36}^{+}~ e_{56}^{-}~ e_{55}^{+}~ e_{45}^{-})~,~
&
\nn\\
&
f_5 = (e_{26}^{+}~ e_{36}^{-}~ e_{32}^{+}~ e_{22}^{-})~,~
f_6 = (e_{15}^{+}~ e_{55}^{-}~ e_{52}^{+}~ e_{32}^{-}~ e_{31}^{+}~ e_{11}^{-})~,~
&
\nn\\
&
f_7 = (e_{33}^{+}~ e_{23}^{-}~ e_{21}^{+}~ e_{31}^{-})~,~
f_8 = (e_{64}^{+}~ e_{14}^{-}~ e_{11}^{+}~ e_{21}^{-}~ e_{22}^{+}~ e_{62}^{-})~,~
&
\eea
satisfying the following constraints, 
\beal{es13c05}
&
f_6 f_7 f_8 = z_1 z_2 z_3 z_8~,~
f_5 f_6^{-1} f_8^{-1} = z_4 z_5 z_6 z_8^{-1}~,~ 
f_4 f_8^{-1} = z_1 z_4 z_5 z_7~,~ 
&
\nn\\
&
f_3 f_7^{-1} = z_4 z_6 z_7 z_8~,~ 
f_2 f_6^{-1}= z_2 z_5 z_6 z_7~,~
f_1 f_7^{-1} = z_1^{-1} z_2^{-1}~,~
&
\nn\\
&
f_1 f_2 f_3 f_4 f_5 f_6 f_7 f_8=1~.~
&
\eea
The face paths can be written in terms of the canonical variables as follows, 
\beal{es12c05_1}
&
f_1=e^{P} z_1^{-1} z_2^{-1}~,~ 
f_2= e^{Q} z_2 z_5 z_6 z_7~,~ 
f_3= e^{P} z_4 z_6 z_7 z_8~,~  
f_4= e^{-Q-P} z_1 z_6^{-1}~,~ 
&
\nn\\
&
f_5= e^{-P} z_7^{-1} z_8^{-1} ~,~ 
f_6= e^{Q} ~,~ 
f_7= e^{P} ~,~ 
f_8=e^{-Q-P} z_1 z_2 z_3 z_8~.~ 
\eea

The Kasteleyn matrix of the brane tiling for Model 4c in \fref{mf_04C} takes the following form, 
\beal{es13c06}
K = 
\left(
\ba{c|cccccc}
\; & b_1 & b_2 & b_3 & b_4 & b_5 & b_6 
\\
\hline
w_1 & e_{11} & 0 & 0 & e_{14} y^{-1} & e_{15} y^{-1} & 0
\\
w_2 & e_{21} & e_{22} x^{-1} & e_{23} & 0 & 0 & e_{26} x^{-1}
\\
w_3 & e_{31} & e_{32} & e_{33} & 0 & 0 & e_{36}
\\
w_4 & 0 & 0 & e_{43} & e_{44} & e_{45} & 0
\\
w_5 & 0 & e_{52} y& 0 & 0 & e_{55} & e_{56}
\\
w_6 & 0 & e_{62} y & 0 & e_{64} x & 0 & e_{66}
\ea
\right)
~.~
\eea
By taking a permanent of the Kasteleyn matrix, 
we obtain the spectral curve of the dimer integrable system for Model 4c as follows,
\beal{es13c07}
&&
0 = \text{perm}~K=\overline{p}_0
\cdot 
\Big[
\delta_{(-1,-1)} \frac{1}{x y} 
+ \delta_{(-1,1)} \frac{y}{x} 
+ \delta_{(1,-1)} \frac{x}{y} 
+ \delta_{(1,1)} xy
\nn\\
&& 
\hspace{1cm}
+ \delta_{(-1,0)} \frac{1}{x} 
+ \delta_{(0,-1)} \frac{1}{y} 
+ \delta_{(1,0)} x 
+ \delta_{(0,1)} y
+ H
\Big]
~,~
\eea
where $\overline{p}_0= e_{11}^{+} e_{23}^{+} e_{36}^{+} e_{45}^{+} e_{52}^{+} e_{64}^{+}$.
The Casimirs $\delta_{(m,n)}$ in \eref{es13c07} can be expressed in terms of
the zig-zag paths in \eref{es13c03} as follows, 
\beal{es13c08}
&
\delta_{(-1,-1)} =z_1 z_5 z_6 z_8 ~,~
\delta_{(-1,0)} =z_2^{-1} z_3^{-1} z_7^{-1}+z_3^{-1} z_4^{-1} z_7^{-1}~,~
&
\nn\\
&
\delta_{(-1,1)} = z_3^{-1} z_7^{-1}~,~
\delta_{(0,-1)} =z_1 z_6 z_8 + z_1 z_5 z_6~,~
\delta_{(0,1)} = z_3^{-1}+z_7^{-1} ~,~
&
\nn\\
&
\delta_{(1,-1)} = z_1 z_6~,~
\delta_{(1,0)} = z_1 + z_6 ~,~
\delta_{(1,1)} = 1~.~
&
\eea
This leads to the following form of the spectral curve for Model 4c,   
\beal{es13c09}
&&
\Sigma~:~
z_1 z_5 z_6 z_8 \frac{1}{xy}+\Big(\frac{1}{z_2 z_3 z_7}+\frac{1}{z_3 z_4 z_7} \Big) \frac{1}{x}+\frac{1}{z_3 z_7} \frac{y}{x}+(z_1 z_6 z_8+z_1 z_5 z_6)\frac{1}{y}
\nn\\
&&
\hspace{1cm}
+\Big(\frac{1}{z_3}+\frac{1}{z_7} \Big)y+(z_1+z_6)x+z_1 z_6 \frac{x}{y}+ xy+H 
= 0
~.~
\eea

The Hamiltonian is a sum over all 14 1-loops $\gamma_i$,
\beal{es13c10}
H=\sum_{i=1}^{14} \gamma_i~,~
\eea
where the 1-loops $\gamma_i$
can be expressed in terms of zig-zag paths and face paths as follows, 
\beal{es13c11}
&
\gamma_1 = z_1 z_3^{-1} z_7^{-1} z_8^{-1} f_1 f_7^{-1} f_8~,~ 
\gamma_2 = z_1 z_4 z_5 z_6 f_2 f_8~,~ 
\gamma_3 = z_2^{-1} z_7^{-1} f_2 ~,~
&
\nn\\
&
\gamma_4 = z_1 z_7^{-1} f_2 f_7^{-1}~,~ 
\gamma_5 = z_1 z_8 f_5~,~ 
\gamma_6 = z_1 z_7^{-1} f_2~,~ 
&
\nn\\
&
\gamma_7 = z_2^{-1} z_7^{-1} f_2 f_7~,~
\gamma_8 = z_1 z_4 z_6 z_8 f_2~,~
\gamma_9 = z_1 z_4 z_6 z_8 f_2 f_7~,~
&
\nn\\
&
\gamma_{10} = z_1 z_2 z_6 z_8 f_1 f_7^{-1} f_8^{-1}~,~
\gamma_{11} = z_1 z_2 z_6 z_8 f_1~,~
\gamma_{12} = z_1 z_3^{-1} f_1 f_8~,~
&
\nn\\
&
\gamma_{13} =  z_1 z_2 z_6 z_8 f_1 f_8^{-1}~,~
\gamma_{14} = z_1 z_3^{-1} f_1~.~
\eea

The commutation matrix $C$ for Model 4c
is given by, 
\beal{es13c12}
&&
C=
\left(
\ba{c|cccccccccccccc}
\; & \gamma_1
& \gamma_2
& \gamma_3
& \gamma_4 
& \gamma_5  
& \gamma_6 
& \gamma_7 
& \gamma_8 
& \gamma_9 
& \gamma_{10} 
& \gamma_{11} 
& \gamma_{12}
& \gamma_{13}
& \gamma_{14}
\\
\hline
\gamma_1 &     0 & 1 & 1 & 2 & 1 & 1 & 0 & 1 & 0 & 0 & -1 & -1 & -1 & -1 \\
\gamma_2 &     -1 & 0 & 1 & 1 & 0 & 1 & 1 & 1 & 1 & 1 & 0 & -1 & 1 & 0 \\
\gamma_3 &     -1 & -1 & 0 & -1 & -1 & 0 & 1 & 0 & 1 & 1 & 1 & 0 & 2 & 1 \\
\gamma_4 &     -2 & -1 & 1 & 0 & -1 & 1 & 2 & 1 & 2 & 2 & 1 & -1 & 3 & 1 \\
\gamma_5 &     -1 & 0 & 1 & 1 & 0 & 1 & 1 & 1 & 1 & 1 & 0 & -1 & 1 & 0 \\
\gamma_6 &     -1 & -1 & 0 & -1 & -1 & 0 & 1 & 0 & 1 & 1 & 1 & 0 & 2 & 1 \\
\gamma_7 &     0 & -1 & -1 & -2 & -1 & -1 & 0 & -1 & 0 & 0 & 1 & 1 & 1 & 1 \\
\gamma_8 &     -1 & -1 & 0 & -1 & -1 & 0 & 1 & 0 & 1 & 1 & 1 & 0 & 2 & 1 \\
\gamma_9 &     0 & -1 & -1 & -2 & -1 & -1 & 0 & -1 & 0 & 0 & 1 & 1 & 1 & 1 \\
\gamma_{10} & 0 & -1 & -1 & -2 & -1 & -1 & 0 & -1 & 0 & 0 & 1 & 1 & 1 & 1 \\
\gamma_{11} & 1 & 0 & -1 & -1 & 0 & -1 & -1 & -1 & -1 & -1 & 0 & 1 & -1 & 0 \\
\gamma_{12} & 1 & 1 & 0 & 1 & 1 & 0 & -1 & 0 & -1 & -1 & -1 & 0 & -2 & -1 \\
\gamma_{13} & 1 & -1 & -2 & -3 & -1 & -2 & -1 & -2 & -1 & -1 & 1 & 2 & 0 & 1 \\
\gamma_{14} & 1 & 0 & -1 & -1 & 0 & -1 & -1 & -1 & -1 & -1 & 0 & 1 & -1 & 0 \\
\ea
\right)
~,~
\eea
where the 1-loops
satisfying the commutation relations 
can be written in terms of the canonical variables as follows, 
\beal{es13c13}
&
\gamma_1 = e^{-Q-P} z_1 z_7^{-1}~,~
\gamma_2 =e^{-P} z_1 z_2 z_5 z_6~,~ 
\gamma_3 = e^{Q}  z_5 z_6 ~,~ 
&
\nn\\
&
\gamma_4 =  e^{Q-P} z_1 z_2 z_5 z_6~,~
\gamma_5 = e^{-P} z_1 z_7^{-1}~,~ 
\gamma_6 =e^{Q} z_1 z_2 z_5 z_6~,~ 
&
\nn\\
&
\gamma_7 = e^{Q+P} z_5 z_6~,~
\gamma_8 = e^{Q} z_3^{-1} z_6~,~
\gamma_9 =  e^{Q+P} z_3^{-1} z_6 ~,~ 
&
\nn\\
&
\gamma_{10} = e^{Q+P} z_1^{-1} z_2^{-1} z_3^{-1} z_6~,~
\gamma_{11} = e^{P}  z_6 z_8~,~
\gamma_{12} = e^{-Q}  z_1 z_8~,~ 
&
\nn\\
&
\gamma_{13} = e^{Q+2P}  z_1^{-1} z_2^{-1} z_3^{-1} z_6~,~
\gamma_{14} = e^{P}  z_2^{-1} z_3^{-1}~.~
&
\eea
\\

%=================================================================
\subsection{Model 4d}
%=================================================================
%------------------------------------------------------------------------------------------------------------------
\begin{figure}[H]
\begin{center}
\resizebox{0.9\hsize}{!}{
\includegraphics{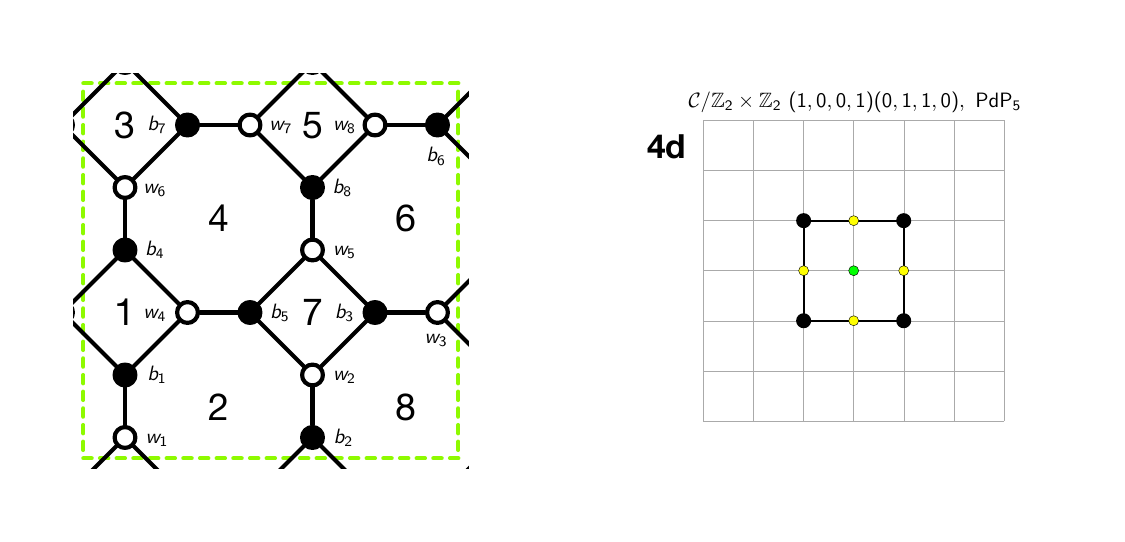}
}
\vspace{-0.5cm}
\caption{The brane tiling and toric diagram of Model 4d.}
\label{mf_04D}
 \end{center}
 \end{figure}
%------------------------------------------------------------------------------------------------------------------

The brane tiling for Model 4d can be expressed in terms of the following pair of permutation tuples
\beal{es13d01}
\sigma_B &=& (e_{11}\ e_{41}\ e_{31})\ (e_{22}\ e_{72}\ e_{82})\ (e_{23}\ e_{33}\ e_{53})\ (e_{34}\ e_{44}\ e_{64})
\nn\\
&&
 (e_{25}\ e_{55}\ e_{45})\ (e_{16}\ e_{86}\ e_{66})\ (e_{17}\ e_{67}\ e_{77})\ (e_{58}\ e_{88}\ e_{78})~,~
\nn\\
\sigma_W^{-1} &=& (e_{11}\ e_{17}\ e_{16})\ (e_{22}\ e_{25}\ e_{23})\ (e_{31}\ e_{33}\ e_{34})\ (e_{41}\ e_{44}\ e_{45})
\nn\\
&&
(e_{53}\ e_{55}\ e_{58})\ (e_{64}\ e_{66}\ e_{67})\ (e_{72}\ e_{78}\ e_{77})\ (e_{82}\ e_{86}\ e_{88})~,~
\eea
which correspond to the black and white nodes in the brane tiling, respectively.\\

The brane tiling for Model 4d has 8 zig-zag paths given by, 
\beal{es13d03}
&
z_1 = (e_ {34}^{+}~ e_ {44}^{-}~ e_ {45}^{+}~ e_ {25}^{-}~ e_ {23}^{+}~ e_{33}^{-})~,~
z_2 = (e_ {41}^{+}~ e_ {31}^{-}~ e_ {33}^{+}~ e_ {53}^{-}~ e_ {55}^{+}~ e_{45}^{-})~,~
&
\nn\\
&
z_3 = (e_ {44}^{+}~ e_ {64}^{-}~ e_ {66}^{+}~ e_ {16}^{-}~ e_ {11}^{+}~ e_{41}^{-})~,~
z_4 = (e_ {77}^{+}~ e_ {17}^{-}~ e_ {16}^{+}~ e_ {86}^{-}~ e_ {88}^{+}~ e_{78}^{-})~,~
&
\nn\\
&
z_5 = (e_ {22}^{+}~ e_ {72}^{-}~ e_ {78}^{+}~ e_ {58}^{-}~ e_ {53}^{+}~ e_{23}^{-})~,~
z_6 = (e_ {31}^{+}~ e_ {11}^{-}~ e_ {17}^{+}~ e_ {67}^{-}~ e_ {64}^{+}~ e_{34}^{-})~,~
&
\nn\\
&
z_7 = (e_ {72}^{+}~ e_ {82}^{-}~ e_ {86}^{+}~ e_ {66}^{-}~ e_ {67}^{+}~ e_{77}^{-})~,~
z_8 = (e_ {25}^{+}~ e_ {55}^{-}~ e_ {58}^{+}~ e_ {88}^{-}~ e_ {82}^{+}~ e_{22}^{-})~,~
&
\eea
and 8 face paths given by, 
\beal{es13d04}
&
f_1 = (e_ {44}^{+}~ e_ {34}^{-}~ e_ {31}^{+}~ e_ {41}^{-})~,~
f_2 = (e_ {41}^{+}~ e_ {11}^{-}~ e_ {17}^{+}~ e_ {77}^{-}~ e_ {72}^{+}~ e_{22}^{-}~ e_ {25}^{+}~ e_ {45}^{-})~,~
&
\nn\\
&
f_3 = (e_ {67}^{+}~ e_ {17}^{-}~ e_ {16}^{+}~ e_ {66}^{-})~,~
f_4 = (e_ {45}^{+}~ e_ {55}^{-}~ e_ {58}^{+}~ e_ {78}^{-}~ e_ {77}^{+}~ e_{67}^{-}~ e_ {64}^{+}~ e_ {44}^{-})~,~
&
\nn\\
&
f_5 = (e_ {78}^{+}~ e_ {88}^{-}~ e_ {82}^{+}~ e_ {72}^{-})~,~
f_6 = (e_ {34}^{+}~ e_ {64}^{-}~ e_ {66}^{+}~ e_ {86}^{-}~ e_ {88}^{+}~ e_{58}^{-}~ e_ {53}^{+}~ e_ {33}^{-})~,~
&
\nn\\
&
f_7 = (e_ {55}^{+}~ e_ {25}^{-}~ e_ {23}^{+}~ e_ {53}^{-})~,~
f_8 = (e_{11}^{+}~ e_{31}^{-}~ e_{33}^{+}~ e_{23}^{-}~ e_{22}^{+}~e_{82}^{-}~ e_{86}^{+}~ e_{16}^{-})~,~
&
\eea
which satisfy the following constraints, 
\beal{es13d05}
&
f_6 f_7^2 f_8 = z_1 z_2 z_3 z_8^{-1}~,~
f_5 f_7^{-1}  = z_5 z_8~,~ 
f_4 f_8^{-1} = z_1 z_4 z_6 z_8~,~ 
f_3 f_7^{-1} = z_4 z_5 z_7 z_8~,~ 
&
\nn\\
&
f_2 f_6^{-1}= z_2 z_6 z_7 z_8~,~
f_1 f_7^{-1} =z_1^{-1} z_2^{-1}~,~
f_1 f_2 f_3 f_4 f_5 f_6 f_7 f_8=1~.~
&
\eea
The face paths can be written in terms of the canonical variables as follows, 
\beal{es13d05_1}
&
f_1=e^{P} z_1^{-1} z_2^{-1}~,~ 
f_2= e^{Q} z_2 z_6 z_7 z_8~,~ 
f_3= e^{P} z_4 z_5 z_7 z_8~,~  
&
\nn\\
&
f_4= e^{-Q-2P} z_1 z_5^{-1} z_7^{-1} z_8^{-1}~,~ 
f_5= e^{P} z_5 z_8 ~,~ 
f_6= e^{Q} ~,~ 
f_7= e^{P} ~,~ 
&
\nn\\
&
f_8=e^{-Q-2P} z_1 z_2 z_3 z_8^{-1}~.~ 
\eea

The Kasteleyn matrix of 
the brane tiling for Model 4d in \fref{mf_04D} is given by, 
\beal{es13d06}
K = 
\left(
\ba{c|cccccccc}
\; & b_1 & b_2 & b_3 & b_4 & b_5 & b_6  & b_7 & b_8
\\
\hline
w_1 & e_{11} & 0 & 0 & 0 & 0 & e_{16} x^{-1} y^{-1} & e_{17} y^{-1} & 0
\\
w_2 & 0 & e_{22} & e_{23} & 0 & e_{25} & 0 & 0 & 0
\\
w_3 & e_{31} x & 0 & e_{33} & e_{34}x & 0 & 0 & 0 & 0
\\
w_4 & e_{41} & 0 & 0 & e_{44} & e_{45} & 0 & 0 & 0
\\
w_5 & 0 & 0 & e_{53} & 0 & e_{55} & 0 & 0 & e_{58}
\\
w_6 & 0 & 0 & 0 & e_{64} & 0 & e_{66} x^{-1} & e_{67} & 0
\\
w_7 & 0 & e_{72} y & 0 & 0 & 0 & 0 & e_{77} & e_{78} 
\\
w_8 & 0 & e_{82} y & 0 & 0 & 0 & e_{86} &0 & e_{88} 
\ea
\right)
~.~
\eea
The permanent of the Kasteleyn matrix in \eref{es13d06}
gives the spectral curve of the dimer integrable system for Model 4d as follows,
\beal{es13d07}
&&
0 = \text{perm}~K=\overline{p}_0
\cdot 
\Big[
\delta_{(-1,-1)} \frac{1}{x y} 
+ \delta_{(-1,1)} \frac{y}{x} 
+ \delta_{(1,-1)} \frac{x}{y} 
+ \delta_{(1,1)} xy
\nn\\
&& 
\hspace{1cm}
+ \delta_{(-1,0)} \frac{1}{x} 
+ \delta_{(0,-1)} \frac{1}{y} 
+ \delta_{(1,0)} x 
+ \delta_{(0,1)} y
+ H
\Big]
~,~
\eea
where $\overline{p}_0= e_{11}^{+} e_{23}^{+} e_{34}^{+} e_{45}^{+} e_{58}^{+} e_{67}^{+} e_{72}^{+} e_{86}^{+}$.
The Casimirs $\delta_{(m,n)}$ in \eref{es13d07} can be expressed in terms of the zig-zag paths in \eref{es13d03} as follows, 
\beal{es13d08}
&
\delta_{(-1,-1)} =z_2 z_4 z_5 z_6 ~,~
\delta_{(-1,0)} = z_1^{-1} z_3^{-1} z_7^{-1} + z_1^{-1} z_7^{-1} z_8^{-1}~,~
&
\nn\\
&
\delta_{(-1,1)} = z_1^{-1} z_7^{-1} ~,~
\delta_{(0,-1)} =z_2 z_5 z_6 + z_4 z_5 z_6~,~
\delta_{(0,1)} = z_1^{-1}+z_7^{-1} ~,~
&
\nn\\
&
\delta_{(1,-1)} = z_5 z_6~,~
\delta_{(1,0)} = z_5 + z_6 ~,~
\delta_{(1,1)} = 1~.~
&
\eea
Accordingly, 
we can express the spectral curve for Model 4d as follows,
\beal{es13d09}
&&
\Sigma~:~
z_2 z_4 z_5 z_6\frac{1}{xy}+\Big( \frac{1}{z_1 z_3 z_7}+\frac{1}{z_1 z_7 z_8} \Big) \frac{1}{x}+\frac{1}{z_1 z_7} \frac{y}{x}
+(z_2 z_5 z_6 + z_4 z_5 z_6)\frac{1}{y}
\nn\\
&&
\hspace{1cm}
+\Big(\frac{1}{z_1}+\frac{1}{z_7} \Big)y+(z_5+z_6)x+z_5 z_6 \frac{x}{y}+xy+ H 
= 0 
~.~
\eea

The Hamiltonian is a sum over all 21 1-loops $\gamma_i$,
\beal{es13d10}
H=\sum_{i=1}^{21} \gamma_i~,~
\eea
where the 1-loops $\gamma_i$
can be expressed in terms of zig-zag paths and face paths as shown below, 
\beal{es13d11}
&
\gamma_1 = z_3^{-1} z_7^{-1} f_1 f_2 f_8~,~
\gamma_2 = z_2 z_5 f_1 f_4~,~
\gamma_3 = z_2 z_5 f_1~,~
\gamma_4 = z_3^{-1}  z_7^{-1} f_1~,~
&
\nn\\
&
\gamma_5 = z_5 z_7^{-1} f_2 f_3~,~
\gamma_6 = z_1^{-1}  z_5 f_8^{-1}~,~
\gamma_7 = z_2 z_5 f_8^{-1}~,~
\gamma_8 = z_1^{-1}  z_5 f_7 f_8^{-1}~,~
&
\nn\\
&
\gamma_9 = z_2 z_3 z_5 z_6 f_1^{-1} f_7^{-1} f_8^{-1}~,~
\gamma_{10} = z_1^{-1}  z_3 z_5 z_6 f_1^{-1} f_8^{-1}~,~
\gamma_{11} = z_2 z_3 z_5 z_6 f_1^{-1} f_8^{-1}~,~
&
\nn\\
&
\gamma_{12} = z_1^{-1}  z_3 z_5 z_6 f_1^{-1} f_7 f_8^{-1}~,~
\gamma_{13} = z_1^{-1}  z_8^{-1} f_1^{-1} f_7^{-1} f_8^{-1}~,~
\gamma_{14} = z_1^{-2}  z_2^{-1} z_8^{-1} f_1^{-1} f_8^{-1}~,~
&
\nn\\
&
\gamma_{15} = z_1^{-1} z_8^{-1} f_1^{-1} f_8^{-1}~,~
\gamma_{16} = z_1^{-2}  z_2^{-1} z_8^{-1} f_1^{-1} f_7 f_8^{-1}~,~
\gamma_{17} = z_1^{-1}  z_5 f_2 f_4 f_6 f_7~,~
&
\nn\\
&
\gamma_{18} = z_1^{-1}  z_2^{-1}  z_7^{-1}  z_8^{-1} f_2 ~,~
\gamma_{19} = z_7^{-1}  z_8^{-1} f_2~,~
\gamma_{20} = z_3 z_4 z_5 z_6 f_2 f_7~,~
&
\nn\\
&
\gamma_{21} = z_2^{-1}  z_4 z_5 z_7^{-1} f_2 f_7 f_8~.~
\eea

The commutation matrix $C$ for Model 4d is given by, 
\beal{es13d12}
&&
C=
\resizebox{0.75\textwidth}{!}{$
\left(
\ba{c|ccccccccccccccccccccc}
\; & \gamma_1
& \gamma_2
& \gamma_3
& \gamma_4 
& \gamma_5  
& \gamma_6 
& \gamma_7 
& \gamma_8 
& \gamma_9 
& \gamma_{10} 
& \gamma_{11} 
& \gamma_{12}
& \gamma_{13}
& \gamma_{14}
& \gamma_{15}
& \gamma_{16}
& \gamma_{17}
& \gamma_{18}
& \gamma_{19}
& \gamma_{20}
& \gamma_{21}
\\
\hline
\gamma_1 &     0 & -1 & 0 & 0 & 1 & 1 & 1 & 1 & 1 & 1 & 1 & 1 & 1 & 1 & 1 & 1 & 1 & 1 & 1 & 1 & 0 \\
\gamma_2 &      1 & 0 & -1 & -1 & 0 & -1 & -1 & -2 & 1 & 0 & 0 & -1 & 1 & 0 & 0 & -1 & 2 & 1 & 1 & 0 & 1 \\
\gamma_3 &      0 & 1 & 0 & 0 & -1 & -1 & -1 & -1 & -1 & -1 & -1 & -1 & -1 & -1 & -1 & -1 & -1 & -1 & -1 & -1 & 0 \\
\gamma_4 &      0 & 1 & 0 & 0 & -1 & -1 & -1 & -1 & -1 & -1 & -1 & -1 & -1 & -1 & -1 & -1 & -1 & -1 & -1 & -1 & 0 \\
\gamma_5 &      -1 & 0 & 1 & 1 & 0 & 1 & 1 & 2 & -1 & 0 & 0 & 1 & -1 & 0 & 0 & 1 & -2 & -1 & -1 & 0 & -1 \\
\gamma_6 &      -1 & 1 & 1 & 1 & -1 & 0 & 0 & 1 & -2 & -1 & -1 & 0 & -2 & -1 & -1 & 0 & -3 & -2 & -2 & -1 & -1 \\
\gamma_7 &      -1 & 1 & 1 & 1 & -1 & 0 & 0 & 1 & -2 & -1 & -1 & 0 & -2 & -1 & -1 & 0 & -3 & -2 & -2 & -1 & -1 \\
\gamma_8 &      -1 & 2 & 1 & 1 & -2 & -1 & -1 & 0 & -3 & -2 & -2 & -1 & -3 & -2 & -2 & -1 & -4 & -3 & -3 & -2 & -1 \\
\gamma_9 &      -1 & -1 & 1 & 1 & 1 & 2 & 2 & 3 & 0 & 1 & 1 & 2 & 0 & 1 & 1 & 2 & -1 & 0 & 0 & 1 & -1 \\
\gamma_{10} &  -1 & 0 & 1 & 1 & 0 & 1 & 1 & 2 & -1 & 0 & 0 & 1 & -1 & 0 & 0 & 1 & -2 & -1 & -1 & 0 & -1 \\
\gamma_{11} &  -1 & 0 & 1 & 1 & 0 & 1 & 1 & 2 & -1 & 0 & 0 & 1 & -1 & 0 & 0 & 1 & -2 & -1 & -1 & 0 & -1 \\
\gamma_{12} &  -1 & 1 & 1 & 1 & -1 & 0 & 0 & 1 & -2 & -1 & -1 & 0 & -2 & -1 & -1 & 0 & -3 & -2 & -2 & -1 & -1 \\
\gamma_{13} &  -1 & -1 & 1 & 1 & 1 & 2 & 2 & 3 & 0 & 1 & 1 & 2 & 0 & 1 & 1 & 2 & -1 & 0 & 0 & 1 & -1 \\
\gamma_{14} &  -1 & 0 & 1 & 1 & 0 & 1 & 1 & 2 & -1 & 0 & 0 & 1 & -1 & 0 & 0 & 1 & -2 & -1 & -1 & 0 & -1 \\
\gamma_{15} &  -1 & 0 & 1 & 1 & 0 & 1 & 1 & 2 & -1 & 0 & 0 & 1 & -1 & 0 & 0 & 1 & -2 & -1 & -1 & 0 & -1 \\
\gamma_{16} &  -1 & 1 & 1 & 1 & -1 & 0 & 0 & 1 & -2 & -1 & -1 & 0 & -2 & -1 & -1 & 0 & -3 & -2 & -2 & -1 & -1 \\
\gamma_{17} &  -1 & -2 & 1 & 1 & 2 & 3 & 3 & 4 & 1 & 2 & 2 & 3 & 1 & 2 & 2 & 3 & 0 & 1 & 1 & 2 & -1 \\
\gamma_{18} &  -1 & -1 & 1 & 1 & 1 & 2 & 2 & 3 & 0 & 1 & 1 & 2 & 0 & 1 & 1 & 2 & -1 & 0 & 0 & 1 & -1 \\
\gamma_{19} &  -1 & -1 & 1 & 1 & 1 & 2 & 2 & 3 & 0 & 1 & 1 & 2 & 0 & 1 & 1 & 2 & -1 & 0 & 0 & 1 & -1 \\
\gamma_{20} &  -1 & 0 & 1 & 1 & 0 & 1 & 1 & 2 & -1 & 0 & 0 & 1 & -1 & 0 & 0 & 1 & -2 & -1 & -1 & 0 & -1 \\
\gamma_{21} &  0 & -1 & 0 & 0 & 1 & 1 & 1 & 1 & 1 & 1 & 1 & 1 & 1 & 1 & 1 & 1 & 1 & 1 & 1 & 1 & 0 \\
\ea
\right)
$}
~.~
\eea
The 1-loops satisfying the commutation relations
can be written in terms of the canonical variables as follows, 
\beal{es13d13}
&
\gamma_1 = e^{-P} z_2 z_6~,~
\gamma_2 = e^{-Q-P} z_7^{-1} z_8^{-1}~,~
\gamma_3 = e^{P} z_1^{-1} z_5~,~
&
\nn\\
&
\gamma_4 = e^{P} z_4 z_5 z_6 z_8~,~
\gamma_5 =e^{Q+P} z_1^{-1} z_3^{-1} z_5 z_8~,~
\gamma_6 = e^{Q+2P} z_1^{-2} z_2^{-1} z_3^{-1} z_5 z_8~,~
&
\nn\\
&
\gamma_7 =e^{Q+2P} z_1^{-1} z_3^{-1} z_5 z_8~,~
\gamma_8 = e^{Q+3P} z_1^{-2} z_2^{-1} z_3^{-1} z_5 z_8~,~
\gamma_9 = e^{Q} z_2 z_5 z_6 z_8~,~
&
\nn\\
&
\gamma_{10} = e^{Q+P} z_1^{-1} z_5 z_6 z_8~,~
\gamma_{11} = e^{Q+P} z_2 z_5 z_6 z_8~,~
\gamma_{12} =e^{Q+2P} z_1^{-1} z_5 z_6 z_8~,~
&
\nn\\
&
\gamma_{13} = e^{Q} z_1^{-1} z_3^{-1}~,~
\gamma_{14} = e^{Q+P} z_1^{-2} z_2^{-1} z_3^{-1}~,~
\gamma_{15} = e^{Q+P} z_1^{-1} z_3^{-1}~,~
&
\nn\\
&
\gamma_{16} = e^{Q+2P} z_1^{-2} z_2^{-1} z_3^{-1}~,~
\gamma_{17} = e^{Q-P} z_2 z_6~,~
\gamma_{18} = e^Q z_1^{-1} z_6~,~
&
\nn\\
&
\gamma_{19} = e^Q z_2 z_6~,~
\gamma_{20} = e^{Q+P} z_1^{-1} z_6~,~
\gamma_{21} = e^{-P} z_7^{-1} z_8^{-1}~.~
&
\eea
\\

%=================================================================
\section{Model 5: $\text{PdP}_{4b}$}
%=================================================================
%------------------------------------------------------------------------------------------------------------------
\begin{figure}[H]
\begin{center}
\resizebox{0.9\hsize}{!}{
\includegraphics{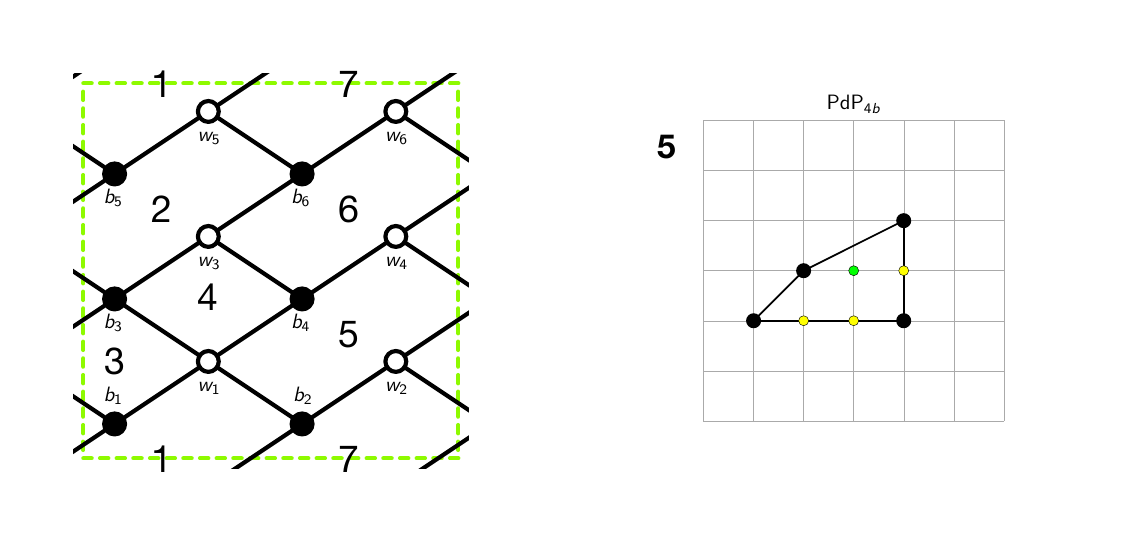}
}
\vspace{-0.5cm}
\caption{The brane tiling and toric diagram of Model 5.}
\label{mf_05}
 \end{center}
 \end{figure}
%------------------------------------------------------------------------------------------------------------------

The brane tiling for Model 5 can be expressed in terms of the following pair of permutation tuples
\beal{es14a01}
\sigma_B &=& (e_{11}\ e_{21}\ e_{61})\ (e_{12}\ e_{52}\ e_{22})\ (e_{13}\ e_{33}\ e_{42}\ e_{22})\ (e_{14}\ e_{44}\ e_{34})
\nn\\
&&
(e_{45}\ e_{55}\ e_{65})\ (e_{36}\ e_{66}\ e_{56})
\nn\\
\sigma_W^{-1} &=& (e_{11}\ e_{13}\ e_{14}\ e_{12})\ (e_{21}\ e_{22}\ e_{23})\ (e_{33}\ e_{36}\ e_{34})\ (e_{43}\ e_{44}\ e_{45})
\nn\\
&&
(e_{52}\ e_{56}\ e_{55})\ (e_{61}\ e_{65}\ e_{66})
\eea
which correspond to the black and white nodes in the brane tiling, respectively.\\

 The brane tiling for Model 5 has 7 zig-zag paths given by, 
\beal{es14a03}
&
z_1 = (e_ {61}^{+}~ e_ {11}^{-}~ e_ {13}^{+}~ e_ {33}^{-}~ e_ {36}^{+}~ e_{66}^{-})~,~
z_2 = (e_ {55}^{+}~ e_ {65}^{-}~ e_ {66}^{+}~ e_ {56}^{-})~,~
&
\nn\\
&
z_3 = (e_ {12}^{+}~ e_ {52}^{-}~ e_ {56}^{+}~ e_ {36}^{-}~ e_ {34}^{+}~ e_{14}^{-})~,~
z_4 = (e_ {52}^{+}~ e_ {22}^{-}~ e_ {23}^{+}~ e_ {13}^{-}~ e_ {14}^{+}~ e_{44}^{-}~ e_ {45}^{+}~ e_ {55}^{-})~,~
&
\nn\\
&
z_5 = (e_ {33}^{+}~ e_ {43}^{-}~ e_ {44}^{+}~ e_ {34}^{-})~,~
z_6 = (e_ {11}^{+}~ e_ {21}^{-}~ e_ {22}^{+}~ e_ {12}^{-})~,~
&
\nn\\
&
z_7 = (e_ {65}^{+}~ e_ {45}^{-}~ e_ {43}^{+}~ e_ {23}^{-}~ e_ {21}^{+}~ e_{61}^{-})~,~
&
\eea
and 7 face paths given by, 
\beal{es14a04}
&
f_1 = (e_ {11}^{+}~ e_ {61}^{-}~ e_ {65}^{+}~ e_ {55}^{-}~ e_ {52}^{+}~ e_{12}^{-})~,~
f_2 = (e_ {55}^{+}~ e_ {45}^{-}~ e_ {43}^{+}~ e_ {33}^{-}~ e_ {36}^{+}~ e_{56}^{-})~,~
&
\nn\\
&
f_3 = (e_ {13}^{+}~ e_ {23}^{-}~ e_ {21}^{+}~ e_ {11}^{-})~,~
f_4 = (e_ {33}^{+}~ e_ {13}^{-}~ e_ {14}^{+}~ e_ {34}^{-})~,~
&
\nn\\
&
f_5 = (e_ {12}^{+}~ e_ {22}^{-}~ e_ {23}^{+}~ e_ {43}^{-}~ e_ {44}^{+}~ e_{14}^{-})~,~
f_6 = (e_{45}^{+}~ e_{65}^{-}~ e_{66}^{+}~ e_{36}^{-}~ e_{34}^{+}~e_{44}^{-})~,~
&
\nn\\
&
f_7 = (e_ {61}^{+}~ e_ {21}^{-}~ e_ {22}^{+}~ e_ {52}^{-}~ e_ {56}^{+}~ e_{66}^{-})~,~
&
\eea
which satisfy the following relations, 
\beal{es14a05}
&
f_5 f_6 f_7 = z_3 z_7^{-1}~,~ f_4 f_7^{-1} = z_2 z_4 z_5 z_7~,~ f_3 f_6^{-1} = z_1 z_5 z_7~,~
&
\nn\\
&
f_2 f_5^{-1} f_7^{-1} = z_2 z_3^{-1} z_5^{-1} z_7~,~f_1 f_5^{-1} f_6^{-1} = z_2^{-1} z_3^{-1} z_6 z_7~,~ f_1 f_2 f_3 f_4 f_5 f_6 f_7=1~.~
&
\eea
The face paths can be written in terms of the canonical variables as follows, 
\beal{es14a05_1}
&
f_1=e^{-P} z_2^{-1} z_6~,~ 
f_2= e^{-Q} z_2 z_5^{-1}~,~ 
f_3= e^{Q} z_1 z_5 z_7~,~  
f_4= e^{P} z_2 z_4 z_5 z_7~,~ 
&
\nn\\
&
f_5= e^{-Q-P} z_3 z_7^{-1} ~,~ 
f_6= e^{Q} ~,~ 
f_7= e^{P} ~.~ 
\eea

The Kasteleyn matrix of the brane tiling for Model 5 
in \fref{mf_05} is given by, 
\beal{es14a06}
K = 
\left(
\ba{c|cccccc}
\; & b_1 & b_2 & b_3 & b_4 & b_5 & b_6  
\\
\hline
w_1 & e_{11} & e_{12} & e_{13} & e_{14} & 0 & 0
\\
w_2 & e_{21} x & e_{22} & e_{23} x & 0 & 0 & 0
\\
w_3 & 0 & 0 & e_{33} & e_{34} & 0 & e_{36}
\\
w_4 & 0 & 0 & e_{43} x & e_{44} & e_{45} x & 0
\\
w_5 & 0 & e_{52} y & 0 & 0 & e_{55} & e_{56}
\\
w_6 & e_{61} xy & 0 & 0 & 0 & e_{65}x & e_{66}
\ea
\right)
~.~
\eea
By taking the permanent of the Kasteleyn matrix,
we obtain the spectral curve of the dimer integrable system for Model 5 as follows,
\beal{es14a07}
0 = \text{perm}~K&=&\overline{p}_0
\cdot x^2 y \cdot
\Big[
\delta_{(-2,-1)} \frac{1}{x^2 y} 
+ \delta_{(-1,-1)} \frac{1}{x y} 
+ \delta_{(-1,0)} \frac{1}{x} 
+ \delta_{(0,-1)} \frac{1}{y}
\nn\\
&& 
+ \delta_{(1,-1)} \frac{x}{y} 
+ \delta_{(1,0)} x 
+ \delta_{(1,1)} x y 
+ H
\Big]
~,~
\eea
where $\overline{p}_0= e_{14}^{+} e_{23}^{+} e_{36}^{+} e_{45}^{+} e_{52}^{+} e_{61}^{+} $.
The Casimirs $\delta_{(m,n)}$ in \eref{es14a07} can be expressed in terms of the zig-zag paths in \eref{es14a03} as follows, 
\beal{es14a08}
&
\delta_{(-2,-1)} =z_2 z_3 z_5 z_6 z_7 ~,~
\delta_{(-1,-1)} =z_3 z_5 z_6 z_7 + z_2 z_3 z_6 z_7 + z_2 z_3 z_5 z_7~,~
&
\nn\\
&
\delta_{(-1,0)} =z_4^{-1} ~,~
\delta_{(0,-1)} =z_3 z_6 z_7 + z_3 z_5 z_7 + z_2 z_3 z_7~,~
\delta_{(1,-1)} = z_3 z_7 ~,~
&
\nn\\
&
\delta_{(1,0)} = z_3 + z_7~,~
\delta_{(1,1)} = 1~,~
&
\eea
such that the spectral curve for Model 5 takes the following form, 
\beal{es14a09}
&&
\Sigma~:~
z_2 z_3 z_5 z_6 z_7 \frac{1}{x^2 y}+ (z_3 z_5 z_6 z_7 + z_2 z_3 z_6 z_7 + z_2 z_3 z_5 z_7)\frac{1}{xy}+
\frac{1}{z_4 x}
\nn\\
&&
\hspace{1cm}
+(z_3 z_6 z_7 + z_3 z_5 z_7 + z_2 z_3 z_7)\frac{1}{y}+z_3 z_7 \frac{x}{y}+(z_3+z_7)x+xy+H 
~.~
\eea

The Hamiltonian is a sum over all 9 1-loops $\gamma_i$,
\beal{es14a10}
H=\sum_{i=1}^{9} \gamma_i~,~
\eea
where the 1-loops $\gamma_i$ can be expressed in terms of zig-zag paths and face paths as follows,
\beal{es14a11}
&
\gamma_1 = z_3 z_5 f_2~,~
\gamma_2 = z_2^{-1} z_3 z_5 z_6 f_2 f_7^{-1}~,~
\gamma_3 = z_3 z_6 f_7^{-1}~,~
&
\nn\\
&
\gamma_4 = z_2^{-1} z_4^{-1} f_7^{-1}~,~
\gamma_5 = z_3 z_6 f_3 f_7^{-1}~,~
\gamma_6 =z_3 z_6 f_3~,~
&
\nn\\
&
\gamma_7 = z_1^{-1} f_3~,~
\gamma_8 = z_3 z_6 f_3 f_4~,~
\gamma_9 = z_4^{-1} z_5^{-1} f_4~.~
\eea

The commutation matrix $C$ for Model 5 takes the following form,
\beal{es14a12}
&&
C=
\left(
\ba{c|ccccccccc}
\; & \gamma_1
& \gamma_2
& \gamma_3
& \gamma_4 
& \gamma_5  
& \gamma_6 
& \gamma_7 
& \gamma_8 
& \gamma_9 
\\
\hline
\gamma_1 &      0 & 1 & 1 & 1 & 1 & 0 & 0 & -1 & -1 \\
\gamma_2 &       -1 & 0 & 1 & 1 & 2 & 1 & 1 & 0 & -1 \\
\gamma_3 &       -1 & -1 & 0 & 0 & 1 & 1 & 1 & 1 & 0 \\
\gamma_4 &       -1 & -1 & 0 & 0 & 1 & 1 & 1 & 1 & 0 \\
\gamma_5 &       -1 & -2 & -1 & -1 & 0 & 1 & 1 & 2 & 1 \\
\gamma_6 &       0 & -1 & -1 & -1 & -1 & 0 & 0 & 1 & 1 \\
\gamma_7 &       0 & -1 & -1 & -1 & -1 & 0 & 0 & 1 & 1 \\
\gamma_8 &       1 & 0 & -1 & -1 & -2 & -1 & -1 & 0 & 1 \\
\gamma_9 &       1 & 1 & 0 & 0 & -1 & -1 & -1 & -1 & 0 \\
\ea
\right)
~,~
\eea
where 
the 1-loops 
satisfying the commutation relations
can be written in terms of the canonical variables as follows, 
\beal{es14a13}
&
\gamma_1 = e^{-Q} z_2 z_3~,~
\gamma_2 = e^{-Q-P} z_3 z_6~,~
\gamma_3 = e^{-P} z_3 z_6~,~
&
\nn\\
&
\gamma_4 = e^{-P} z_2^{-1} z_4^{-1} ~,~
\gamma_5 =e^{Q-P} z_2^{-1} z_4^{-1} ~,~
\gamma_6 = e^{Q} z_2^{-1} z_4^{-1} ~,~
&
\nn\\
&
\gamma_7 =e^{Q} z_5 z_7 ~,~
\gamma_8 = e^{Q+P} z_5 z_7 ~,~
\gamma_9 = e^{P} z_2 z_7~.~
&
\eea
\\

%=================================================================
\section{Model 6: $\text{PdP}_{4a}$}
%=================================================================

%=================================================================
\subsection{Model 6a}
%=================================================================
%------------------------------------------------------------------------------------------------------------------
\begin{figure}[H]
\begin{center}
\resizebox{0.9\hsize}{!}{
\includegraphics{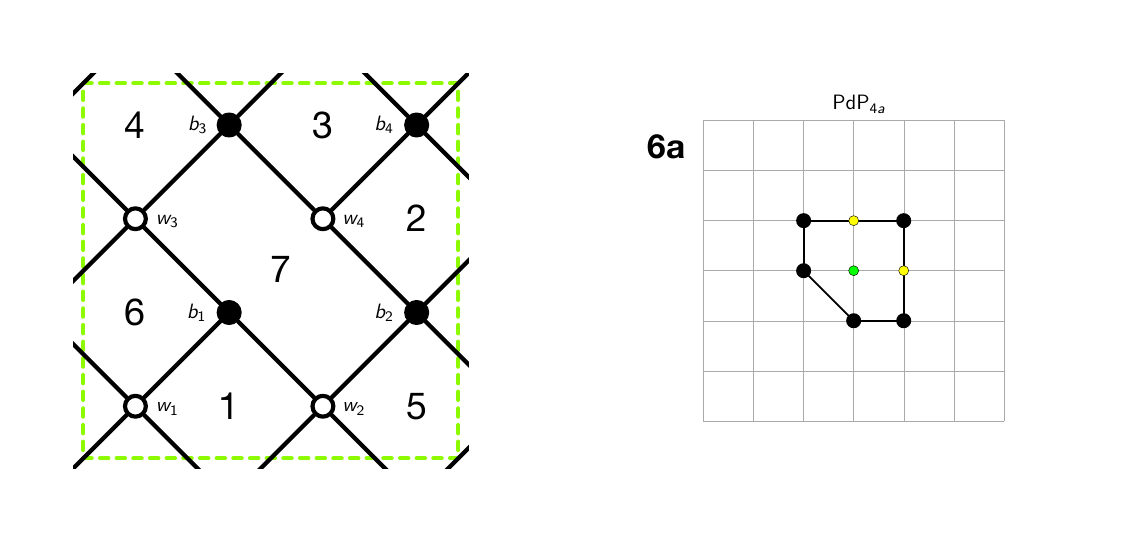}
}
\vspace{-0.5cm}
\caption{The brane tiling and toric diagram of Model 6a.}
\label{mf_06A}
 \end{center}
 \end{figure}
%------------------------------------------------------------------------------------------------------------------

The brane tiling for Model 6a 
can be expressed in terms of the following pair of permutation tuples
\beal{es15a01}
\sigma_B &=& (e_{11}\ e_{21}\ e_{31})\ (e_{12}\ e_{32}\ e_{42}\ e_{22})\ (e_{13}\ e_{33}\ e_{43}\ e_{23})\ (e_{41}\ e_{42}\ e_{44}\ e_{43})
\nn\\
\sigma_W^{-1} &=& (e_{11}\ e_{13}\ e_{14}\ e_{12})\ (e_{21}\ e_{22}\ e_{24}\ e_{23})\ (e_{31}\ e_{32}\ e_{34}\ e_{33})\ (e_{42}\ e_{43}\ e_{44})
\eea
which correspond to black and white nodes in the brane tiling, respectively.\\

The brane tiling for Model 6a has 7 zig-zag paths 
given by, 
\beal{es15a03}
&
z_1 = (e_ {23}^{+}~ e_ {13}^{-}~ e_ {14}^{+}~ e_ {24}^{-})~,~
z_2 = (e_ {11}^{+}~ e_ {21}^{-}~ e_ {22}^{+}~ e_ {12}^{-})~,~
&
\nn\\
&
z_3 = (e_ {44}^{+}~ e_ {34}^{-}~ e_ {33}^{+}~ e_ {43}^{-})~,~
z_4 = (e_ {21}^{+}~ e_ {31}^{-}~ e_ {32}^{+}~ e_ {42}^{-}~ e_ {43}^{+}~ e_{23}^{-})~,~
&
\nn\\
&
z_5 = (e_ {42}^{+}~ e_ {22}^{-}~ e_ {24}^{+}~ e_ {44}^{-})~,~
z_6 = (e_ {34}^{+}~ e_ {14}^{-}~ e_ {12}^{+}~ e_ {32}^{-})~,~
&
\nn\\
&
z_7 = (e_ {13}^{+}~ e_ {33}^{-}~ e_ {31}^{+}~ e_ {11}^{-})~,~
&
\eea
and 7 face paths given by, 
\beal{es15a04}
&
f_1 = (e_ {13}^{+}~ e_ {23}^{-}~ e_ {21}^{+}~ e_ {11}^{-})~,~
f_2 = (e_ {34}^{+}~ e_ {44}^{-}~ e_ {42}^{+}~ e_ {32}^{-})~,~
&
\nn\\
&
f_3 = (e_ {23}^{+}~ e_ {43}^{-}~ e_ {44}^{+}~ e_ {24}^{-})~,~
f_4 = (e_ {14}^{+}~ e_ {34}^{-}~ e_ {33}^{+}~ e_ {13}^{-})~,~
&
\nn\\
&
f_5 = (e_ {24}^{+}~ e_ {14}^{-}~ e_ {12}^{+}~ e_ {22}^{-})~,~
f_6 = (e_ {11}^{+}~ e_ {31}^{-}~ e_ {32}^{+}~ e_ {12}^{-})~,~
&
\nn\\
&
f_7 = (e_ {43}^{+}~ e_ {33}^{-}~ e_ {31}^{+}~ e_ {21}^{-}~ e_ {22}^{+}~ e_{42}^{-})~,~
&
\eea
which satisfy the following relations, 
\beal{es15a05}
&
f_4 f_6 = z_6^{-1} z_7^{-1}~,~ 
f_3 f_6^{-1} =  z_1 z_3 z_6 z_7~,~
f_1 f_2^{-1}=z_3 z_4 z_7~,~
&
\nn\\
&
f_2 f_4^{-1} f_7 = z_2 z_3^{-1} z_6 z_7~,~
f_5 f_6^{-1} f_7^{-1}=z_2^{-1} z_3 z_5 z_6~,~
f_1 f_2 f_3 f_4 f_5 f_6 f_7=1~.~
&
\eea
The face paths can be written in terms of the canonical variables as follows, 
\beal{es15a05_1}
&
f_1=e^{P} ~,~ 
f_2= e^{P} z_1 z_2 z_5 z_6~,~ 
f_3= e^{Q} z_2^{-1} z_4^{-1} z_5^{-1}~,~  
f_4= e^{-Q} z_1 z_2 z_3 z_4 z_5~,~ 
&
\nn\\
&
f_5= e^{-P} z_1^{-1} z_2^{-1} ~,~ 
f_6= e^{Q} ~,~ 
f_7= e^{-Q-P} z_2 z_4 z_7~.~ 
\eea

The Kasteleyn matrix of the brane tiling for Model 6a in \fref{mf_06A} is 
given by, 
\beal{es15a06}
K = 
\left(
\ba{c|cccc}
\; & b_1 & b_2 & b_3 & b_4 
\\
\hline
w_1 & e_{11} & e_{12} x^{-1} & e_{13} y^{-1}& e_{14} x^{-1} y^{-1}
\\
w_2 & e_{21} x & e_{22} & e_{23} y^{-1} & e_{24} y^{-1}
\\
w_3 & e_{31} & e_{32} x^{-1} & e_{33} & e_{34} x^{-1}
\\
w_4 & 0 & e_{42} & e_{43} x & e_{44}
\ea
\right)
~.~
\eea
By taking the permanent of the Kasteleyn matrix, 
we obtain the spectral curve of the dimer integrable system for Model 6a as follows,
\beal{es15a07}
0 = \text{perm}~K&=&\overline{p}_0
\cdot x^{-1} y^{-1} \cdot
\Big[
\delta_{(-1,0)} \frac{1}{x} 
+ \delta_{(-1,1)} \frac{y}{x} 
+ \delta_{(0,-1)} \frac{1}{y} 
+ \delta_{(0,1)} y
\nn\\
&& 
+ \delta_{(1,-1)} \frac{x}{y} 
+ \delta_{(1,0)} x 
+ \delta_{(1,1)} x y 
+ H
\Big]
~,~
\eea
where $\overline{p}_0= e_{11}^{+} e_{22}^{+} e_{33}^{+} e_{44}^{+}$.
The Casimirs $\delta_{(m,n)}$ in \eref{es15a07} can be expressed in terms of the zig-zag paths in \eref{es15a03} as follows, 
\beal{es15a08}
&
\delta_{(-1,0)} =z_1 z_4 z_5 z_7~,~
\delta_{(-1,1)} =z_1 z_4 z_5 z_6 z_7~,~
\delta_{(0,-1)} =z_1 z_5 z_7~,~
&
\nn\\
&
\delta_{(0,1)} =z_2^{-1}+z_3^{-1}~,~
\delta_{(1,-1)} = z_5 z_7 ~,~
\delta_{(1,0)} = z_5 + z_7~,~
\delta_{(1,1)} = 1~,~
&
\eea
which allows us to express the spectral curve of Model 6a
in the following form, 
\beal{es15a09}
\Sigma ~:~
\Big (\frac{1}{z_2}+\frac{1}{z_3} \Big) y+z_1 z_5 z_7 \frac{1}{y}+(y+z_5)(y+z_7) \frac{x}{y}+(1+y z_6)\frac{z_1 z_4 z_5 z_7}{x}+H
= 0 ~.~
\eea

The Hamiltonian is a sum over all 9 1-loops $\gamma_i$,
\beal{es15a10}
H=\sum_{i=1}^{9} \gamma_i~,~
\eea
where the 1-loops $\gamma_i$'s can be expressed in terms of zig-zag paths and face paths as follows,
\beal{es15a11}
&
\gamma_1 = z_3^{-1} z_7 f_4~,~
\gamma_2 = z_1^{-1} z_2^{-1} z_3^{-1} z_7 f_1^{-1} f_4~,~
\gamma_3 = z_1 z_7 f_5~,~
&
\nn\\
&
\gamma_4 = z_3^{-1} z_6^{-1} f_5~,~
\gamma_5 = z_4 z_5 z_7 f_1^{-1} f_3~,~
\gamma_6 = z_4 z_5 z_7 f_3~,~
&
\nn\\
&
\gamma_7 = z_3^{-1} z_5 f_3~,~
\gamma_8 = z_3^{-1} z_5 f_1 f_3~,~
\gamma_9 = z_1 z_5 f_1~.~
\eea

The commutation matrix $C$ for Model 6a is given by, 
\beal{es15a12}
&&
C=
\left(
\ba{c|ccccccccc}
\; & \gamma_1
& \gamma_2
& \gamma_3
& \gamma_4 
& \gamma_5  
& \gamma_6 
& \gamma_7 
& \gamma_8 
& \gamma_9 
\\
\hline
\gamma_1 &       0 & 1 & 1 & 1 & 1 & 0 & 0 & -1 & -1 \\
\gamma_2 &       -1 & 0 & 1 & 1 & 2 & 1 & 1 & 0 & -1 \\
\gamma_3 &       -1 & -1 & 0 & 0 & 1 & 1 & 1 & 1 & 0 \\
\gamma_4 &       -1 & -1 & 0 & 0 & 1 & 1 & 1 & 1 & 0 \\
\gamma_5 &       -1 & -2 & -1 & -1 & 0 & 1 & 1 & 2 & 1 \\
\gamma_6 &       0 & -1 & -1 & -1 & -1 & 0 & 0 & 1 & 1 \\
\gamma_7 &       0 & -1 & -1 & -1 & -1 & 0 & 0 & 1 & 1 \\
\gamma_8 &       1 & 0 & -1 & -1 & -2 & -1 & -1 & 0 & 1 \\
\gamma_9 &       1 & 1 & 0 & 0 & -1 & -1 & -1 & -1 & 0 \\
\ea
\right)
~,~
\eea
where the 1-loops satisfying the commutation relations can be written in terms of the canonical variables as follows, 
\beal{es15a13}
&
\gamma_1 = e^{-Q} z_1 z_2 z_4 z_5 z_7~,~
\gamma_2 = e^{-Q-P} z_4 z_5 z_7~,~
\gamma_3 = e^{-P} z_2^{-1} z_7~,~
&
\nn\\
&
\gamma_4 = e^{-P} z_4 z_5 z_7~,~
\gamma_5 =e^{Q-P} z_2^{-1} z_7~,~
\gamma_6 = e^{Q} z_2^{-1} z_7~,~
&
\nn\\
&
\gamma_7 =e^{Q} z_2^{-1} z_3^{-1} z_4^{-1}~,~
\gamma_8 = e^{Q+P} z_2^{-1} z_3^{-1} z_4^{-1}~,~
\gamma_9 = e^{P} z_1 z_5~.~
&
\eea
\\

%=================================================================
\subsection{Model 6b}
%=================================================================
%------------------------------------------------------------------------------------------------------------------
\begin{figure}[H]
\begin{center}
\resizebox{0.9\hsize}{!}{
\includegraphics{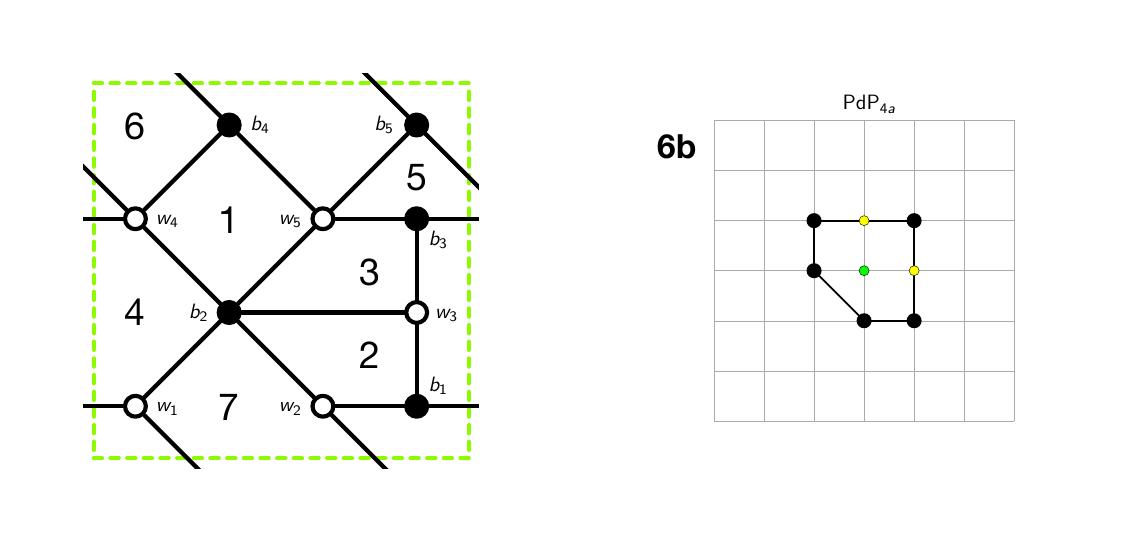}
}
\vspace{-0.5cm}
\caption{The brane tiling and toric diagram of Model 6b.}
\label{mf_06B}
 \end{center}
 \end{figure}
%------------------------------------------------------------------------------------------------------------------

The brane tiling for Model 6b 
can be expressed in terms of the following pair of permutation tuples
\beal{es15b01}
\sigma_B &=& (e_{11}\ e_{31}\ e_{21})\ (e_{12}\ e_{22}\ e_{32}\ e_{52}\ e_{42})\ (e_{33}\ e_{43}\ e_{53})\ (e_{14}\ e_{44}\ e_{54})
\nn\\
&& 
(e_{25}\ e_{55}\ e_{45})
\nn\\
\sigma_W &=& (e_{11}\ e_{12}\ e_{14})\ (e_{21}\ e_{25}\ e_{22})\ (e_{31}\ e_{32}\ e_{33})\ (e_{42}\ e_{43}\ e_{45}\ e_{44})
\nn\\
&&
(e_{52}\ e_{54}\ e_{55}\ e_{53})
\eea
which correspond to black and white nodes in the brane tiling, respectively.\\

The brane tiling for Model 6b has 7 zig-zag paths given by, 
\beal{es15b03}
&
z_1 = (e_ {52}^{+}~ e_ {42}^{-}~ e_ {43}^{+}~ e_ {53}^{-})~,~
z_2 = (e_ {44}^{+}~ e_ {54}^{-}~ e_ {55}^{+}~ e_ {45}^{-})~,~
z_3 = (e_ {22}^{+}~ e_ {32}^{-}~ e_ {33}^{+}~ e_ {43}^{-}~ e_ {45}^{+}~ e_{25}^{-})~,~
&
\nn\\
&
z_4 = (e_ {31}^{+}~ e_ {21}^{-}~ e_ {25}^{+}~ e_ {55}^{-}~ e_ {53}^{+}~ e_{33}^{-})~,~
z_5 = (e_ {54}^{+}~ e_ {14}^{-}~ e_ {11}^{+}~ e_ {31}^{-}~ e_ {32}^{+}~ e_{52}^{-})~,~
&
\nn\\
&
z_6 = (e_ {42}^{+}~ e_ {12}^{-}~ e_ {14}^{+}~ e_ {44}^{-})~,~
z_7 = (e_ {21}^{+}~ e_ {11}^{-}~ e_ {12}^{+}~ e_ {22}^{-})~.~
&
\eea
and 7 face paths given by, 
\beal{es15b04}
&
f_1 = (e_ {13}^{+}~ e_ {23}^{-}~ e_ {21}^{+}~ e_ {11}^{-})~,~
f_2 = (e_ {34}^{+}~ e_ {44}^{-}~ e_ {42}^{+}~ e_ {32}^{-})~,~
&
\nn\\
&
f_3 = (e_ {23}^{+}~ e_ {43}^{-}~ e_ {44}^{+}~ e_ {24}^{-})~,~
f_4 = (e_ {14}^{+}~ e_ {34}^{-}~ e_ {33}^{+}~ e_ {13}^{-})~,~
&
\nn\\
&
f_5 = (e_ {24}^{+}~ e_ {14}^{-}~ e_ {12}^{+}~ e_ {22}^{-})~,~
f_6 = (e_ {11}^{+}~ e_ {31}^{-}~ e_ {32}^{+}~ e_ {12}^{-})~,~
&
\nn\\
&
f_7 = (e_ {43}^{+}~ e_ {33}^{-}~ e_ {31}^{+}~ e_ {21}^{-}~ e_ {22}^{+}~ e_{42}^{-})~,~
&
\eea
which satisfy the following relations, 
\beal{es15b05}
&
f_6 f_7 = z_2 z_7^{-1}~,~
f_4 f_5 f_7^{-1} = z_2^{-1} z_4 z_6^{-1} z_7~,~
f_3 f_6^{-1}= z_1 z_3 z_6 z_7~,~
&
\nn\\
&
f_2 f_5^{-1}= z_3^{-1} z_4^{-1}~,~
f_1 f_4^{-1} f_6^{-1}= z_1^{-1} z_2^{-1} z_4^{-1} z_6~,~
f_1 f_2 f_3 f_4 f_5 f_6 f_7=1~.~
&
\eea
The face paths can be written in terms of the canonical variables as follows, 
\beal{es15b05_1}
&
f_1=e^{-Q} z_5 z_6 z_7~,~ 
f_2= e^{Q} ~,~ 
f_3= e^{P} z_1 z_3 z_6 z_7~,~  
f_4= e^{-Q-P} z_3^{-1} z_6^{-1}~,~ 
&
\nn\\
&
f_5= e^{Q} z_3 z_4 ~,~ 
f_6= e^{P} ~,~ 
f_7= e^{-P} z_2 z_7^{-1}~.~ 
\eea

The Kasteleyn matrix of the brane tiling for Model 6b in \fref{mf_06B} is 
given by, 
\beal{es15b06}
K = 
\left(
\ba{c|ccccc}
\; & b_1 & b_2 & b_3 & b_4 & b_5
\\
\hline
w_1 & e_{11} x^{-1} & e_{12}  & 0 & e_{14} y^{-1} & 0
\\
w_2 & e_{21} x & e_{22} & 0 & 0 & e_{25} y^{-1}
\\
w_3 & e_{31} & e_{32} & e_{33} & 0 & 0
\\
w_4 & 0 & e_{42} & e_{43} x^{-1} & e_{44} & e_{45} x^{-1}
\\
w_5 & 0 & e_{52} & e_{53} x & e_{54} & e_{55}
\ea
\right)
~.~
\eea
The permanent of the Kasteleyn matrix
gives the expression for the spectral curve of the dimer integrable system for Model 6b as follows,
\beal{es15b07}
0 = \text{perm}~K&=&\overline{p}_0
\cdot x^{-1} y^{-1} \cdot
\Big[
\delta_{(-1,0)} \frac{1}{x} 
+ \delta_{(-1,1)} \frac{y}{x} 
+ \delta_{(0,-1)} \frac{1}{y} 
+ \delta_{(0,1)} y
\nn\\
&& 
+ \delta_{(1,-1)} \frac{x}{y} 
+ \delta_{(1,0)} x 
+ \delta_{(1,1)} x y 
+ H
\Big]
~,~
\eea
where $\overline{p}_0= e_{12}^{+} e_{21}^{+} e_{33}^{+} e_{44}^{+} e_{55}^{+}$.
The Casimirs $\delta_{(m,n)}$ in \eref{es15b07} can be expressed in terms of the zig-zag paths in \eref{es15b03} as shown below, 
\beal{es15b08}
&
\delta_{(-1,0)} =z_1 z_4 z_5 z_6~,~
\delta_{(-1,1)} =z_1 z_3 z_4 z_5 z_6~,~
\delta_{(0,-1)} =z_1 z_4 z_6~,~
&
\nn\\
&
\delta_{(0,1)} =z_2^{-1}+z_7^{-1}~,~
\delta_{(1,-1)} = z_4 z_6 ~,~
\delta_{(1,0)} = z_4 + z_6~,~
\delta_{(1,1)} = 1~,~
&
\eea
such that the spectral curve for Model 6b can be written in the following form,  
\beal{es15b09}
&&
\Sigma~:~
\Big (\frac{1}{z_2}+\frac{1}{z_7} \Big) y+z_1 z_4 z_6 \frac{1}{y}+(y+z_4)(y+z_6) \frac{x}{y}+(1+y z_3)\frac{z_1 z_4 z_5 z_6}{x}+H
= 0 ~.~
\nn\\
\eea

The Hamiltonian is a sum over all 9 1-loops $\gamma_i$,
\beal{es15b10}
H=\sum_{i=1}^{9} \gamma_i~,~
\eea
where the 1-loops $\gamma_i$'s can be expressed in terms of zig-zag paths and face paths as follows,
\beal{es15b11}
&
\gamma_1 = z_1 z_4 f_1~,~
\gamma_2 = z_1 z_4 f_1 f_7~,~
\gamma_3 = z_2^{-1} z_4 f_7~,~
&
\nn\\
&
\gamma_4 = z_2^{-1} z_3^{-1} f_7~,~
\gamma_5 = z_2^{-1} z_4 f_2 f_7~,~
\gamma_6 = z_4 z_7^{-1} f_2~,~
&
\nn\\
&
\gamma_7 = z_1 z_3 z_4 z_6 f_2~,~
\gamma_8 = z_4 z_7^{-1} f_2 f_3~,~
\gamma_9 = z_4 z_5 z_6 f_3~.~
\eea

The commutation matrix $C$ for Model 6b is given by, 
\beal{es15b12}
&&
C=
\left(
\ba{c|ccccccccc}
\; & \gamma_1
& \gamma_2
& \gamma_3
& \gamma_4 
& \gamma_5  
& \gamma_6 
& \gamma_7 
& \gamma_8 
& \gamma_9 
\\
\hline
\gamma_1 &       0 & 1 & 1 & 1 & 1 & 0 & 0 & -1 & -1 \\
\gamma_2 &       -1 & 0 & 1 & 1 & 2 & 1 & 1 & 0 & -1 \\
\gamma_3 &       -1 & -1 & 0 & 0 & 1 & 1 & 1 & 1 & 0 \\
\gamma_4 &       -1 & -1 & 0 & 0 & 1 & 1 & 1 & 1 & 0 \\
\gamma_5 &       -1 & -2 & -1 & -1 & 0 & 1 & 1 & 2 & 1 \\
\gamma_6 &       0 & -1 & -1 & -1 & -1 & 0 & 0 & 1 & 1 \\
\gamma_7 &       0 & -1 & -1 & -1 & -1 & 0 & 0 & 1 & 1 \\
\gamma_8 &       1 & 0 & -1 & -1 & -2 & -1 & -1 & 0 & 1 \\
\gamma_9 &       1 & 1 & 0 & 0 & -1 & -1 & -1 & -1 & 0 \\
\ea
\right)
~.~
\eea
The 1-loops satisfying the commutation relations 
can be written in terms of the canonical variables as follows, 
\beal{es15b13}
&
\gamma_1 = e^{-Q} z_2^{-1} z_3^{-1}~,~
\gamma_2 = e^{-Q-P} z_3^{-1} z_7^{-1}~,~
\gamma_3 = e^{-P} z_4 z_7^{-1}~,~
&
\nn\\
&
\gamma_4 =e^{-P} z_3^{-1} z_7^{-1}~,~
\gamma_5 =e^{Q-P} z_4 z_7^{-1}~,~
\gamma_6 = e^{Q} z_4 z_7^{-1}~,~
&
\nn\\
&
\gamma_7 = e^{Q} z_1 z_3 z_4 z_6~,~
\gamma_8 = e^{Q+P} z_1 z_3 z_4 z_6~,~
\gamma_9 = e^{P} z_2^{-1} z_6~.~
&
\eea
\\

%=================================================================
\subsection{Model 6c}
%=================================================================
%------------------------------------------------------------------------------------------------------------------
\begin{figure}[H]
\begin{center}
\resizebox{0.9\hsize}{!}{
\includegraphics{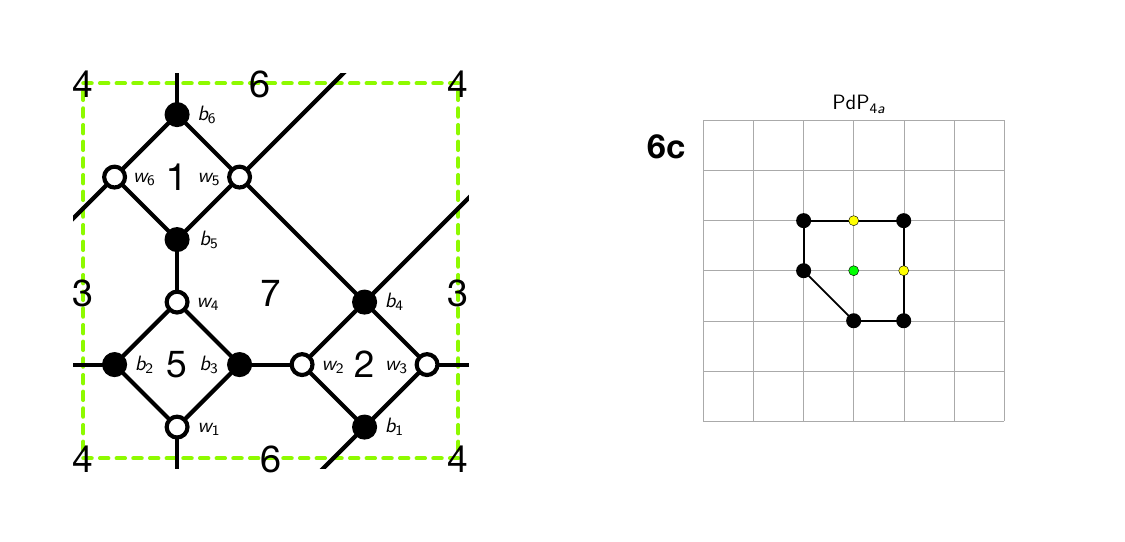}
}
\vspace{-0.5cm}
\caption{The brane tiling and toric diagram of Model 6c.}
\label{mf_06C}
 \end{center}
 \end{figure}
%------------------------------------------------------------------------------------------------------------------

The brane tiling for Model 6c 
can be expressed in terms of the following pair of permutation tuples
\beal{es15c01}
\sigma_B &=& (e_{21}\ e_{51}\ e_{31})\ (e_{12}\ e_{42}\ e_{32})\ (e_{13}\ e_{23}\ e_{43})\ (e_{24}\ e_{34}\ e_{64}\ e_{54})
\nn\\
&&
(e_{45}\ e_{55}\ e_{65})\ (e_{16}\ e_{66}\ e_{56}) 
\nn\\
\sigma_W^{-1} &=& (e_{12}\ e_{13}\ e_{16})\ (e_{21}\ e_{23}\ e_{24})\ (e_{31}\ e_{34}\ e_{32})\ (e_{42}\ e_{45}\ e_{43})\
\nn\\
&&
(e_{51}\ e_{54}\ e_{55}\ e_{56})\ (e_{64}\ e_{66}\ e_{65})
\eea
which correspond to black and white nodes in the brane tiling, respectively.\\
 
The brane tiling for Model 6c has 7 zig-zag paths given by, 
\beal{es15c03}
&
z_1 = (e_ {55}^{+}~ e_ {65}^{-}~ e_ {64}^{+}~ e_ {54}^{-})~,~
z_2 = (e_ {66}^{+}~ e_ {56}^{-}~ e_ {51}^{+}~ e_ {31}^{-}~ e_ {34}^{+}~ e_{64}^{-})~,~
&
\nn\\
&
z_3 = (e_ {24}^{+}~ e_ {34}^{-}~ e_ {32}^{+}~ e_ {12}^{-}~ e_ {13}^{+}~ e_{23}^{-})~,~
z_4 = (e_ {31}^{+}~ e_ {21}^{-}~ e_ {23}^{+}~ e_ {43}^{-}~ e_ {42}^{+}~ e_{32}^{-})~,~
&
\nn\\
&
z_5 = (e_ {65}^{+}~ e_ {45}^{-}~ e_ {43}^{+}~ e_ {13}^{-}~ e_ {16}^{+}~ e_{66}^{-})~,~
z_6 = (e_ {56}^{+}~ e_ {16}^{-}~ e_ {12}^{+}~ e_ {42}^{-}~ e_ {45}^{+}~ e_{55}^{-})~,~
&
\nn\\
&
z_7 = (e_ {21}^{+}~ e_ {51}^{-}~ e_ {54}^{+}~ e_ {24}^{-})~.~
&
\eea
and 7 face paths given by, 
\beal{es15c04}
&
f_1 = (e_ {65}^{+}~ e_ {55}^{-}~ e_ {56}^{+}~ e_ {66}^{-})~,~
f_2 = (e_ {34}^{+}~ e_ {24}^{-}~ e_ {21}^{+}~ e_ {31}^{-})~,~
&
\nn\\
&
f_3 = (e_ {32}^{+}~ e_ {42}^{-}~ e_ {45}^{+}~ e_ {65}^{-}~ e_ {64}^{+}~ e_{34}^{-})~,~
f_4 = (e_ {66}^{+}~ e_ {16}^{-}~ e_ {12}^{+}~ e_ {32}^{-}~ e_ {31}^{+}~ e_{51}^{-}~ e_ {54}^{+}~ e_ {64}^{-})~,~
&
\nn\\
&
f_5 = (e_ {42}^{+}~ e_ {12}^{-}~ e_ {13}^{+}~ e_ {43}^{-})~,~
f_6 = (e_ {51}^{+}~ e_ {21}^{-}~ e_ {23}^{+}~ e_ {13}^{-}~ e_ {16}^{+}~ e_{56}^{-})~,~
&
\nn\\
&
f_7 = (e_ {55}^{+}~ e_ {45}^{-}~ e_ {43}^{+}~ e_ {23}^{-}~ e_ {24}^{+}~ e_{54}^{-})~,~
&
\eea
satisfying the following relations, 
\beal{es15c05}
&
f_5 f_6 f_7=z_6^{-1} z_7^{-1}~,~
f_4 f_6^{-1} f_7^{-2}=z_2 z_3 z_4^2 z_6^2 z_7^3~,~
f_3 f_6^{-1}=z_1 z_3 z_6 z_7~,~
&
\nn\\
&
f_2 f_5^{-1}= z_3^{-1} z_4^{-1}~,~
f_1 f_5^{-1}=z_5 z_6~,~
f_1 f_2 f_3 f_4 f_5 f_6 f_7=1~.~
&
\eea
The face paths can be written in terms of the canonical variables as follows, 
\beal{es15c05_1}
&
f_1=e^{Q} z_5 z_6~,~ 
f_2= e^{Q} z_3^{-1} z_4^{-1}~,~ 
f_3= e^{P} z_2^{-1} z_4^{-1} z_5^{-1}~,~  
f_4= e^{-2Q-P} z_1^{-1} z_4 z_5^{-1} z_6^{-1}~,~ 
&
\nn\\
&
f_5= e^{Q} ~,~ 
f_6= e^{P} ~,~ 
f_7= e^{-Q-P} z_6^{-1} z_7^{-1}~.~ 
\eea

The Kasteleyn matrix of the brane tiling for Model 6c in \fref{mf_06C} 
is given by, 
\beal{es15c06}
K = 
\left(
\ba{c|cccccc}
\; & b_1 & b_2 & b_3 & b_4 & b_5 & b_6  
\\
\hline
w_1 & 0 & e_{12} & e_{13} & 0 & 0 & e_{16} y^{-1}
\\
w_2 & e_{21} & 0 & e_{23} & e_{24} & 0 & 0
\\
w_3 & e_{31} & e_{32} x & 0 & e_{34} & 0 & 0
\\
w_4 & 0 & e_{42} & e_{43} & 0 & e_{45} & 0
\\
w_5 & e_{51} y & 0 & 0 & e_{54} & e_{55} & e_{56}
\\
w_6 & 0 & 0 & 0 & e_{64} x^{-1} & e_{65} & e_{66}
\ea
\right)
~.~
\eea
By taking the permanent of the Kasteleyn matrix in \eref{es15c06} 
with a $GL(2,\mathbb{Z})$ transformation $M : (x,y) \mapsto (x,\frac{1}{y})$, 
we obtain the spectral curve of the dimer integrable system for Model 6c as follows,
\beal{es15c07}
0 &=&\overline{p}_0
\cdot 
\Big[
\delta_{(-1,0)} \frac{1}{x} 
+ \delta_{(-1,1)} \frac{y}{x} 
+ \delta_{(0,-1)} \frac{1}{y} 
+ \delta_{(0,1)} y
\nn\\
&& 
+ \delta_{(1,-1)} \frac{x}{y} 
+ \delta_{(1,0)} x 
+ \delta_{(1,1)} x y 
+ H
\Big]
~,~
\eea
where $\overline{p}_0= e_{13}^{+} e_{24}^{+} e_{32}^{+} e_{45}^{+} e_{51}^{+} e_{66}^{+}$.
The Casimirs $\delta_{(m,n)}$ in \eref{es15c07} can be expressed in terms of the zig-zag paths in \eref{es15c03} as shown below, 
\beal{es15c08}
&
\delta_{(-1,0)} =z_1 z_4 z_5 z_6 z_7~,~
\delta_{(-1,1)} = z_1 z_4 z_5 z_7~,~
\delta_{(0,-1)} = z_3^{-1}~,~
&
\nn\\
&
\delta_{(0,1)} = z_1 z_5 z_7+z_4 z_5 z_7~,~
\delta_{(1,-1)} = 1~,~
\delta_{(1,0)} = z_5 + z_7~,~
\delta_{(1,1)} = z_5 z_7~,~
&
\eea
allowing us to express the spectral curve of Model 6c in the following form,
\beal{es15c09}
&&
\Sigma ~:~
(z_5+z_7)x +\frac{1}{z_3 y}+(x+z_1)(x+z_4)\frac{z_5 z_7}{x} y+z_1 z_4 z_5 z_6 z_7 \frac{1}{x}+\frac{x}{y}+H
= 0 ~.~
\nn\\
\eea

The Hamiltonian is a sum over all 12 1-loops $\gamma_i$,
\beal{es15c10}
H=\sum_{i=1}^{12} \gamma_i~,~
\eea
where the 1-loops $\gamma_i$'s can be expressed in terms of zig-zag paths and face paths as follows, 
\beal{es15c11}
&
\gamma_1 = z_3^{-1} z_5 z_6^{-1} z_7^{-1} f_3 f_4 f_6^{-1} f_7^{-1}~,~
\gamma_2 = z_4 z_7 f_7~,~
\gamma_3 = z_3^{-1} z_6^{-1} f_6^{-1}~,~
&
\nn\\
&
\gamma_4 = z_1 z_5 f_4 ~,~
\gamma_5 = z_3^{-1} z_7 f_7 ~,~
\gamma_6 = z_2^{-1} z_3^{-1} z_4^{-1} f_5 f_6~,~
&
\nn\\
&
\gamma_7 = z_2^{-1} z_3^{-1} z_4^{-1} f_5 ~,~
\gamma_8 = z_3^{-1} z_5 f_5~,~
\gamma_9 = z_4 z_5 z_6 z_7 f_5 f_7~,~
&
\nn\\
&
\gamma_{10} = z_3^{-1} z_5 f_5 f_6^{-1}~,~
\gamma_{11} = z_1^{-1} z_2^{-1} z_3^{-1} f_7~,~
\gamma_{12} = z_3^{-1} z_5 z_6 z_7 f_5 f_7~.~
\eea

The commutation matrix $C$ for Model 6c takes the following form, 
\beal{es15c12}
&&
C=
\left(
\ba{c|cccccccccccc}
\; & \gamma_1
& \gamma_2
& \gamma_3
& \gamma_4 
& \gamma_5  
& \gamma_6 
& \gamma_7
& \gamma_8
& \gamma_9 
& \gamma_{10}  
& \gamma_{11}
& \gamma_{12}
\\
\hline
\gamma_1 &           0 & 1 & 1 & 1 & 1 & -1 & 0 & 0 & 1 & 1 & 1 & 1 \\
\gamma_2 &           -1 & 0 & 1 & -1 & 0 & 0 & 1 & 1 & 1 & 2 & 0 & 1 \\
\gamma_3 &           -1 & -1 & 0 & -2 & -1 & 1 & 1 & 1 & 0 & 1 & -1 & 0 \\
\gamma_4 &           -1 & 1 & 2 & 0 & 1 & -1 & 1 & 1 & 2 & 3 & 1 & 2 \\
\gamma_5 &           -1 & 0 & 1 & -1 & 0 & 0 & 1 & 1 & 1 & 2 & 0 & 1 \\
\gamma_6 &           1 & 0 & -1 & 1 & 0 & 0 & -1 & -1 & -1 & -2 & 0 & -1 \\
\gamma_7 &           0 & -1 & -1 & -1 & -1 & 1 & 0 & 0 & -1 & -1 & -1 & -1 \\\
\gamma_8 &           0 & -1 & -1 & -1 & -1 & 1 & 0 & 0 & -1 & -1 & -1 & -1 \\\
\gamma_9 &           -1 & -1 & 0 & -2 & -1 & 1 & 1 & 1 & 0 & 1 & -1 & 0 \\
\gamma_{10} &      -1 & -2 & -1 & -3 & -2 & 2 & 1 & 1 & -1 & 0 & -2 & -1 \\
\gamma_{11} &       -1 & 0 & 1 & -1 & 0 & 0 & 1 & 1 & 1 & 2 & 0 & 1 \\
\gamma_{12} &      -1 & -1 & 0 & -2 & -1 & 1 & 1 & 1 & 0 & 1 & -1 & 0 \\
\ea
\right)
~.~
\eea
The 1-loops satisfying the commutation relations
can be written in terms of the canonical variables as follows, 
\beal{es15c13}
&
\gamma_1 = e^{-Q} z_4 z_7~,~
\gamma_2 = e^{-P-Q} z_4 z_6^{-1}~,~
\gamma_3 = e^{-P} z_3^{-1} z_6^{-1}~,~
&
\nn\\
&
\gamma_4 = e^{-P-2Q} z_4 z_6^{-1}~,~
\gamma_5 =e^{-P-Q} z_1 z_2 z_4 z_5 z_7 ~,~
\gamma_6 = e^{P+Q} z_2^{-1} z_3^{-1} z_4^{-1}~,~
&
\nn\\
&
\gamma_7 = e^{Q} z_2^{-1} z_3^{-1} z_4^{-1}~,~
\gamma_8 =e^{Q} z_3^{-1} z_5 ~,~
\gamma_9 = e^{-P} z_4 z_5~,~
&
\nn\\
&
\gamma_{10} = e^{-P+Q} z_3^{-1} z_5~,~
\gamma_{11} =e^{-P-Q} z_4 z_5 ~,~
\gamma_{12} = e^{-P} z_3^{-1} z_5 ~.~
&
\eea\\

%=================================================================
\section{Model 7: $\mathbb{C}^3/\mathbb{Z}_6$ $(1,2,3)$, $\text{PdP}_{3a}$}
%=================================================================
%------------------------------------------------------------------------------------------------------------------
\begin{figure}[H]
\begin{center}
\resizebox{0.9\hsize}{!}{
\includegraphics{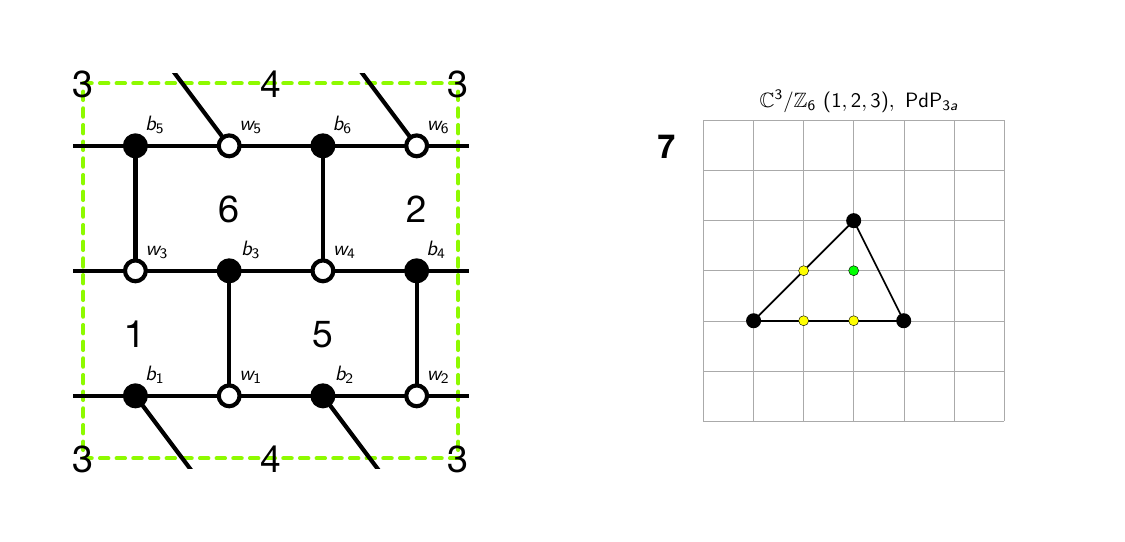}
}
\vspace{-0.5cm}
\caption{The brane tiling and toric diagram of Model 7.}
\label{mf_07}
 \end{center}
 \end{figure}
%------------------------------------------------------------------------------------------------------------------

The brane tiling for Model 7 
can be expressed in terms of the following pair of permutation tuples
\beal{es16a01}
\sigma_B &=& (e_{11}\ e_{21}\ e_{51})\ (e_{12}\ e_{62}\ e_{22})\ (e_{13}\ e_{43}\ e_{33})\ (e_{23}\ e_{34}\ e_{44})\ 
\nn\\
&&
(e_{35}\ e_{55}\ e_{65})\ (e_{46}\ e_{66}\ e_{56})
\nn\\
\sigma_W^{-1} &=& (e_{11}\ e_{13}\ e_{12})\ (e_{21}\ e_{22}\ e_{24})\ (e_{33}\ e_{34}\ e_{35})\ (e_{43}\ e_{46}\ e_{44})
\nn\\
&&
(e_{51}\ e_{56}\ e_{55})\ (e_{62}\ e_{65}\ e_{66})
\eea
which correspond to black and white nodes in the brane tiling, respectively.\\

 The brane tiling for Model 7 has 6 zig-zag paths given by, 
\beal{es16a03}
&
z_1 = (e_ {24}^{+}~ e_ {34}^{-}~ e_ {35}^{+}~ e_ {55}^{-}~ e_ {51}^{+}~ e_{11}^{-}~ e_ {13}^{+}~ e_ {43}^{-}~ e_ {46}^{+}~ e_ {66}^{-}~ e_{62}^{+}~ e_ {22}^{-})~,~
z_2 = (e_ {11}^{+}~ e_ {21}^{-}~ e_ {22}^{+}~ e_ {12}^{-})~,~
&
\nn\\
&
z_3 = (e_ {33}^{+}~ e_ {13}^{-}~ e_ {12}^{+}~ e_ {62}^{-}~ e_ {65}^{+}~ e_{35}^{-})~,~
z_4 = (e_ {34}^{+}~ e_ {44}^{-}~ e_ {43}^{+}~ e_ {33}^{-})~,~
&
\nn\\
&
z_5 = (e_ {66}^{+}~ e_ {56}^{-}~ e_ {55}^{+}~ e_ {65}^{-})~,~
z_6 = (e_ {21}^{+}~ e_ {51}^{-}~ e_ {56}^{+}~ e_ {46}^{-}~ e_ {44}^{+}~ e_{24}^{-})~,~
&
\eea
and 6 face paths given by, 
\beal{es16a04}
&
f_1 = (e_ {34}^{+}~ e_ {24}^{-}~ e_ {21}^{+}~ e_ {11}^{-}~ e_ {13}^{+}~ e_{33}^{-})~,~
f_2 = (e_ {66}^{+}~ e_ {46}^{-}~ e_ {44}^{+}~ e_ {34}^{-}~ e_ {35}^{+}~ e_{65}^{-})~,~
&
\nn\\
&
f_3 = (e_ {65}^{+}~ e_ {55}^{-}~ e_ {51}^{+}~ e_ {21}^{-}~ e_ {22}^{+}~ e_{62}^{-})~,~
f_4 = (e_ {11}^{+}~ e_ {51}^{-}~ e_ {56}^{+}~ e_ {66}^{-}~ e_ {62}^{+}~ e_{12}^{-})~,~
&
\nn\\
&
f_5 = (e_ {24}^{+}~ e_ {44}^{-}~ e_ {43}^{+}~ e_ {13}^{-}~ e_ {12}^{+}~ e_{22}^{-})~,~
f_6 = (e_ {33}^{+}~ e_ {43}^{-}~ e_ {46}^{+}~ e_ {56}^{-}~ e_ {55}^{+}~ e_{35}^{-})~,~
&
\eea
which satisfy the following relations, 
\beal{es16a05}
&
f_3 f_5 f_6= z_3 z_6^{-1} z_7^{-1}~,~
f_4 f_5^{-1} f_6^{-1} = z_1 z_2^2 z_4 z_6^2~,~
&
\nn\\
&
f_2 f_3^{-1} f_5^{-1}=z_1 z_2 z_5^2 z_6^2~,~
f_1 f_3^{-1} f_6^{-1}=z_1 z_4^2 z_5 z_6^2~,~
f_1 f_2 f_3 f_4 f_5 f_6 =1~.~
&
\eea
The face paths can be written in terms of the canonical variables as follows, 
\beal{es16a05_1}
&
f_1=e^{Q} ~,~ 
f_2= e^{P} ~,~ 
f_3= e^{Q+P} z_2 z_3 z_5^{-1} z_6^{-1}~,~  
&
\nn\\
&
f_4= e^{-Q-P} z_3^{-1} z_6~,~ 
f_5= e^{-Q} z_2^{-1} z_4~,~  
f_6= e^{-P} z_4^{-1} z_5 ~.~ 
\eea

The Kasteleyn matrix of the brane tiling for Model 7 in \fref{mf_07} is 
given by, 
\beal{es16a06}
K = 
\left(
\ba{c|cccccc}
\; & b_1 & b_2 & b_3 & b_4 & b_5 & b_6  
\\
\hline
w_1 & e_{11} & e_{12} & e_{13} & 0 & 0 & 0
\\
w_2 & e_{21} x^{-1} & e_{22} & 0 & e_{24} & 0 & 0
\\
w_3 & 0 & 0 & e_{33} & e_{34} x & e_{35} & 0
\\
w_4 & 0 & 0 & e_{43} & e_{44} & 0 & e_{46}
\\
w_5 & e_{51} y & 0 & 0 & 0 & e_{55} & e_{56}
\\
w_6 & 0 & e_{62} y & 0 & 0 & e_{65} x^{-1} & e_{66}
\ea
\right)
~.~
\eea
The permanent of the Kasteleyn matrix 
gives the expression for the spectral curve of the dimer integrable system for Model 7 as follows,
\beal{es16a07}
0 = \text{perm}~K&=&\overline{p}_0
\cdot y \cdot
\Big[
\delta_{(-2,-1)} \frac{1}{x^2 y} 
+ \delta_{(-1,-1)} \frac{1}{x y} 
+ \delta_{(-1,0)} \frac{1}{x} 
\nn\\
&& 
+ \delta_{(0,-1)} \frac{1}{y}
+ \delta_{(0,1)} y
+ \delta_{(1,-1)} \frac{x}{y}
+ H
\Big]
\eea
where $\overline{p}_0= e_{13}^{+} e_{24}^{+} e_{35}^{+} e_{46}^{+} e_{51}^{+} e_{62}^{+}$.
The Casimirs $\delta_{(m,n)}$ in \eref{es16a07} can be expressed in terms of the zig-zag paths in \eref{es16a03} as follows, 
\beal{es16a08}
&
\delta_{(-2,-1)} = z_3 z_6 ~,~
\delta_{(-1,-1)} = z_2 z_3 z_6 +z_3 z_4 z_6 +  z_3 z_5 z_6~,~
\delta_{(-1,0)} = z_3+z_6~,~
&
\nn\\
&
\delta_{(0,-1)} = z_2 z_3 z_4 z_6+z_2 z_3 z_5 z_6+ z_3 z_4 z_5 z_6~,~
\delta_{(0,1)} = 1~,~
\delta_{(1,-1)} = z_2 z_3 z_4 z_5 z_6~,~
&
\eea
such that the spectral curve for Model 7 can be written in the following form,  
\beal{es16a09}
&&
\Sigma~:~
 z_3 z_6 \frac{1}{x^2 y}+ z_3 z_6 (z_2+z_4+z_5) \frac{1}{xy}+(z_3+z_6) \frac{1}{x}
\nn\\
&&
\hspace{1cm}
+z_3 z_6 (z_2 z_4+z_2 z_5+z_4 z_5) \frac{1}{y}+y+z_2 z_3 z_4 z_5 z_6 \frac{x}{y}+H
= 0 ~.~
\eea

The Hamiltonian is a sum over all 6 1-loops $\gamma_i$,
\beal{es16a10}
H=\sum_{i=1}^{6} \gamma_i~,~
\eea
where the 1-loops $\gamma_i$ can be expressed in terms of zig-zag paths and face paths as follows,
\beal{es16a11}
&
\gamma_1 = z_2 z_6 f_5 f_6~,~
\gamma_2 = z_4 z_6 f_6~,~
\gamma_3 = z_2 z_3 f_1~,~
&
\nn\\
&
\gamma_4 = z_2 z_3 f_1 f_2 ~,~
\gamma_5 = z_3 z_4 f_2~,~
\gamma_6 = z_2 z_6 f_5~.~
\eea

The commutation matrix $C$ for Model 7 takes the following form,
\beal{es16a12}
&&
C=
\left(
\ba{c|cccccc}
\; & \gamma_1
& \gamma_2
& \gamma_3
& \gamma_4 
& \gamma_5  
& \gamma_6 
\\
\hline
\gamma_1 &       0 & 1 & 1 & 0 & -1 & -1 \\
\gamma_2 &       -1 & 0 & 1 & 1 & 0 & -1 \\
\gamma_3 &       -1 & -1 & 0 & 1 & 1 & 0 \\
\gamma_4 &       0 & -1 & -1 & 0 & 1 & 1 \\
\gamma_5 &       1 & 0 & -1 & -1 & 0 & 1 \\
\gamma_6 &       1 & 1 & 0 & -1 & -1 & 0 \\
\ea
\right)
~,~
\eea
where the 
1-loops satisfying the commutation relations
can be written in terms of the canonical variables as follows, 
\beal{es16a13}
&
\gamma_1 = e^{-Q-P} z_5 z_6~,~
\gamma_2 = e^{-P} z_5 z_6~,~
\gamma_3 = e^{Q} z_2 z_3~,~
&
\nn\\
&
\gamma_4 = e^{Q+P} z_2 z_3~,~
\gamma_5 =e^{P} z_3 z_4 ~,~
\gamma_6 = e^{-Q} z_4 z_6~.~
&
\eea
\\

%=================================================================
\section{Model 8: $\text{SPP}/\mathbb{Z}_2$ $(0,1,1,1)$, $\text{PdP}_{3c}$}
%=================================================================

%=================================================================
\subsection{Model 8a}
%=================================================================

%------------------------------------------------------------------------------------------------------------------
\begin{figure}[H]
\begin{center}
\resizebox{0.9\hsize}{!}{
\includegraphics{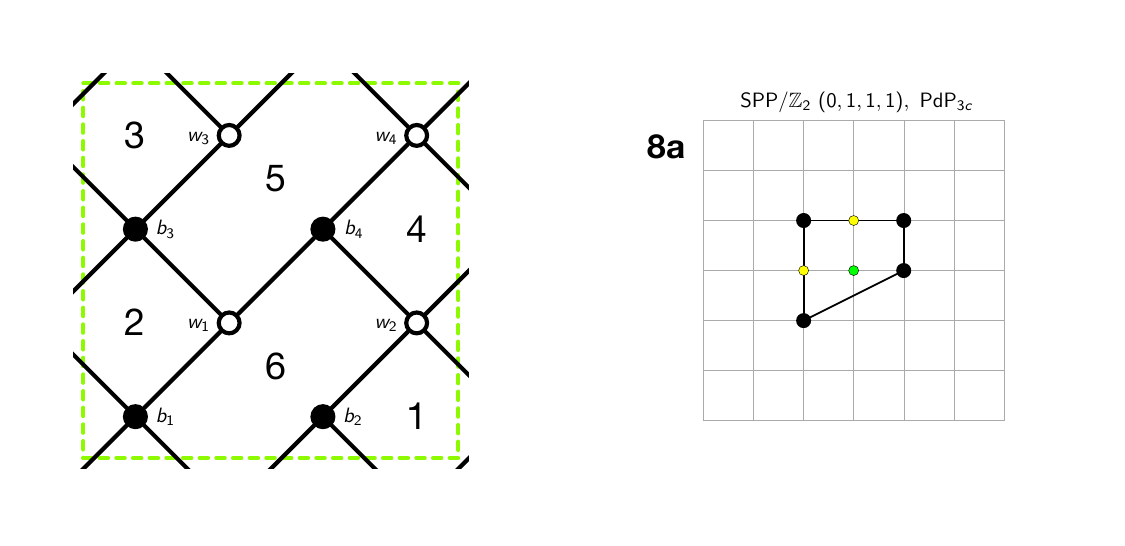}
}
\vspace{-0.5cm}
\caption{The brane tiling and toric diagram of Model 8a.}
\label{mf_08A}
 \end{center}
 \end{figure}
%------------------------------------------------------------------------------------------------------------------

The brane tiling for Model 8a 
can be expressed in terms of the following pair of permutation tuples
\beal{es17a01}
\sigma_B &=& (e_{11}\ e_{21}\ e_{41}\ e_{31})\ (e_{22}\ e_{32}\ e_{42})\ (e_{13}\ e_{33}\ e_{43}\ e_{23})\ (e_{14}\ e_{24}\ e_{44})
\nn\\
\sigma_W^{-1} &=& (e_{11}\ e_{13}\ e_{14})\ (e_{21}\ e_{22}\ e_{24}\ e_{23})\ (e_{31}\ e_{32}\ e_{33})\ (e_{41}\ e_{43}\ e_{44}\ e_{42})
\eea
which correspond to black and white nodes in the brane tiling, respectively.\\
 
The brane tiling for Model 8a has 6 zig-zag paths given by, 
\beal{es17a03}
&
z_1 = (e_{21}^{+}~ e_{41}^{-}~ e_{43}^{+}~ e_{23}^{-})~,~
z_2 = (e_{42}^{+}~ e_{22}^{-}~ e_{24}^{+}~ e_{44}^{-})~,~
&
\nn\\
&
z_3 = (e_{22}^{+}~ e_{32}^{-}~ e_{33}^{+}~ e_{43}^{-}~ e_{44}^{+}~ e_{14}^{-}~ e_{11}^{+}~ e_{21}^{-})~,~
z_4 = (e_{23}^{+}~ e_{13}^{-}~ e_{14}^{+}~ e_{24}^{-})~,~
&
\nn\\
&
z_5 = (e_{41}^{+}~ e_{31}^{-}~ e_{32}^{+}~ e_{42}^{-})~,~
z_6 = (e_{13}^{+}~ e_{33}^{-}~ e_{31}^{+}~ e_{11}^{-})~,~
&
\eea
and 6 face paths given by, 
\beal{es17a04}
&
f_1 = (e_{41}^{+}~ e_{21}^{-}~ e_{22}^{+}~ e_{42}^{-})~,~
f_2 = (e_{21}^{+}~ e_{11}^{-}~ e_{13}^{+}~ e_{23}^{-})~,~
&
\nn\\
&
f_3 = (e_{43}^{+}~ e_{33}^{-}~ e_{31}^{+}~ e_{41}^{-})~,~
f_4 = (e_{23}^{+}~ e_{43}^{-}~ e_{44}^{+}~ e_{24}^{-})~,~
&
\nn\\
&
f_5 = (e_{42}^{+}~ e_{32}^{-}~ e_{33}^{+}~ e_{13}^{-}~ e_{14}^{+}~ e_{44}^{-})~,~
f_6 = (e_{11}^{+}~ e_{31}^{-}~ e_{32}^{+}~ e_{22}^{-}~ e_{24}^{+}~ e_{14}^{-})~,~
&
\eea
which satisfy the following constraints, 
\beal{es17a05}
&
f_5 f_6 = z_2 z_6^{-1}~,~
f_3 f_4 f_6^{-1} = z_2^{-1} z_4 z_5^{-1} z_6~,~
f_2 f_4^{-1} f_5^{-1} = z_1 z_4^{-1} z_5 z_6~,~
&
\nn\\
&
f_1 f_3^{-1} f_5^{-1} =z_3 z_5^2 z_6~,~
f_1 f_2 f_3 f_4 f_5 f_6 =1~.~
&
\eea
The face paths can be written in terms of the canonical variables as follows, 
\beal{es17a05_1}
&
f_1= e^{-P} z_1^{-1} z_2^{-1}~,~ 
f_2= e^{-Q+P} z_1 z_2 z_4^{-1} z_5~,~ 
f_3= e^{Q-P} z_2^{-1} z_4 z_5^{-1} z_6 ~,~ 
&
\nn\\
&
f_4= e^{P} ~,~ 
f_5= e^{-Q} z_2 z_6^{-1}~,~  
f_6= e^{Q} ~.~  
\eea

The Kasteleyn matrix of the brane tiling for Model 8a in \fref{mf_08A} is 
given by,
\beal{es17a06}
K = 
\left(
\ba{c|ccccc}
\; & b_1 & b_2 & b_3 & b_4  
\\
\hline
w_1 & e_{11} & 0 & e_{13} & e_{14} 
\\
w_2 & e_{21}x & e_{22}  & e_{23}x & e_{24}
\\
w_3 & e_{31}y & e_{32}y & e_{33} & 0 
\\
w_4 & e_{41} x y & e_{42} y & e_{43}x & e_{44}
\ea
\right)
~.~
\eea
By taking the permanent of the Kasteleyn matrix, 
we obtain the spectral curve of the dimer integrable system for Model 8a as follows,
\beal{es17a07}
0 = \text{perm}~K&=&\overline{p}_0
\cdot xy \cdot
\Big[
\delta_{(-1,-1)} \frac{1}{x y} 
+ \delta_{(-1,0)} \frac{1}{x} 
+ \delta_{(-1,1)} \frac{y}{x} 
\nn\\
&& 
+ \delta_{(0,1)} y
+ \delta_{(1,0)} x
+ \delta_{(1,1)} x y
+ H
\Big]
~,~
\eea
where $\overline{p}_0= e_{14}^{+} e_{23}^{+} e_{32}^{+} e_{41}^{+} $.
The Casimirs $\delta_{(m,n)}$ in \eref{es17a07} can be expressed in terms of the zig-zag paths in \eref{es17a03} as shown below, 
\beal{es17a08}
&
\delta_{(-1,-1)} = z_1 z_3 ~,~
\delta_{(-1,-0)} = z_1 z_2 z_3+z_1 z_3 z_6~,~
\delta_{(-1,1)} = z_1 z_2 z_3 z_6~,~
&
\nn\\
&
\delta_{(0,1)} = z_1 z_2 z_3 z_4 z_6+z_1 z_2 z_3 z_5 z_6~,~
\delta_{(1,0)} = z_1~,~
\delta_{(1,1)} = 1~.~
&
\eea
Accordingly, we can express the spectral curve of Model 8a as follows, 
\beal{es17a09}
\Sigma~:~
(y+z_1)x+\Big(\frac{1}{z_4}+\frac{1}{z_5}\Big) y+(1+z_2 y)(1+z_6 y)\frac{z_1 z_3}{x y}+H
= 0 ~.~
\eea

The Hamiltonian is a sum over all 6 1-loops $\gamma_i$,
\beal{es17a10}
H=\sum_{i=1}^{6} \gamma_i~,~
\eea
where the 1-loops $\gamma_i$ can be expressed in terms of zig-zag paths and face paths as follows, 
\beal{es17a11}
&
\gamma_1 =  z_1 z_4^{-1} f_4 f_5 ~,~
\gamma_2 = z_2^{-1} z_4^{-1} f_5 ~,~ 
\gamma_3 = z_1 z_5^{-1} f_1~,~
&
\nn\\
&
\gamma_4 = z_1 z_5^{-1} f_1 f_6 ~,~
\gamma_5 = z_1 z_5^{-1} f_1 f_4 f_6 ~,~
\gamma_6 = z_2^{-1} z_4^{-1} f_1^{-1}~.~
\eea

The commutation matrix $C$ for Model 8a is given by, 
\beal{es17a12}
&&
C=
\left(
\ba{c|cccccc}
\; & \gamma_1
& \gamma_2
& \gamma_3
& \gamma_4 
& \gamma_5  
& \gamma_6 
\\
\hline
\gamma_1 &        0 & 1 & 1 & 0 & -1 & -1 \\
\gamma_2 &       -1 & 0 & 1 & 1 & 0 & -1 \\
\gamma_3 &       -1 & -1 & 0 & 1 & 1 & 0 \\
\gamma_4 &       0 & -1 & -1 & 0 & 1 & 1 \\
\gamma_5 &       1 & 0 & -1 & -1 & 0 & 1 \\
\gamma_6 &       1 & 1 & 0 & -1 & -1 & 0 \\
\ea
\right)
~.~
\eea
The 1-loops satisfying the commutation relations
can be written in terms of the canonical variables as follows, 
\beal{es17a13}
&
\gamma_1 = e^{-Q+P} z_1 z_2 z_4^{-1} z_6^{-1}~,~
\gamma_2 =  e^{-Q} z_4^{-1} z_6^{-1}   ~,~
\gamma_3 = e^{-P} z_2^{-1} z_5^{-1}~,~
&
\nn\\
&
\gamma_4 =e^{Q-P} z_2^{-1} z_5^{-1}~,~
\gamma_5 = e^{Q} z_2^{-1} z_5^{-1} ~,~
\gamma_6 = e^{P} z_1 z_4^{-1}~.~
&
\eea
\\
 
 %=================================================================
\subsection{Model 8b}
%=================================================================
%------------------------------------------------------------------------------------------------------------------
\begin{figure}[H]
\begin{center}
\resizebox{0.9\hsize}{!}{
\includegraphics{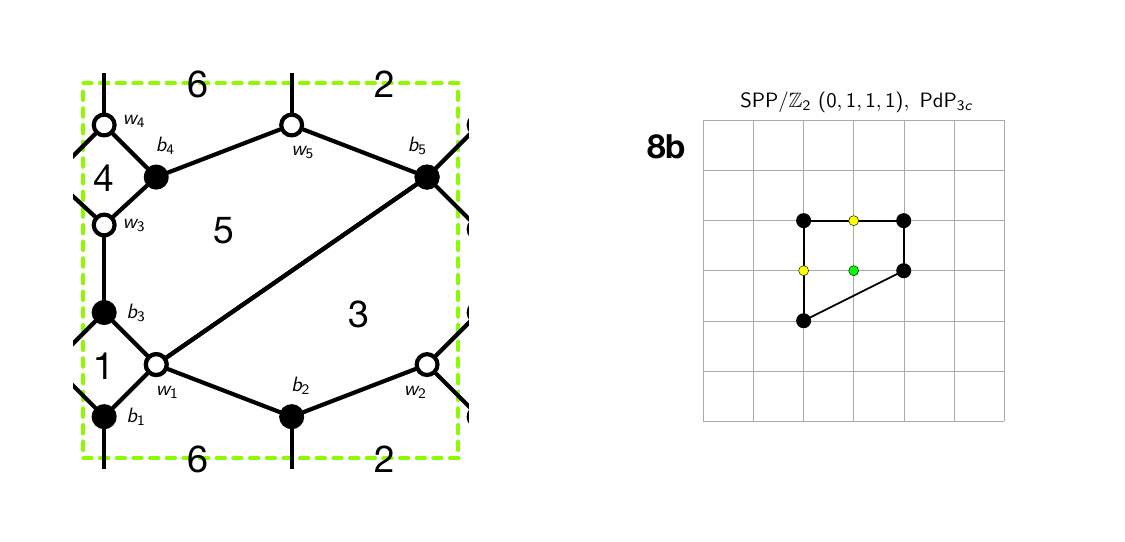}
}
\vspace{-0.5cm}
\caption{The brane tiling and toric diagram of Model 8b.}
\label{mf_08B}
 \end{center}
 \end{figure}
%------------------------------------------------------------------------------------------------------------------

The brane tiling for Model 8b 
can be expressed in terms of the following pair of permutation tuples
\beal{es17b01}
\sigma_B &=& (e_{11}\ e_{21}\ e_{41})\ (e_{12}\ e_{52}\ e_{22})\ (e_{13}\ e_{33}\ e_{23})\ (e_{34}\ e_{54}\ e_{44})\ (e_{15}\ e_{35}\ e_{45}\ e_{55})
\nn\\
\sigma_W^{-1} &=& (e_{11}\ e_{13}\ e_{15}\ e_{12})\ (e_{21}\ e_{22}\ e_{23})\ (e_{33}\ e_{35}\ e_{34})\ (e_{41}\ e_{44}\ e_{45})\ (e_{52}\ e_{55}\ e_{54})
\nn\\
\eea
which correspond to black and white nodes in the brane tiling, respectively.\\
 
The brane tiling for Model 8b has 6 zig-zag paths given by, 
\beal{es17b03}
&
z_1 = (e_{11}^{+}~ e_{21}^{-}~ e_{22}^{+}~ e_{12}^{-})~,~
z_2 = (e_{23}^{+}~ e_{13}^{-}~ e_{15}^{+}~ e_{35}^{-}~ e_{34}^{+}~ e_{54}^{-}~ e_{52}^{+}~ e_{22}^{-})~,~
&
\nn\\
&
z_3 = (e_{45}^{+}~ e_{55}^{-}~ e_{54}^{+}~ e_{44}^{-})~,~
z_4 = (e_{21}^{+}~ e_{41}^{-}~ e_{44}^{+}~ e_{34}^{-}~ e_{33}^{+}~ e_{23}^{-})~,~
&
\nn\\
&
z_5 = (e_{13}^{+}~ e_{33}^{-}~ e_{35}^{+}~ e_{45}^{-}~ e_{41}^{+}~ e_{11}^{-})~,~
z_6 = (e_{55}^{+}~ e_{15}^{-}~ e_{12}^{+}~ e_{52}^{-})~,~
&
\eea
and 6 face paths given by, 
\beal{es17b04}
&
f_1 = (e_{21}^{+}~ e_{11}^{-}~ e_{13}^{+}~ e_{23}^{-})~,~
f_2 = (e_{22}^{+}~ e_{52}^{-}~ e_{55}^{+}~ e_{45}^{-}~ e_{41}^{+}~ e_{21}^{-})~,~
&
\nn\\
&
f_3 = (e_{23}^{+}~ e_{33}^{-}~ e_{35}^{+}~ e_{15}^{-}~ e_{12}^{+}~ e_{22}^{-})~,~
f_4 = (e_{45}^{+}~ e_{35}^{-}~ e_{34}^{+}~ e_{44}^{-})~,~
&
\nn\\
&
f_5 = (e_{15}^{+}~ e_{55}^{-}~ e_{54}^{+}~ e_{34}^{-}~ e_{33}^{+}~ e_{13}^{-})~,~
f_6 = (e_{11}^{+}~ e_{41}^{-}~ e_{44}^{+}~ e_{54}^{-}~ e_{52}^{+}~ e_{12}^{-})~,~
&
\eea
which satisfy the following relations, 
\beal{es17b05}
&
f_1 f_4^{-1} = z_4 z_5~,~
f_4 f_5 f_6=z_5^{-1} z_6^{-1}~,~
f_2 f_4^{-1} f_5^{-1}=z_1 z_3^{-1} z_5 z_6~,~
&
\nn\\
&
f_3 f_5^{-1} f_6^{-2}=z_1^{-1} z_4^{-1} z_3 z_5 z_6^2~,~
f_1 f_2 f_3 f_4 f_5 f_6 =1~.~
&
\eea
The face paths can be expressed in terms of the canonical variables as follows,
\beal{es17b05_1}
&
f_1= e^{Q} z_4 z_5~,~ 
f_2= e^{Q+P} z_1 z_3^{-1} z_5 z_6~,~ 
f_3= e^{-2Q-P} z_2 z_3^2 z_6 ~,~ 
&
\nn\\
&
f_4= e^{Q} ~,~ 
f_5= e^{P}~,~  
f_6= e^{-Q-P} z_5^{-1} z_6^{-1}
~.~  
\eea

The Kasteleyn matrix of the brane tiling for Model 8b in \fref{mf_08B} takes the following form, 
\beal{es17b06}
K = 
\left(
\ba{c|cccccc}
\; & b_1 & b_2 & b_3 & b_4 & b_5 
\\
\hline
w_1 & e_{11} & e_{12} & e_{13} & 0 & e_{15} 
\\
w_2 & e_{21}x & e_{22}  & e_{23}x & 0 & 0
\\
w_3 & 0 & 0 & e_{33} & e_{34} & e_{35} x^{-1} 
\\
w_4 & e_{41} y & 0 & 0 & e_{44} & e_{45} x^{-1}
\\
w_5 & 0 & e_{52} y & 0 & e_{54} & e_{55}
\ea
\right)
~.~
\eea
By taking the permanent of the Kasteleyn matrix in \eref{es17b06} with a $GL(2,\mathbb{Z})$ transformation $M : (x,y) \mapsto (\frac{1}{x},\frac{1}{y})$, 
we obtain the spectral curve of the dimer integrable system for Model 8b as follows,
\beal{es17b07}
0 &=&\overline{p}_0
\cdot y^{-1} \cdot
\Big[
\delta_{(-1,-1)} \frac{1}{x y} 
+ \delta_{(-1,0)} \frac{1}{x} 
+ \delta_{(-1,1)} \frac{y}{x} 
\nn\\
&& 
+ \delta_{(0,1)} y
+ \delta_{(1,0)} x
+ \delta_{(1,1)} x y
+ H
\Big]
~,~
\eea
where $\overline{p}_0= e_{11}^{+} e_{22}^{+} e_{33}^{+} e_{45}^{+} e_{54}^{+}$.
The Casimirs $\delta_{(m,n)}$ in \eref{es17b07} can be written in terms of the zig-zag paths in \eref{es17b03} as follows, 
\beal{es17b08}
&
\delta_{(-1,-1)} = z_2 z_5 ~,~
\delta_{(-1,-0)} = z_2 z_4 z_5 +z_2 z_5 z_6~,~
\delta_{(-1,1)} = z_2 z_4 z_5 z_6 ~,~
&
\nn\\
&
\delta_{(0,1)} =z_1 z_2 z_4 z_5 z_6+ z_2 z_3 z_4 z_5 z_6 ~,~
\delta_{(1,0)} = z_5 ~,~
\delta_{(1,1)} = 1 ~,~
&
\eea
such that the spectral curve for Model 8b takes the following form,  
\beal{es17b09}
&&
\Sigma~:~
(y+z_5)x+\Big(\frac{1}{z_1}+\frac{1}{z_3}\Big) y+(1+z_4 y)(1+z_6 y)\frac{z_2 z_5}{x y}+H
= 0 ~.~
\eea

The Hamiltonian is a sum over all 7 1-loops $\gamma_i$,
\beal{es17b10}
H=\sum_{i=1}^{7} \gamma_i~,~
\eea
where the 1-loops $\gamma_i$ can be expressed in terms zig-zag paths and face paths as follows, 
\beal{es17b11}
&
\gamma_1 = z_3^{-1} z_4^{-1} f_1 ~,~
\gamma_2 = z_3^{-1} z_4^{-1} f_1 f_5 ~,~
\gamma_3 = z_1^{-1} z_4^{-1} f_4^{-1} ~,~
&
\nn\\
&
\gamma_4 = z_1^{-1} z_5 f_1^{-1} f_6 ~,~
\gamma_5 = z_1^{-1} z_4^{-1} f_6 ~,~
\gamma_6 = z_1^{-1} z_5 f_6 ~,~
\gamma_7 = f_1 f_2^{-1} z_3^{-1} z_4^{-1} ~.~
\eea

The commutation matrix $C$ for Model 8b is given by,
\beal{es17b12}
&&
C=
\left(
\ba{c|ccccccc}
\; & \gamma_1
& \gamma_2
& \gamma_3
& \gamma_4 
& \gamma_5  
& \gamma_6 
& \gamma_7 
\\
\hline
\gamma_1 &      0 & 1 & 0 & -1 & -1 & -1 & -1 \\
\gamma_2 &       -1 & 0 & 1 & 1 & 0 & 0 & -1 \\
\gamma_3 &       0 & -1 & 0 & 1 & 1 & 1 & 1 \\
\gamma_4 &       1 & -1 & -1 & 0 & 1 & 1 & 2 \\
\gamma_5 &       1 & 0 & -1 & -1 & 0 & 0 & 1 \\
\gamma_6 &       1 & 0 & -1 & -1 & 0 & 0 & 1 \\
\gamma_7 &       1 & 1 & -1 & -2 & -1 & -1 & 0 \\
\ea
\right)
~.~
\eea
The 1-loops satisfying the commutation relations
can be written in terms of the canonical variables as follows, 
\beal{es17b13}
&
\gamma_1 = e^{Q} z_3^{-1} z_5 ~,~
\gamma_2 = e^{Q+P} z_3^{-1} z_5 ~,~
\gamma_3 = e^{-Q} z_1^{-1} z_4^{-1}~,~
&
\nn\\
&
\gamma_4 = e^{-2Q-P} z_2 z_3 ~,~
\gamma_5 =e^{-Q-P} z_2 z_3 ~,~
\gamma_6 = e^{-Q-P} z_1^{-1} z_6^{-1}~,~
&
\nn\\
&
\gamma_7 = e^{-P} z_1^{-1} z_6^{-1}~.~
&
\eea
\\

%=================================================================
\section{Model 9: $\text{PdP}_{3b}$}
%=================================================================

%=================================================================
\subsection{Model 9a}
%=================================================================
%------------------------------------------------------------------------------------------------------------------
\begin{figure}[H]
\begin{center}
\resizebox{0.9\hsize}{!}{
\includegraphics{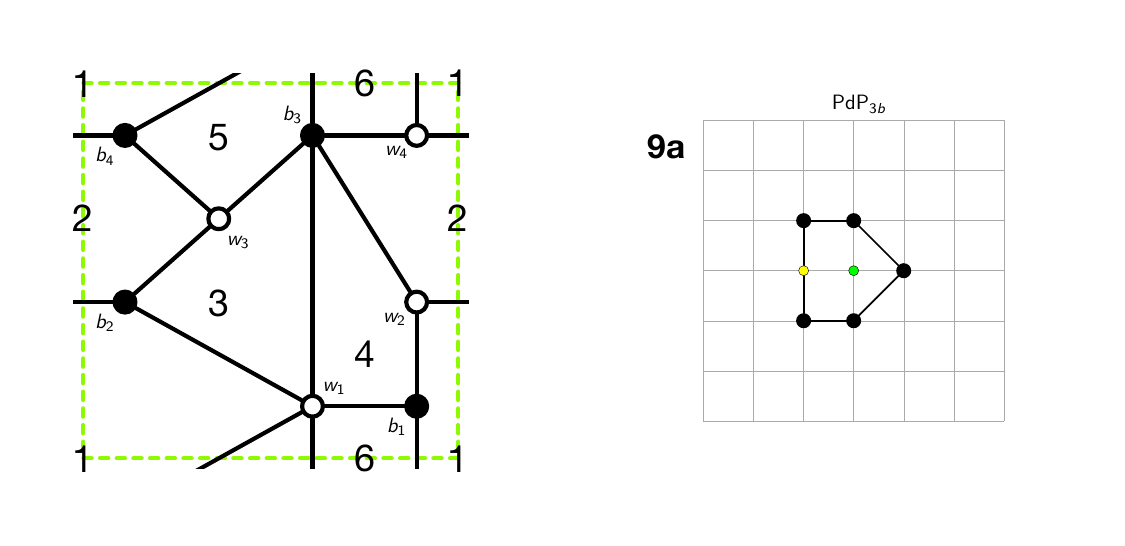}
}
\vspace{-0.5cm}
\caption{The brane tiling and toric diagram of Model 9a.}
\label{mf_09A}
 \end{center}
 \end{figure}
%------------------------------------------------------------------------------------------------------------------

The brane tiling for Model 9a 
can be expressed in terms of the following pair of permutation tuples
\beal{es18a01}
\sigma_B &=& (e_{11}\ e_{41}\ e_{21})\ (e_{12}\ e_{32}\ e_{22})\ (e_{13}^{1}\ e_{23}\ e_{43}\ e_{13}^{2}\ e_{33})\ (e_{14}\ e_{44}\ e_{34})
\nn\\
\sigma_W^{-1} &=& (e_{11}\ e_{13}^{2} \ e_{14}\ e_{12}\ e_{13}^{1})\ (e_{21}\ e_{23}\ e_{22})\ (e_{32}\ e_{34}\ e_{33})\ (e_{41}\ e_{44}\ e_{43})
\eea
which correspond to black and white nodes in the brane tiling, respectively.\\

The brane tiling for Model 9a has 6 zig-zag paths given by,
\beal{es18a03}
&
z_1 = (e_ {14}^{+}~ e_ {44}^{-}~ e_ {43}^{+}~ e_ {13}^{2,-})~,~
z_2 = (e_ {22}^{+}~ e_ {12}^{-}~ e_ {13}^{1,+}, e_ {23}^{-})~,~
&
\nn\\
&
z_3 = (e_ {41}^{+}~ e_ {21}^{-}~ e_ {23}^{+}~ e_ {43}^{-})~,~
z_4 = (e_ {21}^{+}~ e_ {11}^{-}~ e_ {13}^{2,+}, e_ {33}^{-}~ e_{32}^{+}~ e_ {22}^{-})~,~
&
\nn\\
&
z_5 = (e_ {44}^{+}~ e_ {34}^{-}~ e_ {33}^{+}~ e_ {13}^{1,-}, e_{11}^{+}~ e_ {41}^{-})~,~
z_6 = (e_ {12}^{+}~ e_ {32}^{-}~ e_ {34}^{+}~ e_ {14}^{-})~,~
&
\eea
and 6 face paths given by, 
\beal{es18a04}
&
f_1 = (e_ {44}^{+}~ e_ {14}^{-}~ e_ {12}^{+}~ e_ {22}^{-}~ e_{21}^{+}~ e_ {41}^{-})~,~
f_2 = (e_ {22}^{+}~ e_ {32}^{-}~ e_ {34}^{+}~ e_ {44}^{-}~ e_{43}^{+}~ e_ {23}^{-})~,~
&
\nn\\
&
f_3 = (e_ {32}^{+}~ e_ {12}^{-}~ e_ {13}^{1,+}, e_ {33}^{-})~,~
f_4 = (e_ {23}^{+}~ e_ {13}^{1,-}, e_ {11}^{+}~ e_ {21}^{-})~,~
&
\nn\\
&
f_5 = (e_ {14}^{+}~ e_ {34}^{-}~ e_ {33}^{+}~ e_ {13}^{2,-})~,~
f_6 = (e_ {41}^{+}~ e_ {11}^{-}~ e_ {13}^{2,+}, e_ {43}^{-})~,~
&
\eea
satisfying the following relations, 
\beal{es18a05}
&
f_4 f_5 = z_1 z_3 z_5~,~
f_3 f_6 = z_2 z_3 z_4~,~
f_2 f_4^{-1} f_6^{-1}=z_1 z_2 z_3^{-1} z_6~,~
&
\nn\\
&
f_1^{-1} f_3 f_5=z_1 z_2 z_3 z_6^{-1}~,~
f_1 f_2 f_3 f_4 f_5 f_6 =1~.~
&
\eea
The face paths can be written in terms of the canonical variables as follows, 
\beal{es18a05_1}
&
f_1= z_2^{-1} z_5 z_6 e^{Q-P} ~,~ 
f_2= z_2 z_3^{-1} z_5^{-1} e^{-Q+P} ~,~ 
f_3= e^{Q}~,~ 
&
\nn\\
&
f_4= e^{P}~,~ 
f_5= e^{-P} z_1 z_3 z_5~,~  
f_6= e^{-Q} z_2 z_3 z_4 ~.~ 
\eea

The Kasteleyn matrix of the brane tiling for Model 9a in \fref{mf_09A} is 
given by, 
\beal{es18a06}
K = 
\left(
\ba{c|ccccc}
\; & b_1 & b_2 & b_3 & b_4  
\\
\hline
w_1 & e_{11} & e_{12} & e_{13}^{1}+e_{13}^{2} y^{-1} & e_{14} y^{-1}
\\
w_2 & e_{21} & e_{22} x & e_{23} & 0
\\
w_3 & 0 & e_{32} & e_{33} & e_{34} 
\\
w_4 & e_{41} y & 0 & e_{43} & e_{44} x
\ea
\right)
~.~
\eea
By taking a permanent of the Kasteleyn matrix, we obtain the spectral curve of the dimer integrable system for Model 9a as follows,
\beal{es18a07}
0 = \text{perm}~K&=&\overline{p}_0
\cdot x \cdot
\Big[
\delta_{(-1,-1)} \frac{1}{x y} 
+ \delta_{(-1,0)} \frac{1}{x} 
+ \delta_{(-1,1)} \frac{y}{x} 
\nn\\
&& 
+ \delta_{(0,-1)} \frac{1}{y}
+ \delta_{(0,1)} y
+ \delta_{(1,0)} x
+ H
\Big]
~,~
\eea
where $\overline{p}_0= e_{11}^{+} e_{22}^{+} e_{33}^{+} e_{44}^{+} $.
The Casimirs $\delta_{(m,n)}$ in \eref{es18a07} can be written in terms of the zig-zag paths in \eref{es18a03} as shown below, 
\beal{es18a08}
&
\delta_{(-1,-1)} = z_1 z_4 ~,~
\delta_{(-1,-0)} = z_1 z_3 z_4+z_1 z_4 z_6~,~
\delta_{(-1,1)} = z_1 z_3 z_4 z_6~,~
&
\nn\\
&
\delta_{(0,-1)} = z_4~,~
\delta_{(0,1)} = z_5^{-1}~,~
\delta_{(1,0)} = 1~,~
&
\eea
such that the spectral curve for Model 9a takes the following form,
\beal{es18a09}
\Sigma~:~ (1+z_3 y)(1+z_6 y)\frac{z_1 z_4}{x y}+\frac{y}{z_5}+\frac{z_4}{y}+x+H
= 0 ~.~
\eea

The Hamiltonian is a sum over all 6 1-loops $\gamma_i$,
\beal{es18a10}
H=\sum_{i=1}^{6} \gamma_i~,~
\eea
where the 1-loops $\gamma_i$ can be expressed in terms of zig-zag paths and face paths as follows,
\beal{es18a11}
&
\gamma_1 = z_2^{-1} f_2 f_3~,~
\gamma_2 = z_3 z_4 f_2 ~,~
\gamma_3 = z_1 z_2 z_3 z_4 f_3^{-1}~,~
&
\nn\\
&
\gamma_4 = z_1 z_2 z_3 z_4 f_1 f_3^{-1}~,~
\gamma_5 = z_1 z_2 z_3 z_4 f_1~,~
\gamma_6 = z_2^{-1} f_3~.~
\eea

The commutation matrix $C$ for Model 9a takes the following form, 
\beal{es18a12}
&&
C=
\left(
\ba{c|cccccc}
\; & \gamma_1
& \gamma_2
& \gamma_3
& \gamma_4 
& \gamma_5  
& \gamma_6 
\\
\hline
\gamma_1 &       0 & 1 & 1 & 0 & -1 & -1 \\
\gamma_2 &        -1 & 0 & 1 & 1 & 0 & -1 \\
\gamma_3 &        -1 & -1 & 0 & 1 & 1 & 0 \\
\gamma_4 &        0 & -1 & -1 & 0 & 1 & 1 \\
\gamma_5 &        1 & 0 & -1 & -1 & 0 & 1 \\
\gamma_6 &        1 & 1 & 0 & -1 & -1 & 0 \\
\ea
\right)
~.~
\eea
The 1-loops satisfying the commutation relations 
can be written in terms of the canonical variables as follows, 
\beal{es18a13}
&
\gamma_1 = e^{P} z_3^{-1} z_5^{-1}~,~
\gamma_2 =e^{-Q+P} z_2 z_4 z_5^{-1}~,~
\gamma_3 = e^{-Q} z_5^{-1} z_6^{-1}~,~
&
\nn\\
&
\gamma_4 = e^{-P} z_2^{-1}~,~
\gamma_5 = e^{Q-P} z_2^{-1}~,~
\gamma_6 = e^{Q} z_2^{-1}~.~
&
\eea
\\

%=================================================================
\subsection{Model 9b}
%=================================================================
%------------------------------------------------------------------------------------------------------------------
\begin{figure}[H]
\begin{center}
\resizebox{0.9\hsize}{!}{
\includegraphics{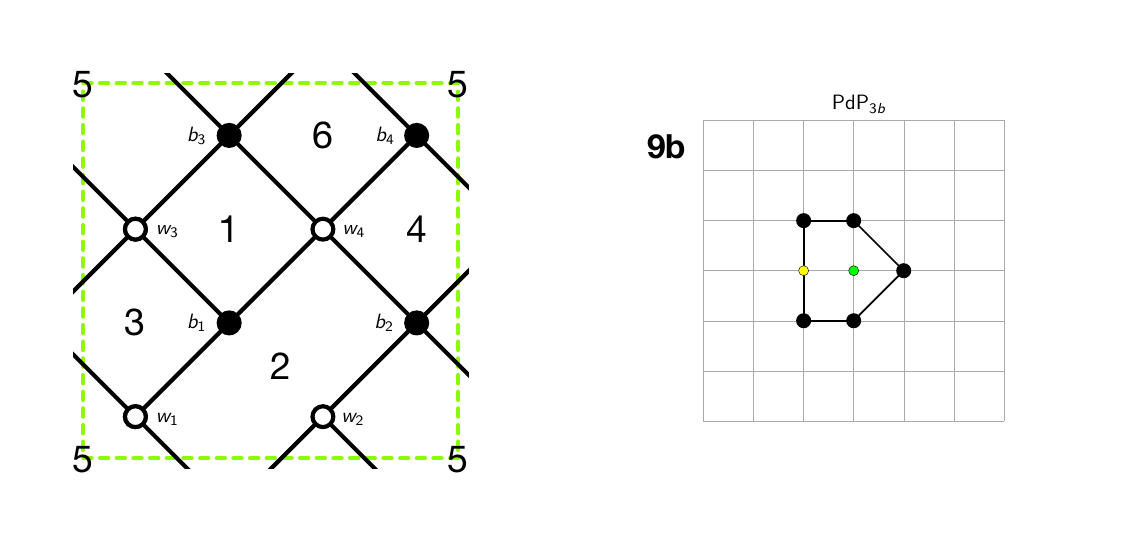}
}
\vspace{-0.5cm}
\caption{The brane tiling and toric diagram of Model 9b.}
\label{mf_09B}
 \end{center}
 \end{figure}
%------------------------------------------------------------------------------------------------------------------

The brane tiling for Model 9b 
can be expressed in terms of the following pair of permutation tuples
\beal{es18b01}
\sigma_B &=& (e_{11}\ e_{41}\ e_{31})\ (e_{12}\ e_{32}\ e_{42}\ e_{32})\ (e_{13}\ e_{33}\ e_{43}\ e_{23})\ (e_{24}\ e_{44}\ e_{34})
\nn\\
\sigma_W^{-1} &=& (e_{11}\ e_{13}\ e_{12})\ (e_{22}\ e_{24}\ e_{23})\ (e_{31}\ e_{32}\ e_{34}\ e_{33})\ (e_{41}\ e_{43}\ e_{44}\ e_{42})
\eea
which correspond to black and white nodes in the brane tiling, respectively.\\

The brane tiling for Model 9b has 6 zig-zag paths given by, 
\beal{es18b03}
&
z_1 = (e_ {41}^{+}~ e_ {31}^{-}~ e_ {32}^{+}~ e_ {42}^{-})~,~
z_2 = (e_ {33}^{+}~ e_ {43}^{-}~ e_ {44}^{+}~ e_ {34}^{-})~,~
&
\nn\\
&
z_3 = (e_ {43}^{+}~ e_ {23}^{-}~ e_ {22}^{+}~ e_ {12}^{-}~ e_ {11}^{+}~ e_{41}^{-})~,~
z_4 = (e_ {42}^{+}~ e_ {22}^{-}~ e_ {24}^{+}~ e_ {44}^{-})~,~
&
\nn\\
&
z_5 = (e_ {31}^{+}~ e_ {11}^{-}~ e_ {13}^{+}~ e_ {33}^{-})~,~
z_6 = (e_ {12}^{+}~ e_ {32}^{-}~ e_ {34}^{+}~ e_ {24}^{-}~ e_ {23}^{+}~ e_{13}^{-})~,~
&
\eea
and 6 face paths given by, 
\beal{es18b04}
&
f_1 = (e_ {31}^{+}~ e_ {41}^{-}~ e_ {43}^{+}~ e_ {33}^{-})~,~
f_2 = (e_ {41}^{+}~ e_ {11}^{-}~ e_ {13}^{+}~ e_ {23}^{-}~ e_ {22}^{+}~ e_{42}^{-})~,~
&
\nn\\
&
f_3 = (e_ {11}^{+}~ e_ {31}^{-}~ e_ {32}^{+}~ e_ {12}^{-})~,~
f_4 = (e_ {42}^{+}~ e_ {32}^{-}~ e_ {34}^{+}~ e_ {44}^{-})~,~
&
\nn\\
&
f_5 = (e_ {33}^{+}~ e_ {13}^{-}~ e_ {12}^{+}~ e_ {22}^{-}~ e_ {24}^{+}~ e_{34}^{-})~,~
f_6 = (e_ {23}^{+}~ e_ {43}^{-}~ e_ {44}^{+}~ e_ {24}^{-})~,~
&
\eea
which satisfy the following relations, 
\beal{es18b05}
&
f_4 f_5 f_6= z_2 z_4 z_6~,~
f_3 f_6^{-1} = z_1 z_3 z_4~,~
f_2 f_4^{-2} f_5^{-1}=z_1^3 z_2^2 z_3^2 z_5^3 z_6~,~
&
\nn\\
&
f_1 f_5^{-1} f_6^{-1}=z_1 z_3^2 z_4 z_5^2 z_6~,~
f_1 f_2 f_3 f_4 f_5 f_6 =1~.~
&
\eea
The face paths can be written in terms of the canonical variables as follows, 
\beal{es18b05_1}
&
f_1= e^{P} ~,~ 
f_2= e^{Q} ~,~ 
f_3= e^{-Q-P} z_1 z_3 z_5~,~ 
&
\nn\\
&
f_4= e^{-P} z_1^{-1} z_2^{-1}~,~ 
f_5= e^{Q+2P} z_2 z_3^{-1} z_4 z_5^{-2}~,~  
f_6= e^{-Q-P} z_4^{-1} z_5 ~.~ 
\eea

The Kasteleyn matrix of the brane tiling for Model 9b in \fref{mf_09B} is 
given by, 
\beal{es18b06}
K = 
\left(
\ba{c|ccccc}
\; & b_1 & b_2 & b_3 & b_4  
\\
\hline
w_1 & e_{11} & e_{12} x^{-1}& e_{13} y^{-1} &0
\\
w_2 & 0 & e_{22} x & e_{23} y^{-1} & e_{24} y^{-1}
\\
w_3 & e_{31} & e_{32} x^{-1}& e_{33} & e_{34} x^{-1} 
\\
w_4 & e_{41} & e_{42} & e_{43} & e_{44}
\ea
\right)
~.~
\eea
By taking the permanent of the Kasteleyn matrix in \eref{es18b06} with a $GL(2,\mathbb{Z})$ transformation $M : (x,y) \mapsto (\frac{1}{x},y)$, 
we obtain the spectral curve of the dimer integrable system for Model 9b as follows,
\beal{es18b07}
0 &=&\overline{p}_0
\cdot x y^{-1} \cdot
\Big[
\delta_{(-1,-1)} \frac{1}{x y} 
+ \delta_{(-1,0)} \frac{1}{x} 
+ \delta_{(-1,1)} \frac{y}{x} 
\nn\\
&& 
+ \delta_{(0,-1)} \frac{1}{y}
+ \delta_{(0,1)} y
+ \delta_{(1,0)} x
+ H
\Big]
~,~
\eea
where $\overline{p}_0= e_{12}^{+} e_{23}^{+} e_{34}^{+} e_{41}^{+} $.
The Casimirs $\delta_{(m,n)}$ in \eref{es18b07} can be written in terms of the zig-zag paths in \eref{es18b03} as follows, 
\beal{es18b08}
&
\delta_{(-1,-1)} = z_1^{-1} z_6^{-1} ~,~
\delta_{(-1,-0)} = z_2 z_3 z_4+z_2 z_3 z_5~,~
\delta_{(-1,1)} = z_2 z_3 ~,~
&
\nn\\
&
\delta_{(0,-1)} = z_6^{-1}~,~
\delta_{(0,1)} = z_3~,~
\delta_{(1,0)} = 1~,~
&
\eea
such that the spectral curve for Model 9b takes the following form,  
\beal{es18b09}
\Sigma~:~ (z_2 z_3 z_4+z_2 z_3 z_5)\frac{1}{x}+\frac{1}{z_6 y}+\frac{1}{z_1 z_6 x y}+z_2 z_3 \frac{y}{x}+z_3 y+x+H
= 0 ~.~
\eea

The Hamiltonian is a sum over all 7 1-loops $\gamma_i$,
\beal{es18b10}
H=\sum_{i=1}^{7} \gamma_i~,~
\eea
where the 1-loops $\gamma_i$ can be expressed in terms of zig-zag paths and face paths as follows,
\beal{es18b11}
&
\gamma_1 = z_2 z_4 z_5^{-1} f_1 f_5^{-1}~,~
\gamma_2 = z_1^{-1} f_2 f_3~,~
\gamma_3 = z_5^{-1} z_6^{-1} f_3^{-1}~,~
&
\nn\\
&
\gamma_4 = z_5^{-1} z_6^{-1} f_1 f_3^{-1}~,~
\gamma_5 = z_5^{-1} z_6^{-1} f_1~,~
\gamma_6 = z_2 f_1 ~,~
\gamma_7 = z_2 f_1 f_3~.~
\eea

The commutation matrix $C$ for Model 9b is given by,
\beal{es18b12}
&&
C=
\left(
\ba{c|ccccccc}
\; & \gamma_1
& \gamma_2
& \gamma_3
& \gamma_4 
& \gamma_5  
& \gamma_6 
& \gamma_7 
\\
\hline
\gamma_1 &        0 & 1 & 0 & -1 & -1 & -1 & -1 \\
\gamma_2 &        -1 & 0 & 1 & 1 & 0 & 0 & -1 \\
\gamma_3 &        0 & -1 & 0 & 1 & 1 & 1 & 1 \\
\gamma_4 &        1 & -1 & -1 & 0 & 1 & 1 & 2 \\
\gamma_5 &        1 & 0 & -1 & -1 & 0 & 0 & 1 \\
\gamma_6 &        1 & 0 & -1 & -1 & 0 & 0 & 1 \\
\gamma_7 &        1 & 1 & -1 & -2 & -1 & -1 & 0 \\
\ea
\right)
~,~
\eea
where the
1-loops satisfying the commutation relations
can be written in terms of the canonical variables as follows, 
\beal{es18b14}
&
\gamma_1 = e^{-Q-P} z_3 z_5~,~
\gamma_2 = e^{-P} z_3 z_5~,~
\gamma_3 = e^{Q+P} z_2 z_4 z_5^{-1}~,~
&
\nn\\
&
\gamma_4 = e^{Q+2P} z_2 z_4 z_5^{-1}~,~
\gamma_5 =e^{P} z_5^{-1} z_6^{-1}~,~
\gamma_6 =e^{P} z_2 ~,~
\gamma_7 = e^{-Q} z_4^{-1} z_6^{-1}~.~
&
\eea
\\

%=================================================================
\subsection{Model 9c}
%=================================================================
%------------------------------------------------------------------------------------------------------------------
\begin{figure}[H]
\begin{center}
\resizebox{0.9\hsize}{!}{
\includegraphics{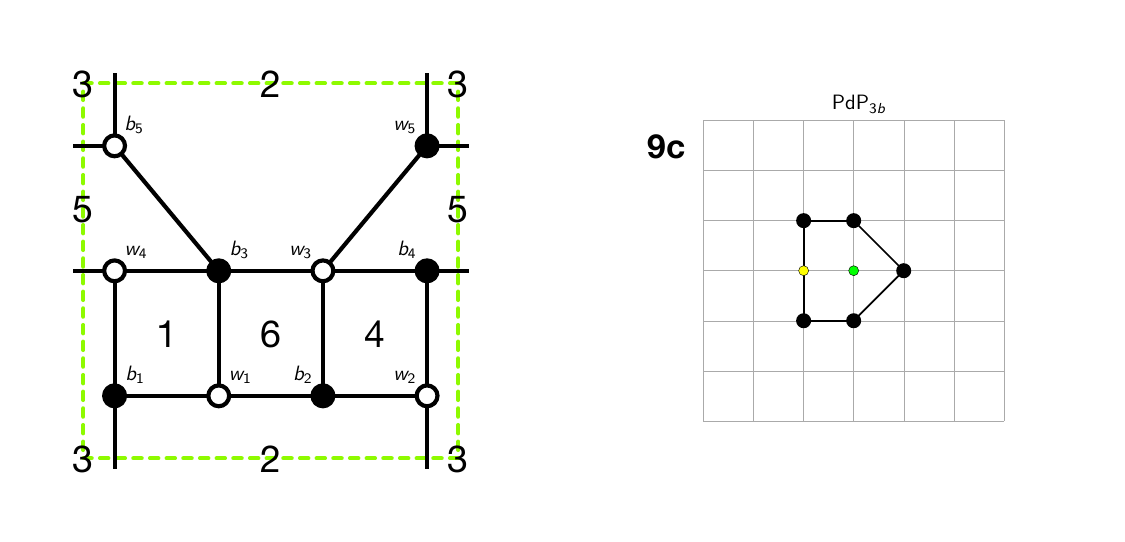}
}
\vspace{-0.5cm}
\caption{The brane tiling and toric diagram of Model 9c.}
\label{mf_09C}
 \end{center}
 \end{figure}
%------------------------------------------------------------------------------------------------------------------

The brane tiling for Model 9c 
can be expressed in terms of the following pair of permutation tuples
\beal{es18c01}
\sigma_B &=& (e_{11}\ e_{41}\ e_{51})\ (e_{12}\ e_{22}\ e_{32})\ (e_{13}\ e_{33}\ e_{53}\ e_{43})\ (e_{24}\ e_{44}\ e_{34})
\nn\\ 
&&
(e_{25}\ e_{35}\ e_{55})
\nn\\
\sigma_W &=& (e_{11}\ e_{13}\ e_{12})\ (e_{22}\ e_{24}\ e_{25})\ (e_{32}\ e_{33}\ e_{35}\ e_{34})\ (e_{41}\ e_{44}\ e_{43})
\nn\\ 
&&
(e_{51}\ e_{53}\ e_{55})
\eea
which correspond to black and white nodes in the brane tiling, respectively.\\

The brane tiling for Model 9c has 6 zig-zag paths given by, 
\beal{es18c03}
&
z_1 = (e_ {11}^{+}~ e_ {41}^{-}~ e_ {44}^{+}~ e_ {34}^{-}~ e_ {32}^{+}~ e_{12}^{-})~,~
z_2 = (e_ {43}^{+}~ e_ {13}^{-}~ e_ {12}^{+}~ e_ {22}^{-}~ e_ {24}^{+}~ e_{44}^{-})~,~
&
\nn\\
&
z_3 = (e_ {13}^{+}~ e_ {33}^{-}~ e_ {35}^{+}~ e_ {55}^{-}~ e_ {51}^{+}~ e_{11}^{-})~,~
z_4 = (e_ {25}^{+}~ e_ {35}^{-}~ e_ {34}^{+}~ e_ {24}^{-})~,~
&
\nn\\
&
z_5 = (e_ {41}^{+}~ e_ {51}^{-}~ e_ {53}^{+}~ e_ {43}^{-})~,~
z_6 = (e_ {22}^{+}~ e_ {32}^{-}~ e_ {33}^{+}~ e_ {53}^{-}~ e_ {55}^{+}~ e_{25}^{-})~,~
&
\eea
and 6 face paths given by, 
\beal{es18c04}
&
f_1 = (e_ {41}^{+}~ e_ {11}^{-}~ e_ {13}^{+}~ e_ {43}^{-})~,~
f_2 = (e_ {11}^{+}~ e_ {51}^{-}~ e_ {53}^{+}~ e_ {33}^{-}~ e_ {35}^{+}~ e_{25}^{-}~ e_ {22}^{+}~ e_ {12}^{-})~,~
&
\nn\\
&
f_3 = (e_ {51}^{+}~ e_ {41}^{-}~ e_ {44}^{+}~ e_ {24}^{-}~ e_ {25}^{+}~ e_{55}^{-})~,~
f_4 = (e_ {24}^{+}~ e_ {34}^{-}~ e_ {32}^{+}~ e_ {22}^{-})~,~
&
\nn\\
&
f_5 = (e_ {43}^{+}~ e_ {53}^{-}~ e_ {55}^{+}~ e_ {35}^{-}~ e_ {34}^{+}~ e_{44}^{-})~,~
f_6 = (e_ {33}^{+}~ e_ {13}^{-}~ e_ {12}^{+}~ e_ {32}^{-})~,~
&
\eea
which satisfy the following relations, 
\beal{es18c05}
&
f_5 f_6=z_2 z_4 z_6~,~
f_3 f_6^{-1} = z_1 z_3 z_4~,~
f_1 f_4^{-1} =z_1^{-1} z_2^{-1}~,~
&
\nn\\
&
f_2 f_4^{2} f_5^{-1}=z_1^3 z_2^2 z_3^2 z_5^3 z_6~,~
f_1 f_2 f_3 f_4 f_5 f_6 =1~.~
&
\eea
The face paths can be written in terms of the canonical variables as follows, 
\beal{es18c05_1}
&
f_1=e^{P} ~,~ 
f_2= e^{Q-2P} z_2^{-1} z_3 z_5~,~ 
f_3= e^{-Q} ~,~  
f_4= e^{P} z_1 z_2~,~ 
&
\nn\\
&
f_5= e^{Q} z_4 z_5^{-1} ~,~ 
f_6= e^{-Q} z_2 z_5 z_6~,~  
\eea

The Kasteleyn matrix of the brane tiling for Model 9b in \fref{mf_09C} is 
given by, 
\beal{es18c06}
K = 
\left(
\ba{c|ccccc}
\; & b_1 & b_2 & b_3 & b_4 & b_5
\\
\hline
w_1 & e_{11} & e_{12} & e_{13}  &0 & 0
\\
w_2 & 0 & e_{22} & 0 & e_{24} & e_{25} y^{-1}
\\
w_3 & 0 & e_{32} & e_{33} & e_{34} & e_{35}
\\
w_4 & e_{41} & 0 & e_{43} & e_{44} x^{-1} & 0
\\
w_5 & e_{51} y& 0 & e_{53} & 0 & e_{55} x^{-1}
\ea
\right)
~.~
\eea
By taking the permanent of the Kasteleyn matrix in
\eref{es18c06} with a $GL(2,\mathbb{Z})$ transformation $M : (x,y) \mapsto (\frac{1}{x},y)$, 
we obtain the spectral curve of the dimer integrable system for Model 9c as follows,
\beal{es18c07}
0 &=&\overline{p}_0
\cdot x  \cdot
\Big[
\delta_{(-1,-1)} \frac{1}{x y} 
+ \delta_{(-1,0)} \frac{1}{x} 
+ \delta_{(-1,1)} \frac{y}{x} 
\nn\\
&& 
+ \delta_{(0,-1)} \frac{1}{y}
+ \delta_{(0,1)} y
+ \delta_{(1,0)} x
+ H
\Big]
~,~
\eea
where $\overline{p}_0= e_{11}^{+} e_{22}^{+} e_{33}^{+} e_{44}^{+} e_{55}^{+}$.
The Casimirs $\delta_{(m,n)}$ in \eref{es18c07} can be written in terms of the zig-zag paths in \eref{es18c03} as shown below, 
\beal{es18c08}
&
\delta_{(-1,-1)} = z_1^{-1} z_6^{-1} ~,~
\delta_{(-1,-0)} = z_2 z_3 z_4+z_2 z_3 z_5~,~
\delta_{(-1,1)} = z_2 z_3 ~,~
&
\nn\\
&
\delta_{(0,-1)} = z_6^{-1}~,~
\delta_{(0,1)} = z_3~,~
\delta_{(1,0)} = 1~,~
&
\eea
such that the spectral curve for Model 9c takes the following form, 
\beal{es18c09}
\Sigma~:~
(z_2 z_3 z_4+z_2 z_3 z_5)\frac{1}{x}+\frac{1}{z_6 y}+\frac{1}{z_1 z_6 x y}+z_2 z_3 \frac{y}{x}+z_3 y+x+H
= 0 ~.~
\eea

The Hamiltonian is a sum over all 8 1-loops $\gamma_i$,
\beal{es18c10}
H=\sum_{i=1}^{8} \gamma_i~,~
\eea
where the 1-loops $\gamma_i$ can be expressed in terms of zig-zag paths and face paths as follows,
\beal{es18c11}
&
\gamma_1 = z_4^{-1} z_6^{-1} f_4^{-1} ~,~
\gamma_2 = z_2 f_4^{-1} f_6^{-1}~,~
\gamma_3 = z_1^{-1} f_6^{-1}~,~
\gamma_4 = z_2 f_6^{-1} ~,~
&
\nn\\
&
\gamma_5 = z_2 f_1 f_6^{-1}~,~
\gamma_6 = z_5^{-1} z_6^{-1} f_1~,~
\gamma_7 = z_2 f_1~,~
\gamma_8 = z_5^{-1} z_6^{-1} f_1 f_6~.~
\eea

The commutation matrix $C$ for Model 9c
is given by, 
\beal{es18c12}
&&
C=
\left(
\ba{c|cccccccc}
\; & \gamma_1
& \gamma_2
& \gamma_3
& \gamma_4 
& \gamma_5  
& \gamma_6 
& \gamma_7 
& \gamma_8 
\\
\hline
\gamma_1 &        0 & 1 & 1 & 1 & 1 & 0 & 0 & -1 \\
\gamma_2 &         -1 & 0 & 1 & 1 & 2 & 1 & 1 & 0 \\
\gamma_3 &         -1 & -1 & 0 & 0 & 1 & 1 & 1 & 1 \\
\gamma_4 &         -1 & -1 & 0 & 0 & 1 & 1 & 1 & 1 \\
\gamma_5 &         -1 & -2 & -1 & -1 & 0 & 1 & 1 & 2 \\
\gamma_6 &         0 & -1 & -1 & -1 & -1 & 0 & 0 & 1 \\
\gamma_7 &         0 & -1 & -1 & -1 & -1 & 0 & 0 & 1 \\
\gamma_8 &   1 & 0 & -1 & -1 & -2 & -1 & -1 & 0 \\
\ea
\right)
~.~
\eea
The 1-loops 
satisfying the commutation relations
can be written in terms of the canonical variables as follows, 
\beal{es18c13}
&
\gamma_1 = e^{-P} z_3 z_5~,~
\gamma_2 = e^{Q-P} z_3 z_4~,~
\gamma_3 = e^{Q} z_3 z_4~,~
\gamma_4 = e^{Q} z_1 z_2 z_3 z_4~,~
&
\nn\\
&
\gamma_5 = e^{Q+P} z_1 z_2 z_3 z_4~,~
\gamma_6 = e^{P} z_5^{-1} z_6^{-1}~,~
\gamma_7 =e^{P} z_2 ~,~
\gamma_8 = e^{-Q+P} z_2 ~.~
&
\eea
\\

%=================================================================
\section{Model 10: $\text{dP}_{3}$}
%=================================================================

%=================================================================
\subsection{Model 10a}
%=================================================================
%------------------------------------------------------------------------------------------------------------------
\begin{figure}[H]
\begin{center}
\resizebox{0.9\hsize}{!}{
\includegraphics{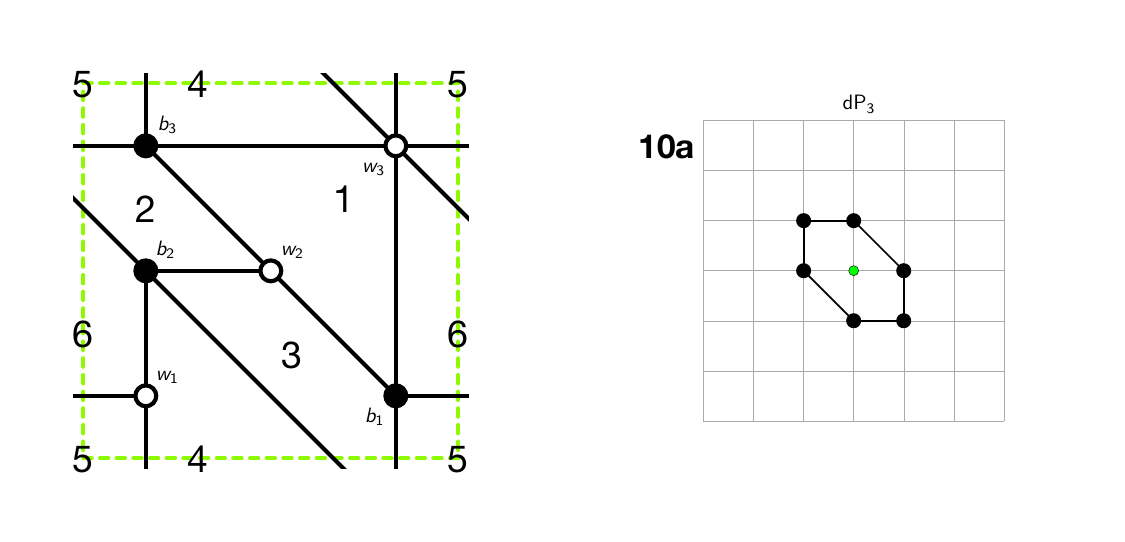}
}
\vspace{-0.5cm}
\caption{The brane tiling and toric diagram of Model 10a.}
\label{mf_10A}
 \end{center}
 \end{figure}
%------------------------------------------------------------------------------------------------------------------

The brane tiling for Model 10a 
can be expressed in terms of the following pair of permutation tuples
\beal{es19a01}
\sigma_B &=& (e_{11} \ e_{31}^{2} \ e_{21}\ e_{31}^{1}) \ (e_{12}\ e_{32}^{2}\ e_{22}\ e_{32}^{1}) \ (e_{13} \ e_{33}^{2} \ e_{23}\ e_{33}^{1}) ~,
\nn\\
\sigma_W^{-1} &=& (e_{11}\ e_{12}\ e_{13}) \ (e_{21}\ e_{22}\ e_{23}) \ (e_{31}^{1}\ e_{33}^{2}\ e_{32}^{1}\ e_{31}^{2}\ e_{33}^{1}\ e_{32}^{2})~,
\eea
which correspond to black and white nodes in the brane tiling, respectively.\\

The brane tiling for Model 10a has 6 zig-zag paths given by, 
\beal{es19a03}
&
z_1 = (e_{31}^{2,+}, e_{21}^{-}~ e_{22}^{+}~ e_{32}^{1,-})~,~
z_2 = (e_{23}^{+}~ e_{33}^{1,-}, e_{32}^{2,+}, e_{22}^{-})~,~
&
\nn\\
&
z_3 = (e_{21}^{+}~ e_{31}^{1,-}, e_{33}^{2,+}, e_{23}^{-})~,~
z_4 = (e_{31}^{1,+}, e_{11}^{-}~ e_{12}^{+}~ e_{32}^{2,-})~,~
&
\nn\\
&
z_5 = (e_{33}^{1,+}, e_{13}^{-}~ e_{11}^{+}~ e_{31}^{2,-})~,~
z_6 = (e_{32}^{1,+}, e_{12}^{-}~ e_{13}^{+}~ e_{33}^{2,-})~,~
\eea
and 6 face paths given by, 
\beal{es19a04}
&
f_1 = (e_{21}^{1,+}, e_{31}^{2,-}, e_{33}^{1,+}, e_{23}^{1,-})~,~
f_2 = (e_{23}^{1,+}, e_{33}^{2,-}, e_{32}^{1,+}, e_{22}^{1,-})~,~
&
\nn\\
&
f_3 = (e_{22}^{1,+}, e_{32}^{2,-}, e_{31}^{1,+}, e_{21}^{1,-})~,~
f_4 = (e_{32}^{2,+}, e_{12}^{1,-}, e_{13}^{1,+}, e_{33}^{1,-})~,~
&
\nn\\
&
f_5 = (e_{33}^{2,+}, e_{13}^{1,-}, e_{11}^{1,+}, e_{31}^{1,-})~,~
f_6 = (e_{31}^{2,+}, e_{11}^{1,-}, e_{12}^{1,+}, e_{32}^{1,-})~,~
\eea
which satisfy the following relations, 
\beal{es19a05}
&
f_4 f_5 f_6=z_1 z_2 z_3 ~,~
f_3 f_6=z_1 z_4~,~
f_2 f_5^{-1} f_6^{-1}=z_2 z_4 z_5 z_6^2~,~
&
\nn\\
&
f_1 f_5=z_3 z_5~,~
f_1 f_2 f_3 f_4 f_5 f_6=1~.~
&
\eea
The face paths can be written in terms of the canonical variables as shown below, 
\beal{es19a05_1}
&
f_1=e^{P},~f_2=e^{Q},~f_3=z_4 z_5 z_6 e^{-Q-P},~f_4=z_2 z_6 e^{-Q},~
&
\nn\\
&
f_5=z_3 z_5 e^{-P},~f_6=z_1 z_5^{-1} z_6^{-1} e^{Q+P}~.~ 
&
\eea

The Kasteleyn matrix of the brane tiling for Model 10a in \fref{mf_10A} 
is given by, 
\beal{es19a06}
K = 
\begin{pmatrix}
e_{11} x^{-1} & e_{12} & e_{13} y^{-1} \\
e_{21} & e_{22} & e_{23} \\
e_{31}^{1} y +e_{31}^{2} & e_{32}^{1} x +e_{32}^{2} y & e_{33}^{1}+e_{33}^{2} x \\
\end{pmatrix} ~.~
\eea
By taking the permanent of the Kasteleyn matrix in \eref{es19a06}, 
we obtain the spectral curve of the dimer integrable system for Model 10a as shown below,
\beal{es19a07}
&&
0 = \text{perm}~K=\overline{p}_0 \cdot \big[\delta_{(1,-1)}\frac{x}{y}+\delta_{(0,-1)}\frac{1}{y}+\delta_{(1,0)}x+\delta_{(-1,0)}\frac{1}{x}+\delta_{(-1,1)}\frac{y}{x}+\delta_{(0,1)}y+H\big]
~,~
\nn\\
\eea
where $\overline{p}_0= e_{12} e_{21} e_{33}^{2}$.

The Casimirs $\delta_{(m,n)}$ in \eref{es19a07} can be written in terms of the zig-zag paths in \eref{es19a03} as follows,
\beal{es19a08}
&
\delta_{(1,0)}=1~,~\delta_{(-1,0)}= z_1 z_5 z_6 ~,~\delta_{(0,1)}= z_3^{-1}~,~
&
\nn\\
&
\delta_{(0,-1)}=z_1 z_6~,~\delta_{(1,-1)}= z_6~,~\delta_{(-1,1)}=z_1 z_2 z_5 z_6~.~
\eea
Accordingly, we can express the spectral curve for Model 10a in the following form, 
\beal{es19a09}
\Sigma~:~
z_6\frac{x}{y}+z_1 z_6 \frac{1}{y}+x+z_1 z_5 z_6 \frac{1}{x}+z_1 z_2 z_5 z_6 \frac{y}{x}+z_3^{-1} y+H
= 0
~.~
\eea

The Hamiltonian is a sum over all 6 1-loops $\gamma_i$,
\beal{es19a10}
H= \gamma_1+\gamma_2+\gamma_3+\gamma_4+\gamma_5+\gamma_6~,~
\eea
where the 1-loops $\gamma_i$ can be expressed in terms of zig-zag paths and face paths as follows, 
\beal{es19a11}
&
\gamma_1 =  z_5 z_6 f_1^{-1}  ~,~
\gamma_2 =  z_5 z_6 f_1^{-1} f_6~,~ 
\gamma_3 =  z_5 z_6 f_6 ~,~
&
\nn\\
&
\gamma_4 = z_1 z_2 z_6 f_1 ~,~
\gamma_5 = z_1 z_2 z_6 f_1 f_3~,~ 
\gamma_6 = z_4^{-1} f_3~.~
&
\eea

The commutation matrix for Model 10a is given by, 
\beal{es19a12}
&&
C =
\left(
\begin{array}{c| c c c c c c}
    & \gamma_1 & \gamma_2 & \gamma_3 & \gamma_4 & \gamma_5 & \gamma_6  \\
    \hline
    \gamma_1      &    0 & 1 & 1 & 0 & -1 & -1 \\
    \gamma_2      &   -1 & 0 & 1 & 1 & 0 & -1 \\
    \gamma_3      &   -1 & -1 & 0 & 1 & 1 & 0 \\
    \gamma_4      &   0 & -1 & -1 & 0 & 1 & 1 \\
    \gamma_5      &   1 & 0 & -1 & -1 & 0 & 1 \\
    \gamma_6      &   1 & 1 & 0 & -1 & -1 & 0 \\
\end{array}
\right)
~.~
\eea
The 1-loops satisfying the commutation relations
can be written in terms of the canonical variables as shown below, 
\beal{es19a13}
&
\gamma_1 =  z_5 z_6 e^{-P}~,~
\gamma_2 =  z_1 e^{Q}~,~ 
\gamma_3 =  z_1 e^{Q+P}~,~
&
\nn\\
&
\gamma_4 = z_1 z_2 z_6 e^{P}~,~
\gamma_5 = z_3^{-1} z_6 e^{-Q}~,~ 
\gamma_6 = z_5 z_6 e^{-Q-P} ~.~
\eea
\\
 
 %=================================================================
\subsection{Model 10b}
%=================================================================
%------------------------------------------------------------------------------------------------------------------
\begin{figure}[H]
\begin{center}
\resizebox{0.9\hsize}{!}{
\includegraphics{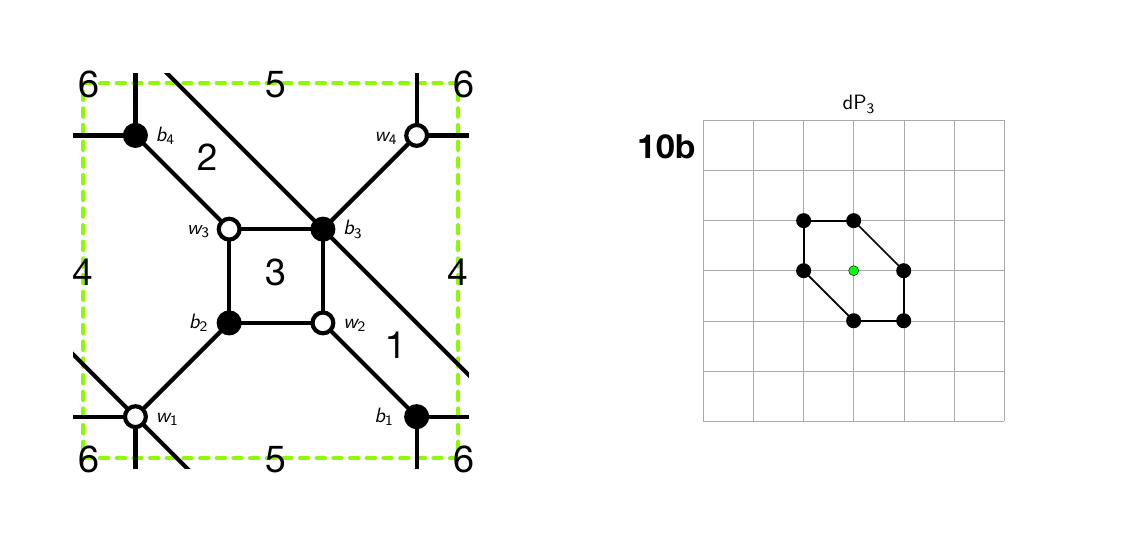}
}
\vspace{-0.5cm}
\caption{The brane tiling and toric diagram of Model 10b.}
\label{mf_10B}
 \end{center}
 \end{figure}
%------------------------------------------------------------------------------------------------------------------

The brane tiling for Model 10b 
can be expressed in terms of the following pair of permutation tuples
\beal{es19b01}
\sigma_B &=& (e_{11}\ e_{21}\ e_{41}) \ (e_{12}\ e_{22}\ e_{32})  \ (e_{23}\ e_{13}^{1}\ e_{43}\ e_{13}^{2}\ e_{33}) \ (e_{14}\ e_{44}\ e_{34}) ~,
\nn\\
\sigma_W^{-1} &=& (e_{11}\ e_{13}^{2}\ e_{12}\ e_{13}^{1}\ e_{14}) \ (e_{21}\ e_{22}\ e_{23}) \ (e_{32}\ e_{34}\ e_{33}) \ (e_{41}\ e_{44}\ e_{43})~,
\eea
which correspond to black and white nodes in the brane tiling, respectively.\\
 
The brane tiling for Model 10b has 6 zig-zag paths given by, 
\beal{es19b03}
&
z_1 = (e_{23}^{+}~ e_{13}^{2,-}, e_{12}^{+}~ e_{22}^{-})~,~
z_2 = (e_{11}^{+}~ e_{21}^{-}~ e_{22}^{+}~ e_{32}^{-}~ e_{34}^{+}~ e_{14}^{-})~,~
&
\nn\\
&
z_3= (e_{13}^{1,+}, e_{33}^{-}~ e_{32}^{+}~ e_{12}^{-})~,~
z_4 = (e_{21}^{+}~ e_{41}^{-}~ e_{44}^{+}~ e_{34}^{-}~ e_{33}^{+}~ e_{23}^{-})~,~
&
\nn\\
&
z_5 = (e_{14}^{+}~ e_{44}^{-}~ e_{43}^{+}~ e_{13}^{1,-})~,~
z_6 = (e_{13}^{2,+}, e_{43}^{-}~ e_{41}^{+}~ e_{11}^{-})~,~
\eea
and 6 face paths given by, 
\beal{es19b04}
&
f_1 = (e_{13}^{2,+}, e_{23}^{-}~ e_{21}^{+}~ e_{11}^{-})~,~
f_2 = (e_{33}^{+}~ e_{13}^{1,-}, e_{14}^{+}~ e_{34}^{-})~,~
&
\nn\\
&
f_3 = (e_{23}^{+}~ e_{33}^{-}~ e_{32}^{+}~ e_{22}^{-})~,~
f_4 = (e_{34}^{+}~ e_{44}^{-}~ e_{43}^{+}~ e_{13}^{2,-}, e_{12}^{+}~ e_{32}^{-})~,~
&
\nn\\
&
f_5 = (e_{13}^{1,+}, e_{43}^{-}~ e_{41}^{+}~ e_{21}^{-}~ e_{22}^{+}~ e_{12}^{-})~,~
f_6 = (e_{11}^{+}~ e_{41}^{-}~ e_{44}^{+}~ e_{14}^{-})~.~
\eea
which satisfy the following constraints, 
\beal{es19b05}
&
f_1 f_2 f_3 f_4 f_5 f_6=1~,~
f_3 f_6^{-1} =  z_1 z_3 z_5 z_6~,~
f_2 f_5^{-1} f_6^{-1} = z_1 z_4 z_5^2 z_6~,~ 
&
\nn\\
&
f_1 f_4^{-1} f_6^{-1} = z_3 z_4 z_5 z_6^2~,~ 
f_4 f_5 f_6^2 = z_1 z_2^2 z_3 z_4~.~ 
&
\eea
The face paths can be written in terms of the canonical variables as follows, 
\beal{es19b05_1}
&
f_1=z_4 z_5 z_6 e^{P},~ f_2 =e^{-P},~ f_3=e^Q,~f_4=z_1 z_5 e^{-Q+P},
&
\nn\\
&
f_5=z_3 z_4^{-1} z_5^{-1} e^{-Q-P},~ f_6=z_2 z_4 e^Q
&
\eea

The Kasteleyn matrix of the brane tiling for Model 10b in \fref{mf_10B} is 
given by, 
\beal{es19b06}
K = 
\begin{pmatrix}
e_{11} x^{-1} & e_{12} & e_{13}^1 y^{-1} + e_{13}^2 x^{-1} & e_{14} y^{-1} \\
e_{21} & e_{22} & e_{23} &0\\
0 & e_{32} & e_{33} & e_{34} \\
e_{41} y & 0 & e_{43} & e_{44} x\\
\end{pmatrix} ~.~
\eea
The permanent of the Kasteleyn matrix gives the spectral curve of the dimer integrable system for Model 10b as follows,
\beal{es19b07}
&&
0 = \text{perm}~K=\overline{p}_0 \cdot \big[\delta_{(1,-1)}\frac{x}{y}+\delta_{(0,-1)}\frac{1}{y}+\delta_{(1,0)}x+\delta_{(-1,0)}\frac{1}{x}+\delta_{(-1,1)}\frac{y}{x}+\delta_{(0,1)}y+H\big]
~,~
\nn\\
\eea
where $\overline{p}_0= e_{12} e_{21} e_{33} e_{44}$.
The Casimirs $\delta_{(m,n)}$ in \eref{es19b07} can be written in terms of the zig-zag paths in \eref{es19b03} as shown below, 
\beal{es19b08}
&
\delta_{(1,0)}=1~,~\delta_{(-1,0)}= z_2 z_3 z_5~,~\delta_{(0,1)}= z_4^{-1}~,~
&
\nn\\
&
\delta_{(0,-1)}=z_3 z_5~,~\delta_{(1,-1)}=z_3 ~,~\delta_{(-1,1)}=z_2 z_3 z_5 z_6
~,~
\eea
such that the spectral curve for Model 10b takes the following form, 
\beal{es19b09}
\Sigma ~:~
z_3 \frac{x}{y}+z_3 z_5\frac{1}{y}+x+z_2 z_3 z_5 \frac{1}{x}+z_2 z_3 z_5 z_6 \frac{y}{x}+z_4^{-1} y+H
= 0 
~.~
\eea

The Hamiltonian is a sum over all 7 1-loops $\gamma_i$,
\beal{es19b10}
H= \gamma_1+\gamma_2+\gamma_3+\gamma_4+\gamma_5+\gamma_6+\gamma_7~,~
\eea
where the 1-loops $\gamma_i$ can be expressed in terms of zig-zag paths and face paths as follows, 
\beal{es19b11}
&
\gamma_1 =  z_5 f_1 f_3 f_5 f_6~,~
\gamma_2 =  z_1 z_2 z_3 z_5 f_1~,~
\gamma_3 = z_3 z_4^{-2} z_5^{-1} z_6^{-1} f_1 f_2 f_3^{-1}~,~ 
&
\nn\\
&
\gamma_4 = z_3 z_4^{-1} f_2 f_3^{-1}~,~
\gamma_5 =  z_5 f_3 f_5 ~,~
\gamma_6 = z_2 z_3 f_2~,~
\gamma_7 = z_5 f_3 f_5 f_6~,~ 
&
\eea

The commutation matrix for Model 10b is given by,
\beal{es19b12}
&&
C =
\left(
\begin{array}{c| c c c c c c c}
    & \gamma_1 & \gamma_2 & \gamma_3 & \gamma_4 & \gamma_5 & \gamma_6 & \gamma_7 \\
    \hline
    \gamma_1      &    0 & 1 & 0 & -1 & -1 & -1 & -1 \\
    \gamma_2      &    -1 & 0 & 1 & 1 & 0 & 0 & -1 \\
    \gamma_3      &    0 & -1 & 0 & 1 & 1 & 1 & 1 \\
    \gamma_4      &    1 & -1 & -1 & 0 & 1 & 1 & 2 \\
    \gamma_5      &    1 & 0 & -1 & -1 & 0 & 0 & 1 \\
    \gamma_6      &    1 & 0 & -1 & -1 & 0 & 0 & 1 \\
    \gamma_7      &1 & 1 & -1 & -2 & -1 & -1 & 0 \\
\end{array}
\right)
~,~
\eea
where
the 1-loops satisfying the commutation relations
can be written in terms of the canonical variables as follows, 
\beal{es19b13}
&
\gamma_1 = z_1^{-1} e^Q ~,~
\gamma_2 = z_5 e^P~,~ 
\gamma_3 = z_3 z_4^{-1} e^{-Q}~,~
&
\nn\\
&
\gamma_4 = z_3 z_4^{-1} e^{-Q-P}~,~
\gamma_5 = z_3 z_4^{-1} e^{-P}~,~ 
\gamma_6 = z_2 z_3 e^{-P}~,~
\gamma_7 = z_2 z_3 e^{Q-P}~,~
\eea
\\
 
%=================================================================
\subsection{Model 10c}
%=================================================================
%------------------------------------------------------------------------------------------------------------------
\begin{figure}[H]
\begin{center}
\resizebox{0.9\hsize}{!}{
\includegraphics{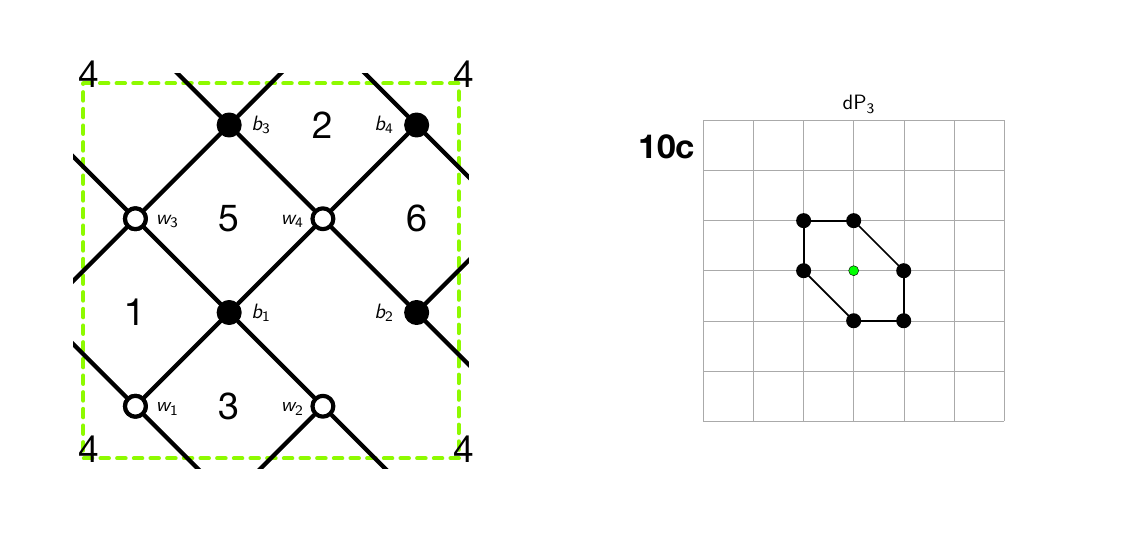}
}
\vspace{-0.5cm}
\caption{The brane tiling and toric diagram of Model 10c.}
\label{mf_10C}
 \end{center}
 \end{figure}
%------------------------------------------------------------------------------------------------------------------

The brane tiling for Model 10c 
can be expressed in terms of the following pair of permutation tuples
\beal{es19c01}
\sigma_B &=& (e_{11}\ e_{21}\ e_{41}\ e_{31})\ (e_{12}\ e_{32}\ e_{42})\ (e_{13}\ e_{33}\ e_{43}\ e_{23})\ (e_{24}\ e_{44}\ e_{34})
\nn\\
\sigma_W^{-1} &=&(e_{11}\ e_{13}\ e_{12})\ (e_{21}\ e_{24}\ e_{23})\ (e_{31}\ e_{32}\ e_{34}\ e_{33})\ (e_{41}\ e_{43}\ e_{44}\ e_{42})~,
\eea
which correspond to black and white nodes in the brane tiling, respectively.\\
 
The brane tiling for Model 10c has 6 zig-zag paths given by, 
\beal{es19c03}
&
z_1 = (e_ {31}^{+}~ e_ {11}^{-}~ e_ {13}^{+}~ e_ {33}^{-})~,~
z_2 = (e_ {12}^{+}~ e_ {32}^{-}~ e_ {34}^{+}~ e_{24}^{-}~ e_{23}^{+}~ e_{13}^{-})~,~
&
\nn\\
&
z_3= (e_{43}^{+}~ e_{23}^{-}~ e_{21}^{+}~ e_{41}^{-})~,~
z_4 = (e_{11}^{+}~ e_{21}^{-}~ e_{24}^{+}~ e_{44}^{-}~ e_{42}^{+}~ e_{12}^{-})~,~
&
\nn\\
&
z_5 = (e_{32}^{+}~ e_{42}^{-}~ e_{41}^{+}~ e_{31}^{-})~,~
z_6 = (e_{44}^{+}~ e_{34}^{-}~ e_{33}^{+}~ e_{43}^{-})~,~
\eea
and 6 face paths given by, 
\beal{es19c04}
&
f_1 = (e_{11}^{+}~ e_{31}^{-}~ e_{32}^{+}~ e_{12}^{-})~,~
f_2 = (e_{23}^{+}~ e_{43}^{-}~ e_{44}^{+}~ e_{24}^{-})~,~
&
\nn\\
&
f_3 = (e_{21}^{+}~ e_{11}^{-}~ e_{13}^{+}~ e_{23}^{-})~,~
f_4 = (e_{12}^{+}~ e_{42}^{-}~ e_{41}^{+}~ e_{21}^{-}~ e_{24}^{+}~ e_{34}^{-}~ e_{33}^{+}~ e_{13}^{-})~,~
&
\nn\\
&
f_5 = (e_{31}^{+}~ e_{41}^{-}~ e_{43}^{+}~ e_{33}^{-})~,~
f_6 = (e_{34}^{+}~ e_{44}^{-}~ e_{42}^{+}~ e_{32}^{-})~.~
\eea
which satisfy the following relations, 
\beal{es19c05}
&
f_5 f_6=z_1 z_2 z_3 z_4~,~
f_3 f_6^{-1}=z_1 z_3 z_5 z_6~,~
f_1 f_2^{-1} = z_3 z_4 z_5 ~,~
&
\nn\\
&
f_2^2 f_4 f_5^{-1}=z_2 z_3^{-1} z_5 z_6^2~,~
f_1 f_2 f_3 f_4 f_5 f_6=1
&
\eea
The face paths can be written in terms of the canonical variables as shown below, 
\beal{es19c05_1}
&
f_1=e^{P}~,~
f_2=e^{P} z_1 z_2 z_6~,~
f_3=e^{-Q} z_1 z_3~,~
&
\nn\\
&
f_4=e^{Q-2P} z_1^{-1} z_4 z_5^2 z_6~,~
f_5=e^{Q}~,~
f_6=e^{-Q} z_1 z_2 z_3 z_4~,~
&
\eea

The Kasteleyn matrix of the brane tiling for Model 10c in \fref{mf_10C} is 
given by,
\beal{es19c06}
K = 
\left(
\ba{c|cccc}
\; & b_1 & b_2 & b_3 & b_4 
\\
\hline
w_1 & e_{11} & e_{12} x^{-1} & e_{13}  y^{-1} &0
\\
w_2 & e_{21} & 0 & e_{23} y^{-1} & e_{24} y^{-1}
\\
w_3 & e_{31} & e_{32} x^{-1} & e_{33} & e_{34} x^{-1}
\\
w_4 & e_{41} & e_{42} & e_{43} & e_{44}
\ea
\right)
~.~
\eea
By taking the permanent of the Kasteleyn matrix, 
we obtain the spectral curve of the dimer integrable system for Model 10c as follows,
\beal{es19c07}
&&
0 = \text{perm}~K=\overline{p}_0 \cdot x^{-1} y^{-1} \cdot \big[\delta_{(-1,0)}\frac{1}{x}+\delta_{(-1,1)}\frac{y}{x}+\delta_{(0,-1)}\frac{1}{y}+\delta_{(0,1)}y
\nn\\
&&
\hspace{1cm}
+\delta_{(1,-1)}\frac{x}{y}+\delta_{(1,0)}x+H\big]
~,~
\eea
where $\overline{p}_0= e_{11}^{+} e_{24}^{+} e_{33}^{+} e_{42}^{+}$.
The Casimirs $\delta_{(m,n)}$ in \eref{es19c07} can be written in terms of the zig-zag paths in \eref{es19c03} as shown below, 
\beal{es19c08}
&
\delta_{(-1,0)} = z_1 z_2 z_5~,~
\delta_{(-1,1)} = z_1 z_2 z_3 z_5~,~
\delta_{(0,-1)} = z_1 z_5 ~,~
&
\nn\\
&
\delta_{(0,1)} = z_1 z_2 z_3 z_5 z_6~,~
\delta_{(1,-1)} = z_1~,~
\delta_{(1,0)} = 1~,~
&
\eea
allows us to express the spectral curve for Model 10c in the following form, 
\beal{es19c09}
\Sigma~:~
\Big(1+\frac{z_1}{y} \Big)x + \frac{y}{z_4}+\frac{z_1 z_5}{y}+(1+z_3 y) \frac{z_1 z_2 z_5}{x}+H
= 0
~.~
\eea

The Hamiltonian is a sum over all 8 1-loops $\gamma_i$,
\beal{es19c10}
H= \gamma_1+\gamma_2+\gamma_3+\gamma_4+\gamma_5+\gamma_6+\gamma_7+\gamma_8~,~
\eea
where the 1-loops $\gamma_i$ can be written in terms of zig-zag paths and face paths as follows,
\beal{es19c11}
&
\gamma_1 =  z_1 z_2 f_1 f_3 f_4~,~
\gamma_2 =  z_1 z_3 z_5 f_2 f_4 f_5 f_6~,~
\gamma_3 = z_5 f_5~,~ 
\gamma_4 = z_1 z_2 z_5 z_6 f_5~,~
&
\nn\\
&
\gamma_5 =  z_1 z_2 z_5 z_6 f_1 f_5~,~
\gamma_6 = z_1 z_4^{-1} f_1~,~
\gamma_7 = z_1 z_2 f_1~,~ 
\gamma_8 = z_1 z_4^{-1} f_1 f_6~,~ 
&
\eea

The commutation matrix $C$ for Model 10c
is given by, 
\beal{es19c12}
&&
C =
\left(
\begin{array}{c| c c c c c c c c}
    & \gamma_1 & \gamma_2 & \gamma_3 & \gamma_4 & \gamma_5 & \gamma_6 & \gamma_7 & \gamma_8 \\
    \hline
    \gamma_1      &  0 & 1 & 1 & 1 & 1 & 0 & 0 & -1 \\
    \gamma_2      &  -1 & 0 & 1 & 1 & 2 & 1 & 1 & 0 \\
    \gamma_3      &  -1 & -1 & 0 & 0 & 1 & 1 & 1 & 1 \\
    \gamma_4      &  -1 & -1 & 0 & 0 & 1 & 1 & 1 & 1 \\
    \gamma_5      &  -1 & -2 & -1 & -1 & 0 & 1 & 1 & 2 \\
    \gamma_6      &  0 & -1 & -1 & -1 & -1 & 0 & 0 & 1 \\
    \gamma_7      & 0 & -1 & -1 & -1 & -1 & 0 & 0 & 1 \\
    \gamma_7      & 1 & 0 & -1 & -1 & -2 & -1 & -1 & 0 \\
    
\end{array}
\right)
~.~
\eea
The 1-loops satisfying the commutation relations
can be written in terms of the canonical variables as follows, 
\beal{es19c13}
&
\gamma_1 = e^{-P} z_5~,~
\gamma_2 = e^{Q-P} z_5~,~ 
\gamma_3 = e^{Q} z_5~,~
\gamma_4 = e^{Q} z_3^{-1} z_4^{-1}~,~
&
\nn\\
&
\gamma_5 = e^{Q+P} z_3^{-1} z_4^{-1}~,~
\gamma_6 = e^{P} z_1 z_4^{-1}~,~
\gamma_7 = e^{P} z_1 z_2~,~
\gamma_8 = e^{-Q+P} z_1^2 z_2 z_3~,~
\eea
\\
 
%=================================================================
\subsection{Model 10d}
%=================================================================
%------------------------------------------------------------------------------------------------------------------
\begin{figure}[H]
\begin{center}
\resizebox{0.9\hsize}{!}{
\includegraphics{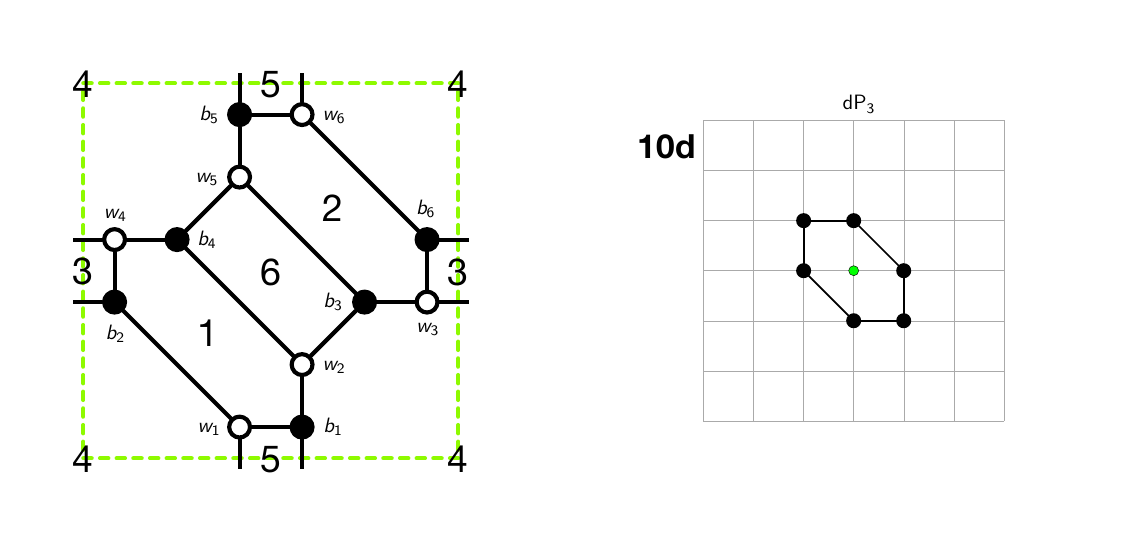}
}
\vspace{-0.5cm}
\caption{The brane tiling and toric diagram of Model 10d.}
\label{mf_10D}
 \end{center}
 \end{figure}
%------------------------------------------------------------------------------------------------------------------

The brane tiling for Model 10d 
can be expressed in terms of the following pair of permutation tuples
\beal{es19d01}
\sigma_B &=& (e_{11}\ e_{61}\ e_{21})\ (e_{12}\ e_{42}\ e_{32})\ (e_{23}\ e_{33}\ e_{53})\ (e_{24}\ e_{54}\ e_{44})
\nn\\
&& 
(e_{15}\ e_{55}\ e_{65})\ (e_{36}\ e_{46}\ e_{66})
\nn\\
\sigma_W^{-1} &=& (e_{11}\ e_{15}\ e_{12})\ (e_{21}\ e_{24}\ e_{23})\ (e_{32}\ e_{33}\ e_{36})\ (e_{42}\ e_{46}\ e_{44})
\nn\\
&& 
(e_{53}\ e_{54}\ e_{55})\ (e_{61}\ e_{66}\ e_{65})
\eea
which correspond to black and white nodes in the brane tiling, respectively.\\
 
The brane tiling for Model 10d has 6 zig-zag paths given by, 
\beal{es19d03}
&
z_1 = (e_{12}^{+}~ e_{42}^{-}~ e_{46}^{+}~ e_{66}^{-}~ e_{65}^{+}~ e_{15}^{-})~,~
z_2 = (e_{21}^{+}~ e_{11}^{-}~ e_{15}^{+}~ e_{55}^{-}~ e_{53}^{+}~ e_{23}^{-})~,~
&
\nn\\
&
z_3= (e_{44}^{+}~ e_{24}^{-}~ e_{23}^{+}~ e_{33}^{-}~ e_{36}^{+}~ e_{46}^{-})~,~
z_4 =(e_{11}^{+}~ e_{61}^{-}~ e_{66}^{+}~ e_{36}^{-}~ e_{32}^{+}~ e_{12}^{-})~,~
&
\nn\\
&
z_5 = (e_{24}^{+}~ e_{54}^{-}~ e_{55}^{+}~ e_{65}^{-}~ e_{61}^{+}~ e_{21}^{-})~,~
z_6 = (e_{42}^{+}~ e_{32}^{-}~ e_{33}^{+}~ e_{53}^{-}~ e_{54}^{+}~ e_{44}^{-})~,~
\eea
and 6 face paths given by, 
\beal{es19d04}
&
f_1 = (e_{42}^{+}~ e_{12}^{-}~ e_{11}^{+}~ e_{21}^{-}~ e_{24}^{+}~ e_{44}^{-})~,~
f_2 = (e_{36}^{+}~ e_{66}^{-}~ e_{65}^{+}~ e_{55}^{-}~ e_{53}^{+}~ e_{33}^{-})~,~
&
\nn\\
&
f_3 = (e_{32}^{+}~ e_{42}^{-}~ e_{46}^{+}~ e_{36}^{-})~,~
f_4 = (e_{44}^{+}~ e_{54}^{-}~ e_{55}^{+}~ e_{15}^{-}~ e_{12}^{+}~ e_{32}^{-}~ e_{33}^{+}~ e_{23}^{-}~ e_{21}^{+}~ e_{61}^{-}~ e_{66}^{+}~ e_{46}^{-})~,~
&
\nn\\
&
f_5 = (e_{15}^{+}~ e_{65}^{-}~ e_{61}^{+}~ e_{11}^{-})~,~
f_6 = (e_{23}^{+}~ e_{53}^{-}~ e_{54}^{+}~ e_{24}^{-})~.~
\eea
which satisfy the following relations, 
\beal{es19d05}
&
f_1^{-1} f_2=z_1 z_2 z_3~,~
f_3 f_5^{-1}=z_1 z_4~,~
f_5^{-1} f_6=z_2^{-1} z_5^{-1}~,~
&
\nn\\
&
f_1^2 f_4 f_5^3=z_1^{-1} z_2 z_5^2 z_6~,~
f_1 f_2 f_3 f_4 f_5 f_6 =1~.~
&
\eea
 The face paths can be written in terms of the canonical variables as follows, 
\beal{es19d05_1}
&
f_1=e^{Q}~,~ 
f_2=e^{Q} z_1 z_2 z_3~,~ 
f_3=e^{P} z_1 z_4 ~,~  
&
\nn\\
&
f_4= e^{-2Q-3P} z_1^{-1} z_2 z_5^2 z_6~,~   
f_5= e^{P}~,~
f_6= e^{P} z_2^{-1} z_5^{-1}~.~
&
\eea

The Kasteleyn matrix of the brane tiling for Model 10d in \fref{mf_10D} is 
given by, 
\beal{es19d06}
K = 
\begin{pmatrix}
e_{11} & e_{12} & 0 & 0 & e_{15} y^{-1} & 0 \\
e_{21} & 0 & e_{23} & e_{24} & 0 & 0 \\
0 & e_{32} x & e_{33} & 0 & 0 & e_{36} \\
0 & e_{42} & 0 & e_{44} & 0 & e_{46} x^{-1} \\
0 & 0 & e_{53} & e_{54} & e_{55} & 0 \\
e_{61} y & 0 & 0 & 0 & e_{65} & e_{66}
\end{pmatrix} ~.~
\eea
By taking the permanent of the Kasteleyn matrix in \eref{es19d06}, we obtain the spectral curve of the dimer integrable system for Model 10d as shown below,
\beal{es19d07}
0 &=& \text{perm}~K=\overline{p}_0  \cdot \big[\delta_{(-1,0)}\frac{1}{x}+\delta_{(-1,1)}\frac{y}{x}+\delta_{(0,-1)}\frac{1}{y}+\delta_{(0,1)}y
\nn\\
&&
+\delta_{(1,-1)}\frac{x}{y}+\delta_{(1,0)}x+H\big]~,~
\eea
where $\overline{p}_0= e_{11}^{+} e_{23}^{+} e_{32}^{+} e_{44}^{+} e_{55}^{+} e_{66}^{+}$.
The Casimirs $\delta_{(m,n)}$ in \eref{es19d07} can be expressed in terms of the zig-zag paths in \eref{es19d03} as follows, 
\beal{es19d08}
&
\delta_{(-1,0)} = z_1 z_2 z_6~,~
\delta_{(-1,1)} = z_1 z_2 z_5 z_6~,~
\delta_{(0,-1)} = z_2 z_6 ~,~
&
\nn\\
&
\delta_{(0,1)} = z_4^{-1}~,~
\delta_{(1,-1)} = z_2~,~
\delta_{(1,0)} = 1~,~
&
\eea
such that the spectral curve for Model 10d takes the following form,  
\beal{es19d09}
\Sigma~:~
\Big(1+\frac{z_2}{y} \Big)x + \frac{y}{z_4}+\frac{z_2 z_6}{y}+(1+z_5 y) \frac{z_1 z_2 z_6}{x}+H
= 0
~.~
\eea

The Hamiltonian is a sum over all 11 1-loops $\gamma_i$,
\beal{es19d10}
H= \sum_{i=1}^{11} \gamma_{i}~,~
\eea
where the 1-loops $\gamma_i$'s can be expressed in terms of zig-zag paths and face paths as follows, 
\beal{es19d11}
&
\gamma_1 = z_3^{-1} f_1^{-1} f_3^{-1}~,~
\gamma_2 = z_3^{-1} f_3^{-1}~,~
\gamma_3 = z_2 z_4^{-1} f_1 ~,~ 
&
\nn\\
&
\gamma_4 = z_2 z_4^{-1} f_1 f_3 ~,~
\gamma_5 = z_2 z_3 z_4^{-1} z_6 f_1 f_3~,~
\gamma_6 = z_1 z_2 f_1 f_5 f_6 ~,~
&
\nn\\
&
\gamma_7 = z_1 z_2 f_3^{-1}~,~
\gamma_8 = z_1 z_2 f_1 f_3^{-1}~,~
\gamma_9 = z_1 z_2 f_1~,~
&
\nn\\
&
\gamma_{10} = z_1 z_2 z_3 z_6 f_1 ~,~
\gamma_{11} = z_1 z_2 z_3 z_6 f_1 f_3 ~.~
&
\eea

The commutation matrix $C$ for Model 10d is given by, 
\beal{es19d12}
&&
C=
\left(
\ba{c|ccccccccccc}
\; & \gamma_1
& \gamma_2
& \gamma_3
& \gamma_4 
& \gamma_5  
& \gamma_6 
& \gamma_7 
& \gamma_8 
& \gamma_9 
& \gamma_{10} 
& \gamma_{11} 
\\
\hline
\gamma_1     & 0 & 1 & 1 & 0 & 0 & -1 & 1 & 2 & 1 & 1 & 0 \\
\gamma_2     &-1 & 0 & 1 & 1 & 1 & 1 & 0 & 1 & 1 & 1 & 1 \\
\gamma_3     &-1 & -1 & 0 & 1 & 1 & 2 & -1 & -1 & 0 & 0 & 1 \\
\gamma_4     &0 & -1 & -1 & 0 & 0 & 1 & -1 & -2 & -1 & -1 & 0 \\
\gamma_5     &0 & -1 & -1 & 0 & 0 & 1 & -1 & -2 & -1 & -1 & 0 \\
\gamma_6     &1 & -1 & -2 & -1 & -1 & 0 & -1 & -3 & -2 & -2 & -1 \\
\gamma_7     &-1 & 0 & 1 & 1 & 1 & 1 & 0 & 1 & 1 & 1 & 1 \\
\gamma_8     &-2 & -1 & 1 & 2 & 2 & 3 & -1 & 0 & 1 & 1 & 2 \\
\gamma_9     &-1 & -1 & 0 & 1 & 1 & 2 & -1 & -1 & 0 & 0 & 1 \\
\gamma_{10} &-1 & -1 & 0 & 1 & 1 & 2 & -1 & -1 & 0 & 0 & 1 \\
\gamma_{11} &0 & -1 & -1 & 0 & 0 & 1 & -1 & -2 & -1 & -1 & 0 \\
\ea
\right)
~.~
\eea
The 1-loops satisfying the commutation relations
can be written in terms of the canonical variables as follows, 
\beal{es19d13}
&
\gamma_1 = e^{-Q-P} z_2 z_5 z_6~,~
\gamma_2 = e^{-P} z_2 z_5 z_6~,~
\gamma_3 = e^{Q} z_2 z_4^{-1}~,~
\gamma_4 = e^{Q+P} z_1 z_2~,~
&
\nn\\
&
\gamma_5 = e^{Q+P} z_4^{-1} z_5^{-1}~,~
\gamma_6 = e^{Q+2P} z_1 z_5^{-1}~,~
\gamma_7 = e^{-P} z_2 z_4^{-1}~,~
\gamma_8 = e^{Q-P} z_2 z_4^{-1}~,~
&
\nn\\
&
\gamma_9 = e^{Q} z_1 z_2 ~,~
\gamma_{10} = e^{Q} z_4^{-1} z_5^{-1}~,~
\gamma_{11} = e^{Q+P} z_1 z_5^{-1}~.~
\eea
\\

%=================================================================
\section{Model 11: $\text{PdP}_2$}
%=================================================================
%------------------------------------------------------------------------------------------------------------------
\begin{figure}[H]
\begin{center}
\resizebox{0.9\hsize}{!}{
\includegraphics{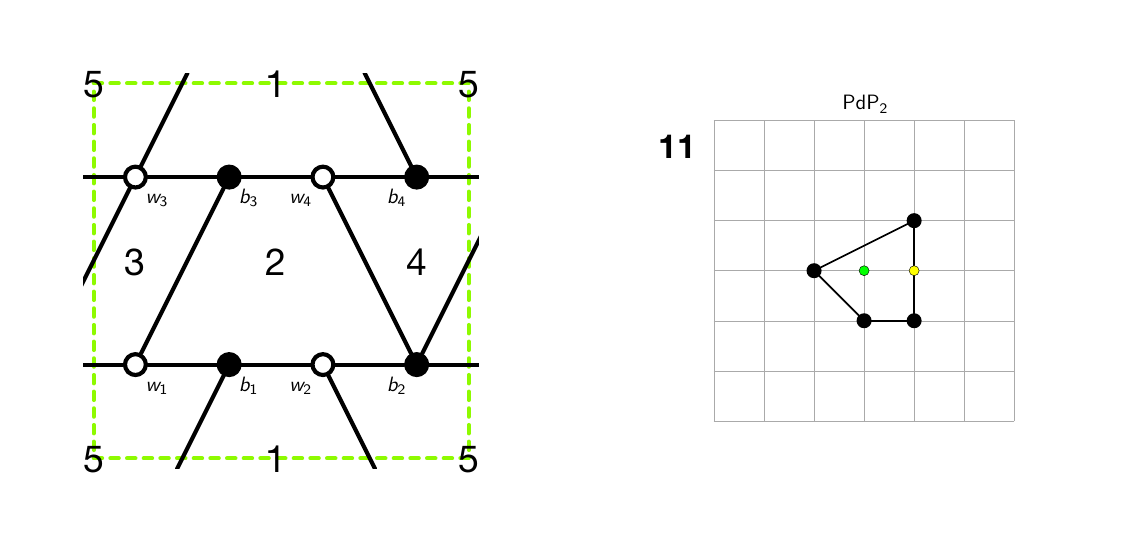}
}
\vspace{-0.5cm}
\caption{The brane tiling and toric diagram of Model 11.}
\label{mf_11}
 \end{center}
 \end{figure}
%------------------------------------------------------------------------------------------------------------------

The brane tiling for Model 11 
can be expressed in terms of the following pair of permutation tuples
\beal{es20a01}
\sigma_B &=& (e_{11}\ e_{31}\ e_{21}) \ (e_{22}\ e_{12}\ e_{32} e_{42})  \ (e_{13}\ e_{43}\ e_{33}) \ (e_{44}\ e_{34}\ e_{24}) ~,
\nn\\
\sigma_W^{-1} &=& (e_{11}\ e_{12}\ e_{13}) \ (e_{21}\ e_{22}\ e_{24}) \ (e_{31}\ e_{33}\ e_{32}~e_{34}) \ (e_{42}\ e_{43}\ e_{44})~,
\eea
which correspond to black and white nodes in the brane tiling, respectively.\\
 
The brane tiling for Model 11 has 5 zig-zag paths given by, 
\beal{es20a03}
&
z_1 = (e_{31}^{+}~ e_{21}^{-}~ e_{22}^{+}~ e_{12}^{-}~ e_{13}^{+}~ e_{43}^{-}~ e_{44}^{+}~ e_{34}^{-})~,~
z_2 = (e_{43}^{+}~ e_{33}^{-}~ e_{32}^{+}~ e_{42}^{-})~,~
&
\nn\\
&
z_3= (e_{24}^{+}~ e_{44}^{-}~ e_{42}^{+}~ e_{22}^{-})~,~
z_4 = (e_{33}^{+}~ e_{13}^{-}~ e_{11}^{+}~ e_{31}^{-})~,~
&
\nn\\
&
z_5 = (e_{21}^{+}~ e_{11}^{-}~ e_{12}^{+}~ e_{32}^{-}~ e_{34}^{+}~ e_{24}^{-})~,~
\eea
and 5 face paths given by, 
\beal{es20a04}
&
f_1 = (e_{21}^{+}~ e_{31}^{-}~ e_{33}^{+}~ e_{43}^{-}~ e_{44}^{+}~ e_{24}^{-})~,~
f_2 = (e_{43}^{+}~ e_{13}^{-}~ e_{11}^{+}~ e_{21}^{-}~ e_{22}^{+}~ e_{42}^{-})~,~
&
\nn\\
&
f_3 = (e_{13}^{+}~ e_{33}^{-}~ e_{32}^{+}~ e_{12}^{-})~,~
f_4 = (e_{42}^{+}~ e_{32}^{-}~ e_{34}^{+}~ e_{44}^{-})~,~
&
\nn\\
&
f_5 = (e_{31}^{+}~ e_{11}^{-}~ e_{12}^{+}~ e_{22}^{-}~ e_{24}^{+}~ e_{34}^{-})~,~
\eea
which satisfy the following relations, 
\beal{es20a05}
&
f_1 f_2 f_3 f_4 f_5 =1~,~
f_3 f_4 f_5 = z_3 z_4^{-1}~,~
&
\nn\\
&
f_2 f_4^{-2} f_5^{-1}  = z_1 z_2^2 z_3^{-1} z_4^2 ~,~ 
f_1 f_3^{-2} f_5^{-1} = z_1 z_4^3 z_5^2~.~ 
&
\eea
The face paths can be written in terms of the canonical variables as shown below, 
\beal{es20a05_1}
&
f_1=z_2^{-1} z_4 z_5 e^{-2Q+P},~ f_2 =z_1 z_2^2 z_4 e^{2Q-P},~ f_3=e^{-Q}
&
\nn\\
&
f_4= e^{Q-P},~ f_5=z_3 z_4^{-1} e^P
&
\eea
The Kasteleyn matrix of the brane tiling for Model 11 in \fref{mf_11} is 
given by,
\beal{es20a06}
K = 
\begin{pmatrix}
e_{11}  & e_{12} x^{-1} & e_{13} & 0 \\
e_{21} & e_{22} & 0 & e_{24} y^{-1}\\
e_{31} y & e_{32} x^{-1} & e_{33} & e_{34} x^{-1} \\
0 & e_{42} & e_{43} & e_{44}\\
\end{pmatrix} ~.~
\eea
By taking the permanent of the Kasteleyn matrix in \eref{es20a06}, we obtain the spectral curve of the dimer integrable system for Model 11 as follows,
\beal{es20a07}
&&
0 = \text{perm}~K=\overline{p}_0 \cdot x^{-1} \cdot \big[\delta_{(-1,0)} x^{-1}+\delta_{(1,0)}x+\delta_{(0,-1)}y^{-1}
\nn\\
&&
\hspace{1cm}
+\delta_{(1,-1)} x y^{-1}+\delta_{(1,1)} x y+H\big]
~,~
\eea
where $\overline{p}_0= e_{13} e_{24} e_{31} e_{42}$.
The Casimirs $\delta_{(m,n)}$ in \eref{es20a07} can be written in terms of the zig-zag paths in \eref{es20a03} as follows,
\beal{es20a08}
&
\delta_{(-1,0)}=z_1^{-1} z_3^{-1}~,~\delta_{(1,0)}= \delta_{(1,0)}^1+\delta_{(1,0)}^2=1+z_3^{-1} z_4~,~\delta_{(0,-1)}= z_2 z_4~,~
&
\nn\\
&
\delta_{(1,-1)}=z_4 ~,~\delta_{(1,1)}=z_3^{-1}~,~
\eea
such that the spectral curve for Model 11 takes the following form,  
\beal{es20a09}
\Sigma~:~
z_1^{-1} z_3^{-1} x^{-1}+(1+z_3^{-1} z_4)x+z_2 z_4 y^{-1}+z_4 x y^{-1}+z_3^{-1} x y+H
= 0
~.~
\eea

The Hamiltonian is a sum over all 5 1-loops $\gamma_i$,
\beal{es20a10}
H= \gamma_1+\gamma_2+\gamma_3+\gamma_4+\gamma_5~,~
\eea
where the 1-loops $\gamma_i$ can be expressed in terms of zig-zag paths and face paths as follows,
\beal{es20a11}
&
\gamma_1 = z_2 f_3^{-1}~,~
\gamma_2 = z_3^{-1} z_4^2 z_5 f_3 f_5 ~,~
\gamma_3 = z_2 f_1~,~ 
&
\nn\\
&
\gamma_4 = z_4 z_5 f_3~,~
\gamma_5 = z_2 f_3^{-1} f_5^{-1}~,~
&
\eea

The commutation matrix $C$ for Model 11
is given by, 
\beal{es20a12}
&&
C =
\left(
\begin{array}{c| c c c c c}
    & \gamma_1 & \gamma_2 & \gamma_3 & \gamma_4 & \gamma_5  \\
    \hline
    \gamma_1      &  0 & 1 & 1 & 0 & -1 \\
    \gamma_2      &  -1 & 0 & 1 & 1 & 0 \\
    \gamma_3      &  -1 & -1 & 0 & 1 & 1 \\
    \gamma_4      &  0 & -1 & -1 & 0 & 1 \\
    \gamma_5.     &  1 & 0 & -1 & -1 & 0 \\
\end{array}
\right)
~,~
\eea
where 
the 1-loops satisfying the commutation relations
can be written in terms of the canonical variables as follows, 
\beal{es20a13}
&
\gamma_1 = e^{Q} z_2 ~,~
\gamma_2 = e^{-Q+P} z_4 z_5~,~
\gamma_3 = e^{-2Q+P} z_4 z_5~,~
&
\nn\\
&
\gamma_4 = e^{-Q} z_4 z_5~,~
\gamma_5 = e^{Q-P} z_2 z_3^{-1} z_4 ~.~
\eea
\\

%=================================================================
\section{Model 12: $\text{dP}_2$}
%=================================================================

%=================================================================
\subsection{Model 12a}
%=================================================================
%------------------------------------------------------------------------------------------------------------------
\begin{figure}[H]
\begin{center}
\resizebox{0.9\hsize}{!}{
\includegraphics{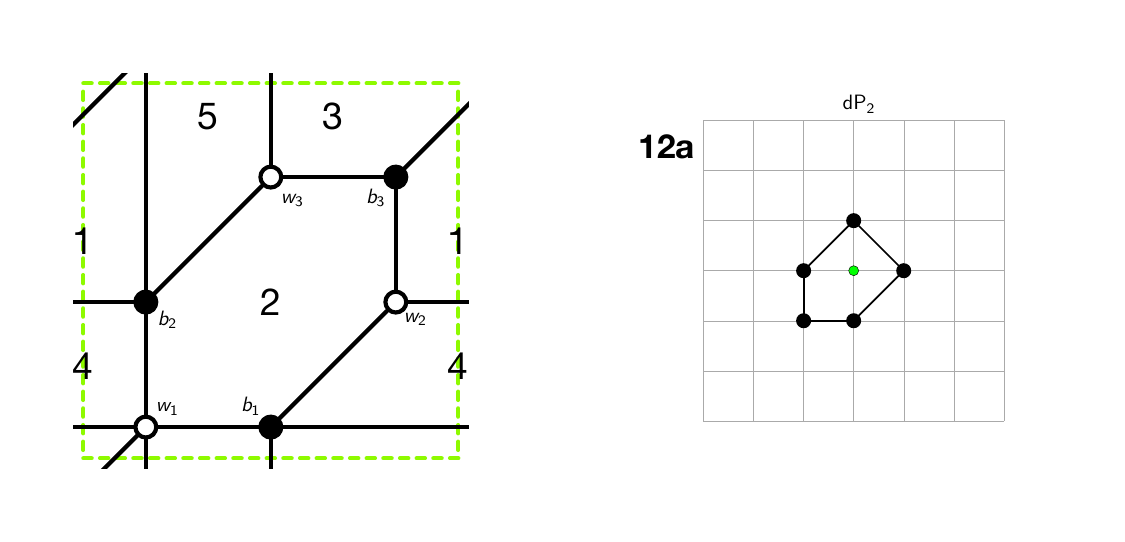}
}
\vspace{-0.5cm}
\caption{The brane tiling and toric diagram of Model 12a.}
\label{mf_12A}
 \end{center}
 \end{figure}
%------------------------------------------------------------------------------------------------------------------

The brane tiling for Model 12a 
can be expressed in terms of the following pair of permutation tuples
\beal{es21a01}
\sigma_B &=& (e_{11}^{1}\ e_{31}\ e_{11}^{2}\ e_{21}) \ (e_{12}^{1} \ e_{22}  \ e_{12}^{2} \ e_{32})  \ (e_{13}\ e_{33}\ e_{23}) ~,
\nn\\
\sigma_W^{-1} &=& (e_{13}\ e_{11}^{2}\ e_{12}^{2}\ e_{11}^{1} \ e_{12}^{1}) \ (e_{21}\ e_{23}\ e_{22}) \ (e_{31}\ e_{33}\ e_{32}) ~,
\eea
which correspond to black and white nodes in the brane tiling, respectively.\\
 
The brane tiling for Model 12a has 5 zig-zag paths 
given by, 
\beal{es21a03}
&
z_1 = (e_ {23}^{+}~ e_ {13}^{-}~ e_ {11}^{2,+}, e_ {21}^{-})~,~
z_2 = (e_ {12}^{1,+}, e_ {22}^{-}~ e_ {21}^{+}~ e_ {11}^{1,-})~,~
&
\nn\\
&
z_3= (e_ {22}^{+}~ e_ {12}^{2,-}, e_ {11}^{1,+}, e_ {31}^{-}~ e_{33}^{+}~ e_ {23}^{-})~,~
z_4 = (e_ {13}^{+}~ e_ {33}^{-}~ e_ {32}^{+}~ e_ {12}^{1,-})~,~
&
\nn\\
&
z_5 = (e_ {31}^{+}~ e_ {11}^{2,-}, e_ {12}^{2,+}, e_ {32}^{-})~,~
\eea
and 5 face paths given by, 
\beal{es21a04}
&
f_1 = (e_ {13}^{+}~ e_ {23}^{-}~ e_ {22}^{+}~ e_ {12}^{1,-})~,~
f_2 = (e_ {23}^{+}~ e_ {33}^{-}~ e_ {32}^{+}~ e_ {12}^{2,-}, e_{11}^{1,+}, e_ {21}^{-})~,~
&
\nn\\
&
f_3 = (e_ {33}^{+}~ e_ {13}^{-}~ e_ {11}^{2,+}, e_ {31}^{-})~,~
f_4 = (e_ {21}^{+}~ e_ {11}^{2,-}, e_ {12}^{2,+}, e_ {22}^{-})~,~
&
\nn\\
&
f_5=(e_ {12}^{1,+}, e_ {32}^{-}~ e_ {31}^{+}~ e_ {11}^{1,-}) ~,~
\eea
which satisfy the following relations, 
\beal{es21a05}
&
f_1 f_2 f_3 f_4 f_5=1~,~
f_4 f_5=z_2 z_5~,~
f_2 f_3^2 f_4^{-1}=z_1^3 z_2 z_3^2 z_4~,~
f_1 f_3^{-1} f_5^{-1}=z_3 z_4^2 z_5~.~
&
\eea
The face paths can be written in terms of the canonical variables as follows, 
\beal{es21a05_1}
&
f_1=e^{Q-P} z_1^{-1} z_4 z_5 ,~
f_2=e^{-2Q+P} z_1^2 z_3 z_5^{-1} ,~
f_3=e^Q,~
f_4=e^P,~
f_5=e^{-P} z_2 z_5 ~.~
&
\eea

The Kasteleyn matrix of the brane tiling for Model 12a in \fref{mf_12A} is 
given by,
\beal{es21a06}
K = 
\begin{pmatrix}
e_{11}^{1}+e_{11}^{2} x^{-1}  & e_{12}^{1} y^{-1}+e_{12}^{2} & e_{13} x^{-1} y^{-1} \\
e_{21} & e_{22} x & e_{23} \\
e_{31} y & e_{32} & e_{33} \\
\end{pmatrix} ~.~
\eea
The permanent of the Kasteleyn matrix
gives the spectral curve of the dimer integrable system, which for Model 12a takes the following form,
\beal{es21a07}
&&
0 = \text{perm}~K=\overline{p}_0 \cdot \big[\delta_{(-1,0)} x^{-1}+\delta_{(1,0)}x+\delta_{(0,-1)}y^{-1}
\nn\\
&& 
\hspace{1cm}
+\delta_{(-1,-1)} x^{-1} y^{-1}+\delta_{(0,1)} y+H\big]
~,~
\eea
where $\overline{p}_0 = e_{11} e_{22} e_{33}$.
The Casimirs $\delta_{(m,n)}$ in \eref{es21a07} can be expressed in terms of the zig-zag paths in \eref{es21a03} as follows, 
\beal{es21a08}
&
\delta_{(-1,0)}=z_1 z_2 z_4~,~\delta_{(1,0)}=1~,~\delta_{(0,-1)}= z_2 ~,~
&
\nn\\
&
\delta_{(-1,-1)}= z_2 z_4~,~\delta_{(0,1)}=z_3^{-1}~,~
\eea
such that the spectral curve for Model 12a takes the following form, 
\beal{es21a09}
\Sigma~:~
z_1 z_2 z_4 x^{-1}+x+z_2 y^{-1}+z_2 z_4 x^{-1} y^{-1}+z_3^{-1} y+H
= 0
~.~
\eea

The Hamiltonian is a sum over all 5 1-loops $\gamma_i$,
\beal{es21a10}
H= \gamma_1+\gamma_2+\gamma_3+\gamma_4+\gamma_5~,~
\eea
where the 1-loops $\gamma_i$ can be expressed in terms of zig-zag paths and face paths as follows,
\beal{es21a11}
&
\gamma_1 =  z_4 f_1^{-1} ~,~
\gamma_2 =  z_4 f_1^{-1} f_5 ~,~
\gamma_3 =  z_4 f_5 ~,~
&
\nn\\
&
\gamma_4 =  z_1 z_2 f_1~,~ 
\gamma_5 =  z_1 z_2 f_1 f_4. 
&
\eea

The commutation matrix $C$
for Model 12a is given by, 
\beal{es21a12}
&&
C =
\left(
\begin{array}{c| c c c c c}
    & \gamma_1 & \gamma_2 & \gamma_3 & \gamma_4 & \gamma_5  \\
    \hline
    \gamma_1      &  0 & 1 & 1 & 0 & -1 \\
    \gamma_2      &  -1 & 0 & 1 & 1 & 0 \\
    \gamma_3      &  -1 & -1 & 0 & 1 & 1 \\
    \gamma_4      &  0 & -1 & -1 & 0 & 1 \\
    \gamma_5.     &  1 & 0 & -1 & -1 & 0 \\
\end{array}
\right)
~,~
\eea
where
the 1-loops satisfying the commutation relations
can be written in terms of the canonical variables as follows, 
\beal{es21a13}
&
\gamma_1 = e^{-Q+P} z_1 z_5^{-1} ~,~
\gamma_2 = e^{-Q} z_1 z_2 ~,~
\gamma_3 = e^{-P} z_2 z_4 z_5 ~,~
&
\nn\\
&
\gamma_4 = e^{Q-P} z_2 z_4 z_5 ~,~
\gamma_5 = e^{Q} z_2 z_4 z_5 ~.~
\eea
\\

 %=================================================================
\subsection{Model 12b}
%=================================================================
%------------------------------------------------------------------------------------------------------------------
\begin{figure}[H]
\begin{center}
\resizebox{0.9\hsize}{!}{
\includegraphics{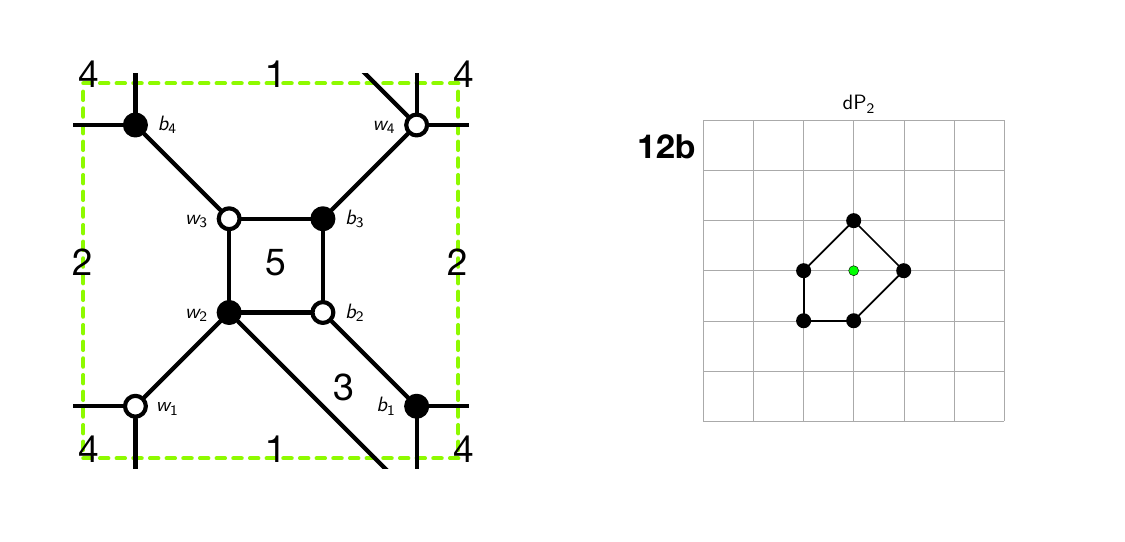}
}
\vspace{-0.5cm}
\caption{The brane tiling and toric diagram of Model 12b.}
\label{mf_12B}
 \end{center}
 \end{figure}
%------------------------------------------------------------------------------------------------------------------

The brane tiling for Model 12b
can be expressed in terms of the following pair of permutation tuples
\beal{es21b01}
\sigma_B &=& (e_{11}\ e_{21}\ e_{41}) \ (e_{12}\ e_{42} \ e_{22} \ e_{32})  \ (e_{23} \ e_{43} \ e_{33}) \ (e_{14} \ e_{44} \ e_{34}) ~,
\nn\\
\sigma_W^{-1} &=& (e_{11}\ e_{12}\ e_{14}) \ (e_{21}\ e_{22}\ e_{23}) \ (e_{32}\ e_{34}\ e_{33}) \ (e_{41}\ e_{44}\ e_{43}\ e_{42})~,
\eea
which correspond to black and white nodes in the brane tiling, respectively.\\

The brane tiling for Model 12b has 5 zig-zag paths given by, 
\beal{es21b03}
&
z_1 = (e_{12}^{+}~ e_{42}^{-}~ e_{41}^{+}~ e_{11}^{-})~,~
z_2 = (e_{34}^{+}~ e_{14}^{-}~ e_{11}^{+}~ e_{21}^{-}~ e_{22}^{+}~ e_{32}^{-})~,~
&
\nn\\
&
z_3= (e_{14}^{+}~ e_{44}^{-}~ e_{43}^{+}~ e_{33}^{-}~ e_{32}^{+}~ e_{12}^{-})~,~
z_4 = (e_{33}^{+}~ e_ {23}^{-}~ e_ {21}^{+}~ e_ {41}^{-}~ e_ {44}^{+}~ e_{34}^{-})~,~
&
\nn\\
&
z_5 = (e_{42}^{+}~ e_{22}^{-}~ e_{23}^{+}~ e_{43}^{-})~,~
\eea
and 5 face paths given by, 
\beal{es21b04}
&
f_1 = (e_{42}^{+}~ e_{12}^{-}~ e_{14}^{+}~ e_{34}^{-}~ e_{33}^{+}~ e_{43}^{-})~,~
f_2 = (e_{12}^{+}~ e_{32}^{-}~ e_{34}^{+}~ e_{44}^{-}~ e_{43}^{+}~ e_{23}^{-}~ e_{21}^{+}~ e_{11}^{-})~,~
&
\nn\\
&
f_3 = (e_{41}^{+}~ e_{21}^{-}~ e_{22}^{+}~ e_{42}^{-})~,~
f_4 = (e_{11}^{+}~ e_{41}^{-}~ e_{44}^{+}~ e_{14}^{-})~,~
&
\nn\\
&
f_5=(e_{23}^{+}~ e_{33}^{-}~ e_{32}^{+}~ e_{22}^{-})
\eea
satisfying the following relations, 
\beal{es21b05}
&
f_1 f_2 f_3 f_4 f_5=1~,~
f_4 f_5^{-1}=z_2 z_4~,~
& 
\nn\\
&
f_2 f_3^2 f_5= z_1 z_4^{-1} z_5^{-1}~,~
f_1 f_2^{-1} f_3^{-3} =z_1^{-1} z_3 z_4^2 z_5^3~.~
&
\eea
The face paths can be written in terms of the canonical variables as follows, 
\beal{es21b05_1}
&
f_1=z_3 z_4 z_5^2 e^{Q-P},~
f_2=z_1 z_4^{-1} z_5^{-1} e^{-2Q-P},~
f_3=e^Q,~
&
\nn\\
&
f_4=z_2 z_4 e^P,~
f_5=e^{P}~.~
&
\eea

The Kasteleyn matrix of the brane tiling for Model 12b in \fref{mf_12B} is 
given by, 
\beal{es21b06}
K = 
\begin{pmatrix}
        e_{11} x^{-1} & e_{12} & 0 & e_{14} y^{-1}\\
        e_{21} & e_{22} & e_{23} & 0\\
        0 & e_{32} & e_{33}  & e_{34}\\
        e_{41} y& e_{42} y & e_{43} & e_{44} x\\
\end{pmatrix} ~.~
\eea
By taking the permanent of the Kasteleyn matrix in \eref{es21b06} with a $GL(2,\mathbb{Z})$ transformation $M : (x,y) \mapsto (x,\frac{1}{y})$, 
we obtain the spectral curve of the dimer integrable system for Model 12b as follows,
\beal{es21b07}
&&
0 =\overline{p}_0 \cdot \big[\delta_{(-1,0)} x^{-1}+\delta_{(1,0)}x+\delta_{(0,-1)}y^{-1}+\delta_{(-1,-1)} x^{-1} y^{-1}+\delta_{(0,1)} y+H\big]
~,~
\nn\\
\eea
where $\overline{p}_0= e_{12} e_{21} e_{33} e_{44}$.
The Casimirs $\delta_{(m,n)}$ in \eref{es21b07} can be written in terms of the zig-zag paths in \eref{es21b03} as shown below, 
\beal{es21b08}
&
\delta_{(-1,0)}= z_2 z_3~,~\delta_{(1,0)}=1~,~\delta_{(0,-1)}=z_4^{-1}~,~\delta_{(-1,-1)}=z_2 z_3 z_5~,~\delta_{(0,1)}= z_3
~,~
\eea
such that the spectral curve for Model 12b takes the following form, 
\beal{es21b09}
\Sigma~:~
x+z_2 z_3 \frac{1}{x}+z_2 z_3 z_5\frac{1}{x y}+z_3 y+\frac{1}{z_4 y}+H
= 0
~.~
\eea

The Hamiltonian is a sum over all 6 1-loops $\gamma_i$,
\beal{es21b10}
H= \gamma_1+\gamma_2+\gamma_3+\gamma_4+\gamma_5+\gamma_6~,~
\eea
where the 1-loops $\gamma_i$ can be expressed in terms of zig-zag paths and face paths as follows,
\beal{es21b11}
&
\gamma_1 = z_1^{-1} f_3 ~,~
\gamma_2 = z_3 z_5 f_3 f_4 ~,~ 
\gamma_3 =  z_1 z_2 z_3 f_1 ~,~
&
\nn\\
&
\gamma_4 =  z_1 z_3 z_4^{-1} f_1 f_4~,~ 
\gamma_5 =  z_1 z_2 z_3 f_1 f_3^{-1} ~,~
\gamma_6 =  z_1^{-1} f_1^{-1} f_4^{-1} ~.~ 
&
\eea

The commutation matrix $C$ for Model 12b is given by, 
\beal{es21b12}
&&
C =
\left(
\begin{array}{c| c c c c c c}
    & \gamma_1 & \gamma_2 & \gamma_3 & \gamma_4 & \gamma_5 & \gamma_6  \\
    \hline
    \gamma_1      &  0 & 1 & -1 & 0 & -1 & 0 \\
    \gamma_2      &  -1 & 0 & -2 & -1 & -1 & 1 \\
    \gamma_3      &  1 & 2 & 0 & 1 & -1 & -1 \\
    \gamma_4      &  0 & 1 & -1 & 0 & -1 & 0 \\
    \gamma_5      &  1 & 1 & 1 & 1 & 0 & -1 \\
    \gamma_6      &  0 & -1 & 1 & 0 & 1 & 0 \\
\end{array}
\right)
~.~
\eea
The 1-loops satisfying the commutation relations
can be written in terms of the canonical variables as shown below, 
\beal{es21b13}
&
\gamma_1 = e^{Q} z_1^{-1} ~,~
\gamma_2 = e^{Q+P} z_1^{-1} ~,~
\gamma_3 = e^{Q-P} z_3 z_5 ~,~
&
\nn\\
&
\gamma_4 = e^{Q} z_3 z_5 ~,~
\gamma_5 = e^{-P} z_3 z_5 ~,~
\gamma_5 = e^{-Q} z_1 z_2 z_3 ~.~
\eea
\\

%=================================================================
\section{Model 13: $\mathbb{C}^3/\mathbb{Z}_4$ $(1,2,2)$, $Y^{2,2}$}
%=================================================================
%------------------------------------------------------------------------------------------------------------------
\begin{figure}[H]
\begin{center}
\resizebox{0.9\hsize}{!}{
\includegraphics{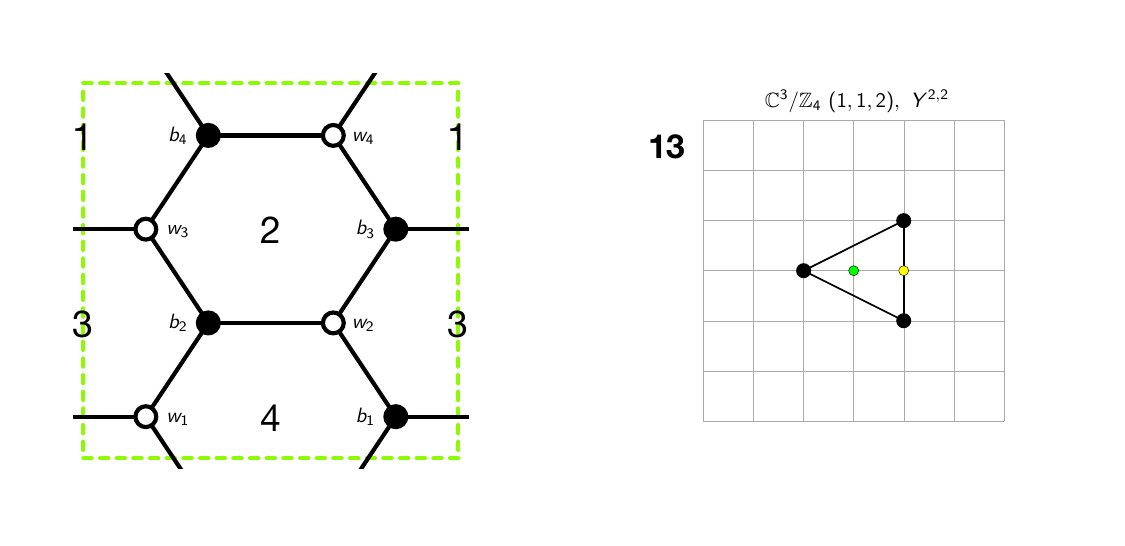}
}
\vspace{-0.5cm}
\caption{The brane tiling and toric diagram of Model 13.}
\label{mf_13}
 \end{center}
 \end{figure}
%------------------------------------------------------------------------------------------------------------------

The brane tiling for Model 13 
can be expressed in terms of the following pair of permutation tuples
\beal{es22a01}
\sigma_B &=& (e_{11}\ e_{21}\ e_{41}) \ (e_{12} \ e_{22} \ e_{32})  \ (e_{23} \ e_{33} \ e_{43}) \ (e_{14} \ e_{34} \ e_{44}) ~,
\nn\\
\sigma_W^{-1} &=& (e_{11}\ e_{12}\ e_{14}) \ (e_{21}\ e_{22}\ e_{23}) \ (e_{32}\ e_{33}\ e_{34}) \ (e_{41}\ e_{43}\ e_{44})~,
\eea
which correspond to black and white nodes in the brane tiling, respectively.\\
 
The brane tiling for Model 13 has 4 zig-zag paths given by, 
\beal{es22a03}
&
z_1 = (e_{33}^{+}~ e_{43}^{-}~ e_{44}^{+}~ e_{14}^{-}~ e_{11}^{+}~ e_{21}^{-}~ \
e_{22}^{+}~ e_{32}^{-})~,~
z_2 = (e_{41}^{+}~ e_{11}^{-}~ e_{12}^{+}~ e_{22}^{-}~ e_{23}^{+}~ e_{33}^{-}~ \
e_{34}^{+}~ e_{44}^{-})~,~
&
\nn\\
&
z_3= (e_{43}^{+}~ e_{23}^{-}~ e_{21}^{+}~ e_{41}^{-})~,~
z_4 = (e_{32}^{+}~ e_{12}^{-}~ e_{14}^{+}~ e_{34}^{-})~,~
\eea
and 4 face paths given by, 
\beal{es22a04}
&
f_1 = (e_{43}^{+}~ e_{33}^{-}~ e_{34}^{+}~ e_{14}^{-}~ e_{11}^{+}~ e_{41}^{-})~,~
f_2 = (e_{23}^{+}~ e_{43}^{-}~ e_{44}^{+}~ e_{34}^{-}~ e_{32}^{+}~ e_{22}^{-})~,~
&
\nn\\
&
f_3 = (e_{33}^{+}~ e_{23}^{-}~ e_{21}^{+}~ e_{11}^{-}~ e_{12}^{+}~ e_{32}^{-})~,~
f_4 = (e_{41}^{+}~ e_{21}^{-}~ e_{22}^{+}~ e_{12}^{-}~ e_{14}^{+}~ e_{44}^{-})~,~
\eea
which satisfy the following relations, 
\beal{es22a05}
&
f_1 f_3=z_3 z_4^{-1}~,~
f_2 f_4=z_3^{-1} z_4 ~,~
f_1 f_2 f_3 f_4=1
&
\eea
The face paths can be written in terms of the canonical variables as shown below, 
\beal{es22a05_1}
&
f_1=e^{-P},~f_2=e^{2Q},~f_3=z_3 z_4^{-1} e^{P},~f_4=z_3^{-1} z_4 e^{-2Q}~.~
&
\eea

The Kasteleyn matrix of the brane tiling for Model 13 in \fref{mf_13}
is given by,
\beal{es22a06}
K = 
\begin{pmatrix}
        e_{11} x^{-1} & e_{12} & 0 & e_{14} y^{-1}\\
        e_{21} & e_{22} & e_{23} & 0\\
        0 & e_{32} & e_{33} x^{-1} & e_{34}\\
        e_{41} y& 0 & e_{43} & e_{44}\\
\end{pmatrix} ~.~
\eea
By taking the permanent of the Kasteleyn matrix in \eref{es22a06}, we obtain the spectral curve of the dimer integrable system for Model 13 as follows,
\beal{es22a07}
&&
0 = \text{perm}~K=\overline{p}_0 \cdot x^{-1} \cdot \big[\delta_{(1,0)}x+\delta_{(-1,0)}\frac{1}{x}+\delta_{(1,1)}x y+\delta_{(1,-1)}\frac{x}{y}+H\big]
~,~
\eea
where $\overline{p}_0= e_{11} e_{22} e_{33} e_{44}$.
The Casimirs $\delta_{(m,n)}$ in \eref{es22a07} can be expressed in terms of the zig-zag paths in \eref{es22a03} as follows, 
\beal{es22a08}
&
\delta_{(-1,0)}=1~,~\delta_{(1,0)}= z_2 z_3+z_2 z_4~,~\delta_{(1,-1)}= z_2 z_3 z_4~,~\delta_{(1,1)}=z_2~,~
\eea
allowing us to express the spectral curve for Model 13 
in the following form,
\beal{es22a09}
\Sigma~:~
\frac{1}{x}+z_2(z_3+z_4)x+z_2 xy+z_2 z_3 z_4\frac{x}{y}+H
= 0
~.~
\eea

The Hamiltonian is a sum over all 4 1-loops $\gamma_i$,
\beal{es22a10}
H= \gamma_1+\gamma_2+\gamma_3+\gamma_4~,~
\eea
where the 1-loops $\gamma_i$ can be expressed in terms of zig-zag paths and face paths as follows,
\beal{es22a11}
&
\gamma_1 =z_2^{1/2} z_4^{1/2} f_1^{1/2} f_2^{1/2} ~,~ 
\gamma_2 =  z_2^{1/2} z_4^{1/2} f_1^{1/2} f_2^{-1/2} ~,~
&
\nn\\
&
\gamma_3 =  z_2^{1/2} z_4^{1/2} f_3^{1/2} f_4^{1/2} ~,~ 
\gamma_4 = z_2^{1/2} z_4^{1/2} f_3^{1/2} f_4^{-1/2} ~.~
&
\eea
The commutation matrix $C$
for Model 13 is given by, 
\beal{es22a12}
&&
C =
\left(
\begin{array}{c| c c c c}
    & \gamma_1 & \gamma_2 & \gamma_3 & \gamma_4  \\
    \hline
    \gamma_1      &   0 & -1 & 0 & 1 \\
    \gamma_2      &   1 & 0 & -1 & 0 \\
    \gamma_3      &   0 & 1 & 0 & -1 \\
    \gamma_4      &   -1 & 0 & 1 & 0 \\
\end{array}
\right)
~,~
\eea
where the
1-loops satisfying the commutation relations
can be written in terms of the canonical variables as shown below, 
\beal{es22a13}
&
\gamma_1=e^{Q-P/2} z_2^{1/2} z_4^{1/2}~,~ \gamma_2=e^{-Q-P/2} z_2^{1/2} z_4^{1/2}~,~ 
&
\nn\\
&
\gamma_3=e^{-Q+P/2} z_2^{1/2} z_4^{1/2} ~,~\gamma_4=e^{Q+P/2} z_2^{1/2} z_3 z_4^{-1/2}~.~
\eea
\\
 
%=================================================================
\section{Model 14: $\text{dP}_1$}
%=================================================================
%------------------------------------------------------------------------------------------------------------------
\begin{figure}[H]
\begin{center}
\resizebox{0.9\hsize}{!}{
\includegraphics{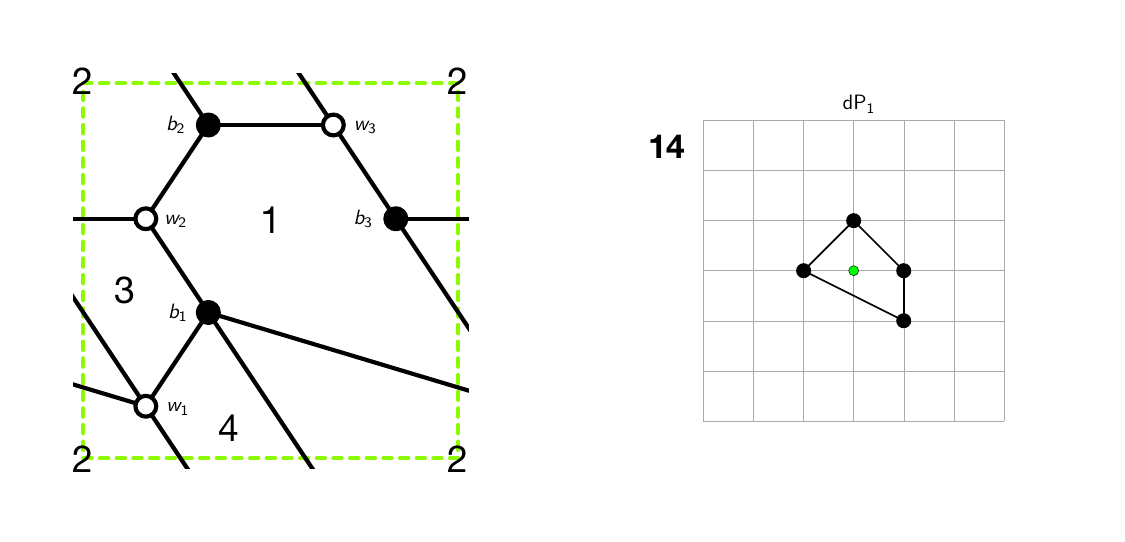}
}
\vspace{-0.5cm}
\caption{The brane tiling and toric diagram of Model 14.}
\label{mf_14}
 \end{center}
 \end{figure}
%------------------------------------------------------------------------------------------------------------------

The brane tiling for Model 14 
can be expressed in terms of the following pair of permutation tuples
\beal{es23a01}
\sigma_B &=& (e_{11}^{1}\ e_{31}\ e_{11}^{2}\ e_{21}) \ (e_{12} \ e_{22} \ e_{32})  \ (e_{13} \ e_{23} \ e_{33}) ~,
\nn\\
\sigma_W^{-1} &=& (e_{11}^{1}\ e_{12}\  e_{11}^{2}\ e_{13}) \ (e_{21}\ e_{23}\ e_{22}) \ (e_{31}\ e_{33}\ e_{32})~,
\eea
which correspond to black and white nodes in the brane tiling, respectively.\\

The brane tiling for Model 14 has 4 zig-zag paths 
given by,
\beal{es23a03}
&
z_1 = (e_{33}^{+}~ e_{13}^{-}~ e_{11}^{1,+}, e_{31}^{-})~,~
z_2 = (e_{13}^{+}~ e_{23}^{-}~ e_{22}^{+}~ e_{32}^{-}~ e_{31}^{+}~ 
e_{11}^{2,-})~,~
&
\nn\\
&
z_3= (e_{32}^{+}~ e_{12}^{-}~ e_{11}^{2,+}, e_{21}^{-}~ e_{23}^{+}~ 
e_{33}^{-})~,~
z_4 = (e_{21}^{+}~ e_{11}^{1,-}, e_{12}^{+}~ e_{22}^{-})~,~
\eea
and
4 face paths given by,
\beal{es23a04}
&
f_1 = (e_{13}^{+}~ e_{33}^{-}~ e_{32}^{+}~ e_{22}^{-}~ e_{21}^{+}~ \
e_{11}^{2,-})~,~
f_2 = (e_{33}^{+}~ e_{23}^{-}~ e_{22}^{+}~ e_{12}^{-}~ e_{11}^{2,+}, \
e_{31}^{-})~,~
&
\nn\\
&
f_3 = (e_{23}^{+}~ e_{13}^{-}~ e_{11}^{1,+}, e_{21}^{-})~,~
f_4 = (e_{31}^{+}~ e_{11}^{1,-}, e_{12}^{+}~ e_{32}^{-})~,~
\eea
which satisfy the following relations, 
\beal{es23a05}
&
f_1^2 f_2 f_4^3=z_2^2 z_3 z_4^3~,~
f_1 f_2^2 f_3^3=z_1 z_2^{-1} z_4^{-2}~,~
f_1 f_2 f_3 f_4=1
&
\eea
The face paths can be written in terms of the canonical variables as follows, 
\beal{es23a05_1}
&
f_1=z_2 e^{2Q-P},~f_2=z_3 e^{-Q -P},~f_3=z_1 e^{P},~f_4=z_4 e^{-Q+P}~.~
&
\eea

The Kasteleyn matrix of the brane tiling for Model 14 in \fref{mf_14} is 
given by, 
\beal{es23a06}
K = 
\begin{pmatrix}
        e_{11}^{1}+e_{11}^{2} x^{-1} & e_{12} y^{-1} & e_{13} x^{-1}\\
        e_{21} & e_{22} & e_{23} x^{-1}\\
        e_{31} y& e_{32} & e_{33}\\
\end{pmatrix} ~.~
\eea
By taking a permanent of the Kasteleyn matrix in \eref{es23a06}, we obtain the spectral curve of the dimer integrable system for Model 14 as shown below,
\beal{es23a07}
&&
0 = \text{perm}~K=\overline{p}_0 \cdot x^{-1} \cdot \big[\delta_{(1,0)}x+\delta_{(-1,0)}\frac{1}{x}+\delta_{(0,1)}y+\delta_{(1,-1)}\frac{x}{y}+H\big]
~,~
\eea
where $\overline{p}_0= e_{11}^{1} e_{22} e_{33}$.
The Casimirs $\delta_{(m,n)}$ in \eref{es23a07} can be written in terms of the zig-zag paths in \eref{es23a03} as shown below, 
\beal{es23a08}
&
\delta_{(1,0)}=1~,~\delta_{(-1,0)}= z_3 z_4~,~\delta_{(0,1)}= z_2 z_3 z_4~,~\delta_{(1,-1)}=z_4~,~
\eea
such that the spectral curve for Model 14 can be expressed in the following form,  
\beal{es23a09}
\Sigma~:~
z_4 \frac{x}{y}+x+z_3 z_4 \frac{1}{x}+z_2 z_3 z_4 y+H
= 0
~.~
\eea

The Hamiltonian is a sum over all 4 1-loops $\gamma_i$,
\beal{es23a10}
H= \gamma_1+\gamma_2+\gamma_3+\gamma_4~,~
\eea
where the 1-loops $\gamma_i$'s can be expressed in terms of zig-zag paths and face paths as follows, 
\beal{es23a11}
&
\gamma_1 =  z_1 z_3 f_1 f_4 ~,~
\gamma_2 =  z_1^{-1} z_4 f_3 ~,~ 
\gamma_3 =  z_1 z_3 f_3^{-1} f_4 ~,~
\gamma_4 =  z_1^{-1} z_4 f_4^{-1}
~.~ 
&
\eea

The commutation matrix $C$ for Model 14 is given by,
\beal{es23a12}
&&
C =
\left(
\begin{array}{c| c c c c}
    & \gamma_1 & \gamma_2 & \gamma_3 & \gamma_4  \\
    \hline
    \gamma_1      &  0 & 1 & 0 & -1 \\
    \gamma_2      &  -1 & 0 & 1 & -1 \\
    \gamma_3      &  0 & -1 & 0 & 1 \\
    \gamma_4      &  1 & 1 & -1 & 0 \\
\end{array}
\right)
~,~\nn\\
\eea
where
the 1-loops satisfying the commutation relations
can be written in terms of the canonical variables as follows, 
\beal{es23a13}
&
\gamma_1=e^{Q}~,~ \gamma_2=z_4 e^{P}~,~ \gamma_3=z_3 z_4 e^{-Q}~,~\gamma_4=z_2 z_3 z_4 e^{Q-P}~.~
\eea
\\

%=================================================================
\section{Model 15: $\mathcal{C}/\mathbb{Z}_2$ $(1,1,1,1)$, $F_0$}
%=================================================================

%=================================================================
\subsection{Model 15a}
%=================================================================
%------------------------------------------------------------------------------------------------------------------
\begin{figure}[H]
\begin{center}
\resizebox{0.9\hsize}{!}{
\includegraphics{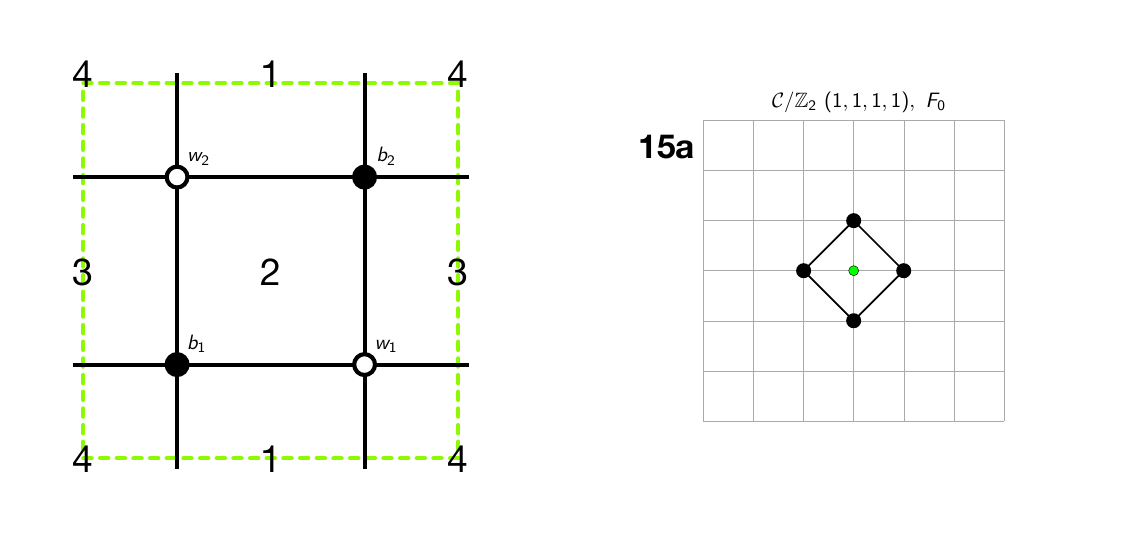}
}
\vspace{-0.5cm}
\caption{The brane tiling and toric diagram of Model 15a.}
\label{mf_15A}
 \end{center}
 \end{figure}
%------------------------------------------------------------------------------------------------------------------

The brane tiling for Model 15a 
can be expressed in terms of the following pair of permutation tuples
\beal{es24a01}
\sigma_B &=& (e_{11}^{1}\ e_{21}^{1}\ e_{11}^{2}\ e_{21}^{2}) \ (e_{12}^{1}\ e_{22}^{1}\ e_{12}^{2}\ e_{22}^{2})~,
\nn\\
\sigma_W^{-1} &=& (e_{11}^{1}\ e_{12}^{1}\ e_{11}^{2}\ e_{12}^{2}) \ (e_{21}^{1}\ e_{22}^{1}\ e_{21}^{2}\ e_{22}^{2})~,
\eea
which correspond to black and white nodes in the brane tiling, respectively.\\
 
The brane tiling for Model 15a has 4 zig-zag paths given by, 
\beal{es24a03}
&
z_1 = (e_{12}^{2,+}, e_{22}^{2,-}, e_{21}^{1,+}, e_{11}^{2,-})~,~
z_2 = (e_{11}^{1,+}, e_{21}^{1,-}, e_{22}^{1,+}, e_{12}^{2,-})~,~
&
\nn\\
&
z_3= (e_{21}^{2,+}, e_{11}^{1,-}, e_{12}^{1,+}, e_{22}^{1,-})~,~
z_4 = (e_{22}^{2,+}, e_{12}^{1,-}, e_{11}^{2,+}, e_{21}^{2,-})~,~
\eea
and 4 face paths given by, 
\beal{es24a04}
&
f_1 = (e_{22}^{2,+}, e_{12}^{2,-}, e_{11}^{1,+}, e_{21}^{2,-})~,~
f_2 = (e_{21}^{1,+}, e_{11}^{1,-}, e_{12}^{1,+}, e_{22}^{2,-})~,~
&
\nn\\
&
f_3 = (e_{11}^{2,+}, e_{21}^{1,-}, e_{22}^{1,+}, e_{12}^{1,-})~,~
f_4 = (e_{12}^{2,+}, e_{22}^{1,-}, e_{21}^{2,+}, e_{11}^{2,-})~,~
\eea
which satisfy the following constraints, 
\beal{es24a05}
&
f_2 f_4=z_1 z_3,~f_1 f_3=z_2 z_4,~f_1 f_2 f_3 f_4=1~.~
&
\eea
The face paths can be written in terms of the canonical variables as follows, 
\beal{es24a05_1}
&
f_1=e^Q,~f_2=e^{2P},~f_3=z_2 z_4 e^{-Q},~f_4=z_1 z_3 e^{-2P}
~.~
&
\eea

The Kasteleyn matrix of the brane tiling for Model 15a in \fref{mf_15A} is 
given by, 
\beal{es24a06}
K = 
\begin{pmatrix}
        e_{11}^{1}+e_{11}^{2} x& e_{21}^{1} +e_{21}^{2} y \\
        e_{12}^{1}+e_{12}^{2} \frac{1}{y} & e_{22}^{1} \frac{1}{x}+e_{22}^{2} \\
\end{pmatrix} ~.~
\eea
The permanent of the Kasteleyn matrix gives the expression for the
spectral curve of the dimer integrable system for Model 15a as follows,
\beal{es24a07}
&&
0 = \text{perm}~K=\overline{p}_0 \cdot  \big[\delta_{(1,0)}x+\delta_{(-1,0)}\frac{1}{x}+\delta_{(0,1)}y+\delta_{(0,-1)}\frac{1}{y}+H\big]
~,~
\eea
where $\overline{p}_0= e_{11}^{2,+} e_{22}^{2,+}$.
The Casimirs $\delta_{(m,n)}$ in \eref{es24a07} can be written in terms of the zig-zag paths in \eref{es24a03} as shown below, 
\beal{es24a08}
&
\delta_{(1,0)}=1~,~\delta_{(-1,0)}= z_1 z_2~,~\delta_{(0,1)}= z_1 z_2 z_3~,~\delta_{(0,-1)}=z_1~,~
\eea
such that the spectral curve for Model 15a can be written in the following form,  
\beal{es24a09}
\Sigma~:~
z_1 \frac{1}{y}+x+z_1 z_2 \frac{1}{x}+z_1 z_2 z_3 y+H
= 0
~.~
\eea

The Hamiltonian is a sum over all 4 1-loops $\gamma_i$,
\beal{es24a10}
H= \gamma_1+\gamma_2+\gamma_3+\gamma_4~,~
\eea
where the 1-loops $\gamma_i$ can be expressed in terms of zig-zag paths and face paths as follows, 
\beal{es24a11}
&
\gamma_1 =  z_1^{1/2} z_4^{-1/2} f_1^{1/2} f_2^{1/2} ~,~
\gamma_2 =  z_1^{1/2} z_4^{-1/2} f_1^{1/2} f_2^{-1/2}~,~
&
\nn\\
&
\gamma_3 =  z_1^{1/2} z_4^{-1/2} f_3^{1/2} f_4^{1/2} ~,~
\gamma_4 =  z_1^{1/2} z_4^{-1/2} f_3^{1/2} f_4^{-1/2} ~.~
&
\eea

The commutation matrix $C$ for Model 15a is given by, 
\beal{es24a12}
&&
C =
\left(
\begin{array}{c| c c c c}
    & \gamma_1 & \gamma_2 & \gamma_3 & \gamma_4  \\
    \hline
    \gamma_1      &   0 & -1 & 0 & 1 \\
    \gamma_2      &  1 & 0 & -1 & 0 \\
    \gamma_3      &  0 & 1 & 0 & -1 \\
    \gamma_4      &  -1 & 0 & 1 & 0 \\
\end{array}
\right)
~.~
\eea
The 1-loops satisfying the commutation relations
can be written in terms of the canonical variables as follows, 
\beal{es24a13}
&
 \gamma_1=z_1^{1/2} z_4^{-1/2}e^{Q/2+P}~,~ \gamma_2=z_1^{1/2} z_4^{-1/2} e^{Q/2-P}~,~
&
\nn\\
&
\gamma_3=z_1^{1/2} z_4^{-1/2}  e^{-Q/2-P}~,~\gamma_4=z_2^{1/2} z_3^{-1/2}e^{-Q/2+P}
~.~
\eea
\\
 
%=================================================================
\subsection{Model 15b}
%=================================================================
%------------------------------------------------------------------------------------------------------------------
\begin{figure}[H]
\begin{center}
\resizebox{0.9\hsize}{!}{
\includegraphics{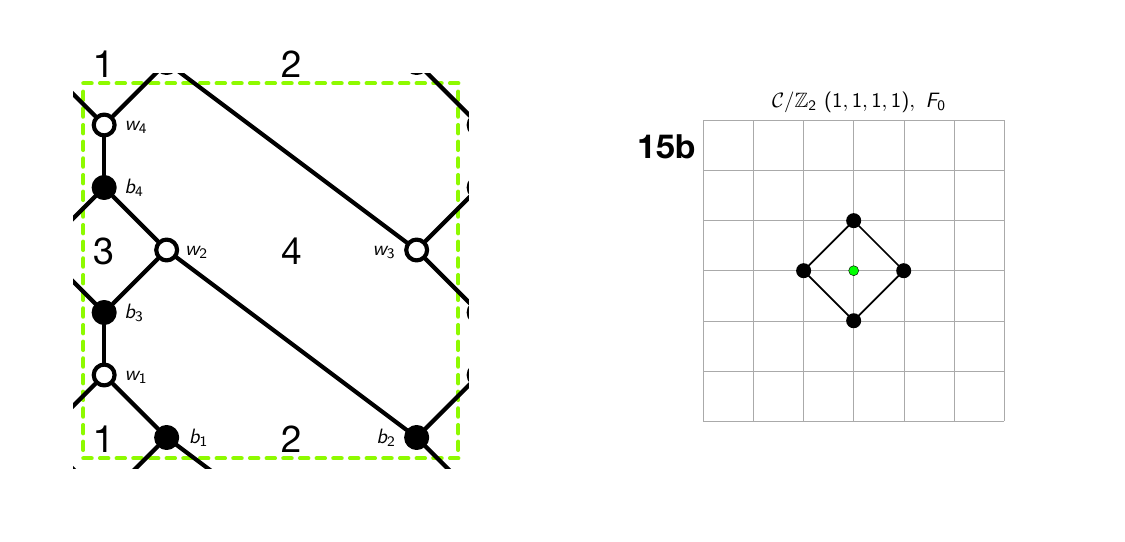}
}
\vspace{-0.5cm}
\caption{The brane tiling and toric diagram of Model 15b.}
\label{mf_15B}
 \end{center}
 \end{figure}
%------------------------------------------------------------------------------------------------------------------

The brane tiling for Model 15b 
can be expressed in terms of the following pair of permutation tuples
\beal{es24b01}
\sigma_B &=& (e_{11}\ e_{41}\ e_{31})\ (e_{22}\ e_{42}\ e_{12} )\ (e_{13}\ e_{23}\ e_{33})\ (e_{24}\ e_{44}\ e_{34}) ~,
\nn\\
\sigma_W^{-1} &=& (e_{11}\ e_{12}\ e_{13}) \ (e_{22}\ e_{23}\ e_{24}) \ (e_{31}\ e_{34}\ e_{33}) \ (e_{41}\ e_{44}\ e_{42}) ~,
\eea
which correspond to black and white nodes in the brane tiling, respectively.\\
 
The brane tiling for Model 15b has 4 zig-zag paths
given by,
\beal{es24b03}
&
z_1 = (e_{12}^{+}~ e_{22}^{-}~ e_{23}^{+}~ e_{33}^{-}~ e_{31}^{+}~ e_{11}^{-})~,~
z_2 = (e_{42}^{+}~ e_{12}^{-}~ e_{13}^{+}~ e_{23}^{-}~ e_{24}^{+}~ e_{44}^{-})~,~
&
\nn\\
&
z_3= (e_{11}^{+}~ e_{41}^{-}~ e_{44}^{+}~ e_{34}^{-}~ e_{33}^{+}~ e_{13}^{-})~,~
z_4= (e_{22}^{+}~ e_{42}^{-}~ e_{41}^{+}~ e_{31}^{-}~ e_{34}^{+}~ e_{24}^{-})~,~
\eea
and 4 face paths given by, 
\beal{es24b04}
&
f_1 = (e_{12}^{+}~ e_{42}^{-}~ e_{41}^{+}~ e_{11}^{-})~,~
f_2 = (e_{11}^{+}~ e_{31}^{-}~ e_{34}^{+}~ e_{44}^{-}~ e_{42}^{+}~ e_{22}^{-}~ e_{23}^{+}~ e_{13}^{-})~,~
&
\nn\\
&
f_3 = (e_{24}^{+}~ e_{34}^{-}~ e_{33}^{+}~ e_{23}^{-})~,~
f_4 = (e_{22}^{+}~ e_{12}^{-}~ e_{13}^{+}~ e_{33}^{-}~ e_{31}^{+}~ e_{41}^{-}~ e_{44}^{+}~ e_{24}^{-})~,~
\eea
satisfying the following relations, 
\beal{es24b05}
&
f_2 f_3^2 f_4=z_2 z_3~,~f_1 f_3^{-1}=z_1 z_4~,~f_1 f_2 f_3 f_4=1~.~
&
\eea
The face paths can be written in terms of the canonical variables as shown below,
\beal{es24b05_1}
&
f_1=e^{2P},~f_2=e^{Q},~f_3=z_2 z_3 e^{2P},~f_4=z_1 z_4 e^{-Q-4P}
&
~.~
\eea

The Kasteleyn matrix of the brane tiling for Model 15b in \fref{mf_15B} is 
given by, 
\beal{es24b06}
K = 
\begin{pmatrix}
        e_{11} & e_{12} x^{-1} & e_{13} & 0 \\
        0 & e_{22}  & e_{23} & e_{24} \\
        e_{31} y & 0 & e_{33} x & e_{34} x \\
        e_{41} y & e_{42} x^{-1} y & 0 & e_{44} \\
\end{pmatrix} ~.~
\eea
The permanent of the Kasteleyn matrix in \eref{es24b06} 
gives the following expression,
\beal{es24b07_1}
&&
0 =
\text{perm}~K=\overline{p}_0 \cdot x y^{-1} \cdot \big[\delta_{(1,0)}x+\delta_{(-1,0)}\frac{1}{x}+\delta_{(1,-1)} \frac{x}{y}+\delta_{(-1,1)}\frac{y}{x}+H\big]~,~ \nn
\eea
where $\overline{p}_0=e_{13} e_{22} e_{34} e_{41}$.
Under a $GL(3,\mathbb{Z})$ transformation $(x,y) \mapsto (x,\frac{x}{y})$, 
we obtain the following form of the
spectral curve of the dimer integrable system for Model 15b,
\beal{es24b07}
&&
\Sigma~:~
\delta_{(1,0)}x+\delta_{(-1,0)}\frac{1}{x}+\delta_{(0,1)}y+\delta_{(0,-1)}\frac{1}{y}+H
= 0 
~.~
\eea
The Casimirs $\delta_{(m,n)}$ in \eref{es24b07} can be written in terms of the 4 zig-zag paths in \eref{es24b03} as follows, 
\beal{es24b08}
&
\delta_{(1,0)}=1~,~\delta_{(-1,0)}= z_1 z_3~,~\delta_{(0,1)}= z_3~,~\delta_{(0,-1)}=z_4^{-1}~,~
\eea
allowing us to express the spectral curve for Model 15b in the following form,
\beal{es24b09}
\Sigma~:~
\frac{1}{z_4 y}+x+z_1 z_3 \frac{1}{x}+z_3 y+H
= 0
~.~
\eea

The Hamiltonian is a sum over all 5 1-loops $\gamma_i$,
\beal{es24b10}
H= \gamma_1+\gamma_2+\gamma_3+\gamma_4+\gamma_5~,~
\eea
where the 1-loops $\gamma_i$ can be expressed in terms of zig-zag paths and face paths as follows,
\beal{es24b11}
&
\gamma_1 = z_1^{1/2} z_3^{1/2} f_1^{1/2} f_2^{1/2}~,~
\gamma_2 = z_1^{1/2} z_3^{1/2} f_3^{1/2} f_4^{-1/2} ~,~
\gamma_3 = z_3^{1/2} z_4^{-1/2} f_2^{1/2} f_3^{-1/2}~,~
&
\nn\\
&
\gamma_4 = z_3^{1/2} z_4^{-1/2} f_2^{1/2} f_3^{1/2}~,~
\gamma_5 = z_1^{1/2} z_3^{1/2} f_3^{1/2} f_4^{1/2}~,~
&
\eea

The commutation matrix $C$
for Model 15b is given by, 
\beal{es24b12}
&&
C =
\left(
\begin{array}{c| c c c c c}
    & \gamma_1 & \gamma_2 & \gamma_3  & \gamma_4 & \gamma_5 \\
    \hline
    \gamma_1   &      0 & 1 & -1 & 0 & 0 \\
    \gamma_2   &      -1 & 0 & -2 & -1 & 1 \\
    \gamma_3   &     1 & 2 & 0 & 1 & -1 \\
    \gamma_4   &      0 & 1 & -1 & 0 & 0 \\
    \gamma_5   &      0 & -1 & 1 & 0 & 0 \\
\end{array}
\right)
~.~
\eea
The 1-loops satisfying the commutation relations
can be written in terms of the canonical variables as follows, 
\beal{es24b13}
&
\gamma_1 = z_1^{1/2} z_3^{1/2} e^{Q/2+P}~,~
\gamma_2 = z_2^{1/2} z_3 z_4^{-1/2} e^{Q/2+3P} ~,~
\gamma_3 = z_1^{1/2} z_3^{1/2} e^{Q/2-P}~,~
&
\nn\\
&
\gamma_4 = z_2^{1/2} z_3 z_4^{-1/2} e^{Q/2+P}~,~
\gamma_5 = z_1^{1/2} z_3^{1/2} e^{-Q/2-P}~.~
\eea
\\

%=================================================================
\section{Model 16: $\mathbb{C}^3/\mathbb{Z}_3$ $(1,1,1)$, $\text{dP}_0$ \label{sec:18}}
%=================================================================
%------------------------------------------------------------------------------------------------------------------
\begin{figure}[H]
\begin{center}
\resizebox{0.9\hsize}{!}{
\includegraphics{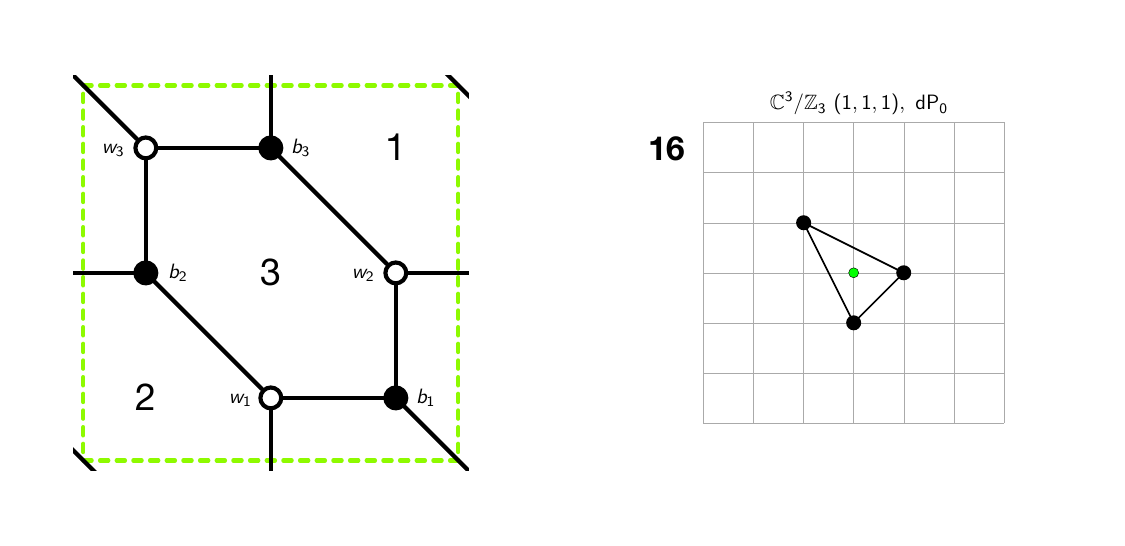}
}
\vspace{-0.5cm}
\caption{The brane tiling and toric diagram of Model 16.}
\label{mf_16}
 \end{center}
 \end{figure}
%------------------------------------------------------------------------------------------------------------------

The brane tiling for Model 16 
can be expressed in terms of the following pair of permutation tuples
 \beal{es25a01}
\sigma_B &=& (e_{11}\ e_{31}\ e_{21})\ (e_{12}\ e_{32}\ e_{22})\ (e_{13}\ e_{33}\ e_{23})~,
\nn\\
\sigma_W^{-1} &=& (e_{11}\ e_{13}\ e_{12})\ (e_{21}\ e_{23}\ e_{22})\ (e_{31}\ e_{33}\ e_{32})~,
\eea
which correspond to black and white nodes in the brane tiling, respectively.\\

The brane tiling for Model 16 has 3 zig-zag paths given by,
\beal{es25a03}
&
z_1 = (e_{13}^{+}~e_{33}^{-}~e_{32}^{+}~e_{22}^{-}~e_{21}^{+}~e_{11}^{-})~,~
z_2 = (e_{11}^{+}~e_{31}^{-}~e_{33}^{+}~e_{23}^{-}~e_{22}^{+}~e_{12}^{-})~,~
&
\nn\\
&
z_3 = (e_{12}^{+}~e_{32}^{-}~e_{31}^{+}~e_{21}^{-}~e_{23}^{+}~e_{13}^{-})~,~
\eea
and 3 face paths given by,
\beal{es25a04}
&
f_1 = (e_{13}^{+}~e_{23}^{-}~e_{22}^{+}~e_{32}^{-}~e_{31}^{+}~e_{11}^{-})~,~
f_2 = (e_{12}^{+}~e_{22}^{-}~e_{21}^{+}~e_{31}^{-}~e_{33}^{+}~e_{13}^{-})~,~
&
\nn\\
&
f_3 = (e_{11}^{+}~e_{21}^{-}~e_{23}^{+}~e_{33}^{-}~e_{32}^{+}~e_{12}^{-})~,
\eea
which satisfy the following relation,
\beal{es25a05}
&
f_1 f_2 f_3=1~.~
&
\eea
The face paths can be written in terms of the canonical variables as follows, 
\beal{es25a05_1}
&
f_1=e^Q,~f_2=e^{-Q+3P},~f_3=e^{-3P}~.~
&
\eea

The Kasteleyn matrix of the brane tiling for Model 16 in \fref{mf_16} is 
given by,
\beal{es25a06}
K = 
\begin{pmatrix}
        e_{11} & e_{12} & e_{13} y^{-1} \\
        e_{21} & e_{22} x & e_{23} \\
        e_{31} \frac{y}{x} & e_{32} & e_{33} \\
  
\end{pmatrix} ~.~
\eea
By taking the permanent of the Kasteleyn matrix in \eref{es25a06}, we obtain the spectral curve of the dimer integrable system for Model 16 as follows,
\beal{es25a07}
&&
0 = \text{perm}~K=\overline{p}_0 \cdot \big[\delta_{(0,-1)}\frac{1}{y}+\delta_{(1,0)}x+\delta_{(-1,1)}\frac{y}{x}+H\big]
~,~
\eea
where $\overline{p}_0= e_{11}e_{22} e_{33}$.
The Casimirs $\delta_{(m,n)}$ in \eref{es25a07} can be written in terms of the 3 zig-zag paths in \eref{es20a03} as shown below,
\beal{es25a08}
&
\delta_{(0,-1)}= z_1~,~\delta_{(1,0)}=1~,~\delta_{(-1,1)}=z_1 z_3~,~
\eea
allowing us to express the spectral curve of Model 16 in the following form, 
\beal{es25a09}
\Sigma~:~
z_1 \frac{1}{y}+x+z_1 z_3 \frac{y}{x}+H
= 0
~.~
\eea

The Hamiltonian is a sum over all 3 1-loops $\gamma_i$,
\beal{es25a10}
H= \gamma_1+\gamma_2+\gamma_3~,~
\eea
where the 1-loops $\gamma_i$ can be expressed in terms of zig-zag paths and face paths as follows,
\beal{es25a11}
&
\gamma_1 = z_1^{1/3} z_2^{-1/3} f_1^{1/3} f_2^{-1/3}~,~
\gamma_2 = z_1^{1/3} z_2^{-1/3} f_1^{1/3} f_2^{2/3} ~,~
\gamma_3 = z_1^{1/3} z_2^{-1/3} f_1^{-1/3} f_3^{1/3} ~.~
\nn\\
&
\eea

The commutation matrix $C$
for Model 16 is given by,
\beal{es25a12}
&&
C =
\left(
\begin{array}{c| c c c }
    & \gamma_1 & \gamma_2 & \gamma_3  \\
    \hline
    \gamma_1      &  0 & 1 & -1   \\
    \gamma_2      & -1 & 0 & 1  \\
    \gamma_3      & 1 & -1 & 0  \\
\end{array}
\right)
~,~
\eea
where the 1-loops 
satisfying the commutation relations
can be written in terms of the canonical variables as follows, 
\beal{es25a13}
\gamma_1=z_1^{1/3} z_2^{-1/3} e^{\frac{2}{3}Q-P}~,~ \gamma_2=z_1^{1/3} z_2^{-1/3}  e^{-\frac{Q}{3}+2P}~,~ \gamma_3=z_1^{1/3} z_2^{-1/3} e^{-\frac{Q}{3}-P}~.~
\eea
\\

%=================================================================
%=================================================================
\section{Bucket 1 \label{sec:19}}
%=================================================================

%---------------------------------------------------- 
\begin{figure}[H]
\begin{center}
\resizebox{0.75\hsize}{!}{
\includegraphics{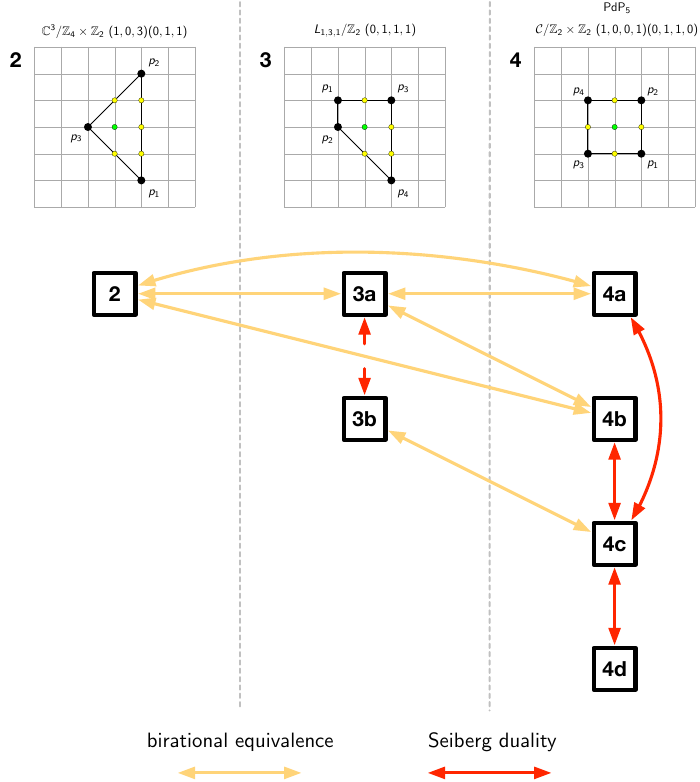}
}
\caption{Brane tilings and toric diagrams in Bucket 1.}
\label{fig_bucket01}
 \end{center}
 \end{figure}
%---------------------------------------------------- 

%=================================================================
\subsection{Hilbert series and generators of the mesonic moduli spaces}
%=================================================================
 
\fref{fig_bucket01} summarizes the brane tilings
related by birational transformations in bucket 1.
From the results in \cite{Hanany:2012hi}, 
we have the refined Hilbert series of the mesonic moduli spaces of these models 
in terms of fugacities $t_a$ corresponding to GLSM fields $p_a$ 
as follows, 
\beal{es40a00}
g(t_a; \mathcal{M}^{mes}_{\text{Model 2}})
&=&
\frac{
(1 - t_1^4 t_2^4) (1 - t_1^2 t_2^2 t_3^2)
}{
(1 -  t_1^4) (1 - t_1^2 t_2^2) (1 - t_2^4) (1 - t_1 t_2 t_3) (1 - t_3^2)
}
~,~
\nn
\\
g(t_a; \mathcal{M}^{mes}_{\text{Model 3a, 3b}})
&=&
\frac{
(1 - t_1^2 t_2^2 t_3^2 t_4^2) (1 - t_1 t_2 t_3^3 t_4^3)
}{
(1 - t_1^2 t_2^2) (1 - t_1 t_3^3) (1 - t_1 t_2 t_3 t_4) 
(1 - t_3^2 t_4^2) (1 - t_2 t_4^3)
}
~,~
\nn
\\
g(t_a; \mathcal{M}^{mes}_{\text{Model 4a, 4b, 4c, 4d}})
&=&
\frac{
(1 - t_1^2 t_2^2 t_3^2 t_4^2)^2
}{
(1 - t_1^2 t_2^2) (1 - t_1^2 t_3^2) (1 - t_1 t_2 t_3 t_4) (1 - t_2^2 t_4^2) (1 - t_3^2 t_4^2)
}
~.~
\nn
\\
\eea
We note here that brane tilings related by Seiberg duality have the same mesonic moduli space
and therefore have the same associated Hilbert series.
\\

%-------------------------
\begin{table}[H]
\begin{center}
\begin{minipage}[t]{0.8\linewidth}
\centering
\begin{tabular}{|c|c|l|}
\hline
\multicolumn{3}{|c|}{Model 2}
\\
\hline
GLSM & $U(1)_R$ & fugacity \\
\hline
$p_1$ & $r$
&
$t_1 = \bar{t}$
\\
$p_2$ & $r$
&
$t_2 = \bar{t}$
\\
$p_3$ & $2r$
&
$t_3 = \bar{t}^2$
\\
\hline
\end{tabular}
\hspace{1cm}
\begin{tabular}{|c|c|l|}
\hline
\multicolumn{3}{|c|}{Model 3a, 3b}
\\
\hline
GLSM & $U(1)_R$ & fugacity \\
\hline
$p_1$ & $r$
&
$t_1 = \bar{t}$
\\
$p_2$ & $r$
&
$t_2 = \bar{t}$
\\
$p_3$ & $r$
&
$t_3 = \bar{t}$
\\
$p_4$ & $r$
&
$t_4 = \bar{t}$
\\
\hline
\end{tabular}
\end{minipage}
\\
\vspace{0.5cm}
\begin{minipage}[t]{0.8\linewidth}
\centering
\begin{tabular}{|c|c|l|}
\hline
\multicolumn{3}{|c|}{Model 4a, 4b, 4c, 4d}
\\
\hline
GLSM & $U(1)_R$ & fugacity \\
\hline
$p_1$ & $r$
&
$t_1 = \bar{t}$
\\
$p_2$ & $r$
&
$t_2 = \bar{t}$
\\
$p_3$ & $r$
&
$t_3 = \bar{t}$
\\
$p_4$ & $r$
&
$t_4 = \bar{t}$
\\
\hline
\end{tabular}
\end{minipage}
\caption{$U(1)_R$ charge assignment on GLSM fields of birationally related brane tilings in bucket 1 such that the $U(1)_R$ charge of the superpotentials is $4r = 2$ and that the generators of the mesonic moduli spaces have all $U(1)_R$ charge $4r$.}
\label{tab_buck1}
\end{center}
\end{table}
%-------------------------

\tref{tab_buck1}
summarizes the $U(1)_R$ charge assignment on the GLSM fields for the brane tilings in bucket 1
in terms of a $U(1)_R$ charge $r$, 
ensuring that the superpotentials of the brane tilings have all $U(1)_R$ charge $4r = 2$
and the generators of the mesonic moduli spaces have all $U(1)_R$ charge $4r$.
In terms of the fugacity $\bar{t}$
corresponding to $U(1)_R$ charge $r$, 
the refined Hilbert series in \eref{es40a00} all become, 
\beal{es40a01}
g(\bar{t};\mathcal{M}^{mes}_{\text{bucket 1}})
= 
\frac{
(1-\bar{t}^8)^2
}{
(1-\bar{t}^4)^5
}
~,~
\eea
confirming that the birational transformations relating the brane tilings in bucket 1 
leave the $U(1)_R$-refined Hilbert series of the associated mesonic moduli spaces invariant.

We also note here based on the results in \cite{Hanany:2012hi} that the brane tilings in bucket 1 have all mesonic moduli spaces with 5 generators 
confirming that birational transformations also leave the number of generators invariant.
This can be seen by taking the plethystic logarithm \cite{Benvenuti:2006qr, Hanany:2006uc, Butti:2007jv, Feng:2007ur, Hanany:2007zz} of the Hilbert series in \eref{es40a01}, which takes the form, 
\beal{es40a02}
PL[
g(\bar{t};\mathcal{M}^{mes}_{\text{bucket 1}})
]
= 5 \bar{t} - 2 \bar{t}^8 ~,~
\eea
confirming the number of mesonic moduli space generators to be $5$.
\\

In the following sections, 
we illustrate how birational transformations in bucket 1
map between birationally equivalent dimer integrable systems
defined by the corresponding brane tilings. 
\\

%=================================================================
\subsection{Model 2 to Model 3a}
%=================================================================
Let us refer to
the spectral curve in \eref{es11a09} for Model 2 as $\Sigma^{(2)}$
and the spectral curve in \eref{es12a09} for Model 3a as $\Sigma^{(3a)}$.

Under the following birational transformation,
\beal{es27a01}
&&
\varphi_{A;M;N}=M \circ \varphi_{A} \circ N~:~ 
(x,y) \mapsto \Big(\frac{1}{xy (1+\frac{z_5^{(2)}}{y})(1+\frac{z_6^{(2)}}{y})(1+\frac{z_8^{(2)}}{y})}, y \Big)
~,~
\eea
where 
\beal{es27a02}
&
{M} ~:~ (x,y) \mapsto \Big(\frac{1}{x}, y \Big)~,~N ~:~ (x,y) \mapsto \Big(xy,y \Big)~,~
&
\nn\\
&
\varphi_{A} ~:~ (x,y) \mapsto \Big(\Big(1+\frac{z_5^{(2)}}{y}\Big)\Big(1+\frac{z_6^{(2)}}{y}\Big)\Big(1+\frac{z_8^{(2)}}{y}\Big)x, y  \Big)~,~
\eea
we discover that the spectral curve $\Sigma^{(2)}$ in \eref{es11a09} is mapped to $\Sigma^{(3a)}$ in \eref{es12a09},
\beal{es27a03}
\varphi_{A;M;N} \Sigma^{(2)} = \Sigma^{(3a)} ~.~
\eea
Based on this map,
we have the following identifications between
the zig-zag paths,
\beal{es27a04}
&
z_1^{(2)} = \frac{1}{z_3^{(3a)}} ~,~
z_2^{(2)} = \frac{1}{z_4^{(3a)}} ~,~
z_3^{(2)} = z_1^{(3a)} z_3^{(3a)} z_4^{(3a)} z_7^{(3a)}~,~
\nn\\
&
z_4^{(2)} = z_3^{(3a)} z_4^{(3a)} z_5^{(3a)} z_7^{(3a)}~,~ 
z_5^{(2)} = z_2^{(3a)} ~,~
z_6^{(2)} = z_6^{(3a)} ~,~
\nn\\
&
z_7^{(2)} = \frac{1}{z_7^{(3a)}} ~,~
z_8^{(2)} = z_8^{(3a)} ~,~
&
\eea
as well as between the face paths,
\beal{es27a05}
&
f_1^{(2)} = f_4^{(3a)} ~,~ 
f_2^{(2)} = f_6^{(3a)} ~,~ 
f_3^{(2)} = f_2^{(3a)} ~,~ 
f_4^{(2)} = f_5^{(3a)} ~,~ 
\nn\\
& 
f_5^{(2)} = f_3^{(3a)} ~,~ 
f_6^{(2)} = f_7^{(3a)} ~,~ 
f_7^{(2)} = f_8^{(3a)} ~,~ 
f_8^{(2)} = f_1^{(3a)} ~.~ 
&
\eea
Moreover, the 1-loops of the two dimer integrable systems are identified as follows,
\beal{es27a06}
\gamma_u^{(2)} = \gamma_u^{(3a)} ~,~
\eea
for all $u=1, \dots, 12$. 
This implies that the two Hamiltonians of the dimer integrable systems are identical under the birational transformation in \eref{es27a01},
\beal{es27a07}
H^{(2)} = H^{(3a)} ~.~
\eea
By identifying \eref{es11a05_1} with \eref{es12a05_1}, we also obtain the following canonical transformation, 
\beal{es27a08}
e^{Q^{(2)}} = \frac{1}{z_2^{(3a)} z_4^{(3a)} z_5^{(3a)}} e^{-Q^{(3a)}}~,~
e^{P^{(2)}} = \frac{z_2^{(3a)}}{z_6^{(3a)}} e^{-P^{(3a)}} ~.~
\eea
We conclude that the dimer integrable systems for Model 2 and Model 3a are birationally equivalent to each other. 
\\

%=================================================================
\subsection{Model 2 to Model 4a}
%=================================================================

Let us refer to the spectral curve in 
\eref{es11a09} for Model 2 as
$\Sigma^{(2)}$ and the spectral curve in \eref{es13a09} for Model 4a as
 $\Sigma^{(4a)}$.

Under the following birational transformation,
\beal{es27b01}
&&
\varphi_A ~:~ 
(x,y) \mapsto 
\nn\\
&&
\hspace{0.5cm}
\left(
\left(
\frac{y}{1+\frac{y}{z_6^{(2)}}}
\right)
\left(
\frac{y}{1+\frac{y}{z_8^{(2)}}}
\right)
\frac{1}{z_6^{(2)} y}
x
~,~
y
\right)
~,~
\eea
we discover that the spectral curve $\Sigma^{(2)}$ in \eref{es11a09} 
is mapped to $\Sigma^{(4a)}$ in \eref{es13a09},
\beal{es27b02}
\varphi_A \Sigma^{(2)} = \Sigma^{(4a)} ~.~
\eea
Based on this map,
we have the following identifications between
the zig-zag paths,
\beal{es27b03}
&
z_1^{(2)} =z_1^{(4a)} z_8^{(4a)} ~,~
z_2^{(2)} = z_3^{(4a)} z_8^{(4a)}  ~,~
z_{3}^{(2)}= z_{4}^{(4a)} z_6^{(4a)}  ~,~
&
\nn\\
&
z_{4}^{(2)} =z_{2}^{(4a)} z_6^{(4a)} ~,~
z_5^{(2)} =z_5^{(4a)}  ~,~
z_6^{(2)}=\frac{1}{z_6^{(4a)}}  ~,~
&
\nn\\
&
z_7^{(2)} = z_7^{(4a)}  ~,~
z_8^{(2)}=\frac{1}{z_8^{(4a)}} ~,~
&
\eea
as well as between the face paths,
\beal{es27b04}
&
f_1^{(2)} =f_8^{(4a)}~,~ 
f_2^{(2)} = f_6^{(4a)}~,~ 
f_3^{(2)} = f_3^{(4a)} ~,~ 
&
\nn\\
&
f_4^{(2)} = f_1^{(4a)} ~,~
f_5^{(2)}  = f_2^{(4a)}~,~ 
f_6^{(2)} = f_4^{(4a)} ~,~ 
&
\nn\\
&
f_7^{(2)} = f_5^{(4a)} ~,~ 
f_8^{(2)} = f_7^{(4a)} ~.~
&
\eea
Moreover, the 1-loops of the two dimer integrable systems are identified as follows,
\beal{es27b05}
\gamma_u^{(2)} = \gamma_u^{(4a)} ~,~
\eea
for all $u=1, \dots, 12$. 
This implies that the two Hamiltonians of the dimer integrable systems are identical under the birational transformation in 
\eref{es27b01},
\beal{es27b06}
H^{(2)} = H^{(4a)} ~.~
\eea
By identifying \eref{es11a05_1} with \eref{es13a05_1}, we also obtain the following canonical transformation, 
\beal{es27b07}
e^{Q^{(2)}} = e^{Q^{(4a)}} z_3^{(4a)} z_4^{(4a)} z_6^{(4a)} z_7^{(4a)}
~,~
e^{P^{(2)}} = e^{P^{(4a)}} 
\eea
We conclude that the dimer integrable systems for Model 2 and Model 4a are birationally equivalent to each other. 
\\

%=================================================================
\subsection{Model 2 to Model 4b}
%=================================================================
Let us refer to the spectral curve in \eref{es11a09} for Model 2 as
$\Sigma^{(2)}$ and the spectral curve in \eref{es13b09} for Model 4b as
 $\Sigma^{(4b)}$.

Under the following birational transformation,
\beal{es27c01}
&&
\varphi_{A:N} =\varphi_A \circ N~:~ 
(x,y) \mapsto 
\left(
\frac{1}{(1+\frac{1}{z_6^{(2)} y})(1+\frac{1}{z_7^{(2)} y})}\frac{1}{xy}
~,~
\frac{1}{y}
\right)
~,~
\eea
where 
\beal{es27c01_01}
&
\varphi_{A} ~:~ (x,y) \mapsto \Big( \Big(\frac{x}{1+\frac{y}{z_6^{(2)}}} \Big) \Big(\frac{x}{1+\frac{y}{z_7^{(2)}}} \Big) x ,y\Big)~,~N ~:~ (x,y) \mapsto \Big(\frac{y}{x}, \frac{1}{y} \Big)~,~
\eea
we discover that the spectral curve $\Sigma^{(2)}$ in \eref{es11a09} 
is mapped to $\Sigma^{(4b)}$ in \eref{es13b09},
\beal{es27c02}
\varphi_A \Sigma^{(2)} = \Sigma^{(4b)} ~.~
\eea
Based on this map,
we have the following identifications between
the zig-zag paths,
\beal{es27c03}
&
z_1^{(2)} =z_2^{(4b)} z_4^{(4b)} z_8^{(4b)} ~,~
z_2^{(2)} = z_2^{(4b)} z_3^{(4b)} z_4^{(4b)}  ~,~
z_{3}^{(2)}= z_5^{(4b)}  ~,~
&
\nn\\
&
z_{4}^{(2)} = z_6^{(4b)}  ~,~
z_5^{(2)} = z_7^{(4b)}  ~,~
z_6^{(2)}=\frac{1}{z_4^{(4b)}}  ~,~
z_7^{(2)} = \frac{1}{z_2^{(4b)}}  ~,~
z_8^{(2)} = z_1^{(4b)}  ~,~
&
\eea
as well as between the face paths,
\beal{es27c04}
&
f_1^{(2)} =f_7^{(4b)}~,~ 
f_2^{(2)} = f_1^{(4b)}~,~ 
f_3^{(2)} = f_8^{(4b)} ~,~ 
&
\nn\\
&
f_4^{(2)} = f_2^{(4b)} ~,~
f_5^{(2)}  = f_3^{(4b)}~,~ 
f_6^{(2)} = f_5^{(4b)} ~,~ 
&
\nn\\
&
f_7^{(2)} = f_6^{(4b)} ~,~ 
f_8^{(2)} = f_4^{(4b)} ~.~
&
\eea
Moreover, the 1-loops of the two dimer integrable systems are identified as follows,
\beal{es27c05}
\gamma_u^{(2)} = \gamma_u^{(4b)} ~,~
\eea
for all $u=1, \dots, 12$. 
This implies that the two Hamiltonians of the dimer integrable systems are identical under the birational transformation in \eref{es27c01},
\beal{es27c06}
H^{(2)} = H^{(4b)} ~.~
\eea
By identifying \eref{es11a05_1} with \eref{es13a05_1}, we also obtain the following canonical transformation, 
\beal{es27c07}
e^{Q^{(2)}} = e^{Q^{(4b)}} 
~,~
e^{P^{(2)}} = e^{P^{(4b)}} ~.~
\eea
We conclude that the dimer integrable systems for Model 2 and Model 4b are birationally equivalent to each other. 
\\

%=================================================================
\subsection{Model 3a to Model 4a}
%=================================================================
Let us refer to the spectral curve in \eref{es12a09} for Model 3a as
$\Sigma^{(3a)}$ and the spectral curve in \eref{es13a09} for Model 4a as
 $\Sigma^{(4a)}$.

Under the following birational transformation,
\beal{es27d01}
&&
\varphi_{A;M;N}=M \circ \varphi_{A} \circ N~:~ 
\nn\\&&
(x,y) \mapsto 
\hspace{0.5cm}
\left(
\frac{1}{(y+z_2^{(3a)})z_8^{(3a)}} \frac{y}{x}
~,~
y
\right)
~,~
\eea
where 
\beal{es27d01_01}
&
{M} ~:~ (x,y) \mapsto \Big(\frac{x}{y}, y \Big)~,~N ~:~ (x,y) \mapsto \Big(\frac{1}{x},y \Big)~,~
& 
\nn\\
&
\varphi_{A} ~:~ (x,y) \mapsto \Big(\frac{y^2}{(y+z_2^{(3a)}) z_8^{(3a)}}x  , y\Big)~,~
\eea
we discover that the spectral curve $\Sigma^{(3a)}$ in \eref{es12a09} 
is mapped to $\Sigma^{(4a)}$ in \eref{es13a09},
\beal{es27d02}
\varphi_A \Sigma^{(3a)} = \Sigma^{(4a)} ~.~
\eea
Based on this map,
we have the following identifications between
the zig-zag paths,
\beal{es27d03}
&
z_1^{(3a)} = \frac{z_8^{(4a)}}{z_2^{(4a)} z_5^{(4a)}} ~,~
z_2^{(3a)} = z_5^{(4a)} ~,~
z_3^{(3a)}= \frac{1}{z_1^{(4a)} z_8^{(4a)}}  ~,~
z_4^{(3a)} = \frac{1}{z_3^{(4a)} z_8^{(4a)}}  ~,~
&
\nn\\
&
z_5^{(3a)} = \frac{z_8^{(4a)}}{z_4^{(4a)} z_5^{(4a)}}  ~,~
z_6^{(3a)}=\frac{1}{z_6^{(4a)}}  ~,~
z_7^{(3a)} =\frac{1}{z_7^{(4a)}}  ~,~
z_8^{(3a)} =\frac{1}{z_8^{(4a)}}  ~,~
&
\eea
as well as between the face paths,
\beal{es27d04}
&
f_1^{(3a)} =f_7^{(4a)}~,~ 
f_2^{(3a)} = f_3^{(4a)}~,~ 
f_3^{(3a)} = f_2^{(4a)} ~,~ 
&
\nn\\
&
f_4^{(3a)} = f_8^{(4a)} ~,~
f_5^{(3a)}  = f_1^{(4a)}~,~ 
f_6^{(3a)} = f_6^{(4a)} ~,~ 
&
\nn\\
&
f_7^{(3a)} = f_4^{(4a)} ~,~ 
f_8^{(3a)} = f_5^{(4a)} ~.~
&
\eea
Moreover, the 1-loops of the two dimer integrable systems are identified as follows,
\beal{es27d05}
\gamma_u^{(3a)} = \gamma_u^{(4a)} ~,~
\eea
for all $u=1, \dots, 12$. 
This implies that the two Hamiltonians of the dimer integrable systems are identical under the birational transformation in \eref{es27d01},
\beal{es27d06}
H^{(3a)} = H^{(4a)} ~.~
\eea
By identifying \eref{es12a05_1} with \eref{es13a05_1}, we also obtain the following canonical transformation, 
\beal{es27d07}
e^{Q^{(3a)}} = \frac{1}{z_6^{(4a)} z_7^{(4a)}}e^{-Q^{(4a)}} 
~,~
e^{P^{(3a)}} = z_5^{(4a)} z_6^{(4a)} e^{-P^{(4a)}} ~.~
\eea
We conclude that the dimer integrable systems for Model 3a and Model 4a are birationally equivalent to each other. 
\\

%=================================================================
\subsection{Model 3a to Model 4b}
%=================================================================
Let us refer to the spectral curve in \eref{es12a09} for Model 3a as
$\Sigma^{(3a)}$ and the spectral curve in \eref{es13b09} for Model 4b as
 $\Sigma^{(4b)}$.

Under the following birational transformation,
\beal{es27e01}
&&
\varphi_{A;M;N}=M \circ \varphi_{A} \circ N~:~ 
\nn\\
&&
\hspace{0.5cm}
(x,y) \mapsto 
\left(
\frac{(y+z_7^{(3a)})}{(1+z_2^{(3a)} y)(1+z_8^{(3a)} y)z_6^{(3a)}}x
~,~
\frac{1}{y}
\right)
~,~
\eea
where 
\beal{es27e01_01}
&
{M} ~:~ (x,y) \mapsto \Big(\frac{x}{y}, y \Big)~,~N ~:~ (x,y) \mapsto \Big(\frac{x}{y},\frac{1}{y} \Big)~,~
& 
\nn\\
&
\varphi_{A} ~:~ (x,y) \mapsto \Big(\frac{1+z_7^{(3a)} y}{(y+z_2^{(3a)}) (y+z_8^{(3a)})z_6^{(3a)}}x y , y\Big)
~,~
\eea
we discover that the spectral curve $\Sigma^{(3a)}$ in \eref{es12a09} 
is mapped to $\Sigma^{(4b)}$ in \eref{es13b09},
\beal{es27e02}
\varphi_A \Sigma^{(3a)} = \Sigma^{(4b)} ~.~
\eea
Based on this map,
we have the following identifications between
the zig-zag paths,
\beal{es27e03}
&
z_1^{(3a)} = z_2^{(4b)} z_3^{(4b)} {z_4^{(4b)}}^2 z_5^{(4b)} z_8^{(4b)} ~,~
z_2^{(3a)} = z_7^{(4b)} ~,~
z_3^{(3a)}= \frac{1}{z_2^{(4b)} z_4^{(4b)} z_8^{(4b)}}  ~,~
&
\nn\\
&
z_4^{(3a)} = \frac{1}{z_2^{(4b)} z_3^{(4b)} z_4^{(4b)}}  ~,~
z_5^{(3a)} = z_2^{(4b)} z_3^{(4b)} {z_4^{(4b)}}^2 z_6^{(4b)} z_8^{(4b)} ~,~
&
\nn\\
&
z_6^{(3a)}=\frac{1}{z_4^{(4b)}}  ~,~
z_7^{(3a)} =z_2^{(4b)}  ~,~
z_8^{(3a)} =z_1^{(4b)}  ~,~
&
\eea
as well as between the face paths,
\beal{es27e04}
&
f_1^{(2)} =f_4^{(4b)}~,~ 
f_2^{(2)} = f_8^{(4b)}~,~ 
f_3^{(2)} = f_3^{(4b)} ~,~ 
&
\nn\\
&
f_4^{(2)} = f_7^{(4b)} ~,~
f_5^{(2)}  = f_2^{(4b)}~,~ 
f_6^{(2)} = f_1^{(4b)} ~,~ 
&
\nn\\
&
f_7^{(2)} = f_5^{(4b)} ~,~ 
f_8^{(2)} = f_6^{(4b)} ~.~
&
\eea
Moreover, the 1-loops of the two dimer integrable systems are identified as follows,
\beal{es27e05}
\gamma_u^{(3a)} = \gamma_u^{(4b)} ~,~
\eea
for all $u=1, \dots, 12$. 
This implies that the two Hamiltonians of the dimer integrable systems are identical under the birational transformation in \eref{es27e01},
\beal{es27e06}
H^{(3a)} = H^{(4b)} ~.~
\eea
By identifying \eref{es12a05_1} with \eref{es13b05_1}, we also obtain the following canonical transformation, 
\beal{es27e07}
e^{Q^{(3a)}} = z_1^{(4b)} z_2^{(4b)} z_3^{(4b)} z_5^{(4b)} e^{-Q^{(4b)}} 
~,~
e^{P^{(3a)}} = z_4^{(4b)} z_7^{(4b)} e^{-P^{(4b)}} ~.~
\eea
We conclude that the dimer integrable systems for Model 3a and Model 4b are birationally equivalent to each other. 
\\

%=================================================================
\subsection{Model 3b to Model 4c}
%=================================================================
Let us refer to the spectral curve in \eref{es12b09} for Model 3b as
$\Sigma^{(3b)}$ and the spectral curve in \eref{es13c09} for Model 4c as
 $\Sigma^{(4c)}$.

Under the following birational transformation,
\beal{es27f01}
&&
\varphi_{A;M;N}=M \circ \varphi_{A} \circ N~:~ 
\nn\\
&&
\hspace{0.5cm}
(x,y) \mapsto 
\left(
\frac{x}{(1+\frac{z_2^{(3b)}}{y})y}
~,~
\frac{1}{y}
\right)
~,~
\eea
where 
\beal{es27f01_01}
&
{M} ~:~ (x,y) \mapsto \Big(\frac{x}{y}, y \Big)~,~N ~:~ (x,y) \mapsto \Big(\frac{x}{y},\frac{1}{y} \Big)~,~
& 
\nn\\
&
\varphi_{A} ~:~ (x,y) \mapsto \Big(\frac{x}{(1+z_2^{(3b)} y)}x , y\Big)~,~
\eea
we discover that the spectral curve $\Sigma^{(3b)}$ in \eref{es12b09} 
is mapped to $\Sigma^{(4c)}$ in \eref{es13c09},
\beal{es27f02}
\varphi_A \Sigma^{(3b)} = \Sigma^{(4c)} ~.~
\eea
Based on this map,
we have the following identifications between
the zig-zag paths,
\beal{es27f03}
&
z_1^{(3b)} = z_4^{(4c)} z_8^{(4c)}~,~
z_2^{(3b)} = \frac{1}{z_4^{(4c)}}~,~
z_3^{(3b)}= z_7^{(4c)}~,~
&
\nn\\
&
z_4^{(3b)} = z_2^{(4c)}~,~
z_5^{(3b)} = z_1^{(4c)} ~,~
&
\nn\\
&
z_6^{(3b)}=z_4^{(4c)} z_5^{(4c)}~,~
z_7^{(3b)} =z_6^{(4c)} ~,~
z_8^{(3b)} =z_3^{(4c)}  ~.~
&
\eea
as well as between the face paths,
\beal{es27f04}
&
f_1^{(3b)} =f_5^{(4c)}~,~ 
f_2^{(3b)} = f_1^{(4c)}~,~ 
f_3^{(3b)} = f_8^{(4c)} ~,~ 
&
\nn\\
&
f_4^{(3b)} = f_6^{(4c)} ~,~
f_5^{(3b)}  = f_7^{(4c)}~,~ 
f_6^{(3b)} = f_4^{(4c)} ~,~ 
&
\nn\\
&
f_7^{(3b)} = f_2^{(4c)} ~,~ 
f_8^{(3b)} = f_3^{(4c)} ~.~
&
\eea
Moreover, the 1-loops of the two dimer integrable systems are identified as follows,
\beal{es27f05}
\gamma_u^{(3b)} = \gamma_u^{(4c)} ~,~
\eea
for all $u=1, \dots, 14$. 
This implies that the two Hamiltonians of the dimer integrable systems are identical under the birational transformation in \eref{es27f01},
\beal{es27f06}
H^{(3b)} = H^{(4c)} ~.~
\eea
By identifying \eref{es12a05_1} with \eref{es13b05_1}, we also obtain the following canonical transformation, 
\beal{es27f07}
e^{Q^{(3b)}} = z_2^{(4c)} z_5^{(4c)} z_6^{(4c)} z_7^{(4c)} e^{Q^{(4c)}} 
~,~
e^{P^{(3b)}} = z_4^{(4c)} z_6^{(4c)} z_7^{(4c)} z_8^{(4c)} e^{P^{(4c)}} ~.~
\eea
We conclude that the dimer integrable systems for Model 3b and Model 4c are birationally equivalent to each other. 
\\

%=================================================================
%=================================================================
\section{Bucket 2 \label{sec:20}}
%=================================================================

%---------------------------------------------------- 
\begin{figure}[H]
\begin{center}
\resizebox{0.5\hsize}{!}{
\includegraphics{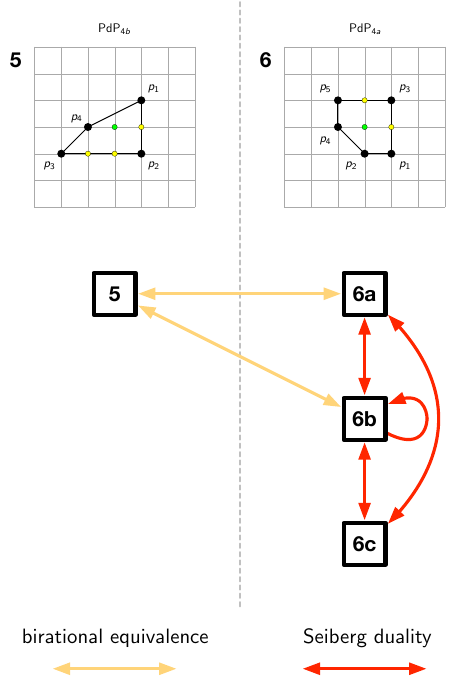}
}
\caption{Brane tilings and toric diagrams in Bucket 2.}
\label{fig_bucket02}
 \end{center}
 \end{figure}
 %---------------------------------------------------- 
 
%=================================================================
\subsection{Hilbert series and generators of the mesonic moduli spaces}
%=================================================================
 
\fref{fig_bucket02}
summarizes the brane tilings in bucket 2 that are related by birational transformations.
Based on the results in \cite{Hanany:2012hi}, 
the refined Hilbert series of the mesonic moduli spaces
of these brane tilings
in terms of fugacities $t_a$ corresponding to GLSM fields $p_a$
are given as follows, 
\beal{es41a00}
g(t_a; \mathcal{M}^{mes}_{\text{Model 5}})
&=&
\frac{
1 + t_1 t_2 t_3 t_4 - t_1^2 t_2^4 t_3 t_4 + t_2^2 t_3^2 t_4 - t_1 t_2^5 t_3^2 t_4 -  t_1^2 t_2^6 t_3^3 t_4^2
}{
(1 - t_1 t_2^3) (1 - t_2^4 t_3) (1 - t_1^2 t_4) (1 - t_3^3 t_4^2)
}
~,~
\nn\\
g(t_a; \mathcal{M}^{mes}_{\text{Model 6a, 6b, 6c}})
&=&
\frac{
1
}{
(1 -  t_1^2 t_2 t_3^2) (1 -  t_1^2 t_2^2 t_4) (1 -  t_1 t_3^3 t_5) (1 -   t_3^2 t_4 t_5^2) (1 -  t_2 t_4^2 t_5^2)
}
\nn\\
&&
\times
(1 + t_1 t_2 t_3 t_4 t_5 - t_1^3 t_2^2 t_3^3 t_4 t_5 - t_1^2 t_2 t_3^4 t_4 t_5^2 
\nn\\
&&
\hspace{0.5cm}
-  t_1^2 t_2^2 t_3^2 t_4^2 t_5^2 - t_1 t_2 t_3^3 t_4^2 t_5^3 +  t_1^3 t_2^2 t_3^5 t_4^2 t_5^3 + t_1^4 t_2^3 t_3^6 t_4^3 t_5^4)
~,~
\eea
where we note that brane tilings related by Seiberg duality have the same mesonic moduli space and therefore have the same corresponding Hilbert series. 
\\

%-------------------------
\begin{table}[H]
\begin{center}
\begin{minipage}[t]{0.8\linewidth}
\centering
\begin{tabular}{|c|c|l|}
\hline
\multicolumn{3}{|c|}{Model 5}
\\
\hline
GLSM & $U(1)_R$ & fugacity \\
\hline
$p_1$ & $2r$
&
$t_1 = \bar{t}^2$
\\
$p_2$ & $r$
&
$t_2 = \bar{t}$
\\
$p_3$ & $r$
&
$t_3 = \bar{t}$
\\
$p_4$ & $r$
&
$t_4 = \bar{t}$
\\
\hline
\end{tabular}
\hspace{1cm}
\begin{tabular}{|c|c|l|}
\hline
\multicolumn{3}{|c|}{Model 6a, 6b, 6c}
\\
\hline
GLSM & $U(1)_R$ & fugacity \\
\hline
$p_1$ & $r$
&
$t_1 = \bar{t}$
\\
$p_2$ & $r$
&
$t_2 = \bar{t}$
\\
$p_3$ & $r$
&
$t_3 = \bar{t}$
\\
$p_4$ & $r$
&
$t_4 = \bar{t}$
\\
$p_5$ & $r$
&
$t_5 = \bar{t}$
\\
\hline
\end{tabular}
\end{minipage}
\caption{$U(1)_R$ charge assignment on GLSM fields of birationally related brane tilings in bucket 2 such that the $U(1)_R$ charge of the superpotentials is $5r = 2$ and that the generators of the mesonic moduli spaces have all $U(1)_R$ charge $5r$.}
\label{tab_buck2}
\end{center}
\end{table}
%-------------------------

\tref{tab_buck2}
summarizes the $U(1)_R$ charge assignment on the GLSM fields
in terms of a $U(1)_R$ charge $r$,
ensuring that the superpotentials of the brane tilings in bucket 2
have all $U(1)_R$ charge $5r = 2$
and the generators of the mesonic moduli spaces have all $U(1)_R$ charge $5r$.
Based on this $U(1)_R$ charge assignment, 
in terms of a fugacity $\bar{t}$ corresponding to $U(1)_R$ charge $r$, 
the refined Hilbert series in \eref{es41a00} all become,
\beal{es40a11}
g(\bar{t};\mathcal{M}^{mes}_{\text{bucket 2}})
=
\frac{1 + 3 \bar{t}^5 + \bar{t}^{10}}{
(1 - \bar{t}^5)^3
}
~.~
\eea
This confirms that
the birational transformations relating the brane tilings in bucket 2 keep the $U(1)_R$-refined Hilbert series of the associated mesonic moduli spaces invariant.

Based on the results in \cite{Hanany:2012hi},
we also note here that the brane tilings in bucket 2 have all mesonic moduli spaces with 6 generators.
This can also be seen by taking the plethystic logarithm \cite{Benvenuti:2006qr, Hanany:2006uc, Butti:2007jv, Feng:2007ur, Hanany:2007zz}
of the Hilbert series in \eref{es40a11}, giving us,
\beal{es41a12}
PL[g(\bar{t};\mathcal{M}^{mes}_{\text{bucket 2}})]
=
6 \bar{t}^5 - 5 \bar{t}^{10} + 5 \bar{t}^{15} + \dots 
~,~
\eea
which 
confirms the number of generators to be 6 for all brane tilings in bucket 2. 
\\

In the following sections, 
we illustrate how the brane tilings in bucket 2 define dimer integrable systems that are equivalent under birational transformations.
 \\

%=================================================================
\subsection{Model 5 to Model 6a}
%=================================================================
   Let us refer to the spectral curve in \eref{es14a09} for Model 5 as
$\Sigma^{(5)}$ and the spectral curve in \eref{es15a09} for Model 6a as
 $\Sigma^{(6a)}$.

Under the following birational transformation,
\beal{es28a01}
&&
\varphi_{A;N}=\varphi_{A} \circ N~:~ 
(x,y) \mapsto \Big(x, z_3^{(5)}  z_7^{(5)} \frac{(x+z_2^{(5)} )}{xy} \Big)
~,~
\eea
where 
\beal{es28a02}
\varphi_{A} ~:~ (x,y) \mapsto \Big(x,z_3^{(5)}  z_7^{(5)}  \frac{(x+z_2^{(5)} )}{x} y\Big)~,~N ~:~ (x,y) \mapsto \Big(x, \frac{1}{y} \Big)~,~ 
\eea
we discover that the spectral curve $\Sigma^{(5)}$ in \eref{es14a09} is mapped to $\Sigma^{(6a)}$ in \eref{es15a09},
\beal{es28a03}
\varphi_{A;N} \Sigma^{(5)} = \Sigma^{(6a)} ~.~
\eea
Based on this map,
we have the following identifications between
the zig-zag paths,
\beal{es28a04}
&
z_1^{(5)} = z_2^{(6a)} z_3^{(6a)} z_4^{(6a)} ~,~
z_2^{(5)} = z_1^{(6a)} ~,~
z_3^{(5)} = z_7^{(6a)}~,~
z_4^{(5)} = z_2^{(6a)} z_3^{(6a)} z_6^{(6a)}~,~
\nn\\
&
z_5^{(5)} = \frac{1}{z_3^{(6a)}}~,~
z_6^{(5)} = \frac{1}{z_2^{(6a)}}~,~
z_7^{(5)} = z_5^{(6a)}~,~
&
\eea
as well as between the face paths,
\beal{es28a05}
&
f_1^{(5)} = f_5^{(6a)} ~,~ 
f_2^{(5)} = f_4^{(6a)} ~,~ 
f_3^{(5)} = f_6^{(6a)} ~,~ 
f_4^{(5)} = f_2^{(6a)} ~,~ 
\nn\\
&
f_5^{(5)} = f_7^{(6a)} ~,~ 
f_6^{(5)} = f_3^{(6a)} ~,~ 
f_7^{(5)} = f_1^{(6a)} ~.~ 
&
\eea
Moreover, the 1-loops of the two dimer integrable systems are identified as follows,
\beal{es28a06}
\gamma_u^{(5)} = \gamma_u^{(6a)} ~,~
\eea
for all $u=1, \dots,9$. 
This implies that the two Hamiltonians of the dimer integrable systems are identical under the birational transformation in \eref{es28a01},
\beal{es28a07}
H^{(5)} = H^{(6a)} ~.~
\eea
By identifying \eref{es14a05_1} with \eref{es15a05_1}, we also obtain the following canonical transformation, 
\beal{es28a08}
e^{Q^{(5)}} &=& \frac{1}{z_2^{(6a)} z_4^{(6a)} z_5^{(6a)}}e^{Q^{(6a)}} ~,~ e^{P^{(5)}} = e^{P^{(6a)}}
~.~
\eea
We conclude that the dimer integrable systems for Model 5 and Model 6a are birationally equivalent to each other. 
\\

%=================================================================
\subsection{Model 5 to Model 6b}
%=================================================================
  Let us refer to the spectral curve in \eref{es14a09} for Model 5 as
$\Sigma^{(5)}$ and the spectral curve in \eref{es15b09} for Model 6b as
 $\Sigma^{(6b)}$.

Under the following birational transformation,
\beal{es28b01}
&&
\varphi_{A;N}=\varphi_{A} \circ N~:~ 
(x,y) \mapsto \Big(x, z_3^{(5)}  z_7^{(5)} \frac{(x+z_5^{(5)} )}{xy} \Big)
~,~
\eea
where 
\beal{es28b02}
\varphi_{A} ~:~ (x,y) \mapsto \Big(x, z_3^{(5)}  z_7^{(5)} \frac{(x+z_5^{(5)} )}{x} y\Big)~,~N ~:~ (x,y) \mapsto \Big(x, \frac{1}{y} \Big)~,~ 
\eea
we discover that the spectral curve $\Sigma^{(5)}$ in \eref{es14a09} is mapped to $\Sigma^{(6b)}$ in \eref{es15b09},
\beal{es28b03}
\varphi_{A;N} \Sigma^{(5)} = \Sigma^{(6b)} ~.~
\eea
Based on this map,
we have the following identifications between
the zig-zag paths,
\beal{es28b04}
&
z_1^{(5)} = z_2^{(6b)} z_5^{(6b)} z_7^{(6b)} ~,~
z_2^{(5)} = \frac{1}{z_2^{(6b)}} ~,~
z_3^{(5)} = z_4^{(6b)}~,~
z_4^{(5)} = z_2^{(6b)} z_3^{(6b)} z_7^{(6b)}~,~
\nn\\
&
z_5^{(5)} = z_1^{(6b)}~,~
z_6^{(5)} = \frac{1}{z_7^{(6b)}}~,~
z_7^{(5)} = z_6^{(6b)}~,~
&
\eea
as well as between the face paths,
\beal{es28b05}
&
f_1^{(5)} = f_7^{(6b)} ~,~ 
f_2^{(5)} = f_1^{(6b)} ~,~ 
f_3^{(5)} = f_2^{(6b)} ~,~ 
f_4^{(5)} = f_3^{(6b)} ~,~ 
\nn\\
& 
f_5^{(5)} = f_4^{(6b)} ~,~ 
f_6^{(5)} = f_5^{(6b)} ~,~ 
f_7^{(5)} = f_6^{(6b)} ~.~ 
&
\eea
Moreover, the 1-loops of the two dimer integrable systems are identified as follows,
\beal{es28b06}
\gamma_u^{(5)} = \gamma_u^{(6b)} ~,~
\eea
for all $u=1, \dots,9$. 
This implies that the two Hamiltonians of the dimer integrable systems are identical under the birational transformation in \eref{es28b01},
\beal{es28b07}
H^{(5)} = H^{(6b)} ~.~
\eea
By identifying \eref{es14a05_1} with \eref{es15b05_1}, we also obtain the following canonical transformation, 
\beal{es28b08}
e^{Q^{(5)}} &=&z_3^{(6b)} z_4^{(6b)} e^{Q^{(6b)}} ~,~ e^{P^{(5)}} = e^{P^{(6b)}}
~.~
\eea
We conclude that the dimer integrable systems for Model 5 and Model 6b are birationally equivalent to each other. 
\\

%=================================================================
%=================================================================
\section{Bucket 3 \label{sec:21}}
%=================================================================

%---------------------------------------------------- 
\begin{figure}[H]
\begin{center}
\resizebox{1\hsize}{!}{
\includegraphics{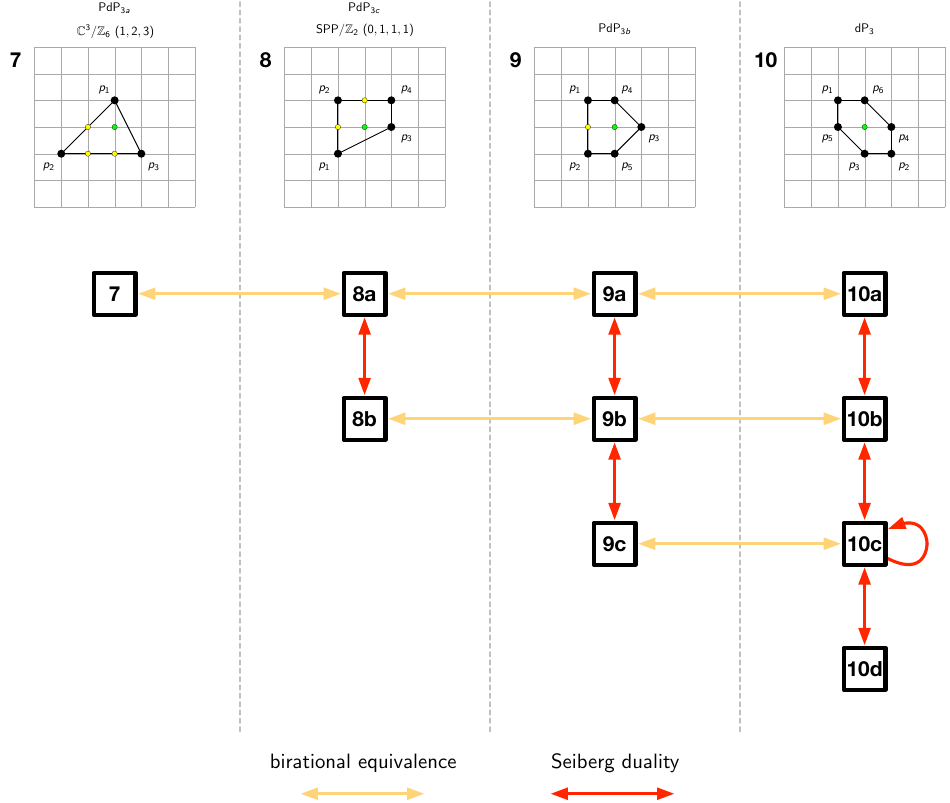}
}
\caption{Brane tilings and toric diagrams in Bucket 3.}
\label{fig_bucket03}
 \end{center}
 \end{figure}
%---------------------------------------------------- 
 
%=================================================================
\subsection{Hilbert series and generators of the mesonic moduli spaces}
%=================================================================
 
\fref{fig_bucket03}
summarizes the brane tilings
in bucket 3 that are related by birational transformations.
Taking the results in \cite{Hanany:2012hi}, 
the refined Hilbert series of the mesonic moduli spaces of these brane tilings 
in terms of fugacities $t_a$ corresponding to GLSM fields $p_a$
are given by,
\beal{es42a00}
g(t_a; \mathcal{M}^{mes}_{\text{Model 7}})
&=&
\frac{
1 + t_1 t_2^3 + t_1 t_2 t_3 + t_2^4 t_3 + t_2^2 t_3^2 + t_1 t_2^5 t_3^2
}{
(1 -  t_1^2) (1 -  t_2^6) (1 -  t_3^3)
}
~,~
\nn\\
g(t_a; \mathcal{M}^{mes}_{\text{Model 8a, 8b}})
&=&
\frac{
1
}{
(1 -  t_1^2 t_2^2) (1 -  t_1^2 t_3^4) (1 -  t_2 t_4^2) (1 -  t_3^2 t_4^2)
}
\nn\\
&&
\times 
(1 + t_1^2 t_2 t_3^2 + t_1 t_2 t_3 t_4 + t_1 t_3^3 t_4 - t_1^2 t_2^2 t_3^2 t_4^2 - t_1^2 t_2 t_3^4 t_4^2 
\nn\\
&&
\hspace{0.5cm}
- t_1 t_2 t_3^3 t_4^3 - t_1^3 t_2^2 t_3^5 t_4^3)
~,~
\nn\\
g(t_a; \mathcal{M}^{mes}_{\text{Model 9a, 9b, 9c}})
&=&
\frac{
1
}{
(1 -  t_1^3 t_2 t_4^2) (1 -  t_1^2 t_3 t_4^2) (1 -  t_3^2 t_4 t_5) (1 -  t_1 t_2^3 t_5^2) (1 -  t_2^2 t_3 t_5^2)
}
\nn\\
&&
\times (
1 + t_1^2 t_2^2 t_4 t_5 + t_1 t_2 t_3 t_4 t_5 - t_1^4 t_2^2 t_3 t_4^3 t_5 -  t_1^3 t_2 t_3^2 t_4^3 t_5 
\nn\\
&&
\hspace{0.5cm}
- t_1^3 t_2^3 t_3 t_4^2 t_5^2 -  t_1^2 t_2^2 t_3^2 t_4^2 t_5^2 - t_1^2 t_2^4 t_3 t_4 t_5^3 -  t_1 t_2^3 t_3^2 t_4 t_5^3 
\nn\\
&&
\hspace{0.5cm}
+ t_1^4 t_2^4 t_3^2 t_4^3 t_5^3 +  t_1^3 t_2^3 t_3^3 t_4^3 t_5^3 + t_1^5 t_2^5 t_3^3 t_4^4 t_5^4
)
~,~
\nn\\
\nn\\
g(t_a; \mathcal{M}^{mes}_{\text{Model 10a, 10b, 10c, 10d}})
&=&
\frac{
(1 - t_1 t_2 t_3 t_4 t_5 t_6)
}{
(1 -  t_2^2 t_3^2 t_4 t_5) (1 -  t_1 t_2 t_3^2 t_5^2) (1 -  t_2^2 t_3 t_4^2 t_6) (1 -  t_1^2 t_3 t_5^2 t_6) 
}
\nn\\
&&
\times
\frac{1}{
(1 -  t_1 t_2 t_4^2 t_6^2) (1 -  t_1^2 t_4 t_5 t_6^2)
}
\times
(
1 + 2 t_1 t_2 t_3 t_4 t_5 t_6 
\nn\\
&&
\hspace{0.5cm}
- t_1 t_2^3 t_3^3 t_4^2 t_5^2 t_6 -  t_1^2 t_2^2 t_3^3 t_4 t_5^3 t_6 - t_1 t_2^3 t_3^2 t_4^3 t_5 t_6^2 
\nn\\
&&
\hspace{0.5cm}
-  t_1^3 t_2 t_3^2 t_4 t_5^3 t_6^2 
- t_1^2 t_2^2 t_3 t_4^3 t_5 t_6^3 -  t_1^3 t_2 t_3 t_4^2 t_5^2 t_6^3 
\nn\\
&&
\hspace{0.5cm}
+ 2 t_1^3 t_2^3 t_3^3 t_4^3 t_5^3 t_6^3 +  t_1^4 t_2^4 t_3^4 t_4^4 t_5^4 t_6^4
)
~.~
\eea
Here, we note that brane tilings related by Seiberg duality have the same mesonic moduli space and associated Hilbert series. 
\\

%-------------------------
\begin{table}[http!!]
\begin{center}
\begin{minipage}[t]{0.8\linewidth}
\centering
\begin{tabular}{|c|c|l|}
\hline
\multicolumn{3}{|c|}{Model 7}
\\
\hline
GLSM & $U(1)_R$ & fugacity \\
\hline
$p_1$ & $3r$
&
$t_1 = \bar{t}^3$
\\
$p_2$ & $r$
&
$t_2 = \bar{t}$
\\
$p_3$ & $2r$
&
$t_3 = \bar{t}^2$
\\
\hline
\end{tabular}
\hspace{1cm}
\begin{tabular}{|c|c|l|}
\hline
\multicolumn{3}{|c|}{Model 8a, 8b}
\\
\hline
GLSM & $U(1)_R$ & fugacity \\
\hline
$p_1$ & $r$
&
$t_1 = \bar{t}$
\\
$p_2$ & $2r$
&
$t_2 = \bar{t}^2$
\\
$p_3$ & $r$
&
$t_3 = \bar{t}$
\\
$p_4$ & $2r$
&
$t_4 = \bar{t}^2$
\\
\hline
\end{tabular}
\end{minipage}
\\
\vspace{0.5cm}
\begin{minipage}[t]{0.8\linewidth}
\centering
\begin{tabular}{|c|c|l|}
\hline
\multicolumn{3}{|c|}{Model 9a, 9b, 9c}
\\
\hline
GLSM & $U(1)_R$ & fugacity \\
\hline
$p_1$ & $r$
&
$t_1 = \bar{t}$
\\
$p_2$ & $r$
&
$t_2 = \bar{t}$
\\
$p_3$ & $2r$
&
$t_3 = \bar{t}^2$
\\
$p_4$ & $r$
&
$t_4 = \bar{t}$
\\
$p_5$ & $r$
&
$t_5 = \bar{t}$
\\
\hline
\end{tabular}
\hspace{1cm}
\begin{tabular}{|c|c|l|}
\hline
\multicolumn{3}{|c|}{Model 10a, 10b, 10c, 10d}
\\
\hline
GLSM & $U(1)_R$ & fugacity \\
\hline
$p_1$ & $r$
&
$t_1 = \bar{t}$
\\
$p_2$ & $r$
&
$t_2 = \bar{t}$
\\
$p_3$ & $r$
&
$t_3 = \bar{t}$
\\
$p_4$ & $r$
&
$t_4 = \bar{t}$
\\
$p_5$ & $r$
&
$t_5 = \bar{t}$
\\
$p_6$ & $r$
&
$t_6 = \bar{t}$
\\
\hline
\end{tabular}
\end{minipage}
\caption{$U(1)_R$ charge assignment on GLSM fields of birationally related brane tilings in bucket 3 such that the $U(1)_R$ charge of the superpotentials is $6r = 2$ and that the generators of the mesonic moduli spaces have all $U(1)_R$ charge $6r$.}
\label{tab_buck3}
\end{center}
\end{table}
%-------------------------

\tref{tab_buck3} summarizes
the $U(1)_R$ charge assignment on the GLSM fields in terms of a $U(1)_R$ charge $r$,
which ensures that the superpotentials of the brane tilings in bucket 3 have all $U(1)_R$ charge $6r = 2$
and the generators of the mesonic moduli spaces have all $U(1)_R$ charge $6r$.
Using this $U(1)_R$ charge assignment, 
we see that the refined Hilbert series in \eref{es42a00}
expressed in terms of a single fugacity $\bar{t}$ corresponding to $U(1)_R$ charge $r$ all become,
\beal{es42a01}
g(\bar{t};\mathcal{M}^{mes}_{\text{bucket 3}})
=
\frac{
1 + 4 \bar{t}^6 + \bar{t}^{12}}{
(1 - \bar{t}^6)^3
}
~.~
\eea
This confirms that the birational transformations relating the brane tilings in bucket 3
preserve the Hilbert series of the mesonic moduli spaces when refined only under the $U(1)_R$ symmetry.

Moreover, by further referring to the results in \cite{Hanany:2012hi}, 
we note here that the brane tilings in bucket 3 all have mesonic moduli spaces with 7 generators.
This can also be seen through the plethystic logarithm \cite{Benvenuti:2006qr, Hanany:2006uc, Butti:2007jv, Feng:2007ur, Hanany:2007zz}
of the Hilbert series in \eref{es42a01}, which takes the following form,
 \beal{es42a02}
PL[g(\bar{t};\mathcal{M}^{mes}_{\text{bucket 3}})]
=
7 \bar{t}^6 - 9 \bar{t}^{12} + 16 \bar{t}^{18}
+ \dots 
~.~
 \eea
 This confirms the mesonic moduli spaces in bucket 3 all have 7 generators.
\\

The following sections illustrate how the brane tilings in bucket 3 define dimer integrable systems that are birationally equivalent to each other.
\\

 %=================================================================
 \subsection{Model 7 to Model 8a}
 %=================================================================
   Let us refer to the spectral curve in \eref{es16a09} for Model 7 as
$\Sigma^{(7)}$ and the spectral curve in \eref{es17a09} for Model 8a as
 $\Sigma^{(8a)}$.

Under the following birational transformation,
\beal{es30a01}
&&
\varphi_{A;N}=\varphi_{A} \circ N~:~ 
(x,y) \mapsto \Big(\frac{1}{y}, z_3^{(7)} z_6^{(7)} \frac{(y+z_4^{(7)})(y+z_5^{(7)})}{xy} \Big)
~,~
\eea
where 
\beal{es30a02}
\varphi_{A} ~:~ (x,y) \mapsto \Big(x,z_3^{(7)} z_6^{(7)} (x+z_4^{(7)})(x+z_5^{(7)})y \Big)~,~N ~:~ (x,y) \mapsto \Big(\frac{1}{y}, \frac{y}{x} \Big)~,~ 
\eea
we discover that the spectral curve $\Sigma^{(7)}$ in \eref{es16a09} is mapped to $\Sigma^{(8a)}$ in \eref{es17a09},
\beal{es30a03}
\varphi_{M;N} \Sigma^{(7)} = \Sigma^{(8a)} ~.~
\eea
Based on this map,
we have the following identifications between
the zig-zag paths,
\beal{es30a04}
&
z_1^{(7)} = \frac{1}{{z_1^{(8a)}}^2 z_3^{(8a)}} ~,~
z_2^{(7)} = z_1^{(8a)} ~,~
z_3^{(7)} = \frac{1}{z_5^{(8a)}} ~,~
z_4^{(7)} = \frac{1}{z_2^{(8a)}} ~,~
z_5^{(7)} = \frac{1}{z_6^{(8a)}} ~,~
z_6^{(7)} = \frac{1}{z_4^{(8a)}} ~,~
&
\nn \\
\eea
as well as between the face paths,
\beal{es30a05}
&
f_1^{(7)} = f_1^{(8a)} ~,~ 
f_2^{(7)} = f_6^{(8a)} ~,~ 
f_3^{(7)} = f_3^{(8a)} ~,~ 
\nn\\
&
f_4^{(7)} = f_2^{(8a)} ~,~ 
f_5^{(7)} = f_4^{(8a)} ~,~
f_6^{(7)} = f_5^{(8a)} ~.~  
&
\eea
Moreover, the 1-loops of the two dimer integrable systems are identified as follows,
\beal{es30a06}
\gamma_u^{(7)} = \gamma_u^{(8a)} ~,~
\eea
for all $u=1, \dots,6$. 
This implies that the two Hamiltonians of the dimer integrable systems are identical under the birational transformation in \eref{es30a01},
\beal{es30a07}
H^{(7)} = H^{(8a)} ~.~
\eea
By identifying \eref{es16a05_1} with \eref{es17a05_1}, we also obtain the following canonical transformation, 
\beal{es30a08}
e^{Q^{(7)}} &=& \frac{1}{z_1 z_2} e^{-P^{(8a)}}  ~,~ e^{P^{(7)}} = e^{Q^{(8a)}} 
~.~
\eea
We conclude that the dimer integrable systems for Model 7 and Model 8a are birationally equivalent to each other. 
\\

%=================================================================
\subsection{Model 8a to Model 9a}
%=================================================================
     Let us refer to the spectral curve in \eref{es17a09} for Model 8a as
$\Sigma^{(8a)}$ and the spectral curve in \eref{es18a09} for Model 9a as
 $\Sigma^{(9a)}$.

Under the following birational transformation,
\beal{es30b01}
&&
\varphi_{M;A;N}=M \circ \varphi_{A} \circ N~:~ 
(x,y) \mapsto \Big(\frac{x}{z_1^{(8a)}},\frac{1}{y+\frac{z_4^{(8a)} x y}{z_1^{(8a)}}} \Big)
~,~
\eea
where 
\beal{es30b02}
&
M ~:~ (x,y) \mapsto \Big(x, \frac{1}{y} \Big)~,~\varphi_{A} ~:~ (x,y) \mapsto \Big(x,(1+z_4^{(8a)} x)y \Big)~,~
&
\nn\\
&
N ~:~ (x,y) \mapsto \Big(\frac{x}{z_1^{(8a)}}, y \Big)~,~ 
&
\eea
we discover that the spectral curve $\Sigma^{(8a)}$ in \eref{es17a09} is mapped to $\Sigma^{(9a)}$ in \eref{es18a09},
\beal{es30b03}
\varphi_{M;A;N} \Sigma^{(8a)} = \Sigma^{(9a)} ~.~
\eea
Based on this map,
we have the following identifications between
the zig-zag paths,
\beal{es30b04}
&
z_1^{(8a)} = \frac{1}{z_2^{(9a)} z_4^{(9a)}} ~,~
z_2^{(8a)} = \frac{1}{z_3^{(9a)}} ~,~
z_3^{(8a)} = \frac{z_2^{(9a)} {z_4^{(9a)}}^2}{z_5^{(9a)}} ~,~
\nn\\
&
z_4^{(8a)} = \frac{1}{z_4^{(9a)}} ~,~
z_5^{(8a)} = \frac{1}{z_1^{(9a)} z_2^{(9a)} z_4^{(9a)}} ~,~
z_6^{(8a)} = \frac{1}{z_6^{(9a)}} ~,~
&
\eea
as well as between the face paths,
\beal{es30b05}
&
f_1^{(8a)} = f_6^{(9a)} ~,~ 
f_2^{(8a)} = f_4^{(9a)} ~,~ 
f_3^{(8a)} = f_5^{(9a)} ~,~ 
\nn\\
&
f_4^{(8a)} = f_3^{(9a)} ~,~ 
f_5^{(8a)} = f_2^{(9a)} ~,~
f_6^{(8a)} = f_1^{(9a)} ~.~  
&
\eea
Moreover, the 1-loops of the two dimer integrable systems are identified as follows,
\beal{es30b06}
\gamma_u^{(8a)} = \gamma_u^{(9a)} ~,~
\eea
for all $u=1, \dots,6$. 
This implies that the two Hamiltonians of the dimer integrable systems are identical under the birational transformation in \eref{es30b01},
\beal{es30b07}
H^{(8a)} = H^{(9a)} ~.~
\eea
By identifying \eref{es17a05_1} with \eref{es18a05_1}, we also obtain the following canonical transformation, 
\beal{es30b32}
e^{Q^{(8a)}} &=&\frac{z_5^{(9a)} z_6^{(9a)}}{z_2^{(9a)}}e^{Q^{(9a)}-P^{(9a)}}  ~,~ e^{P^{(8a)}} =e^{Q^{(9a)}} 
~.~
\eea
We conclude that the dimer integrable systems for Model 8a and Model 9a are birationally equivalent to each other. 
\\

%=================================================================
\subsection{Model 8b to Model 9b}
%=================================================================

Let us refer to the spectral curve in \eref{es17b09} for Model 8b as
$\Sigma^{(8b)}$ and the spectral curve in \eref{es18b09} for Model 9b as
 $\Sigma^{(9b)}$.

Under the following birational transformation,
\beal{es30d01}
&&
\varphi_{M;A;N}= M \circ \varphi_{A} \circ N~:~ 
(x,y) \mapsto \Big(\frac{x}{z_5^{(8b)}}, \frac{1}{(1+\frac{z_1^{(8b)}}{z_5^{(8b)}}x) y}\Big)
~,~
\eea
where 
\beal{es30d02}
&
M ~:~ (x,y) \mapsto \Big(x, \frac{1}{y} \Big)~,~\varphi_{A} ~:~ (x,y) \mapsto \Big(x,(1+z_1^{(8a)} x)y \Big)~,~
&
\nn\\
&
N ~:~ (x,y) \mapsto \Big(\frac{x}{z_5^{(8a)}}, y \Big)~,~
&
\eea
we discover that the spectral curve $\Sigma^{(8b)}$ in \eref{es17b09} is mapped to $\Sigma^{(9b)}$ in \eref{es18b09},
\beal{es30d03}
\varphi_{M;A;N} \Sigma^{(8b)} = \Sigma^{(9b)} ~.~
\eea
Based on this map,
we have the following identifications between
the zig-zag paths,
\beal{es30d04}
&
z_1^{(8b)} = z_6^{(9b)} ~,~
z_2^{(8b)} = {z_2^{(9b)}}^{-1} z_3^{(9b)} {z_6^{(9b)}}^{-2} ~,~
z_3^{(8b)} = z_1^{(9b)} z_2^{(9b)} z_6^{(9b)} ~,~
\nn\\
&
z_4^{(8b)} = z_5^{(9b)} ~,~
z_5^{(8b)} = z_2^{(9b)} z_6^{(9b)} ~,~
z_6^{(8b)} = z_4^{(9b)} ~,~
&
\eea
as well as between the face paths,
\beal{es30d05}
&
f_1^{(8b)} = f_6^{(9b)} ~,~ 
f_2^{(8b)} = f_4^{(9b)} ~,~ 
f_3^{(8b)} = f_5^{(9b)} ~,~ 
\nn\\
& 
f_4^{(8b)} = f_3^{(9b)} ~,~ 
f_5^{(8b)} = f_2^{(9b)} ~,~
f_6^{(8b)} = f_1^{(9b)} ~.~  
&
\eea
Moreover, the 1-loops of the two dimer integrable systems are identified as follows,
\beal{es30d06}
\gamma_u^{(8b)} = \gamma_u^{(9b)} ~,~
\eea
for all $u=1, \dots,7$. 
This implies that the two Hamiltonians of the dimer integrable systems are identical under the birational transformation in \eref{es30d01},
\beal{es30d07}
H^{(8b)} = H^{(9b)} ~.~
\eea
By identifying \eref{es17b05_1} with \eref{es18b05_1}, we also obtain the following canonical transformation, 
\beal{es30d08}
e^{Q^{(8b)}} &=&z_1^{(9b)} z_3^{(9b)} z_5^{(9b)} e^{-Q^{(9b)}-P^{(9b)}}  ~,~ e^{P^{(8b)}} =e^{Q^{(9b)}}
~.~
\eea
We conclude that the dimer integrable systems for Model 8b and Model 9b are birationally equivalent to each other. 
\\

%=================================================================
\subsection{Model 9a to Model 10a}
%=================================================================
Let us refer to the spectral curve in \eref{es18a09} for Model 9a as
$\Sigma^{(9a)}$ and the spectral curve in \eref{es19a09} for Model 10a as
 $\Sigma^{(10a)}$.

Under the following birational transformation,
\beal{es30c01}
&&
\varphi_{A;N}= \varphi_{A} \circ N~:~ 
(x,y) \mapsto \Big(z_1^{(9a)} z_3^{(9a)} z_4^{(9a)} \frac{(1+z_6^{(9a)} y)}{x},y \Big)
~,~
\eea
where 
\beal{es30c02}
\varphi_{A} ~:~ (x,y) \mapsto \Big(z_1^{(9a)} z_3^{(9a)} z_4^{(9a)} (1+z_6^{(9a)} y)x,y\Big)~,~N ~:~ (x,y) \mapsto \Big(\frac{1}{x}, y \Big)~,~
\eea
we discover that the spectral curve $\Sigma^{(9a)}$ in \eref{es18a09} is mapped to $\Sigma^{(10a)}$ in \eref{es19a09},
\beal{es30c03}
\varphi_{A;N} \Sigma^{(9a)} = \Sigma^{(10a)} ~.~
\eea
Based on this map,
we have the following identifications between
the zig-zag paths,
\beal{es30c04}
&
z_1^{(9a)} = z_5^{(10a)} z_6^{(10a)} ~,~
z_2^{(9a)} = z_4^{(10a)} ~,~
z_3^{(9a)} = \frac{1}{z_6^{(10a)}} ~,~
\nn\\
&
z_4^{(9a)} = z_1^{(10a)} z_6^{(10a)} ~,~
z_5^{(9a)} = z_3^{(10a)} ~,~
z_6^{(9a)} = z_2^{(10a)} ~,~
&
\eea
as well as between the face paths,
\beal{es30c05}
&
f_1^{(9a)} = f_4^{(10a)} ~,~ 
f_2^{(9a)} = f_2^{(10a)} ~,~ 
f_3^{(9a)} = f_3^{(10a)} ~,~ 
\nn\\
&
f_4^{(9a)} = f_5^{(10a)} ~,~ 
f_5^{(9a)} = f_1^{(10a)} ~,~
f_6^{(9a)} = f_6^{(10a)} ~.~  
&
\eea
Moreover, the 1-loops of the two dimer integrable systems are identified as follows,
\beal{es30c06}
\gamma_u^{(9a)} = \gamma_u^{(10a)} ~,~
\eea
for all $u=1, \dots,6$. 
This implies that the two Hamiltonians of the dimer integrable systems are identical under the birational transformation in \eref{es30c01},
\beal{es30c07}
H^{(9a)} = H^{(10a)} ~.~
\eea
By identifying \eref{es18a05_1} with \eref{es19a05_1}, we also obtain the following canonical transformation, 
\beal{es30c08}
e^{Q^{(9a)}} &=&z_4^{(10a)} z_5^{(10a)} z_6^{(10a)} e^{-Q^{(10a)}-P^{(10a)}}  ~,~ e^{P^{(9a)}} =z_3^{(10a)} z_5^{(10a)} e^{-P^{(10a)}}
~.~
\eea
We conclude that the dimer integrable systems for Model 9a and Model 10a are birationally equivalent to each other. 
\\

%=================================================================
\subsection{Model 9b to Model 10b}
%=================================================================
Let us refer to the spectral curve in \eref{es18b09} for Model 9b as
$\Sigma^{(9b)}$ and the spectral curve in \eref{es19b09} for Model 10b as
 $\Sigma^{(10b)}$.

Under the following birational transformation,
\beal{es31d01}
&&
\varphi_{A;N}= \varphi_{A} \circ N~:~ 
(x,y) \mapsto \Big(\frac{(\frac{1}{y}+z_5^{(9b)})x}{z_5^{(9b)}}, \frac{1}{y}\Big)
~,~
\eea
where 
\beal{es31d02}
\varphi_{A} ~:~ (x,y) \mapsto \Big(x,\frac{(y+z_5^{(9b)})}{z_5^{(9b)}} x \Big)~,~N ~:~ (x,y) \mapsto \Big(x, \frac{1}{y} \Big)~,~ 
\eea
we discover that the spectral curve $\Sigma^{(9b)}$ in \eref{es18b09} is mapped to $\Sigma^{(10b)}$ in \eref{es19b09},
\beal{es31d03}
\varphi_{A;N} \Sigma^{(9b)} = \Sigma^{(10b)} ~.~
\eea
Based on this map,
we have the following identifications between
the zig-zag paths,
\beal{es31d04}
&
z_1^{(9b)} = z_1^{(10b)} ~,~
z_2^{(9b)} = z_2^{(10b)} z_3^{(10b)} ~,~
z_3^{(9b)} = z_3^{(10b)} z_5^{(10b)}  ~,~
\nn\\
&
z_4^{(9b)} = z_6^{(10b)} ~,~
z_5^{(9b)} = \frac{1}{z_3^{(10b)}} ~,~
z_6^{(9b)} = z_4^{(10b)} ~.~
&
\eea
as well as between the face paths,
\beal{es31d05}
&
f_1^{(9b)} = f_2^{(10b)} ~,~ 
f_2^{(9b)} = f_4^{(10b)} ~,~ 
f_3^{(9b)} = f_3^{(10b)} ~,~ 
\nn\\
&
f_4^{(9b)} = f_1^{(10b)} ~,~ 
f_5^{(9b)} = f_5^{(10b)} ~,~
f_6^{(9b)} = f_6^{(10b)} ~.~  
&
\eea
Moreover, the 1-loops of the two dimer integrable systems are identified as follows,
\beal{es31d06}
\gamma_u^{(9b)} = \gamma_u^{(10b)} ~,~
\eea
for all $u=1, \dots,7$. 
This implies that the two Hamiltonians of the dimer integrable systems are identical under the birational transformation in \eref{es30d01},
\beal{es31d07}
H^{(9b)} = H^{(10b)} ~.~
\eea
By identifying \eref{es18b05_1} with \eref{es19b05_1}, we also obtain the following canonical transformation, 
\beal{es31d08}
e^{Q^{(9b)}} &=&z_1^{(10b)} z_5^{(10b)} e^{-Q^{(10b)}+P^{(10b)}}  ~,~ e^{P^{(9b)}} =e^{-P^{(10b)}}
~.~
\eea
We conclude that the dimer integrable systems for Model 9b and Model 10b are birationally equivalent to each other. 
\\

%=================================================================
\subsection{Model 9c to Model 10c}
%=================================================================
 Let us refer to the spectral curve in \eref{es18c09} for Model 9c as
$\Sigma^{(9c)}$ and the spectral curve in \eref{es19c09} for Model 10c as
 $\Sigma^{(10c)}$.

Under the following birational transformation,
\beal{es31e01}
&&
\varphi_{A;N}= \varphi_{A} \circ N~:~ 
(x,y) \mapsto \Big(\frac{(\frac{1}{y}+z_5^{(9c)})x}{z_5^{(9c)}}, \frac{1}{y}\Big)
~,~
\eea
where 
\beal{es31e02}
\varphi_{A} ~:~ (x,y) \mapsto \Big(x,\frac{(y+z_5^{(9c)})}{z_5^{(9c)}} x \Big)~,~N ~:~ (x,y) \mapsto \Big(x, \frac{1}{y} \Big)~,~
\eea
we discover that the spectral curve $\Sigma^{(9c)}$ in \eref{es18c09} is mapped to $\Sigma^{(10c)}$ in \eref{es19c09},
\beal{es31e03}
\varphi_{A;N} \Sigma^{(9c)} = \Sigma^{(10c)} ~.~
\eea
Based on this map,
we have the following identifications between
the zig-zag paths,
\beal{es31e04}
&
z_1^{(9c)} = z_6^{(10c)} ~,~
z_2^{(9c)} = z_1^{(10c)} z_2^{(10c)} ~,~
z_3^{(9c)} = z_1^{(10c)} z_5^{(10c)}  ~,~
\nn\\
&
z_4^{(9c)} = z_3^{(10c)} ~,~
z_5^{(9c)} = \frac{1}{z_1^{(10c)}} ~,~
z_6^{(9c)} = z_4^{(10c)} ~,~
&
\eea
as well as between the face paths,
\beal{es31e05}
&
f_1^{(9c)} = f_1^{(10c)} ~,~ 
f_2^{(9c)} = f_4^{(10c)} ~,~ 
f_3^{(9c)} = f_3^{(10c)} ~,~ 
\\ & \nn
f_4^{(9c)} = f_2^{(10c)} ~,~ 
f_5^{(9c)} = f_5^{(10c)} ~,~
f_6^{(9c)} = f_6^{(10c)} ~.~  
&
\eea
Moreover, the 1-loops of the two dimer integrable systems are identified as follows,
\beal{es31e06}
\gamma_u^{(9c)} = \gamma_u^{(10c)} ~,~
\eea
for all $u=1, \dots,8$. 
This implies that the two Hamiltonians of the dimer integrable systems are identical under the birational transformation in \eref{es30d01},
\beal{es31e07}
H^{(9c)} = H^{(10c)} ~.~
\eea
By identifying \eref{es18c05_1} with \eref{es19c05_1}, we also obtain the following canonical transformation, 
\beal{es30e08}
e^{Q^{(9c)}} &=& \frac{1}{z_1^{(10c)} z_3^{(10c)}}e^{Q^{(10c)}}  ~,~ e^{P^{(9c)}} =e^{P^{(10c)}}
~.~
\eea 
We conclude that the dimer integrable systems for Model 9c and Model 10c are birationally equivalent to each other. 
\\

%=======================================================================
 %=================================================================
\section{Bucket 4 \label{sec:22}}
%=======================================================================

%---------------------------------------------------- 
\begin{figure}[H]
\begin{center}
\resizebox{0.5\hsize}{!}{
\includegraphics{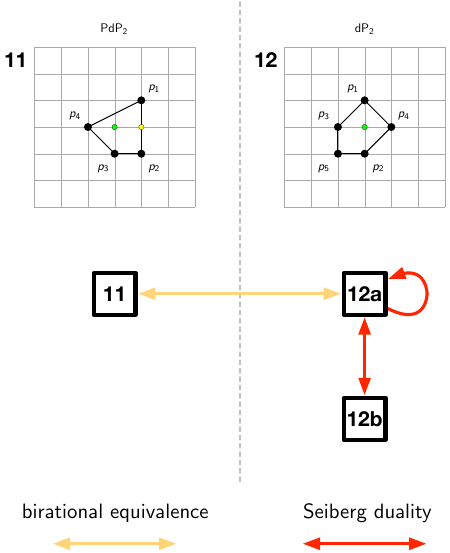}
}
\caption{Brane tilings and toric diagrams in Bucket 4.}
\label{fig_bucket04}
 \end{center}
 \end{figure}
%---------------------------------------------------- 
 
%=================================================================
\subsection{Hilbert series and generators of the mesonic moduli spaces}
%=================================================================

\fref{fig_bucket04}
illustrates how birational transformations relate brane tilings in bucket 4.
From \cite{Hanany:2012hi}, 
we have the refined Hilbert series of the mesonic moduli spaces of these brane tilings in bucket 4
in terms of fugacities $t_a$ corresponding to GLSM fields $p_a$.
These refined Hilbert series take the following form, 
\beal{es43a00}
g(t_a; \mathcal{M}^{mes}_{\text{Model 11}})
&=&
\frac{
1
}{
(1 -  t_1^3 t_2) (1 -  t_2^4 t_3^3) (1 -  t_1^2 t_4) (1 -  t_3 t_4^2)
}
\nn\\
&&
\times
(1 + t_1^2 t_2^2 t_3 + t_1 t_2^3 t_3^2 + t_1 t_2 t_3 t_4 - t_1^4 t_2^2 t_3 t_4 +  t_2^2 t_3^2 t_4 
\nn\\
&&
\hspace{0.5cm}
- t_1^3 t_2^3 t_3^2 t_4 - t_1^3 t_2 t_3 t_4^2 -  t_1^2 t_2^2 t_3^2 t_4^2 - t_1^4 t_2^4 t_3^3 t_4^2)
~,~
\nn\\
g(t_a; \mathcal{M}^{mes}_{\text{Model 12a, 12b}})
&=&
\frac{
1
}{
(1 -  t_1^2 t_3 t_4) (1 -  t_1 t_2 t_4^2) (1 -  t_1^2 t_3^2 t_5) (1 - t_2^2 t_4^2 t_5) (1 -  t_2^2 t_3^2 t_5^3)
}
\nn\\
&&
\times
(1 + t_1 t_2 t_3 t_4 t_5 - t_1^3 t_2 t_3^2 t_4^2 t_5 - t_1^2 t_2^2 t_3 t_4^3 t_5 +  t_1 t_2 t_3^2 t_5^2 + t_2^2 t_3 t_4 t_5^2 
\nn\\
&&
\hspace{0.5cm}
- t_1^3 t_2 t_3^3 t_4 t_5^2 -  2 t_1^2 t_2^2 t_3^2 t_4^2 t_5^2 - t_1 t_2^3 t_3 t_4^3 t_5^2 +  t_1^4 t_2^2 t_3^3 t_4^3 t_5^2 + t_1^3 t_2^3 t_3^2 t_4^4 t_5^2 
\nn\\
&&
\hspace{0.5cm}
-  t_1^2 t_2^2 t_3^3 t_4 t_5^3 - t_1 t_2^3 t_3^2 t_4^2 t_5^3 +  t_1^3 t_2^3 t_3^3 t_4^3 t_5^3 + t_1^4 t_2^4 t_3^4 t_4^4 t_5^4)
~,~
\eea
where we note that brane tilings related by Seiberg duality have the same mesonic moduli space and therefore the same corresponding Hilbert series.
\\

%-------------------------
\begin{table}[H]
\begin{center}
\begin{minipage}[t]{0.8\linewidth}
\centering
\begin{tabular}{|c|c|l|}
\hline
\multicolumn{3}{|c|}{Model 11}
\\
\hline
GLSM & $U(1)_R$ & fugacity \\
\hline
$p_1$ & $2r$
&
$t_1 = \bar{t}^2$
\\
$p_2$ & $r$
&
$t_2 = \bar{t}$
\\
$p_3$ & $r$
&
$t_3 = \bar{t}$
\\
$p_4$ & $3r$
&
$t_4 = \bar{t}^3$
\\
\hline
\end{tabular}
\hspace{1cm}
\begin{tabular}{|c|c|l|}
\hline
\multicolumn{3}{|c|}{Model 12a, 12b}
\\
\hline
GLSM & $U(1)_R$ & fugacity \\
\hline
$p_1$ & $2r$
&
$t_1 = \bar{t}^2$
\\
$p_2$ & $r$
&
$t_2 = \bar{t}$
\\
$p_3$ & $r$
&
$t_3 = \bar{t}$
\\
$p_4$ & $2r$
&
$t_4 = \bar{t}^2$
\\
$p_5$ & $r$
&
$t_5 = \bar{t}$
\\
\hline
\end{tabular}
\end{minipage}
\caption{$U(1)_R$ charge assignment on GLSM fields of birationally related brane tilings in bucket 4 such that the $U(1)_R$ charge of the superpotentials is $7r = 2$ and that the generators of the mesonic moduli spaces have all $U(1)_R$ charge $7r$.}
\label{tab_buck4}
\end{center}
\end{table}
%-------------------------

Under the $U(1)_R$ charge assignment on the GLSM fields
summarized in \tref{tab_buck4}, 
the superpotentials of the brane tilings in bucket 4
have all $U(1)_R$ charge $7r = 2$
and the generators of the mesonic moduli spaces have all $U(1)_R$ charge $7r$.
Using this $U(1)_R$ charge assignment, 
the refined Hilbert series in \eref{es43a00}
can be rewritten in terms of a single fugacity $\bar{t}$ corresponding to $U(1)_R$ charge $r$.
We note here that the Hilbert series in terms of the fugacity $\bar{t}$
takes the following form for all brane tilings in bucket 4, 
\beal{es43a01}
g(\bar{t};\mathcal{M}^{mes}_{\text{bucket 4}})
= 
\frac{
1 + 5 \bar{t}^7 + \bar{t}^{14}
}{
(1 - \bar{t}^7)^3
}
~,~
\eea
confirming that birational transformations relating brane tilings in bucket 4 preserve the Hilbert series when it is refined only under $U(1)_R$.

Using the results in \cite{Hanany:2012hi}, 
we also note that the brane tilings in bucket 4 all have mesonic moduli spaces with 8 generators.
This can be seen by taking the plethystic logarithm \cite{Benvenuti:2006qr, Hanany:2006uc, Butti:2007jv, Feng:2007ur, Hanany:2007zz}
of the Hilbert series in \eref{es43a01}, which gives,
\beal{es43a02}
PL[g(\bar{t};\mathcal{M}^{mes}_{\text{bucket 4}})]
= 
8 \bar{t}^7 - 14 \bar{t}^{14} + 35 \bar{t}^{21}
+ \dots
~.~
\eea
This confirms that the number of generators is 8 for all mesonic moduli spaces in bucket 4. 
\\

In the following sections, we illustrate how brane tilings in bucket 4 define dimer integrable systems that are equivalent under birational transformations.
\\

 %=================================================================
\subsection{Model 11 to Model 12a}
 %=======================================================================
 
  Let us refer to the spectral curve in \eref{es20a09} for Model 11 as
$\Sigma^{(11)}$ and the spectral curve in \eref{es21a09} for Model 12a as
 $\Sigma^{(12a)}$.

Under the following birational transformation,
\beal{es31a01}
&&
\varphi_{A;M;N}=M \circ  \varphi_{A} \circ N~:~ 
(x,y) \mapsto \Big(\frac{z_3^{(11)} x}{z_3^{(11)}+x y}, x y \Big)
~,~
\eea
where 
\beal{es31a02}
M ~:~ (x,y) \mapsto \Big(\frac{1}{x},y \Big)~,~\varphi_{A} ~:~ (x,y) \mapsto \Big(\frac{(y+z_3^{(11)})}{z_3^{(11)}} x,y \Big)~,~N ~:~ (x,y) \mapsto \Big(\frac{1}{x}, x y \Big)~,~ \nn\\
\eea
we discover that the spectral curve $\Sigma^{(11)}$ in \eref{es20a09} is mapped to $\Sigma^{(12a)}$ in \eref{es21a09},
\beal{es31a03}
\varphi_{A;M;N} \Sigma^{(11)} = \Sigma^{(12a)} ~.~
\eea
Based on this map,
we have the following identifications between
the zig-zag paths,
\beal{es31a04}
&
z_1^{(11)} = \frac{z_5^{(12a)}}{z_1^{(12a)} z_2^{(12a)} z_4^{(12a)}}~,~
z_2^{(11)} = z_4^{(12a)} ~,~
z_3^{(11)} = \frac{1}{z_5^{(12a)}}~,~
z_4^{(11)} = z_2^{(12a)}~,~
z_5^{(11)} = z_1^{(12a)}~,~
&
\nn \\
\eea
as well as between the face paths,
\beal{es31a05}
&
f_1^{(11)} = f_5^{(12a)} ~,~ 
f_2^{(11)} = f_4^{(12a)} ~,~ 
f_3^{(11)} = f_1^{(12a)} ~,~ 
f_4^{(11)} = f_3^{(12a)} ~,~ 
f_5^{(11)} = f_2^{(12a)} ~.~ 
&
\nn\\
\eea
Moreover, the 1-loops of the two dimer integrable systems are identified as follows,
\beal{es31a06}
\gamma_u^{(11)} = \gamma_u^{(12a)} ~,~
\eea
for all $u=1, \dots,5$. 
This implies that the two Hamiltonians of the dimer integrable systems are identical under the birational transformation in \eref{es31a01},
\beal{es31a07}
H^{(11)} = H^{(12a)} ~.~
\eea
By identifying \eref{es20a05_1} with \eref{es21a05_1}, we also obtain the following canonical transformation, 
\beal{es31a08}
e^{Q^{(11)}} &=& e^{-Q^{(12a)}+P^{(12a)}} \frac{z_1^{(12a)}}{z_4^{(12a)} z_5^{(12a)}} ~,~ e^{P^{(11)}} = e^{-2Q^{(12a)}+P^{(12a)}} \frac{z_1^{(12a)}}{z_4^{(12a)} z_5^{(12a)}}
~.~
\eea
We conclude that the dimer integrable systems for Model 11 and Model 12a are birationally equivalent to each other. 
\\
  
%=======================================================================
\section{Bucket 5 \label{sec:23}}
%=======================================================================

%---------------------------------------------------- 
\begin{figure}[H]
\begin{center}
\resizebox{0.5\hsize}{!}{
\includegraphics{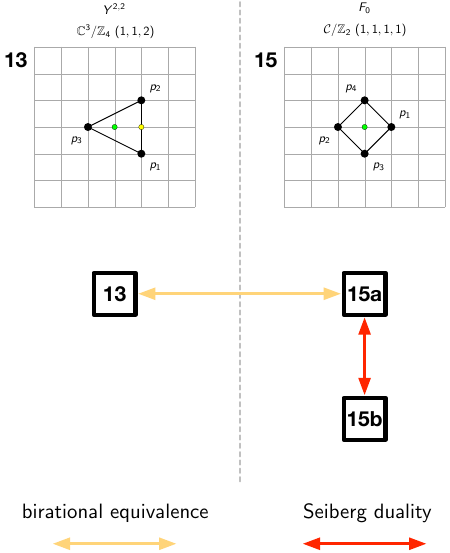}
}
\caption{Brane tilings and toric diagrams in Bucket 5.}
\label{fig_bucket05}
 \end{center}
 \end{figure}
%---------------------------------------------------- 

%=================================================================
\subsection{Hilbert series and generators of the mesonic moduli spaces}
%=================================================================

\fref{fig_bucket05}
illustrates how brane tilings in bucket 5 are related by birational transformations.
Using the results in \cite{Hanany:2012hi}, 
we have the refined Hilbert series of the mesonic moduli spaces of the brane tilings in bucket 4
in terms of fugacities $t_a$ corresponding to GLSM fields $p_a$.
These refined Hilbert series are as follows, 
\beal{es44a00}
g(t_a; \mathcal{M}^{mes}_{\text{Model 13}})
&=&
\frac{
1 + t_1^3 t_2 + t_1^2 t_2^2 + t_1 t_2^3 + t_1^2 t_3 + t_1 t_2 t_3 + t_2^2 t_3 +  t_1^3 t_2^3 t_3
}{
(1 -  t_1^4) (1 -  t_2^4) (1 -  t_3^2)
}
~,~
\nn\\
g(t_a; \mathcal{M}^{mes}_{\text{Model 15a, 15b}})
&=&
\frac{
1 - t_1 t_2 t_3 t_4
}{
(1 -  t_1^2 t_3^2) (1 -  t_2^2 t_3^2) (1 -  t_1^2 t_4^2) (1 -  t_2^2 t_4^2)
}
\nn\\
&&
\times
(1 + t_1 t_2 t_3^2 + t_1^2 t_3 t_4 + 2 t_1 t_2 t_3 t_4 + t_2^2 t_3 t_4 
\nn\\
&&
\hspace{0.5cm}
+  t_1 t_2 t_4^2 + t_1^2 t_2^2 t_3^2 t_4^2)
~,~
\eea
where we note that brane tilings related by Seiberg duality have the same mesonic moduli space and Hilbert series. 
\\

%-------------------------
\begin{table}[H]
\begin{center}
\begin{minipage}[t]{0.8\linewidth}
\centering
\begin{tabular}{|c|c|l|}
\hline
\multicolumn{3}{|c|}{Model 13}
\\
\hline
GLSM & $U(1)_R$ & fugacity \\
\hline
$p_1$ & $r$
&
$t_1 = \bar{t}$
\\
$p_2$ & $r$
&
$t_2 = \bar{t}$
\\
$p_3$ & $2r$
&
$t_3 = \bar{t}^2$
\\
\hline
\end{tabular}
\hspace{1cm}
\begin{tabular}{|c|c|l|}
\hline
\multicolumn{3}{|c|}{Model 15a, 15b}
\\
\hline
GLSM & $U(1)_R$ & fugacity \\
\hline
$p_1$ & $r$
&
$t_1 = \bar{t}$
\\
$p_2$ & $r$
&
$t_2 = \bar{t}$
\\
$p_3$ & $r$
&
$t_3 = \bar{t}$
\\
$p_4$ & $r$
&
$t_4 = \bar{t}$
\\
\hline
\end{tabular}
\end{minipage}
\caption{$U(1)_R$ charge assignment on GLSM fields of birationally related brane tilings in bucket 5 such that the $U(1)_R$ charge of the superpotentials is $4r = 2$ and that the generators of the mesonic moduli spaces have all $U(1)_R$ charge $4r$.}
\label{tab_buck5}
\end{center}
\end{table}
%-------------------------

\tref{tab_buck5}
summarizes a $U(1)_R$ charge assignment in terms of $U(1)_R$ charge $r$
on the GLSM fields such that the superpotentials of the brane tilings in bucket 5 all have $U(1)_R$ charge $4r = 2$
and the generators of the mesonic moduli spaces have all $U(1)_R$ charge $4r$.
In terms of this $U(1)_R$ charge assignment, 
the refined Hilbert series in \eref{es44a00}
can be expressed in terms of a single fugacity $\bar{t}$ corresponding to $U(1)_R$ charge $r$.
We note here that the Hilbert series in terms of $\bar{t}$ all become, 
\beal{es44a01}
g(\bar{t};\mathcal{M}^{mes}_{\text{bucket 5}})
= 
\frac{
1 + 6 \bar{t}^4 + \bar{t}^8
}{
(1 - \bar{t}^4)^3
}
~,~
\eea
which confirms that brane tilings related by birational transformations in bucket 5
share the same Hilbert series refined only under $U(1)_R$.

By further using the results in \cite{Hanany:2012hi}, 
we note that the brane tilings in bucket 5 all have mesonic moduli spaces with 9 generators.
We can see this also by taking the plethystic logarithm \cite{Benvenuti:2006qr, Hanany:2006uc, Butti:2007jv, Feng:2007ur, Hanany:2007zz}
of the Hilbert series in \eref{es44a01}, which takes the form,
\beal{es44a02}
PL[g(\bar{t};\mathcal{M}^{mes}_{\text{bucket 5}})]
= 
9 \bar{t}^4 - 20 \bar{t}^8 + 64 \bar{t}^{12}
+ \dots
~.~
\eea
We note here that the above plethystic logarithm confirms that the number of generators is 9 for all mesonic moduli spaces of brane tilings in bucket 5.
\\

The following sections illustrate how brane tilings in bucket 5 define dimer integrable systems that are equivalent under birational transformations.
\\

%=======================================================================
\subsection{Model 13 to Model 15a}
%=======================================================================

  Let us refer to the spectral curve in \eref{es22a09} for Model 13 as
$\Sigma^{(13)}$ and the spectral curve in \eref{es24a09} for Model 15a as
 $\Sigma^{(15a)}$.

Under the following birational transformation,
\beal{es32a01}
&&
\varphi_{A;M;N}=M \circ  \varphi_{A} \circ N~:~ 
(x,y) \mapsto \Big(\frac{z_3^{(13)}}{(\frac{y}{x}+z_3^{(13)})x},\frac{y}{x}\Big)
~,~
\eea
where 
\beal{es32a02}
&
M ~:~ (x,y) \mapsto \Big(\frac{1}{x},y \Big)~,~N ~:~ (x,y) \mapsto \Big(x, \frac{y}{x} \Big)~,~
&
\nn\\
&
\varphi_{A} ~:~ (x,y) \mapsto \Big(\frac{(y+z_3^{(13)})}{z_3^{(13)}}x,y \Big)~,~
\eea
we discover that the spectral curve $\Sigma^{(13)}$ in \eref{es22a09} is mapped to $\Sigma^{(15a)}$ in \eref{es24a09},
\beal{es32a03}
\varphi_{A;M;N} \Sigma^{(13)} = \Sigma^{(15a)} ~.~
\eea
Based on this map,
we have the following identifications between
the zig-zag paths,
\beal{es32a04}
&
z_1^{(13)} = \frac{1}{z_1^{(15a)}}~,~
z_2^{(13)} = \frac{z_1^{(15a)} z_2^{(15a)}}{z_4^{(15a)}}~,~
z_3^{(13)} = z_4^{(15a)}~,~
z_4^{(13)} = \frac{1}{z_2^{(15a)}} ~,~
&
\eea
as well as between the face paths,
\beal{es32a05}
&
f_1^{(13)} = {f_1^{(15a)}} ~,~ 
f_2^{(13)} = {f_2^{(15a)}} ~,~ 
f_3^{(13)} = {f_3^{(15a)}} ~,~ 
f_4^{(13)} = {f_4^{(15a)}} ~.~ 
&
\eea
Moreover, the 1-loops of the two dimer integrable systems are identified as follows,
\beal{es32a06}
\gamma_u^{(13)} = \gamma_u^{(15a)} ~,~
\eea
for all $u=1, \dots, 4$. 
This implies that the two Hamiltonians of the dimer integrable systems are identical under the birational transformation in \eref{es32a01},
\beal{es32a07}
H^{(13)} = H^{(15a)} ~.~
\eea
By identifying \eref{es22a13} with \eref{es24a13}, we also obtain the following canonical transformation, 
\beal{es32a08}
e^{Q^{(13)}} &=& e^{P^{(15a)}}~,~e^{P^{(13)}} = {e^{-Q^{(15a)}}}~.~
\eea
We conclude that the dimer integrable systems for Model 13 and Model 15a are birationally equivalent to each other. 
\\

%=================================================================
\section{Conclusions and Discussions \label{sec:24}}
%=================================================================

In this work, we present a complete classification of dimer integrable systems that
correspond to the 16 reflexive polygons in 2 dimensions. 
The reflexive polygons are toric diagrams of toric Calabi-Yau 3-folds and each of the dimer integrable systems
in the classification correspond to a brane tiling associated to these toric Calabi-Yau
3-folds.
The classification contains 30 dimer integrable systems and is based on the 30
brane tilings in the classification in \cite{Hanany:2012hi}.
There are more brane tilings and associated dimer integrable systems than toric Calabi-Yau 3-folds because when the associated brane tilings are related by Seiberg duality then they correspond to the same toric
Calabi-Yau 3-fold and the corresponding dimer integrable systems are equivalent under
a canonical transformation \cite{goncharov2012dimersclusterintegrablesystems}.

In our classification, we present for each dimer integrable system the Casimirs, the single Hamiltonian, the spectral curve and the Poisson commutation relations. 
In order to express these, we make use of directed paths along edges in the bipartite periodic graph on the 2-torus given by the associated brane tiling, including zig-zag paths and paths around faces of the brane tiling. 
The dimer integrable systems in our classification contain only a single Hamiltonian because the corresponding toric diagrams are reflexive and have a single internal vertex corresponding to the Hamiltonian.

As part of our classification, we identify 16 pairs of birationally equivalent dimer
integrable systems. 
Equivalence between dimer integrable systems via birational transformations between the associated toric Calabi-Yau 3-folds and brane tilings has been first studied in \cite{Kho:2025fmp}.
In our work, we give explicit expressions for the birational transformations that map the Casimirs, the Hamiltonian, the spectral curve and the Poisson commutation relations between birationally equivalent dimer integrable systems. 
Combined with equivalence due to Seiberg duality of the associated brane tilings, birational equivalence subdivides the dimer integrable systems in our classification into 5 equivalence classes that we call buckets \cite{akhtar2012minkowski}. 

We note here that the results of our work relate to recent developments in the
study of 5$d$ superconformal field theories defined by (p,q)-web diagrams corresponding
to the reflexive toric diagrams in our work \cite{Intriligator:1997pq, Leung:1997tw, Aharony:1997bh, Benini:2009gi, Jefferson:2017ahm, Closset:2018bjz, Closset:2019juk, Franco:2023flw, Franco:2023mkw, Bourget:2023wlb, Arias-Tamargo:2024fjt}.
Furthermore, our find-
ings relate to recent work on birational transformations for higher-dimensional toric
Calabi-Yau 4-folds \cite{Ghim:2024asj, Ghim:2025zhs, Franco:2015tna, Franco:2015tya, Franco:2016nwv, Franco:2016qxh, Franco:2022gvl, Kho:2023dcm}. We plan to investigate these connections further in
future upcoming work.
\\

%=================================================================
\section*{Acknowledgments}

The authors would like to thank S. Franco, D. Ghim, S. Jeong, K. Lee and D. Voloshyn for discussions.
The authors would also like to thank the Simons Center for Geometry and Physics for their hospitality during part of this work.
The work of N.L is supported by IBS project IBS-R003-D1. 
R.-K. S. is supported by an Outstanding Young Scientist Grant (RS-2025-00516583) of the National Research Foundation of Korea (NRF).
He is also partly supported by the BK21 Program (``Next Generation Education Program for Mathematical Sciences'', 4299990414089) funded by the Ministry of Education in Korea and the National Research Foundation of Korea (NRF).

%======================================================================
\bibliographystyle{JHEP}
\bibliography{mybib}
%======================================================================

\end{document}